\documentclass[preprint2]{aastex}
\usepackage{natbib}
\usepackage{float}
\usepackage{times}
\usepackage{graphicx}
\usepackage{amssymb}
\usepackage{color}

\newcommand{\mbh}{$M_{BH}$}

\newcommand{\msig}{$M_{BH}$-$\sigma$}
\newcommand{\msigrho}{$M_{BH}$-$\sigma$-$\rho_h$}
\newcommand{\msigr}{$M_{BH}$-$\sigma$-$r_h$}
\newcommand{\mrho}{$M_{BH}$-$\rho_h$}
\newcommand{\mr}{$M_{BH}$-$r_h$}
\newcommand{\mbu}{$M_{BH}$-$M_{Bu}$}
\newcommand{\mburs}{$M_{Bu}$-$r_h$-$\sigma$}
\newcommand{\mburho}{$M_{BH}$-$M_{Bu}$-$\rho_h$}
\newcommand{\mbus}{$M_{BH}$-$M_{Bu}$-$\sigma$}
\newcommand{\mbur}{$M_{BH}$-$M_{Bu}$-$r_h$}
\newcommand{\mhop}{$M_{BH}$-$M_{Bu}^{0.5}\sigma^2$}
\newcommand{\mfeo}{$M_{BH}$-$M_{Bu}\sigma^2$}
\newcommand{\mbulger}{$M_{Bu}$-$r_h$}
\newcommand{\rhor}{$\rho_h$-$r_h$}
\newcommand{\mbulgerho}{$M_{Bu}$-$\rho_h$}

\shorttitle{Supermassive black holes and bulges.}
\shortauthors{Saglia et al.}

\begin{document}

\title{The SINFONI Black Hole Survey:  The Black Hole Fundamental Plane 
revisited and the paths of (co-) evolution of supermassive black holes 
and bulges.}

\author{R.P. Saglia\altaffilmark{1,2},  M. Opitsch \altaffilmark{1,2,3}, P. Erwin\altaffilmark{1,2}, J. Thomas\altaffilmark{1,2}, A. Beifiori\altaffilmark{1,2}, M. Fabricius\altaffilmark{1,2}, X. Mazzalay \altaffilmark{1,2}, N. Nowak\altaffilmark{4},  S.P. Rusli\altaffilmark{1,2},  R. Bender\altaffilmark{1,2}}

\altaffiltext{1}{Max-Planck-Institut f\"{u}r extraterrestrische Physik, Giessenbachstrasse, D-85748 Garching, Germany}
\altaffiltext{2}{Universit\"{a}ts-Sternwarte M\"{u}nchen, Scheinerstrasse 1, D-81679 M\"{u}nchen, Germany}
\altaffiltext{3}{Exzellenzcluster Universe, Boltzmannstr. 2, D-85748 Garching, Germany}
\altaffiltext{4}{
Stockholm University, Department of Astronomy, Oskar Klein Centre, AlbaNova, SE-10691 Stockholm, Sweden}
\begin{abstract}
 We investigate the correlations between the black hole mass
  $M_{BH}$, the velocity dispersion $\sigma$, the bulge mass $M_{Bu}$,
  the bulge average spherical density $\rho_h$ and its spherical half
  mass radius $r_h$, constructing a database of 97 galaxies (31 core
  ellipticals, 17 power-law ellipticals, 30 classical bulges, 19
  pseudo bulges) by joining 72 galaxies from the literature to 
  25 galaxies observed during our recent SINFONI black hole
  survey. For the first time we discuss the full error covariance
  matrix. We analyse the well known \msig\ and \mbu\ relations and
  establish the existence of statistically significant correlations
  between $M_{Bu}$ and $r_h$ and anti-correlations between $M_{Bu}$
  and $\rho_h$. We establish five significant bivariate correlations
  (\msigrho, \msigr, \mbus, \mburho, \mbur) that predict $M_{BH}$
  of 77 core and power-law ellipticals and classical bulges
  with measured and intrinsic scatter as small as $\approx 0.36$ dex
  and $\approx 0.33$ dex respectively, or 0.26 dex when the subsample of
  45 galaxies defined by \citet{KormendyHo2013} is considered.  
  In contrast, pseudo bulges have
  systematically lower $M_{BH}$, but approach the predictions
  of all the above relations at spherical densities $\rho_h\ge 10^{10}
  M_\odot/kpc^3$ or scale lengths $r_h\le 1$ kpc.  These findings fit
  in a scenario of co-evolution of BH and classical-bulge masses,
  where core ellipticals are the product of dry mergers of power-law
  bulges and power-law Es and bulges the result of (early) gas-rich
  mergers and of disk galaxies. In contrast, the (secular) growth of
  BHs is decoupled from the growth of their pseudo bulge hosts, except
  when (gas) densities are high enough to trigger the feedback
  mechanism responsible for the existence of the correlations between
  $M_{BH}$ and galaxy structural parameters.
\end{abstract}

\keywords{galaxies:bulges;galaxies:fundamental parameters;galaxies:supermassive black holes;galaxies:elliptical and lenticular, cD; galaxies: spiral}

\maketitle
\section{Introduction}
\label{intro}

The last two decades have made clear that supermassive black holes are
ubiquitous at the centres of galaxies with bulges. The galaxy velocity
dispersion $\sigma$ \citep{Ferrarese2000,Gebhardt2000,Gueltekin2009b,
McConnell2011,McConnellMa2013}, luminosity
\citep{Dressler1989,Kormendy1993,KormendyRichstone1995,Kormendy2001,Marconi2003},
bulge mass $M_{Bu}$ \citep{Magorrian1998,Haering2004} and the mass of
the black hole $M_{BH}$ are proportional with a scatter of a factor of 2,
which implies that galaxy bulges and black holes somehow grew in lock
step. Important clues concerning this interconnection are encoded in
the steepness and intrinsic scatter of scaling laws like the \msig\
and the \mbu\ relations. \citet[and references
therein]{KormendyHo2013} rederive the global correlations with
  black hole mass and review the interpretation framework of these
findings. When black holes accrete mass, they shine as quasars or AGN,
and this activity interferes with the star formation which contributes
to bulge growth. Gas can make it to the central region of a galaxy, where 
a black hole (hereafter
BH) might sit, when non-axisymmetric distortions and/or temporal variations 
of the gravitational potential are
strong enough. This can happen through secular evolution of a disk,
possibly related to the formation and dissolution of bars, which also
leads to the build-up of a pseudo bulge. These pseudo bulges
structurally resemble disks, e.g. in their flattening and rotational
support. Mergers are another channel to funnel material towards 
the central region of a galaxy. Mergers
produce classical bulges and elliptical galaxies. Different regimes of
the \msig\ and \mbu\ scaling relations isolate different stages and
modes of BH and/or bulge growth
\citep{Kormendy2011,Mathur2012,KormendyHo2013}. At the low-mass end galaxies are
disk dominated (possibly with pseudo bulges), mergers are unimportant
and those few scaling relations that do exist mostly probe secular
evolution processes in disks. At the high-mass end, gas-poor mergers
dominate and drive the formation of core ellipticals, where the most
massive BHs live. Here, the averaging effect of a succession of major
mergers is expected to reduce the fractional dispersion of the \mbu\
relation \citep{Peng2007}.  Core ellipticals have stellar densities
mildly increasing towards the center \citep{Faber1997,Kormendy2009}, a
result of binary black hole scouring
\citep{Ebisuzaki1991,Milosavljevic2001} that also leaves a dynamical
imprint on the stellar orbits \citep{Thomas2014} and generates a tight
correlation between core radius and BH mass \citep{Kormendy2009,KormendyBender2009,Rusli2013b}. Core
ellipticals are also slow rotators and mildly triaxial
\citep{Nieto1989,Kormendy1996,Faber1997,Emsellem2007,Lauer2012}, a
further clue to their dry merger origin.  In between the two extremes,
early gas-rich, dissipational mergers of disk galaxies are responsible
for the formation of classical bulges and power-law ellipticals and
the lock step accretion on the central BHs mirrored in the \msig\ and
\mbu\ relations \citep{King2003, Hopkins2007a, Hopkins2007b}.
Power-law ellipticals have stellar densities steeply increasing
towards the center, a result of star formation in the high density
central gas concentration originating during a gas rich merger
\citep{Faber1997,Kormendy2009}. They are axisymmetric and fast
rotators \citep{Nieto1991,Faber1997,Emsellem2007}, reminiscent of the
structure and dynamics of disk galaxies
\citep{Bender1988,Kormendy1996}.

Here we reconsider this scenario by discussing the relationship
between the residuals from the \msig\ and \mbu\ relations and the
average spherical stellar-mass density (or scale length) of the
classical or pseudo bulges and pay particular attention to the
families of galaxies discussed above, core and power-law ellipticals,
classical and pseudo bulges and the possible presence of bars.
Attempts to detect a 'second parameter' or ``BH Fundamental Plane''
(BH FP) \footnote{This is different from the Fundamental Plane of
  black hole activity discovered by \citet{Merloni2003}.}  are
numerous and contradictory.  \citet{Feoli2005} ask
 the question whether the black
  hole masses correlate with the kinetic energy of elliptical
  galaxies. \citet{Aller2007} claim based on a sample of $\sim 20$
galaxies that the black hole masses best correlate with $E_g^{0.6}$,
where $E_g\sim M_{Bu}^2/r_h$ is the bulge gravitational binding
energy. With $M_{Bu}\sim r_h \sigma^2$ and $\rho\sim M_{Bu}/r_h^3$
this implies $M_{BH}\sim \rho^{0.2}M_{Bu}\sim
r_h^{-0.6}M_{Bu}^{1.2}\sim
M_{Bu}^{0.6}\sigma^{1.2}\sim\rho^{-0.3}\sigma^3\sim
r_h^{0.6}\sigma^{2.4}$, where $\rho$ is the mean density of the bulge
and $r_h$ its scale-length. A further empirical study of the BH FP is
given by \citet{Barway2007}.

\citet{Hopkins2007a,Hopkins2007b} investigate the BH FP with the help
of hydrodynamical simulations, finding that the empirical relations
$M_{BH}\sim \sigma^{3.0\pm 0.3}R_e^{0.43\pm 0.19}$ and $M_{BH}\sim
M_{Bu}^{0.54\pm 0.17}\sigma^{2.2\pm 0.5}$ can be explained
theoretically by noting that the black hole mass should scale as
$M_{Bu}^{0.5}\sigma^2$.  \citet{Graham2008} reports that the BH FP is
possibly driven by the barred galaxies in the sample. Nevertheless,
\citet{Graham2001} and more recently \citet{Savorgnan2013} argue for a
strong correlation between BH mass and galaxy concentration.
\citet{Feoli2009} and \citet{Mancini2012} investigate the relation
between BH mass and kinetic energy of the bulge $M_{BH}\sim
M_{Bu}\sigma^2$, discussing the existence of a main-sequence like
diagram. \citet{Soker2011} propose that the black hole masses
  should correlate with $M_{Bu}\sigma$. In contrast, \citet{Sani2011}
fail to detect bivariate correlations. \citet{Beifiori2012} find only
weak evidence for bivariate correlations by analysing 49 galaxies from
\citet{Gueltekin2009b} and a large sample of galaxies with upper
limits to BH masses from \citet{Beifiori2009}. Finally,
\citet{Graham2013} claim that ``Sersic galaxies'' follow a quadratic
more than a linear \mbu\ relation.  We will see that to settle the
issue it is important to consider a large database with dynamically
measured BH masses and accurate bulge plus disk decompositions, and to
distinguish between the different families of objects (core and
power-law ellipticals, classical bulges and pseudo bulges, barred
objects), which to some extent obey different residual correlations.

The structure of the paper is the following. In Sect. \ref{sec_data}
we describe the data sample and the methods used to measure the bulge
average densities.  In Sect. \ref{sec_errors} we discuss the error
matrix, exploring the covariances between the parameters.  
In Sect. \ref{sec_correlations} we discuss the method adopted
to investigate multivariate correlations between our parameters.  In
Sect. \ref{correlations} we present the results of our correlation
analysis. In Sect. \ref{discussion}  we investigate which of the quantities
$M_{Bu}^{0.5}\sigma^2$,  $M_{Bu}\sigma^2$ and $M_{Bu}\sigma$ best
correlate with black hole masses
and discuss the implications for
the coevolution of bulges and black holes.  In Sect. \ref{conclusions}
we draw our conclusions. Four appendices discuss how we measure
effective velocity dispersions (App. A), how we determine the
luminosity profiles and the mass-to-light ratios (hereafter $M/L$) of
bulges (App. B) and how we compute simple Jeans $M/L$ for some of
our galaxies (App. C). App. D lists correlation results for a restricted sample of galaxies. 

\section{The data sample}
\label{sec_data}

\subsection{Distances, BH masses and velocity dispersions}
\label{sec_BDdata}

Our sample includes galaxies from \citet{Gueltekin2009b},
\citet{Sani2011}, \citet{McConnell2011}, \citet{McConnellMa2013},
and \citet{KormendyHo2013}.  We tested various combinations of these
  datasets, obtaining compatible results. Here, we start with the database
  (morphological type, distances, black hole masses, velocity
  dispersions and their errors) of \citet{KormendyHo2013}, without those
  galaxies belonging to our SINFONI black hole survey (see below). We
  supplement this list with 8 galaxies (NGC 2974, NGC 3079, NGC 3414,
  NGC 4151, NGC 4552, NGC 4621, NGC 5813, NGC 5846) which are quoted by
  \citet{Sani2011}. References to the original sources can be found in
  these two papers.  We estimate the errors on distances from NED;
  they amount typically to 9\%. We compute symmetrized logarithmic
  errors for black hole masses and velocity dispersions.

  We do not consider three galaxies with upper limits on their black
  hole masses (namely NGC 2778, NGC 4382, IC2560). Furthermore, we
  exclude the following objects.  Cygnus A has an uncertain BH mass
  and velocity dispersion \citep{KormendyHo2013}; moreover the strong
  internal dust absorption prevents the derivation of a reliable
  photometric profile. IC 1481 is undergoing a merger, which makes the
  derivation of a reliable photometric profile difficult. The
  determination of the extremely large BH mass of NGC 1277 
  \citep{vandenBosch2012} has been questioned by \citet{Emsellem2013}.  
  For NGC 4945 the 'binding mass of $\sim
  10^6 M_\odot$ within 0.3 pc' quoted in the abstract and the
  conclusions of \citet{Greenhill1997} comes from maser measurements
  that point to a non-Keplerian rotation.

  We complement these measurements with the 25 determinations from our
  SINFONI black hole survey; 9 of these values are currently
  unpublished.  They are all based on the stellar dynamical analysis
  of our SINFONI kinematics, coupled with extended long-slit or
  integral field stellar kinematics of the outer regions of the
  galaxies. A detailed description of the SINFONI dataset, the methods
  and of some specific cases can be found in \citet{Nowak2007},
  \citet{Nowak2008}, \citet{Nowak2010}, \citet{Rusli2011},
  \citet{Rusli2013a}, \citet{Mazzalay2015}, \citet{Erwin2015b},
  \citet{Thomas2015a}, \citet{Bender2015}. In summary, our SINFONI
  black hole sample consists of 30 galaxies that we observed with the
  Spectrograph for INtegral Field Observations in the Near Infrared
  \citep[][SINFONI,]{Eisenhauer2003,Bonnet2003} at the UT4 of the Very
  Large Telescope under nearly diffraction limited conditions. The
  sample was selected to explore poorly populated regions of the
  \msig\ and \mbu\ correlations, with particular attention to high
  velocity dispersion early-type galaxies, low velocity dispersion and
  pseudo bulge galaxies, mergers and galaxies with low luminosity
  AGNs.  Through our Schwarzschild axisymmetric code
  \citep{Thomas2004,Thomas2005} we determine the best fitting black
  hole mass $M_{BH}$ (and the mass-to-light ratio $M/L$ of the stellar
  component(s)), taking into account the bulge and disk components of
  the galaxies separately when necessary. The appropriate dark matter
  potential is also considered when necessary \citep{Rusli2013a}.  We
  summarize in Fig. \ref{fig_MBHML} the resulting $M/L$ and $M_{BH}$
  for the 9 galaxies that will be discussed in the papers in
  preparation quoted above; see also Appendix B.

  Distances for the
SINFONI sample are directly measured (from Cepheids or surface
brightness fluctuations) or computed from the redshifts using the
standard cosmology ($\Omega=0.3, \Lambda=0.7,H_0=70$). We determine the
$\sigma$ for our SINFONI galaxies using the long-slit and integral
field stellar kinematics used in the modeling (see Appendix A).

\subsection{Bulge masses, sizes and densities}
\label{sec_MBuData}

For all galaxies except the Milky Way, we compute bulge masses,
half-light radii and densities from the photometry of the galaxies,
decomposed into a bulge and further components (a disk, a bar, a ring)
when necessary. For some galaxies we find evidence for composite
(classical plus pseudo) bulges \citep{Erwin2015a}: in these cases we
consider the classical component of the two. We note that some other disk
galaxies identified as having pseudo bulges may prove to have
composite bulges as well, but the necessary data for a proper
assessment is currently lacking for many. We use the dynamically
determined mass-to-light ratios $M/L$ taken from the literature or our
own modeling to convert light into mass. For the Milky Way we use the
axisymmetric bulge density profile of \citet{McMillan2011}, which we
integrate spherically to get $M_{MW}(<r)$. A detailed description of
the methods and procedures is given in the Appendices B  and C; 
here we give a short summary only.

We measure surface brightness profiles from images taken from the
Spitzer archive or SDSS \citep{York2000} and the ESO Key Program
described in \citet{Scorza1998}, or take them from the literature.
Bulge-disk decompositions, when necessary, are taken from
\citet{Fisher2008}, \citet{Gadotti2008}, \citet{Sani2011},
\citet{Beifiori2012}, \citet{Vika2012} or are performed by us (in 16
cases), using the program of \citet{Erwin2015c} or the procedures of
\citet{Fisher2008}. Mass-to-light ratios come from different types of
dynamical modeling, that can be based on spherical distribution
functions \citep{Kronawitter2000}, or Schwarzschild modeling
\citep{Schulze2011}, Jeans equations \citep{Haering2004}, or gas
dynamics \citep{DallaBonta2009}. If none of this is available, they
are computed by matching the stellar kinetic energy profiles
$\sqrt{v^2(R)+\sigma^2(R)}$ or the central velocity dispersion through
spherical Jeans equations (see Appendix B).  When not already done by
the authors, we correct the $M/L$ for galactic extinction following
\citet{Schlegel1998} and we transform them to the band of the
available surface brightness profiles using the galaxy colors from the
Hyperleda database \citep{Paturel2003}. We adopt this approach to
  test whether dynamically determined $M/L$ can deliver bulge masses
  and densities that better correlate (i.e. with smaller scatter) with
  BH masses. \citet{KormendyHo2013} consider $M/L$ determinations from
  colors, deriving a \mbu\ relation with impressive small scatter
  (0.29 dex). In principle, stellar-population $M/L$ determinations
  could be uncertain, because they depend on the proper choice of
  the stellar initial mass function \citep{Thomas2011,Cappellari2013} and
  internal dust corrections, see discussion in \citet{Rusli2013b}.
  We investigate this issue using the bulge masses quoted by
  \citet{KormendyHo2013} for the 45 galaxies used in their fits 
(see below). 

\begin{figure}[h!]
  \begin{center}
    \includegraphics[trim=0 4cm 0 4cm,clip,width=8cm]{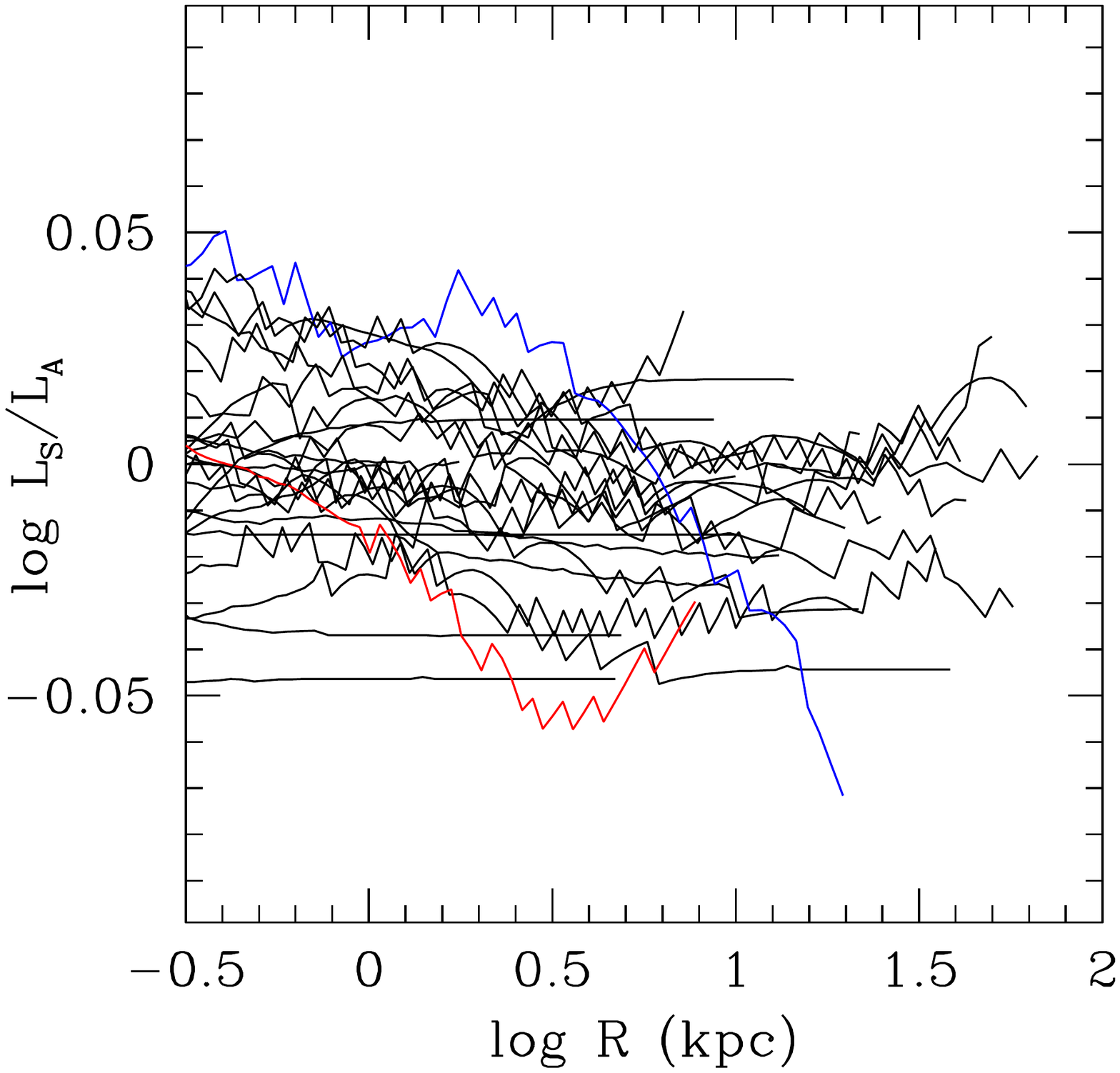}
  \end{center}
  \caption{The comparison between the spherical luminosity profiles
derived using Eqs. \ref{eq_lum} and \ref{eq_lumA} for the SINFONI galaxies.
The red (NGC 4486a) and blue (NGC 4751) lines show 
the most deviant profiles.\label{fig_LumSLumA}}
\end{figure}

We circularize the bulge photometry and deproject it assuming
spherical symmetry to derive the spherical luminosity density
$l_S(r)$.  We then compute the spherical luminosity:
\begin{equation}
L_S(<r)=\int_0^r 4\pi r^2 l_S(r) dr.
\label{eq_lum}
\end{equation}
For the SINFONI sample of galaxies we have also performed an
axisymmetric deprojection of the (decomposed bulge and disk) surface
brightness profiles and derived the axisymmetric luminosity density
$l_A(r,\theta,i)$, where $r$ and $\theta$ are spherical coordinates
and $i$ the assumed inclination angle. We spherically integrate $l_A$ to get 
a second estimate of the spherical luminosity $L_A$:
\begin{equation}
L_A(<r)=\int_{Sphere(r)}l_A dV. 
\label{eq_lumA}
\end{equation}
Fig. \ref{fig_LumSLumA} compares the spherical luminosity profiles
derived using Eqs. \ref{eq_lum} and \ref{eq_lumA}. $L_S$ and $L_A$ are
similar within 0.05 dex, with typical deviations less than 0.02
dex. The most deviant profiles are for flattened galaxies seen nearly
edge-on. 

The spherical mass profile of the bulge is $M(<r)=L(<r)\times M/L$, 
where the mass-to-light ratio $M/L$ is determined dynamically.
We discuss in the Appendix B on a case by
case basis the applied transformations  necessary to 
homogenize the used photometry and $M/L$.

We extend $M(<r)$ to large radii using a spline extrapolation of the
surface brighness, adding a further point to the measured profile at
very large radii (typically several tens to hundreds of arcsec
depending on the size of the bulges) and with a surface brightness of 70
mag/arcsec$^2$. This generates a deprojected density decreasing as
$r^{-3}$ at large radii, see Fig. \ref{fig_densmass}.  Then we get the
bulge mass as:
\begin{equation}
M_{Bu}=M(<\infty)=M(<r_{max})+M_{extrap},
\label{eq_LBu}
\end{equation}
where $r_{max}$ is the distance of the last measured surface brightness
point and $M_{extrap}$ is the light contribution due to the
extrapolation to the last computed deprojected density point.  
We define the half-ligh radius $r_h$ of
the bulge component as the radius where $M(<r_h)=M_{Bu}/2$. 
The bulge's averaged density within $r_h$ is
\begin{equation}
\rho_h=\frac{M_{Bu}/2}{4\pi r_h^3/3}.
\label{eq_rho}
\end{equation}
Fig. \ref{fig_densmass} shows that the bulge masses (to the precision
given in Table \ref{tab_data}) are reached at approximately 20 $r_h$,
where we effectively cut the density profiles, not to have to worry
about the logarithmic divergence of the mass profile implied by the
$r^{-3}$ behaviour of our extrapolation.

\begin{figure*}
  \begin{center}
    \includegraphics[trim=0 4cm 0 4cm,clip,width=8cm]{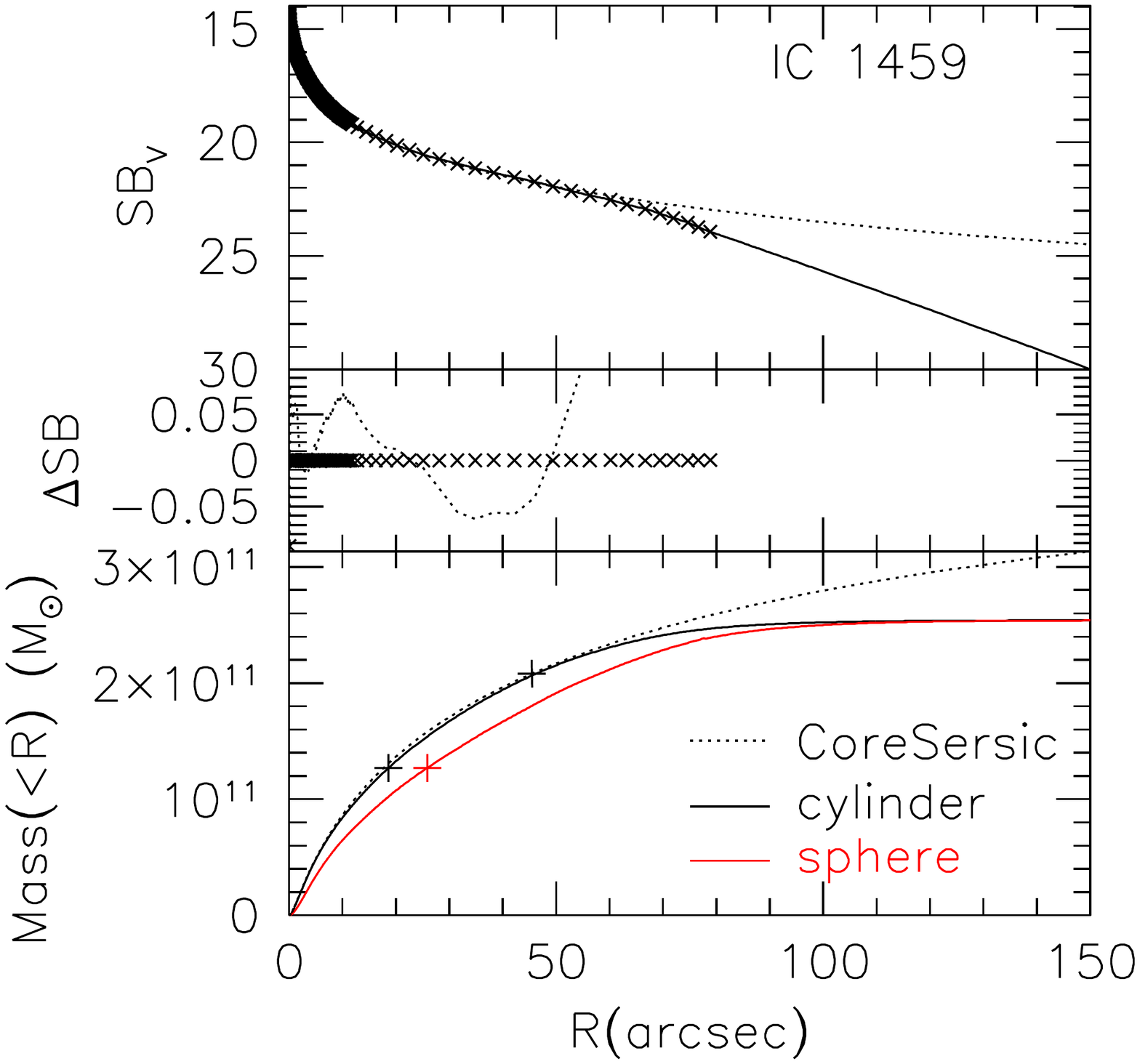}
    \includegraphics[trim=0 4cm 0 4cm,clip,width=8cm]{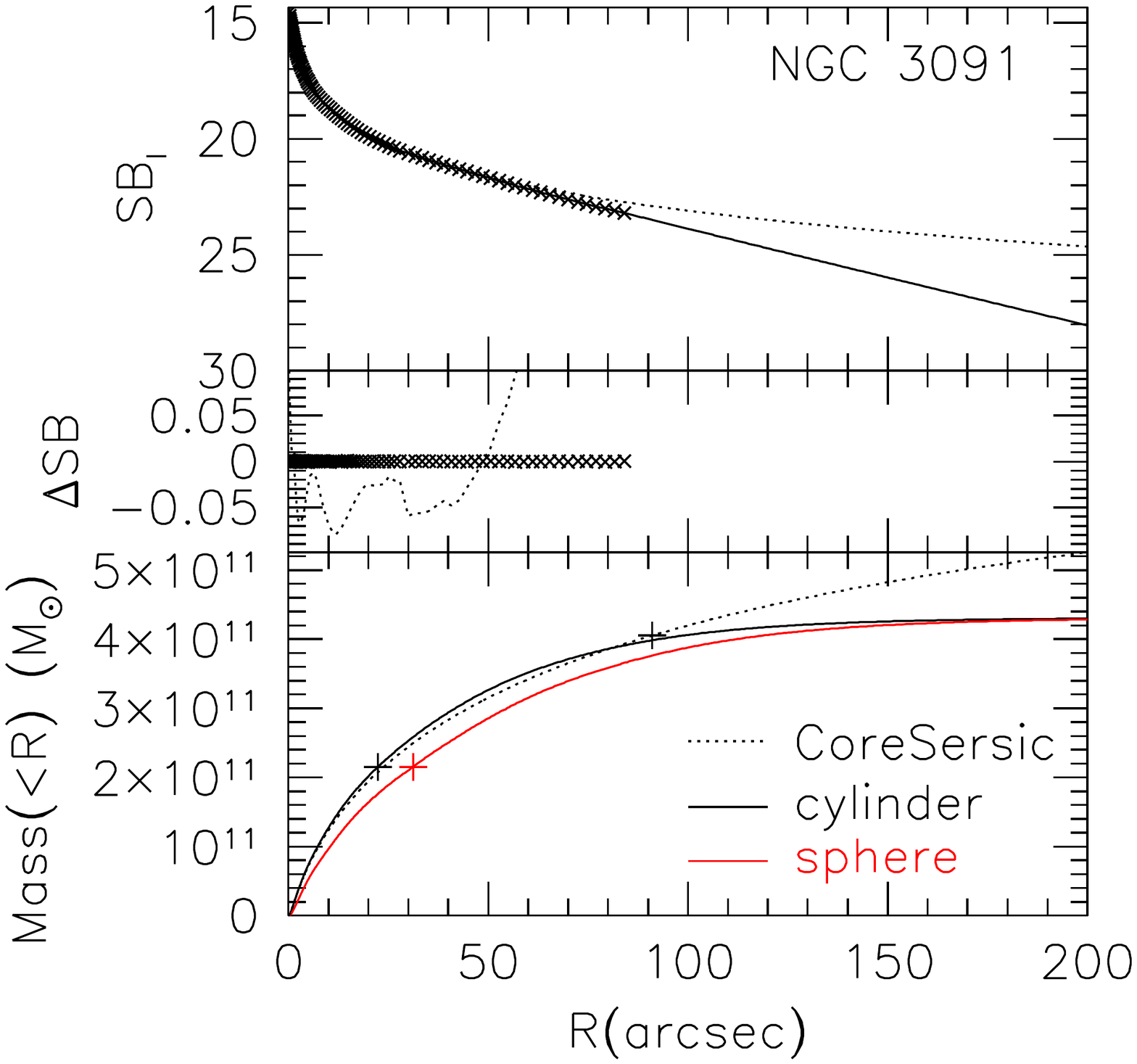}
    \includegraphics[trim=0 4cm 0 4cm,clip,width=8cm]{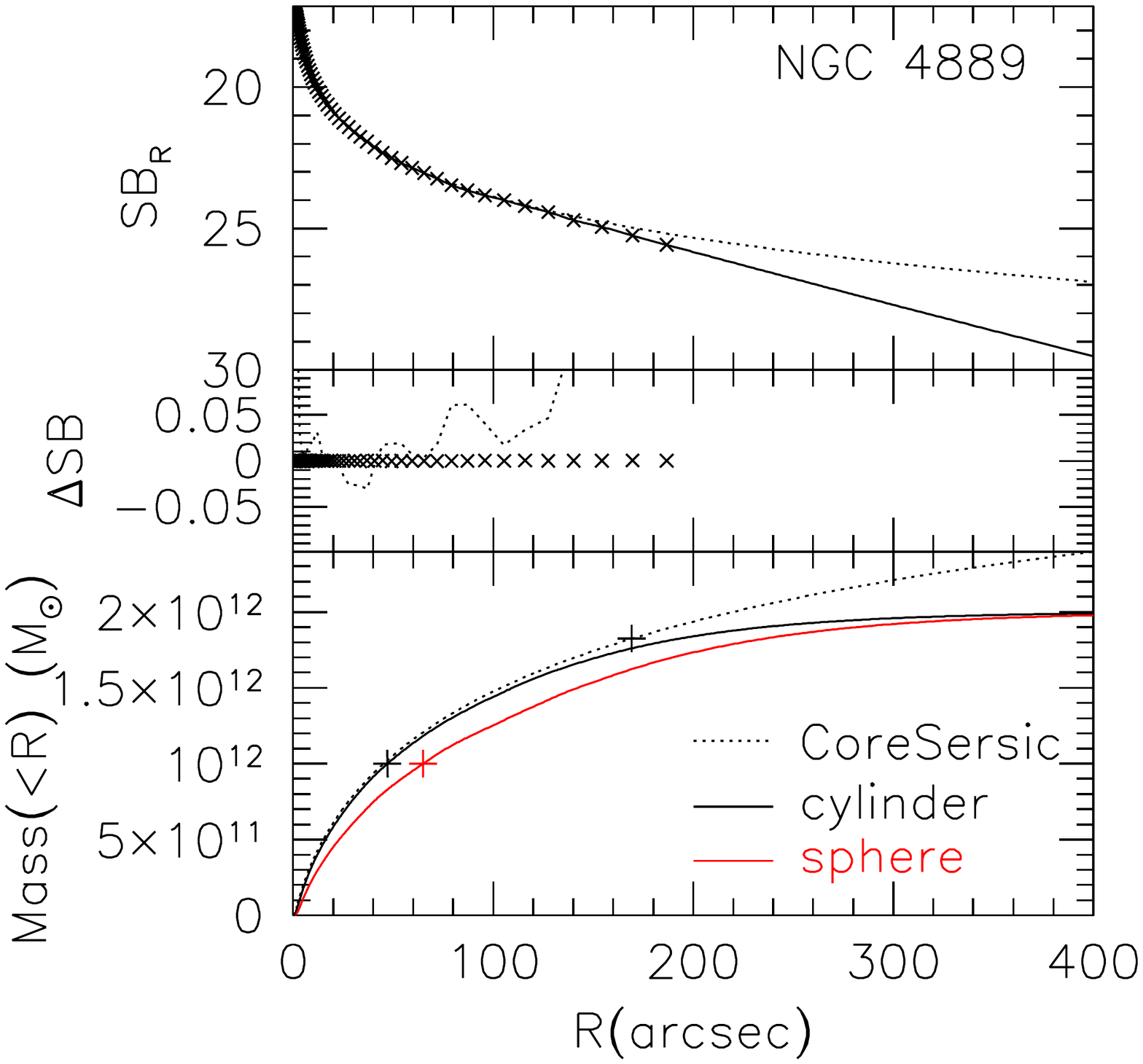}
    \includegraphics[trim=0 4cm 0 4cm,clip,width=8cm]{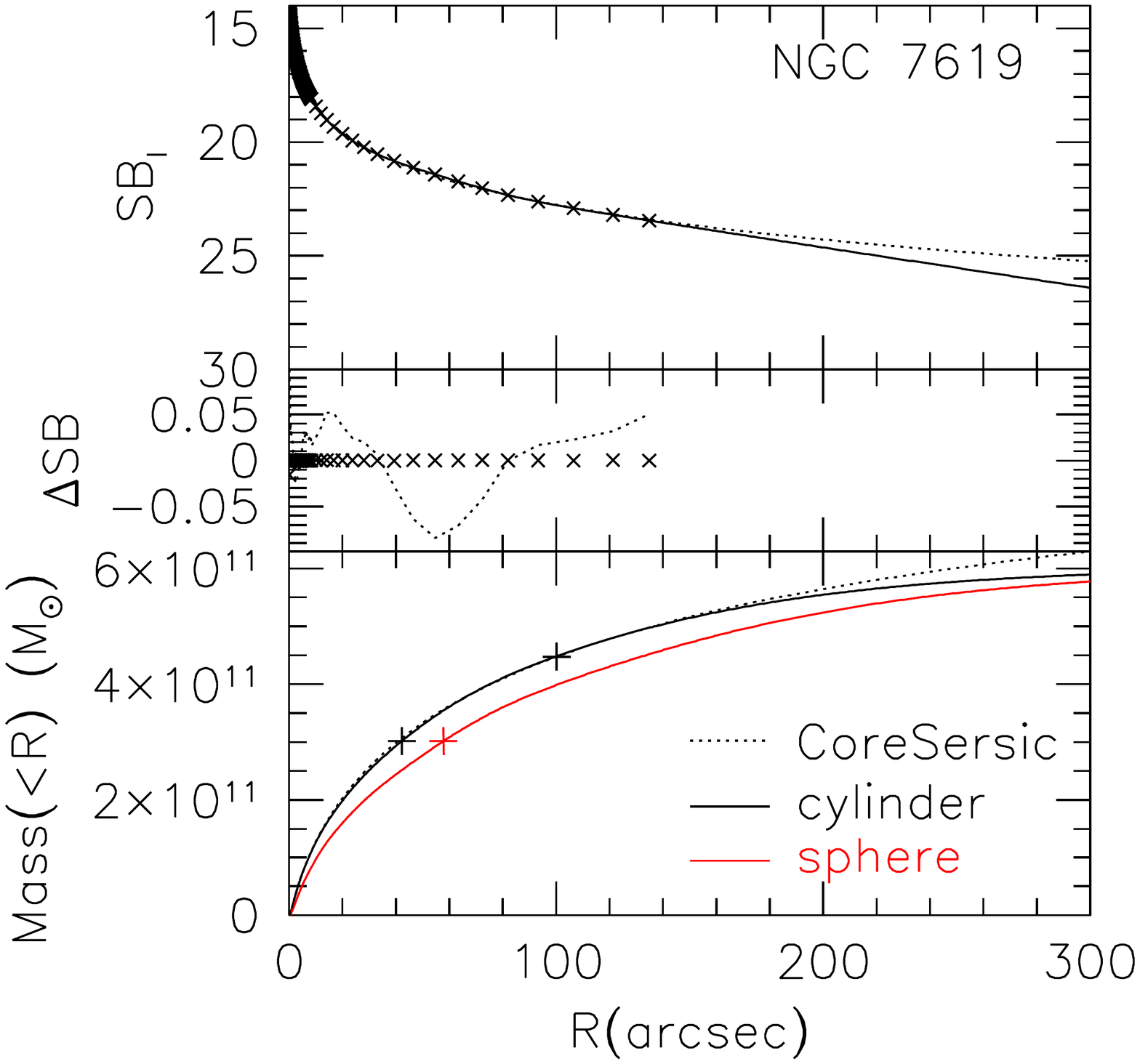}
  \end{center}
  \caption{For each of the core ellipticals IC1459, NGC 3091, NGC 4889, and NGC 7619 we present three plots. 
The plots at the top show the observed circularized surface brightness profiles (crosses), together with the projection along the light of sight 
of the luminosity density we derived (full line). The dotted lines show the best-fitting Core-Sersic profile of  \citet{Rusli2013b}. The plots in 
the middle show the differences in surface brightness between the data and the projected luminosity density profiles (crosses)  and 
the core-Sersic fits (dotted lines) of  \citet{Rusli2013b}. The plots at the bottom show three lines: our spherical (full red lines) and 
cylindrical (full black lines) mass profiles and the (cylindrical) mass profile implied by the Core-Sersic fits of   \citet{Rusli2013b}. 
The three crosses show the positions of $r_h$ and $R_e$ listed in Table \ref{tab_data} and $r_e$ from Table 2 of   \citet{Rusli2013b}.
\label{fig_extrap}}
\end{figure*}

In addition, for all galaxies except the Milky Way we compute the projected circularized
half-luminosity radius $R_e$ of the bulge from the curve of growth of the
projection along the line of sight of the luminosity density profile. 
Examples of the procedure are given in Fig. \ref{fig_extrap} discussed below.
We get $R_e/r_h=0.74$ on average, with $rms=0.01$. 

Fig. \ref{fig_extrap} shows four examples of our surface brightness
profiles, to clarify the role of extrapolation in the determination of
$R_e$ and $r_h$. For these galaxies our derived $R_e$ differ by more
than 0.3 dex from the $r_e$ values quoted by \citet{Rusli2013b}, Table
2, from a Core-Sersic fit (see comments in Appendix B).  By
construction, our procedure reproduces the observed profile
perfectly. This is not always true, when for example Sersic or
Core-Sersic fits are used to derive scale radii, as done in
  \citet{Rusli2013b}. We quantify the amount of extrapolation
involved in our analysis in Sect. \ref{sec_ErrorsData}.

\begin{figure}
  \begin{center}
    \includegraphics[trim=0 4cm 0 4cm,clip,width=8cm]{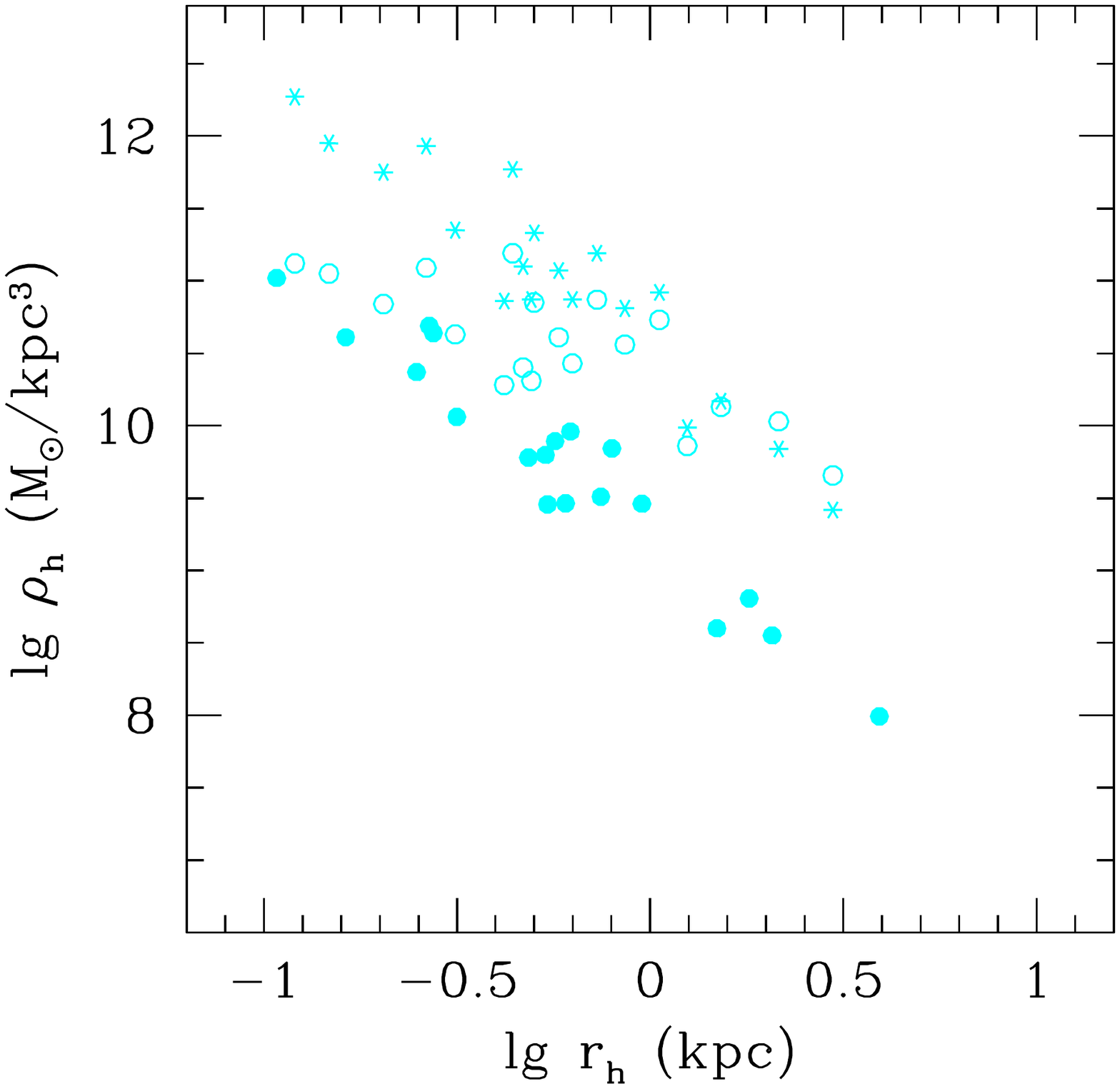}
 \end{center}
  \caption{Cylindrical averaged scalelengths and densities of pseudo bulges, assuming vertical scale height 
 $h_z=0.2$ kpc (open circles) or  $h_z/h=0.2$ (stars); see Eq. \ref{eq_pseudodensity}, case (a) and (b) respectively.
The filled points show the spherical average quantities. 
\label{fig_pseudodensity}}
\end{figure}

We compute spherically averaged densities for all classes of objects
in our sample in order to have a homogeneous data set. But is it
physically meaningful to consider spherical half-mass radii and
spherically averaged densities also for pseudo bulges?  Our current
understanding \citep[][and references therein]{Erwin2015a} is that
these structures are more similar to disks than spheroids. Therefore,
for these objects (except the Milky Way) we also estimate cylindrical
average densities:
\begin{equation}
\rho_{h,c}=\frac{M_{Bu}/2}{\pi a_e^2h_z},
\label{eq_pseudodensity}
\end{equation}
where $a_e=R_e(a/b)^{1/2}$ is the projected half-luminosity radius
along the major axis (the proper scale length of an inclined disk),
$a/b$ the major to minor axis ratio of the bulge (taken from the
decompositions described in Appendix B), and $h_z$ an estimate of the
thickness. We consider the case of (a) a fixed thickness of $h_z=0.2$
kpc or (b) a thickness $h_z=0.2a_e/1.67$, which is 20\% of the exponential
scale-length of the disk $h=a_e/1.67$. Fig. \ref{fig_pseudodensity}
shows the results.  On average, the scale lengths do not change much
($\langle \log a_e/r_h\rangle=-0.016$), because the $a/b$ dependency
compensates the $R_e/r_h$ ratio. However, the cylindrical densities
are one order of magnitude larger ($\langle \log
\rho_{h,c}/\rho_h\rangle=1.31$ for $h_z=0.2a_e/1.67$).

Table \ref{tab_data} lists the galaxy names (Column 1) , the galaxy
type (Column 2), a series of flags (Column 3 to 8, see description in the
footnote of the table), the distance used (Column 9) and the
logarithms of the measured values of the parameters $M_{BH}$ (in
$M_\odot$), $\sigma$ (in km/s), $M_{Bu}$ (in $M_\odot$), $\rho_h$ (in
$M_\odot/kpc^3$), $r_h$ (in $kpc$) with their errors in Columns 10 to 14
(see also Sect. \ref{sec_errors}), plus the values of $R_e$ (in
arcsec) in Column 15. Table \ref{tab_datapseudo} lists the cylindrical
average quantities for pseudo bulges, with galaxy names (Column 1),
the logarithms of $a_e$ (in kpc) and $\rho_{c,h}$ (in $M_\odot/kpc^3$,
computed with $h_z=0.2a_e/1.67$ ).  As discussed above, we also
consider the bulge masses quoted by \citet{KormendyHo2013} for 
  the 45 galaxies used in their fits, here after KH45. They are given in Table
\ref{tab_dataKormendy}, coupled to our sizes $r_h$ to derive
colour-based density estimates. This procedure is uncertain, since we
do not know how the bulge plus disk decompositions of
\citet{KormendyHo2013} were performed. In detail, Table
\ref{tab_dataKormendy} gives the names of the galaxies of the
  KH45 sample  (Column 1), the bulge masses
(Column 2, computed from luminosites using mass-to-light ratios
$M/L_C$ derived from colors), and the spherical bulge densities
(Column 4) computed using the $r_h$ of Table \ref{tab_data}.  The
logarithmic errors (Column 3 and 5) are computed as described in
Sect. \ref{sec_errors}.

\begin{deluxetable}{lllllclllllllll}
\tablecaption{The galaxy sample with the values of the measured parameters.\label{tab_data}}
\rotate
\tabletypesize{\tiny}
\tablewidth{0pt}
\tablehead{
\colhead{Galaxy} & \colhead{Type} & \colhead{S} & \colhead{T} & \colhead{m} & \colhead{b} & \colhead{B} & \colhead{M} &\colhead{$D(Mpc)$} &\colhead{ $\log M_{BH}(M_\odot)$} & \colhead{$\log \sigma(km s^{-1})$} & \colhead{$\log M_{Bu}(M_\odot)$} & \colhead{$\log \rho_h (M_\odot kpc ^{-3})$} & \colhead{$\log r_h(kpc)$}& \colhead{$R_e('')$}}
\startdata
          MW           &            Sbc        &           0 & 3 & 0 &   1 & 1 & 0  &         0.008       $\pm$       0.001           &       6.633       $\pm$       0.048           &       2.021       $\pm$       0.084           &      10.375       $\pm$       0.069           &       9.283       $\pm$       0.111           &       0.056       $\pm$       0.052    &           -          \\ 
    Circinus           &          SABb:        &           0 & 3 & 0 &   1 & 1 & 2  &         2.820       $\pm$       0.470           &       6.057       $\pm$       0.105           &       1.898       $\pm$       0.017           &       9.170       $\pm$       0.089           &      10.611       $\pm$       0.153           &      -0.788       $\pm$       0.072    &         8.9          \\ 
       A1836           &           BCGE        &           0 & 0 & 0 &   0 & 0 & 4  &       152.400       $\pm$       8.426           &       9.573       $\pm$       0.061           &       2.459       $\pm$       0.021           &      12.026       $\pm$       0.135           &       7.392       $\pm$       0.187           &       1.237       $\pm$       0.060    &        17.2          \\ 
      IC1459           &             E4        &           0 & 0 & 0 &   0 & 0 & 2  &        28.920       $\pm$       3.739           &       9.394       $\pm$       0.079           &       2.520       $\pm$       0.007           &      11.401       $\pm$       0.076           &       8.800       $\pm$       0.153           &       0.559       $\pm$       0.069    &        18.6          \\ 
      IC4296           &           BCGE        &           0 & 0 & 0 &   0 & 0 & 1  &        49.200       $\pm$       3.628           &       9.114       $\pm$       0.073           &       2.508       $\pm$       0.022           &      12.026       $\pm$       0.110           &       6.995       $\pm$       0.355           &       1.369       $\pm$       0.146    &        68.1          \\
\enddata
\tablenotetext{\ }{Column 1: galaxy name; Column 2: galaxy type;
  Column 3: Flag S; $S=0$ flag identifies galaxies from the
  literature, $S=1$ identifies galaxies belonging to the SINFONI
  sample. Column 4: Flag T; T=0 for core ellipticals, 1 for power-law
  ellipticals, 2 for classical bulges, 3 for pseudo bulges. Column 5:
  Flag m; m=1 identifies merger remnants. Column 6: The b flag; b=1
  identifies barred galaxies, b=0.5 if a bar cannot be excluded
  because the galaxy is edge-on. Here we follow the strategy of
  classifing a galaxy as barred even when discrepant opinions are
  present, see comments in Appendix B. Column 7 : Flag B; B=0
  identifies galaxies where $M_{Bu}$ comes from the measured
  photometric profile, B=1 identifies galaxies where $M_{Bu}$ is
  computed after a bulge plus disk decomposition, B=2 identifies composite
  galaxies, where both a classical and a pseudo bulge are present. In
  these cases $M_{Bu}$ is the mass of the classical bulge
  component. Column 8: Flag M, M=0 indicates the special case of the
  Milky Way, see text, M=1 indicates that the $M/L$ was computed from
  distribution function or Schwarzschild modeling, M=2 from Jeans
  modeling of stellar kinematics profiles, M=3 from Jeans modeling of
  central stellar velocity dispersions, M=4 from gas dynamics. See
  Appendix B for comments on single galaxies. Column 9: distance and
  its error. Column 10: black hole mass and its error. Column 11:
  velocity dispersion and its error. Column 12: bulge mass and its
  error. Column 13: bulge spherical average density and its
  error. Column 14: bulge spherical half-mass radius and its
  error. Column 15: bulge cylindrical half-mass radius.}

\end{deluxetable}

\begin{deluxetable}{lll}
\tablecaption{The cylindrical parameters of the pseudo bulges.\label{tab_datapseudo}}
\tablewidth{0pt}
\tablehead{
\colhead{Galaxy} & \colhead{$\log a_e(kpc)$} & 
\colhead{$\log \rho_{h,c} (M_\odot kpc ^{-3})$} }
\startdata
    Circinus           &     -0.8313           &       11.95          \\
      IC2560           &    -0.06558           &       10.81          \\
     NGC1300           &     -0.3074           &       10.87          \\
     NGC2273           &     -0.5796           &       11.93          \\
     NGC2748           &     -0.3778           &       10.86          \\
     NGC2960           &     -0.1376           &       11.19          \\
     NGC3079           &     -0.3559           &       11.77          \\
     NGC3227           &     -0.3287           &        11.1          \\
     NGC3384           &     0.02405           &       10.92          \\
     NGC3393           &     -0.3002           &       11.33          \\
     NGC4388           &      0.3326           &        9.84          \\
     NGC4736           &     -0.6902           &       11.75          \\
     NGC4826           &     0.09586           &       9.985          \\
     NGC6264           &       0.183           &       10.17          \\
     NGC6323           &      0.4727           &       9.418          \\
     NGC7582           &     -0.5045           &       11.35          \\
     UGC3789           &     -0.2014           &       10.87          \\
     NGC3627           &     -0.9195           &       12.27          \\
     NGC4501           &     -0.2367           &       11.07          \\
\enddata
\tablenotetext{\ }{Column 1: galaxy name. Column 2: bulge cylindrical major axis half-mass radius. Column 3: bulge cylindrical average density.}
\end{deluxetable}

\begin{deluxetable}{lllll}
\tablecaption{The bulge masses and densities of \citet{KormendyHo2013}.\label{tab_dataKormendy}}
\tablewidth{0pt}
\tablehead{
\colhead{Galaxy} & \colhead{$\log M_{Bu,C}(M_\odot)$} & \colhead{$\delta \log M_{Bu,C}(M_\odot)$} & \colhead{$\log \rho_{h,C} (M_\odot kpc ^{-3})$} & \colhead{$\log \rho_{h,C} (M_\odot kpc ^{-3})$}}
\startdata
      IC1459           &        11.6           &      0.1459           &       8.999           &       0.142          \\
     NGC0221           &        9.05           &      0.1053           &       10.97           &      0.1041          \\
     NGC0224           &       10.35           &     0.09694           &       9.462           &     0.09182          \\
     NGC0524           &       11.26           &      0.1247           &       7.737           &      0.1585          \\
     NGC0821           &       10.98           &       0.144           &       7.864           &      0.3854          \\
     NGC1023           &       10.53           &      0.1105           &       10.45           &     0.09558          \\
     NGC1194           &       10.64           &      0.1305           &       8.044           &      0.1046          \\
     NGC1399           &        11.5           &     0.09493           &       7.064           &      0.1045          \\
\enddata
\tablenotetext{\ }{Column 1: galaxy name; Column 2 and 3: bulge masses
  from \citet{KormendyHo2013} derived from colors with our error
  estimates; Column 4 and 5: bulge densities derived from colors using
  the $r_h$ of Table \ref{tab_data} and errors.}
\end{deluxetable}

\subsection{Errors on $r_h$, $\rho_h$ and  $M_{Bu}$}
\label{sec_ErrorsData}

\begin{figure*}
  \begin{center}
    \includegraphics[trim=0 4cm 0 4cm,clip,width=\textwidth]{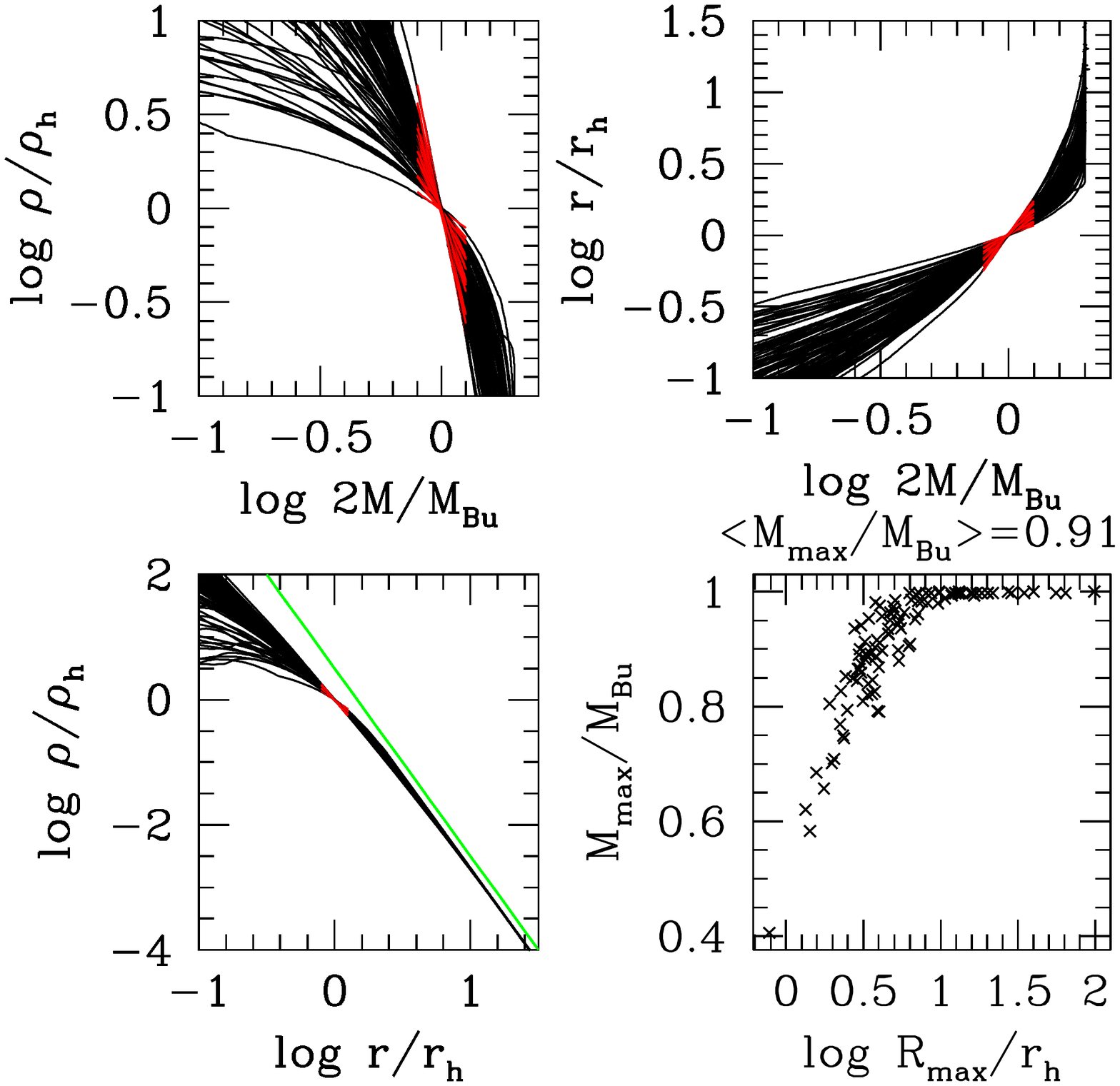}
  \end{center}
  \caption{The bulge density and mass profiles of our galaxy
    sample. The red lines show the region where the logarithmic
    derivatives $a_{\rho M}$ (top left), $a_{rM}$ (top right) and $a_{\rho  r}$ 
(bottom left) are fitted. The green line shows $\rho \propto r^{-3}$. Bottom right:
the correlation between the radial extent of the profile in units of $r_h$ and 
the fractional bulge mass sampled. On average, our profiles probe 
$\approx 91$\% of the bulge mass.
\label{fig_densmass}}
\end{figure*}

\begin{figure*}[h!]
  \begin{center}
    \includegraphics[trim=0 4cm 0 4cm,clip,width=8cm]{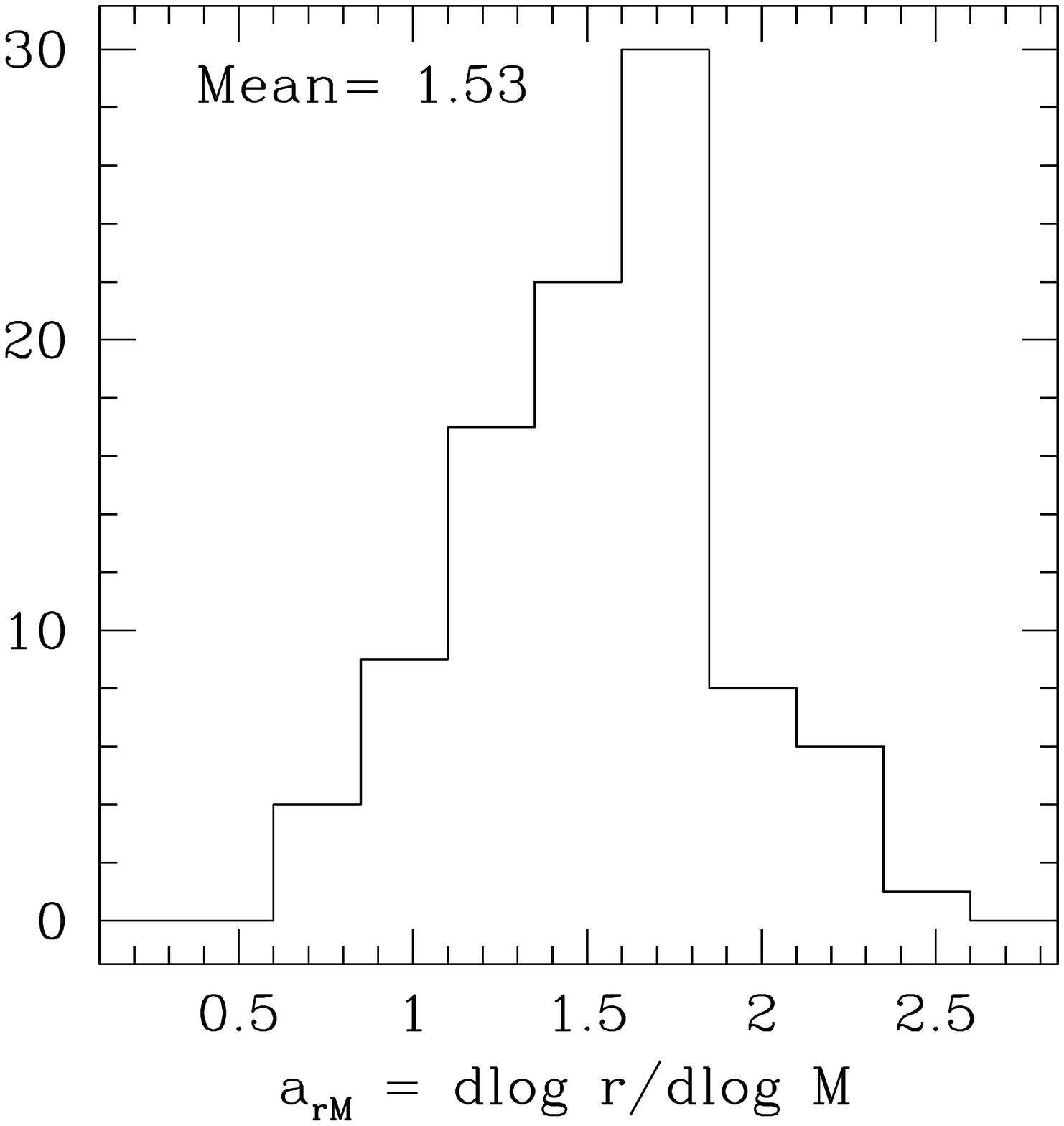}
     \includegraphics[trim=0 4cm 0 4cm,clip,width=8cm]{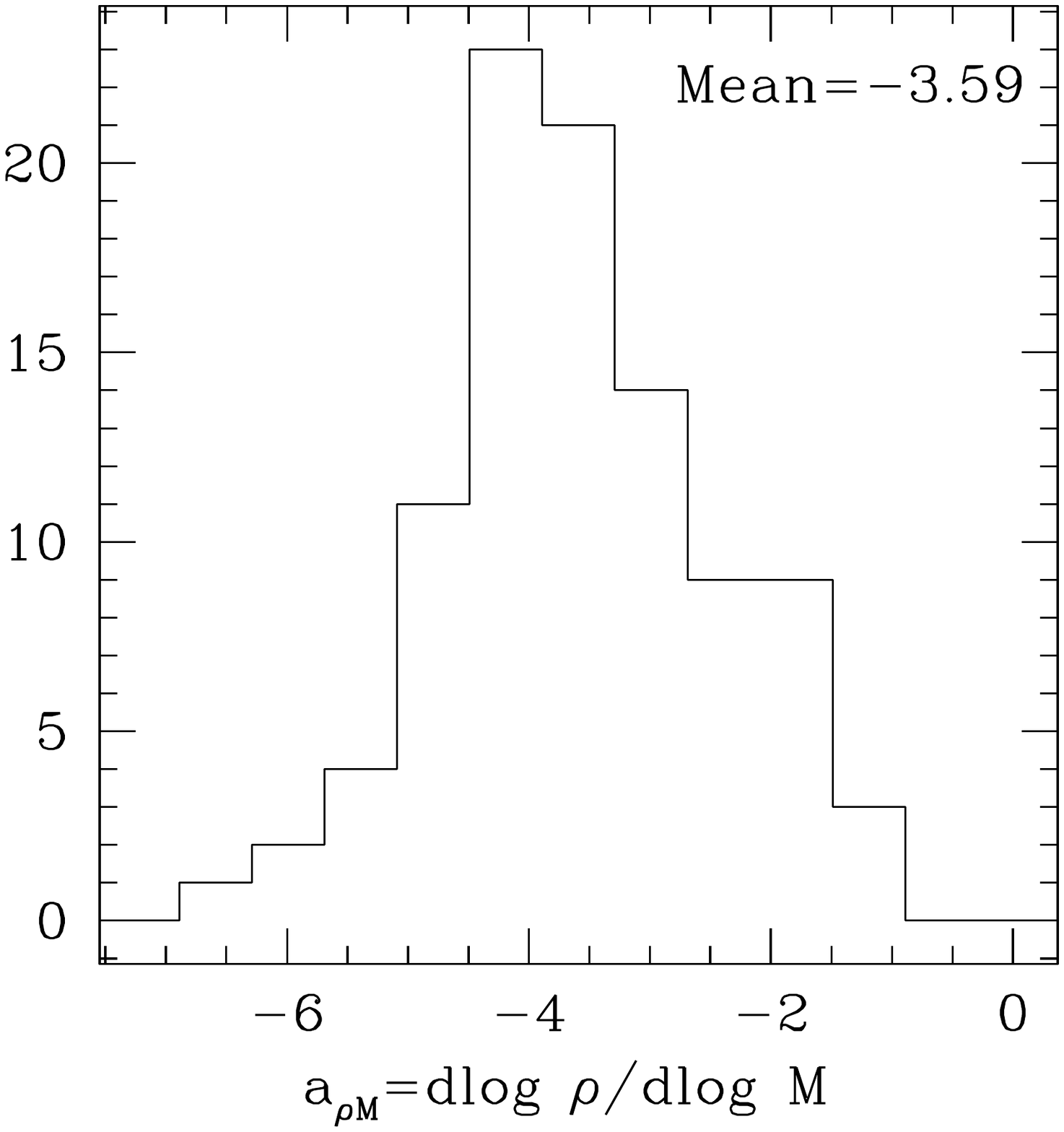}\\
    \includegraphics[trim=0 4cm 0 4cm,clip,width=8cm]{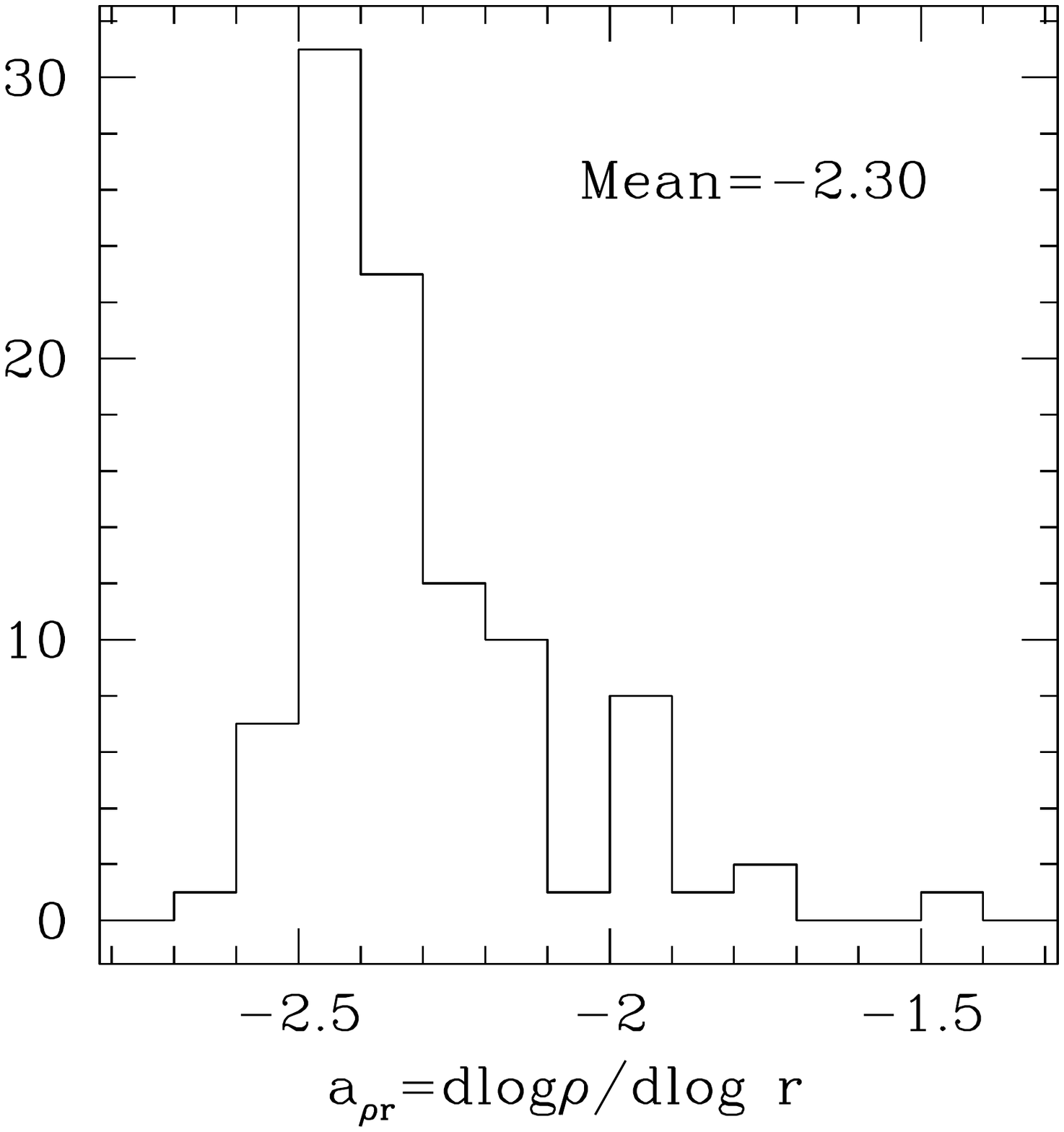}\\
 \end{center}
  \caption{The histograms of the values of the 
logarithmic derivatives $a_{rM}$ (top left), $a_{\rho M}$ (top right) and $a_{\rho  r}$ (bottom). \label{fig_ahist}}
\end{figure*}

We now turn to the errors affecting the quantities
$r_h, \rho_h,$ and $M_{Bu}$. Errors on the distance affect all of them, as well as
the errors coming from the extrapolation to
compute total luminosities. Errors on $M_{Bu}$ and $\rho_h$ have a
further component due to the $M/L$ factor (see
Sect. \ref{sec_errors}) . We estimate the fractional error on the
bulge mass due to extrapolation as $\delta
M_{Bu}/M_{Bu}=M_{extrap}/M_{Bu}$. It is on average 9\%.  We then
determine the logarithmic derivates $a_{rM}=d\log r /d\log M$,
$a_{\rho M}=d\log \rho/ d\log M$, and $a_{\rho r}=d\log \rho/d\log r$
at $r=r_h$ by a least-squares fit in a region $\pm 0.1 dex$ around the
mass point $\log M_{Bu}/2$. The logarithmic errors on $M_{Bu}$, $r_h$,
$\rho_h$ due to extrapolatin are then 
$d\log M_{Bu}=\delta M_{Bu}/M_{Bu}\times\log e$,
$d\log r_h=a_{rM}d\log M_{Bu}$, $d\log \rho_h=a_{\rho M}d\log M_{Bu}$.
Fig. \ref{fig_densmass} shows the bulge density and mass profiles of our
galaxy sample. Fig. \ref{fig_ahist} shows the histograms of the values
of the parameters $a_{rM}$, $a_{\rho M}$ and $a_{\rho r}$. Their
values can be derived from Table \ref{tab_data} using the equations
given in Table \ref{tab_errormatrix} (see Sect. \ref{sec_errors} 
for a full description of this Table). On average, we have $\rho\sim
r^{-2.3}$ near $r\sim r_h$, which implies $M\sim r^{3-2.3}=r^{0.7}$,
or $r\sim M^{1.5}$ and $\rho\sim M^{-2.3\times1.5}=M^{-3.5}$.
Therefore, the logarithmic errors on $r_h$ due to the extrapolation
are on average 1.5 times larger than the ones on $M_{Bu}$, while the
logarithmic errors on $\rho_h$ are 3.5 times larger.

\subsection{External comparisons}
\label{sec_ComparisonData}

\begin{figure*}
  \begin{center}
    \includegraphics[trim=0 4cm 0 4cm,clip,width=8cm]{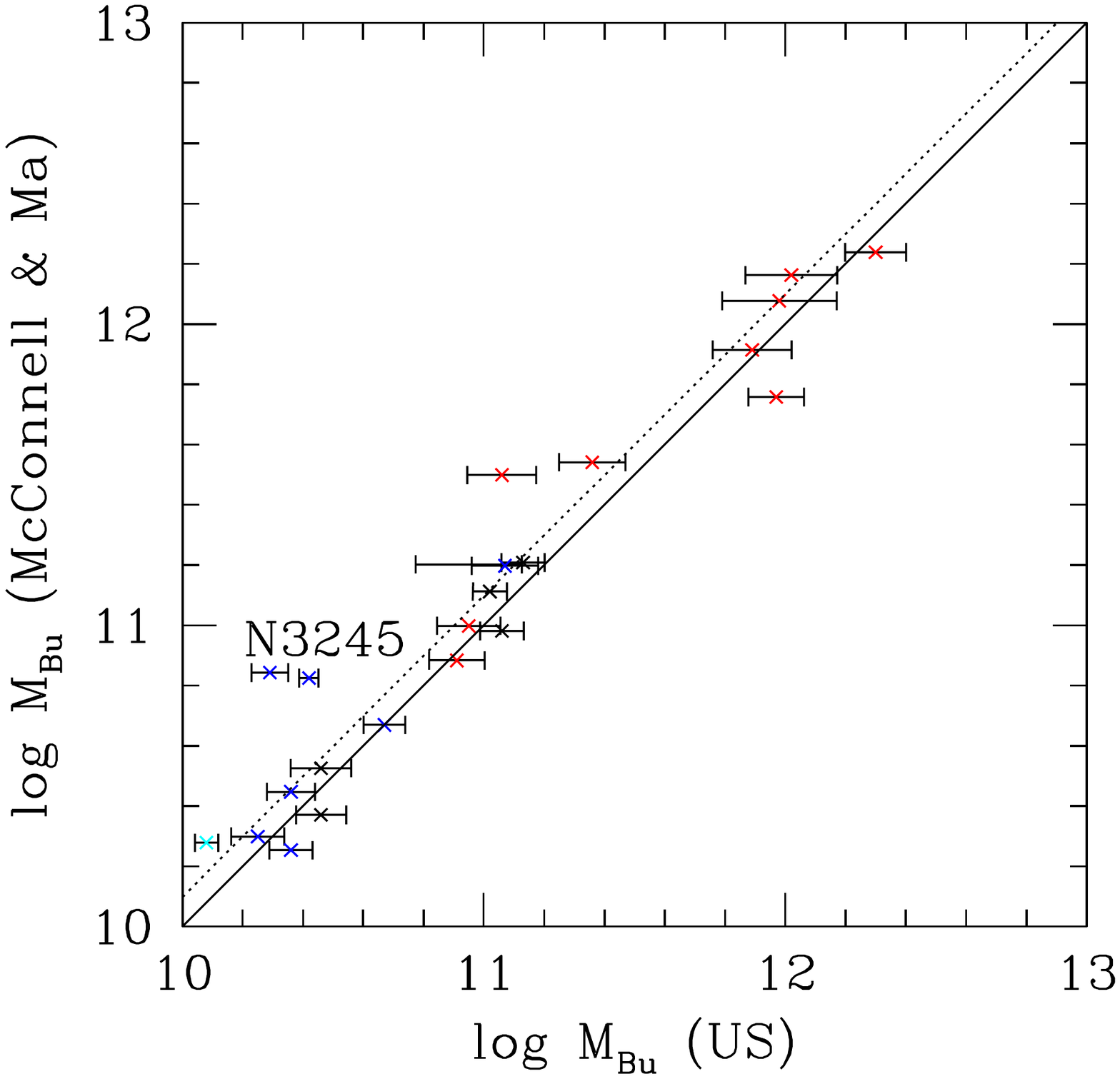}
    \includegraphics[trim=0 4cm 0 4cm,clip,width=8cm]{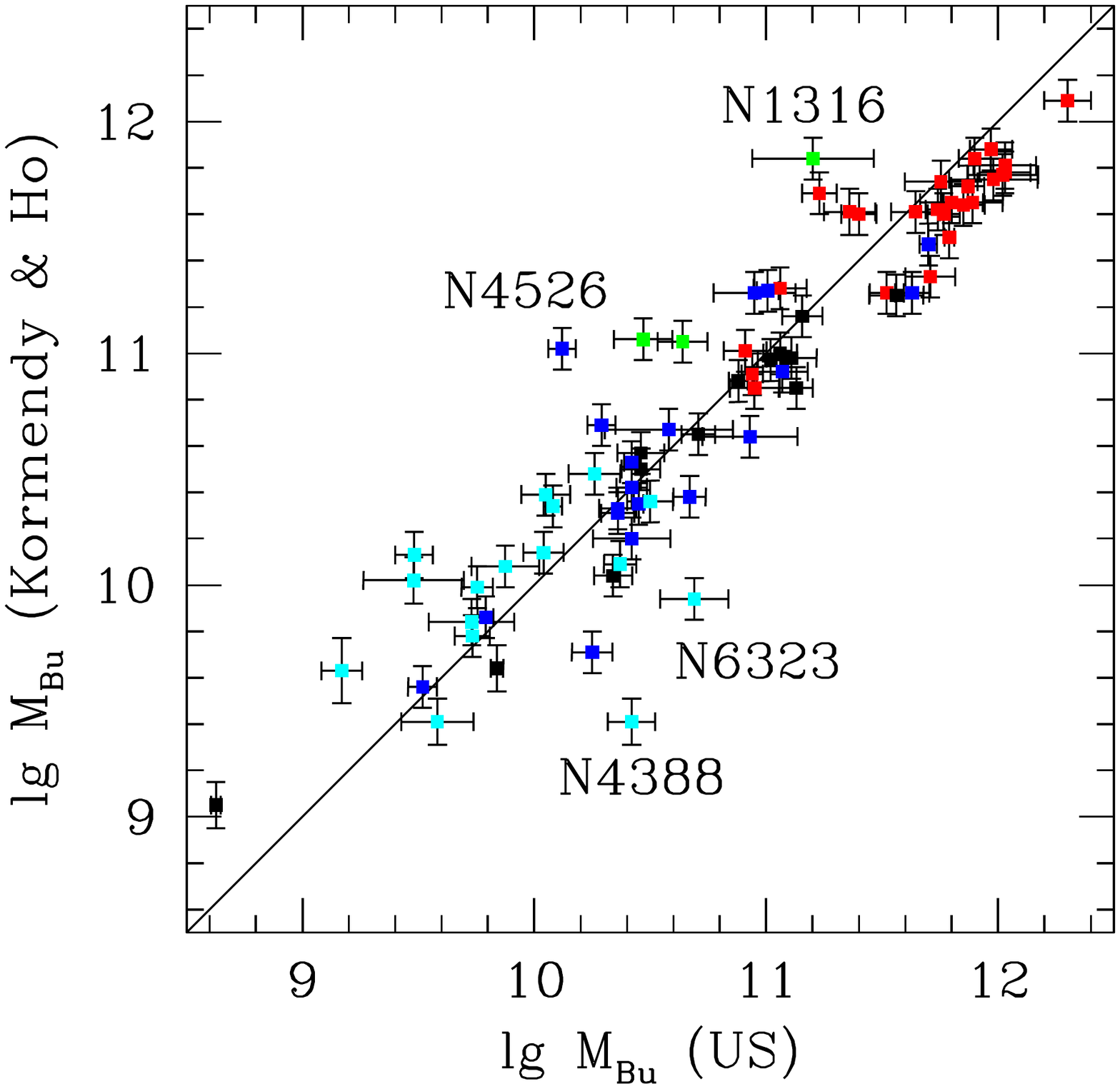}
    \includegraphics[trim=0 4cm 0 4cm,clip,width=8cm]{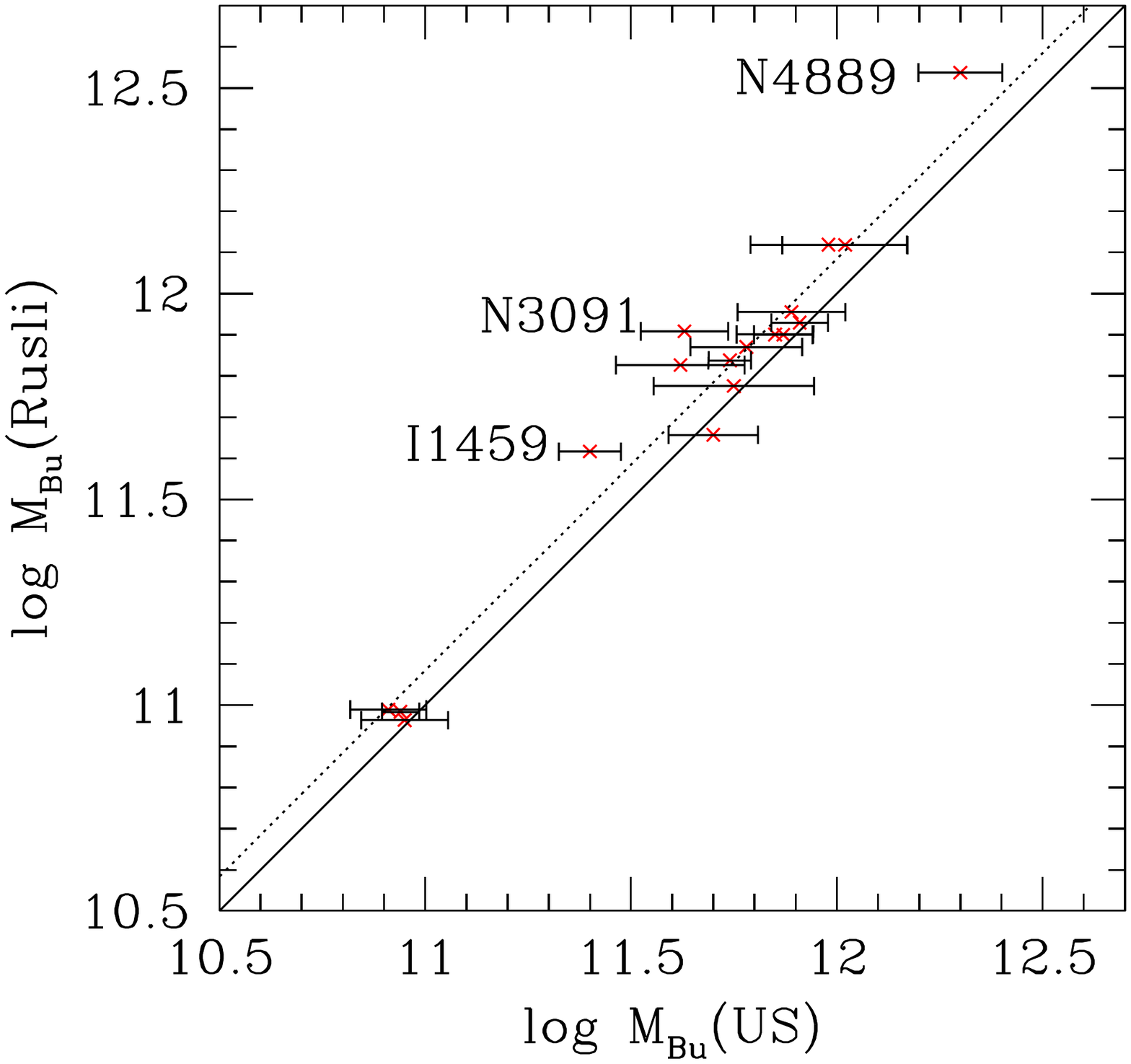}
    \includegraphics[trim=0 4cm 0 4cm,clip,width=8cm]{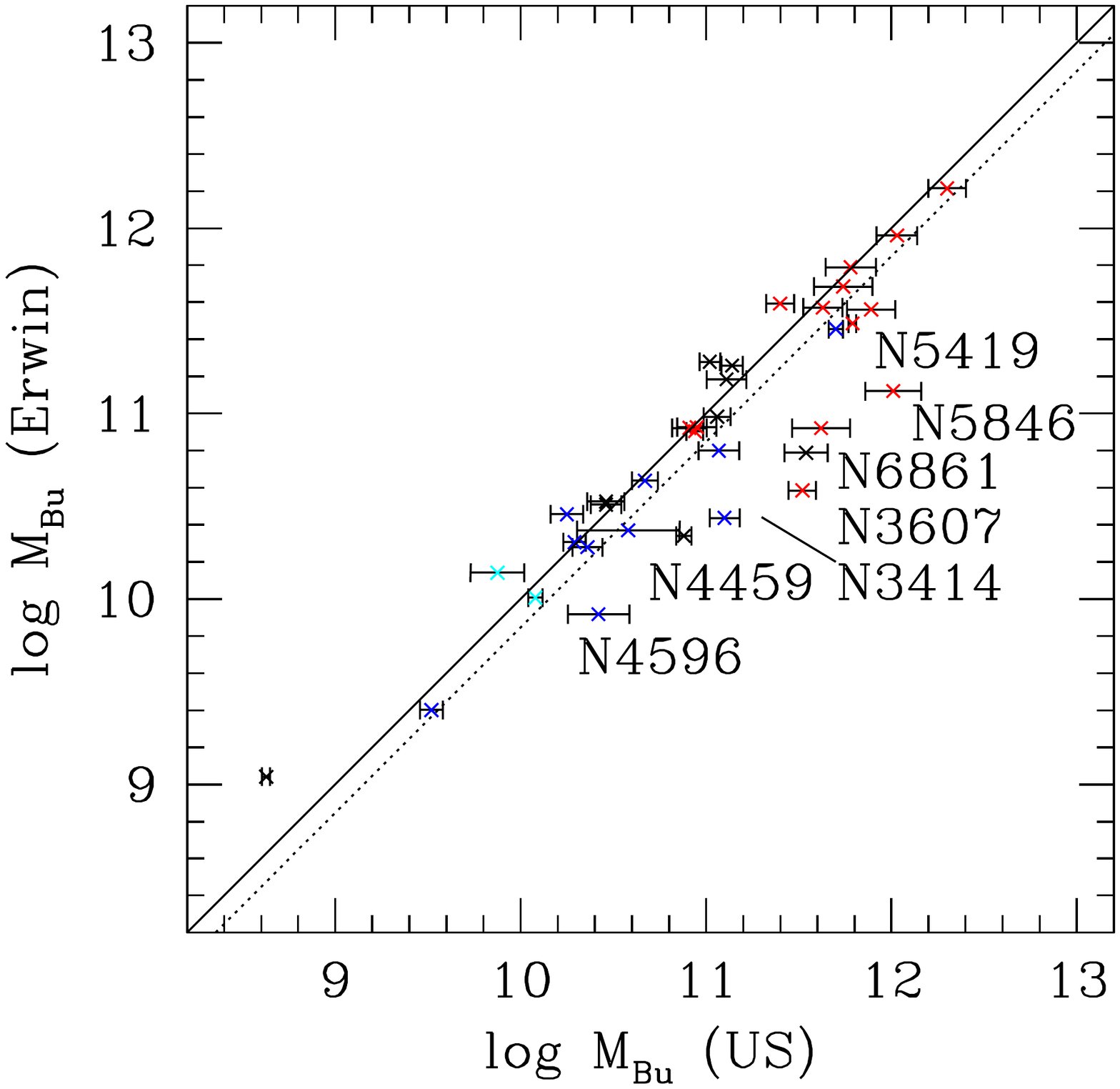}
 \end{center}
 \caption{The bulge masses compared to the values of \citet[][top
   left]{McConnellMa2013}, \citet[][top right]{KormendyHo2013},
   \citet[][bottom left]{Rusli2013b} and \citet[][bottom
   right]{Erwin2015a}.  Red points are core ellipticals, black
     points are power-law ellipticals, blue points are classical
     bulges, cyan points are pseudo bulges, green points are
   mergers. In the upper left plot we exclude the galaxies where we
   consider the classical component of a composite (pseudo plus
   classical) bulge.  In the bottom right plot we show only galaxies
   fitted by one component. The full line shows the one-to-one
   relation, the dotted line is shifted to fit the datapoints on
   average \citep[indistinguishable from the full line for the sample
   of][]{KormendyHo2013}.
\label{fig_bulgecomp}}
\end{figure*}

In Fig. \ref{fig_bulgecomp} we compare our bulge masses to the ones
reported by \citet{McConnellMa2013}, \citet{KormendyHo2013} (where we
exclude pseudo bulges with classical components), \citet{Rusli2013a}
(where we show only galaxies fitted by one component) and
\citet{Erwin2015a}.  They compare reasonably well, with an rms scatter
of $\approx 0.2$ dex and estimated error bars of 0.1 dex. The most
deviant point in the comparison with \citet{McConnellMa2013} is NGC
3245. The difference stems from their large assumed $B/T$ ratio
(0.76). The \citet{KormendyHo2013} points deviating more strongly are
pseudo bulges, in particular NGC 4388 and NGC 6323, and the classical
bulge of NGC 4526, see discussion in Appendix B. The fits discussed in
Sect. \ref{correlations} and presented in the Tables \ref{tab_kormendy}
and  \ref{tab_twokormendy}
show that using the bulge masses 
of \citet{KormendyHo2013}, which are derived using $M/L$ from colors, does not
change our conclusions.  The differences with the masses of
\citet{Rusli2013b} are within the estimated errors due to the
extrapolation. The four most deviant galaxies are discussed in detail
in Fig. \ref{fig_extrap}. Some of the masses of \citet{Erwin2015a}
come from the sample of \citet{Laurikainen2011}, who did not publish
distances; the latter were provided to us by Laurikainen (private
comm.). For five galaxies \citet{Erwin2015a} quote (bulge) masses a
factor 10 smaller than what we get.  The differences originate mainly
from the amount of extrapolation (see below).  In addition,
\citet{Erwin2015a} derive stellar masses based on $M/L$ from colors.

\begin{figure*}
  \begin{center}
    \includegraphics[trim=0 4cm 0 4cm,clip,width=8cm]{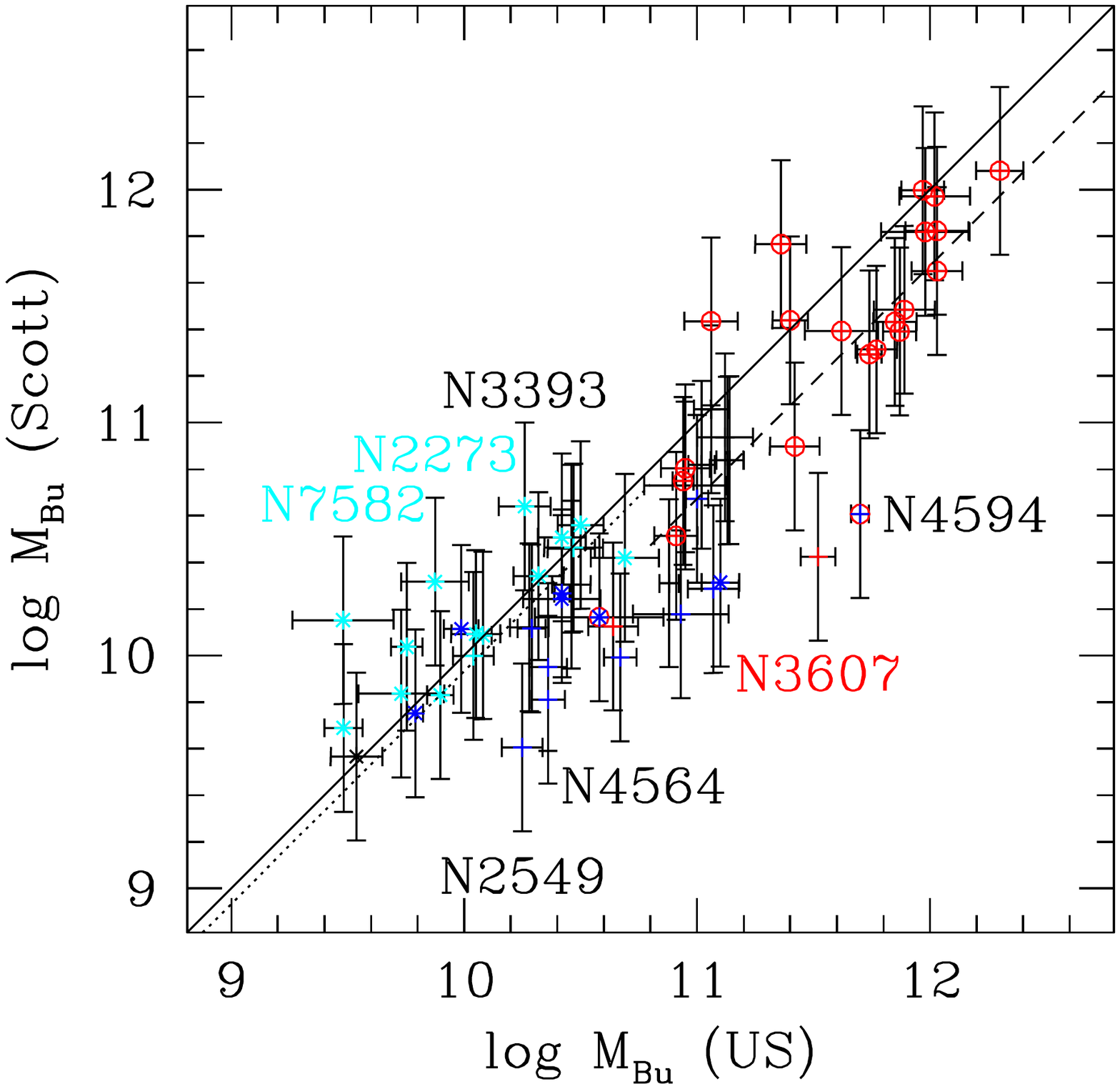}
    \includegraphics[trim=0 4cm 0 4cm,clip,width=8cm]{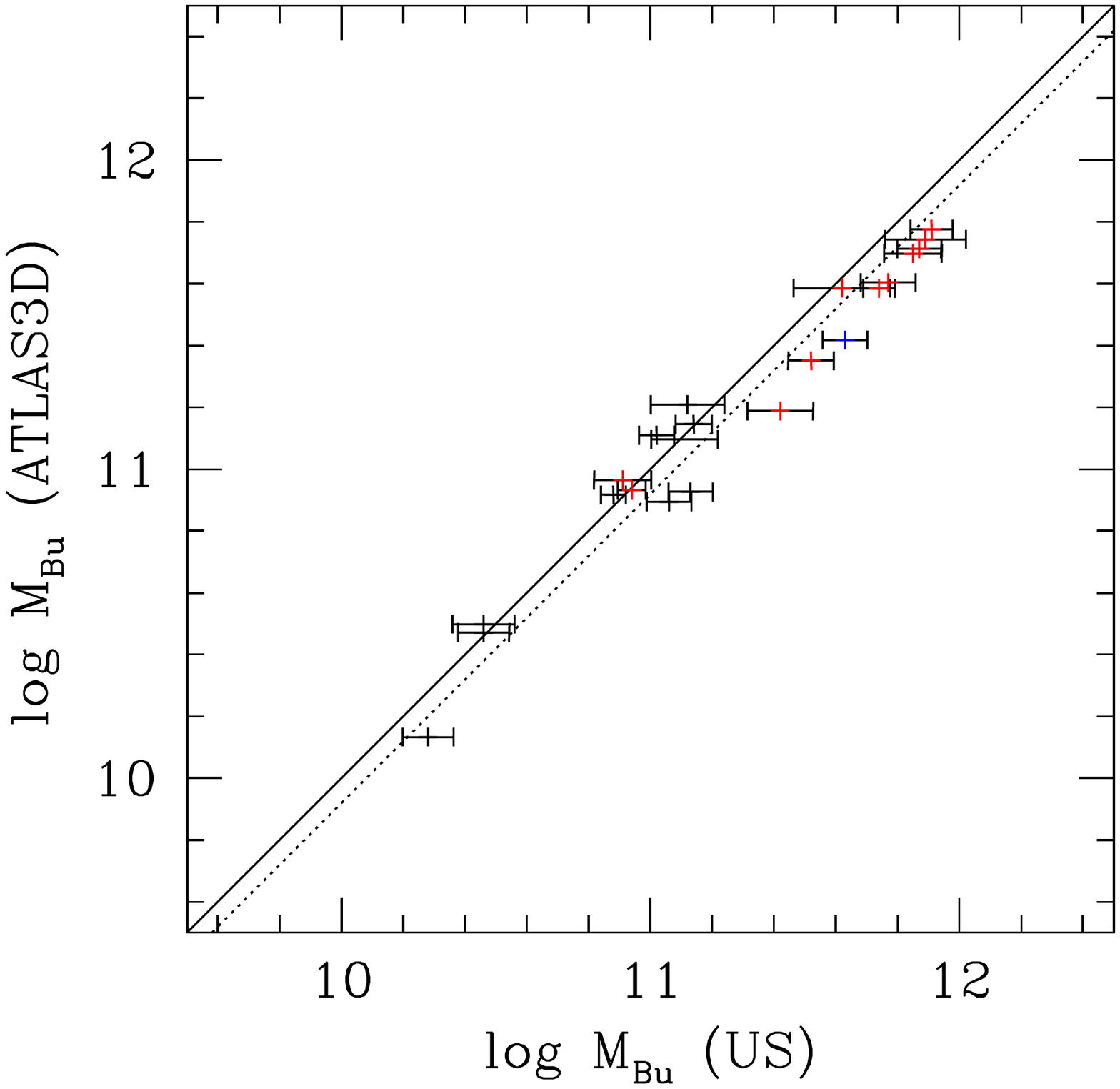}\\
    \includegraphics[trim=0 4cm 0 4cm,clip,width=8cm]{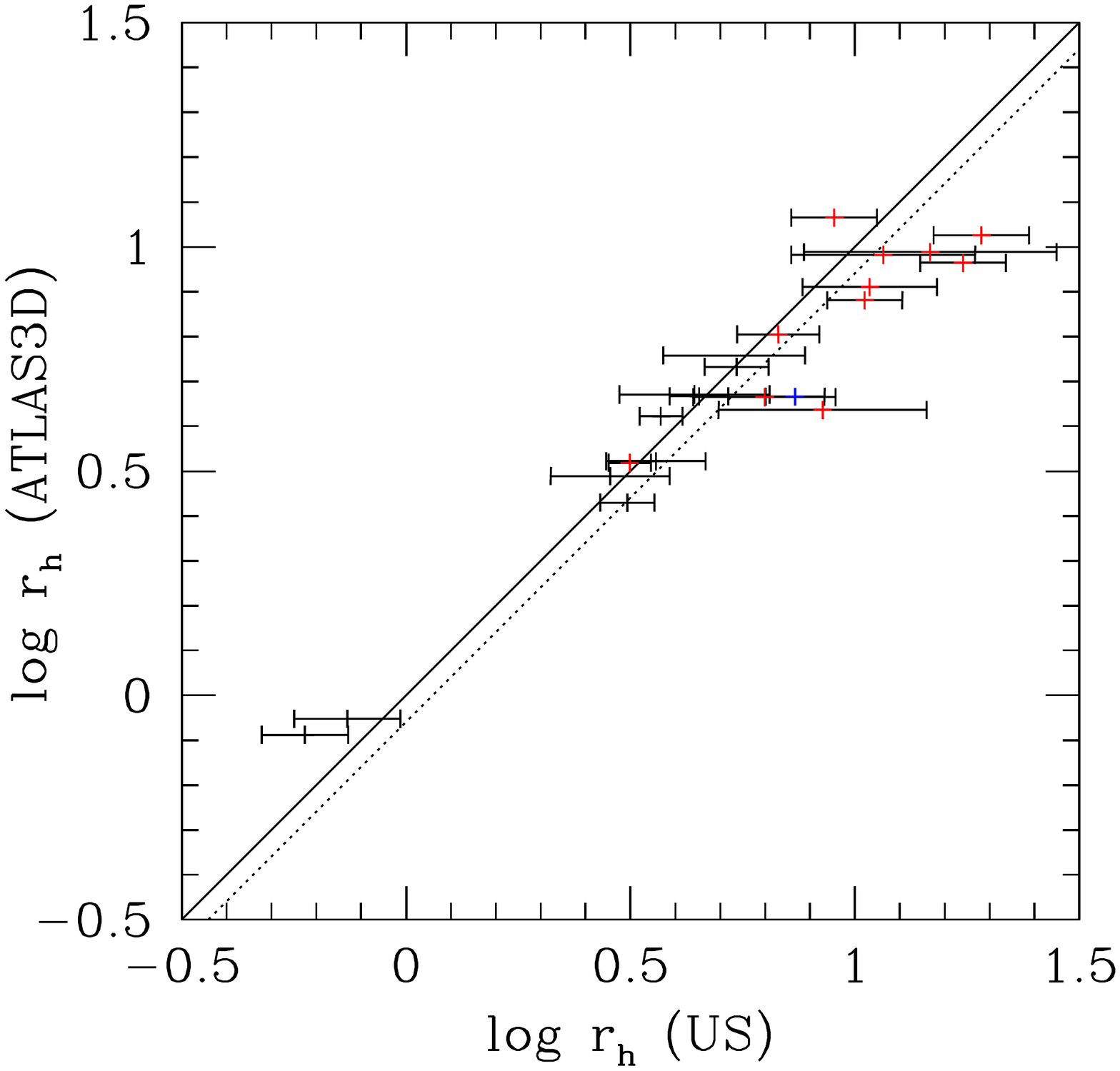}
    \includegraphics[trim=0 4cm 0 4cm,clip,width=8cm]{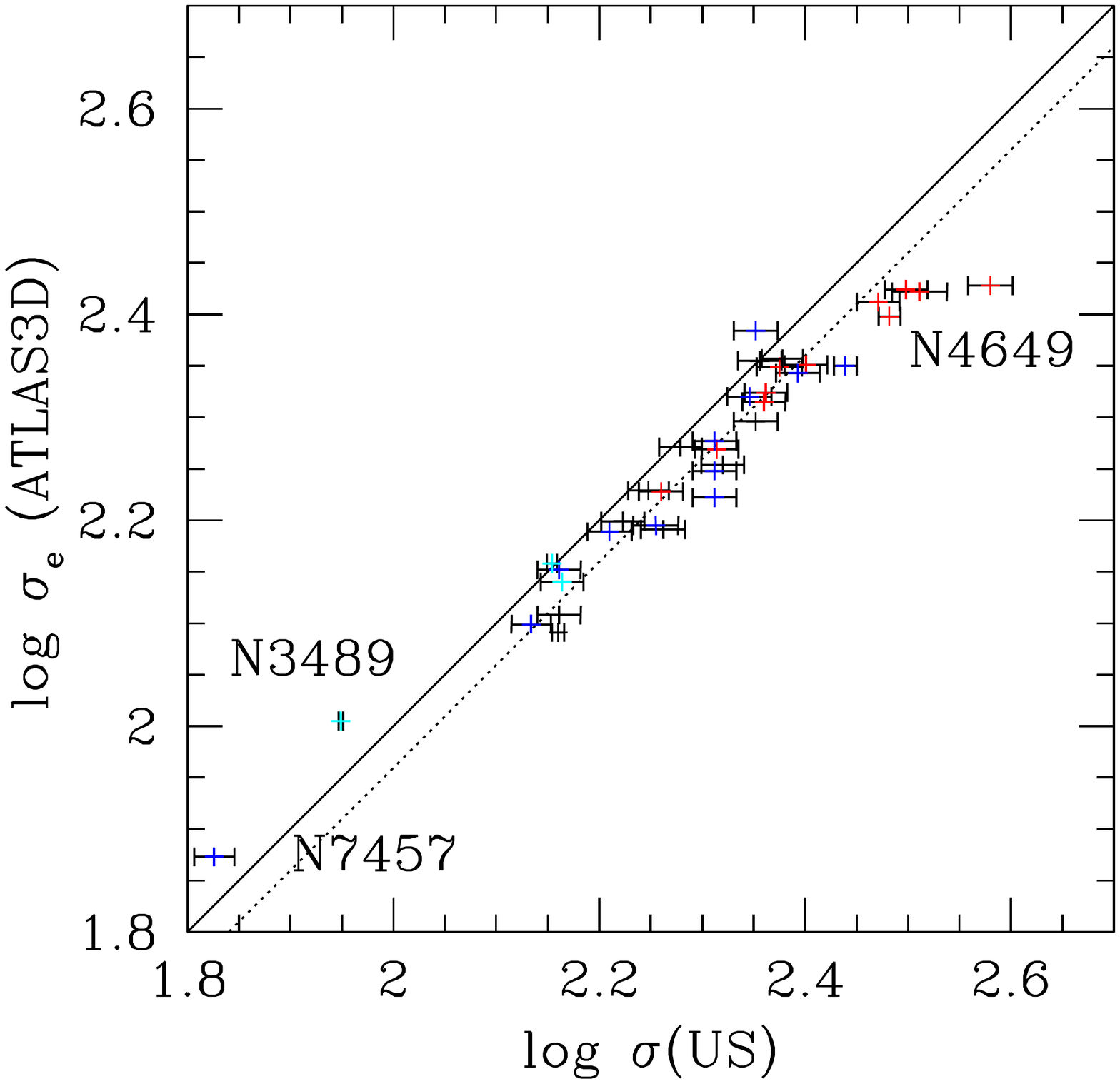}\\
 \end{center}
  \caption{Top left: our bulge masses compared to the values of 
\citet{Scott2013}.  
We exclude NGC 1399 and NGC 6086, where
we subtract the outer halo from the profile. 
Colors and continuous line as in Fig. \ref{fig_bulgecomp}; the dotted and 
dashed lines are shifted to fit the datapoints at $\log M_{Bu}<10.8$ and  
$\log M_{Bu}\ge10.8$ respectively; circles indicate Core-Sersic 
galaxies according to the classification of  \citet{Scott2013}, asterisks 
indicate galaxies that we classified as barred. Top right: the bulge masses 
compared to the values of \citet{Cappellari2013} for the galaxies where we do
 not apply a decomposition. Bottom left: the half-mass radii  compared to the 
values of \citet{Cappellari2013} for the galaxies where we do not apply a decomposition.
Bottom right: the velocity dispersions compared to the $\sigma_e$ values of \citet{Cappellari2013}. 
The three most deviant galaxies are labelled. 
Colors and continuous lines as above. The dotted lines are shifted to the average difference. 
\label{fig_MbulgeScott}}
\end{figure*}

Fig. \ref{fig_MbulgeScott}, upper left, shows the comparison between
our bulge masses and the ones reported by \citet{Scott2013}. The
latter are not based on proper bulge plus disk decompositions, but are
instead derived from total $K_S$-band magnitudes by applying a
statistical bulge-to-disk correction that depends on morphological
type. In addition, their conversion of light into mass involves
(dust-corrected) $M/L$ values derived from $(B-K)$ colors.
The overall scatter in Fig. \ref{fig_MbulgeScott} (0.36 dex) matches the total
uncertainty quoted by \citet{Scott2013}. For galaxies where we measure
$\log M_{Bu}<10.8$ the agreement is fair (the average difference is
0.05 dex). However, at larger masses \citet{Scott2013} derive values
0.34 dex smaller.  This and the
missing distinction between classical and pseudo bulges drive the
steepening of the \mbu\ relation at small bulge masses of ``Sersic
galaxies'' noticed by \citet{Graham2013}.

Fig. \ref{fig_MbulgeScott}, upper right, shows the comparison between
our bulge masses and the ones reported by the ATLAS3D collaboration
\citep{Cappellari2013} for the galaxies where we do not apply a
decomposition.  \citet{Cappellari2013} do not attempt any
extrapolation, this probably explains most of the measured average
shift of -0.11 dex. The measured scatter matches our error
estimates. Similar conclusions are reached when we compare the
half-mass radii (Fig. \ref{fig_MbulgeScott}, bottom left). Finally and
for completeness, Fig. \ref{fig_MbulgeScott}, bottom right, shows the
comparison between the velocity dispersions used here and the
$\sigma_e$ values reported by \citet{Cappellari2013}. The ATLAS3D are
systematically smaller by $\approx 10$\%; the scatter is a bit larger
than the errors.

\begin{figure*}
  \begin{center}
    \includegraphics[trim=0 4cm 0 4cm,clip,width=6cm]{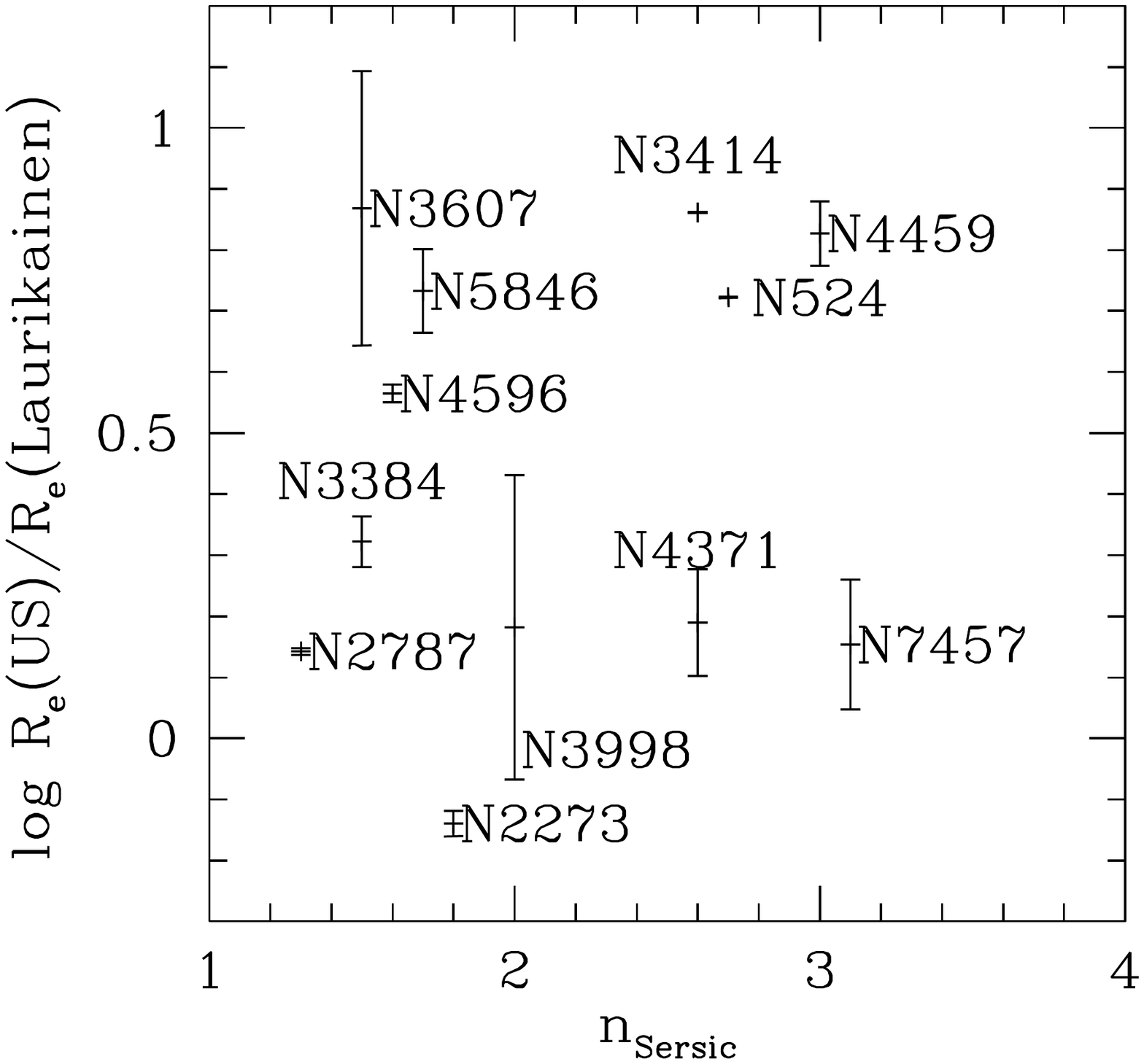}
    \includegraphics[trim=0 4cm 0 4cm,clip,width=6cm]{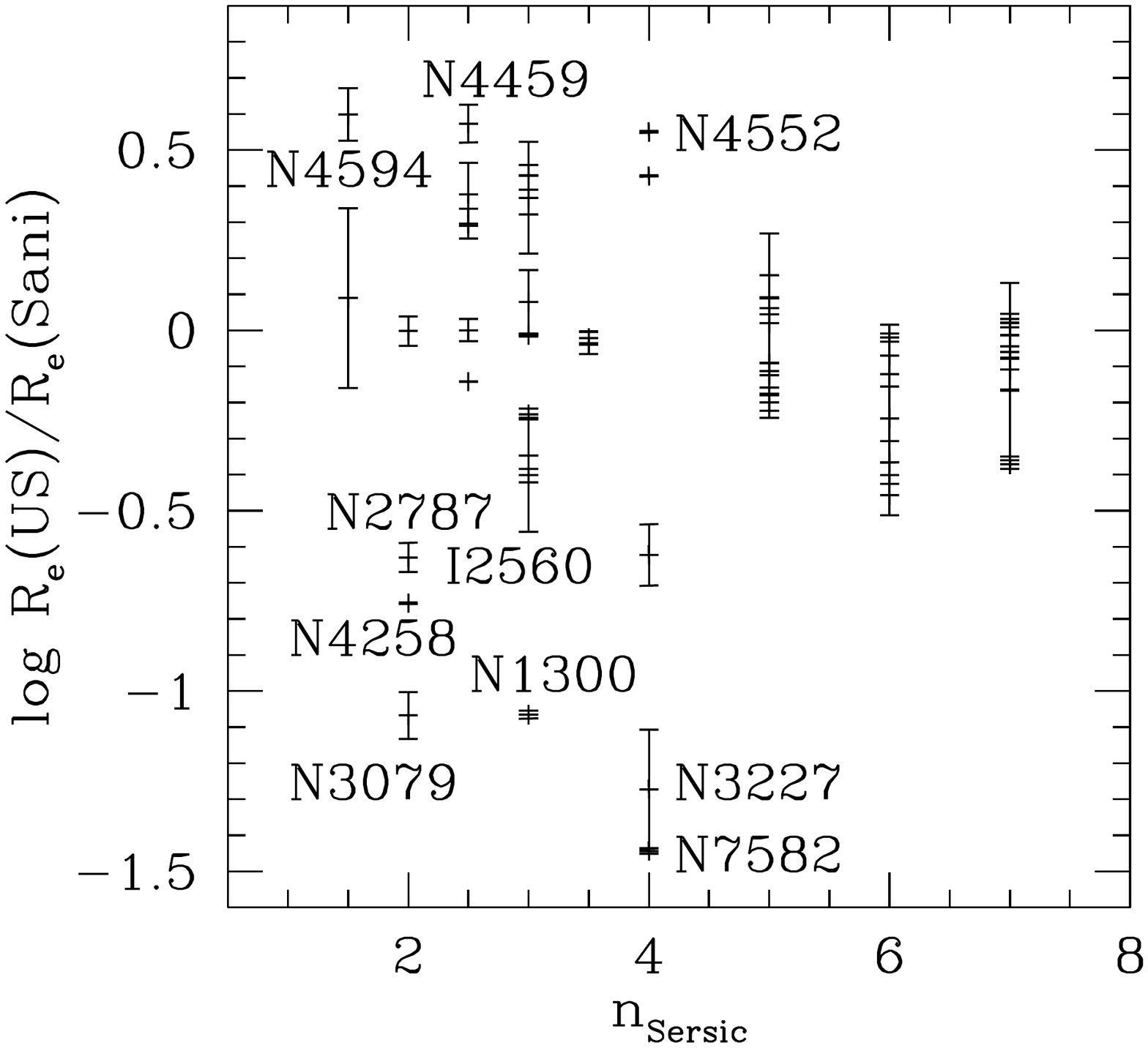}\\
    \includegraphics[trim=0 4cm 0 4cm,clip,width=6cm]{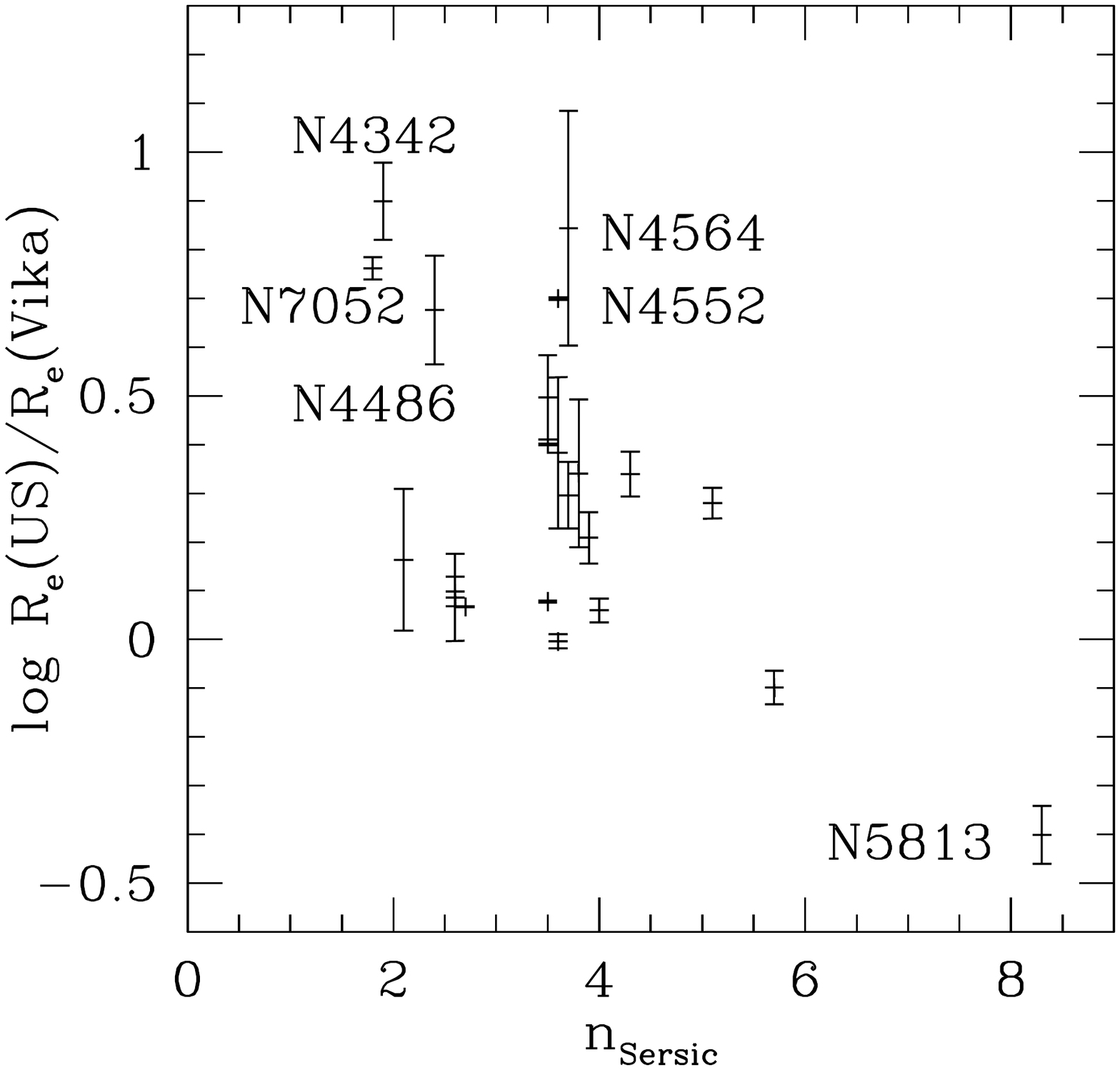}
    \includegraphics[trim=0 4cm 0 4cm,clip,width=6cm]{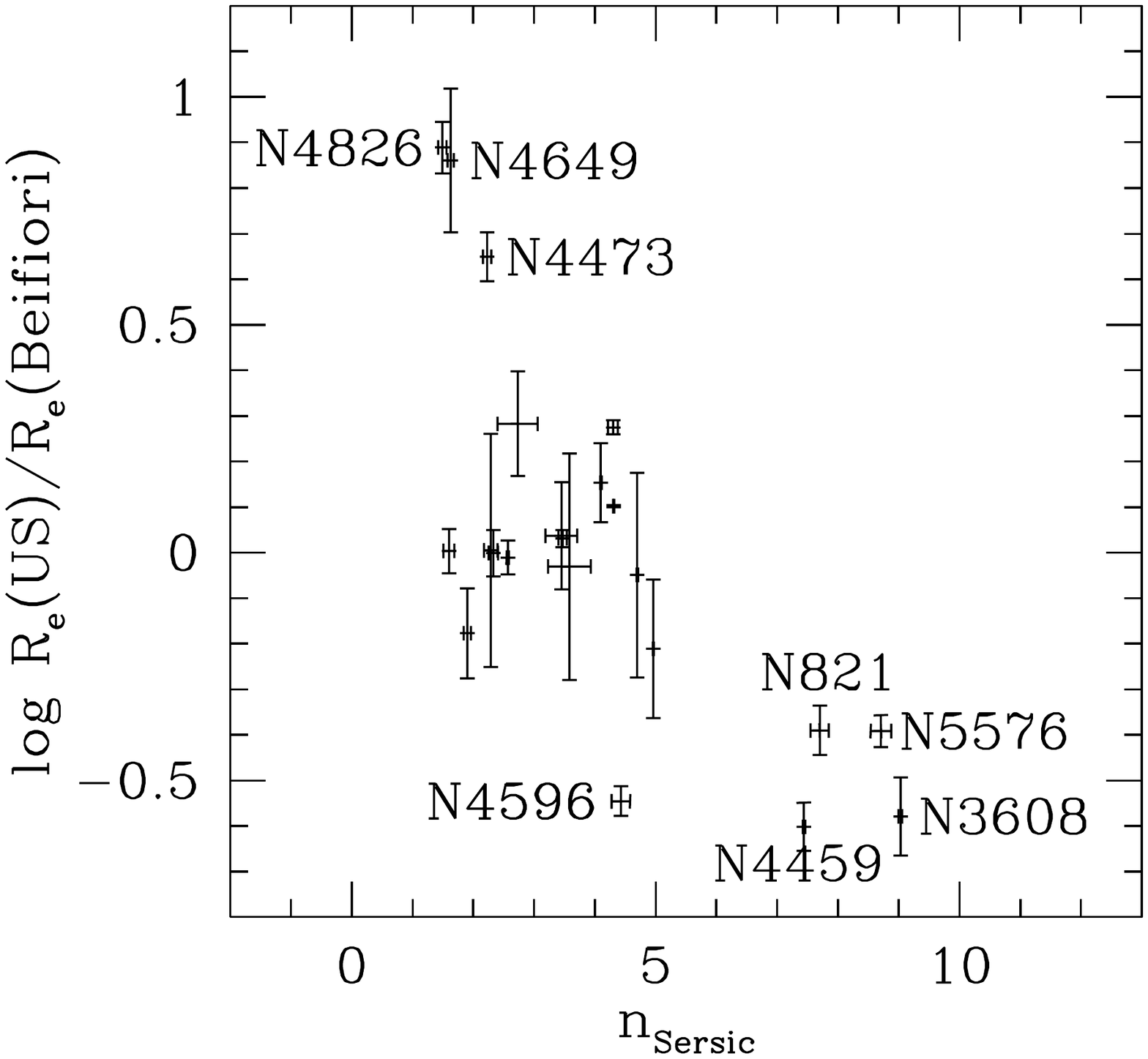}\\
    \includegraphics[trim=0 4cm 0 4cm,clip,width=6cm]{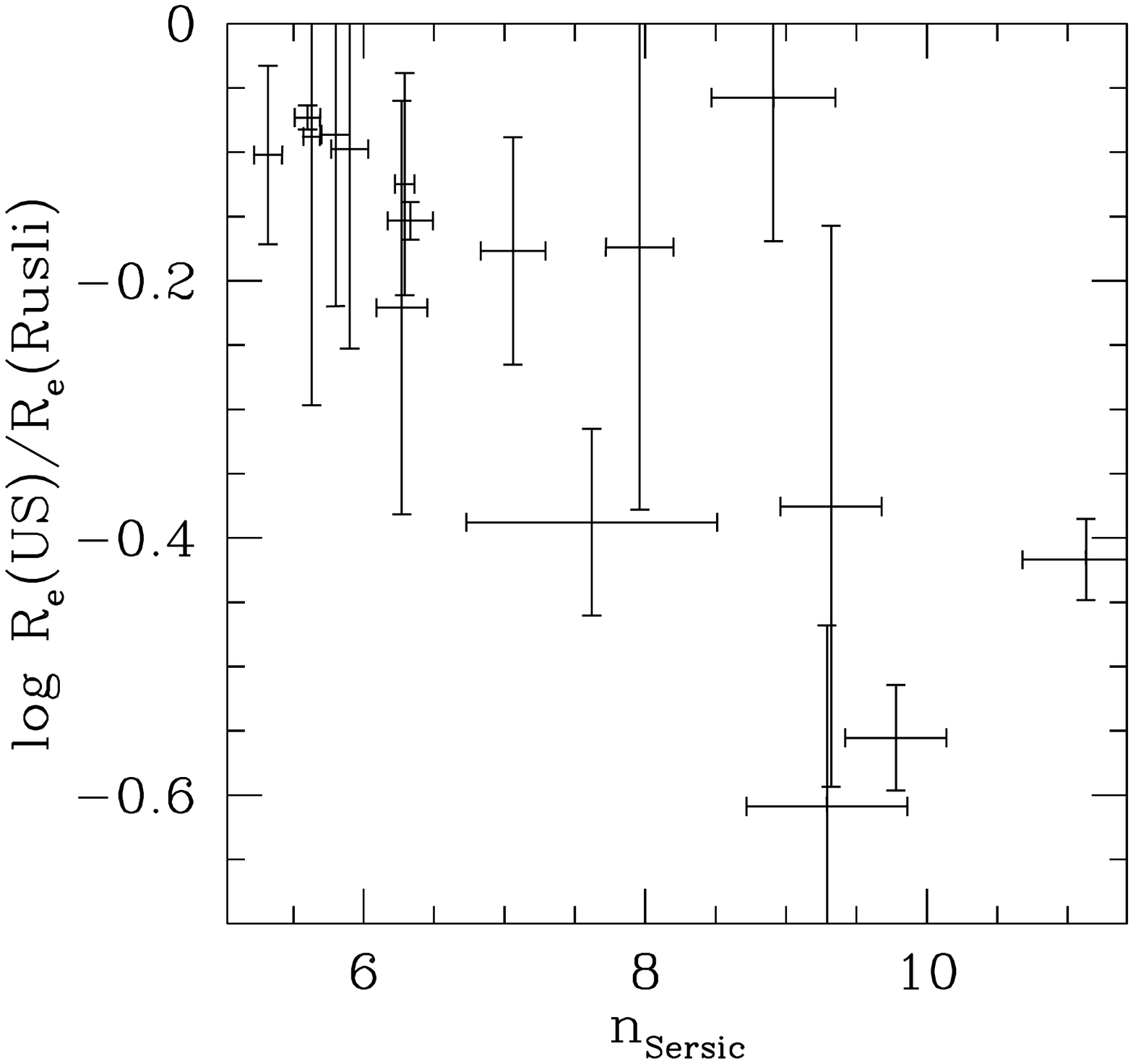}
    \includegraphics[trim=0 4cm 0 4cm,clip,width=6cm]{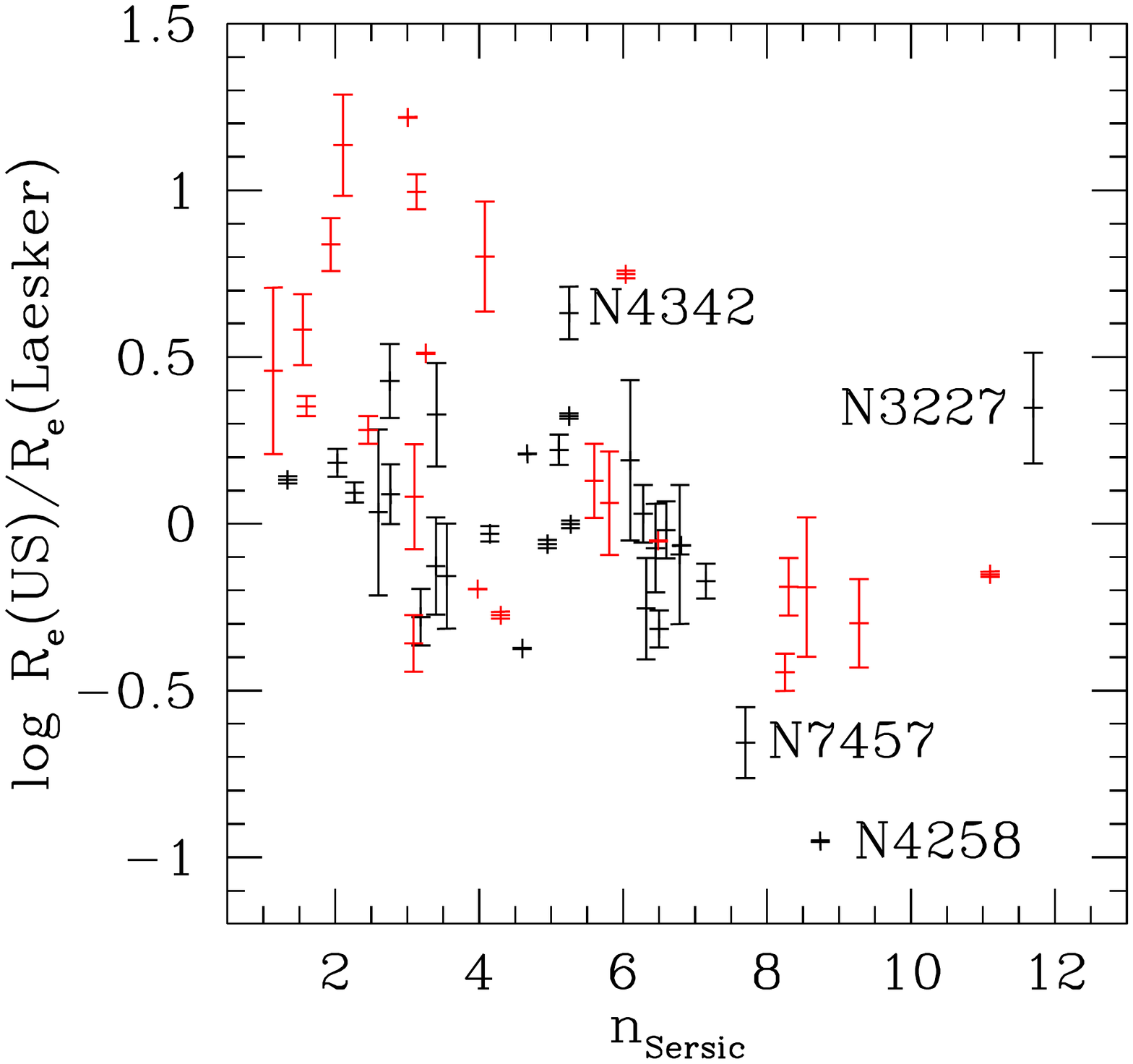}
 \end{center}
 \caption{The bulge circularized half-luminosity radius $R_e$ compared
   to the values of  \citet[][top left]{Laurikainen2010}, \citet[][top right]{Sani2011}, \citet[][middle
   left]{Vika2012},  \citet[][middle
   right]{Beifiori2012}, \citet[][bottom left]{Rusli2013a} and \citet[][bottom
   right]{Laesker2014} as a function of the fitted Sersic index
   $n_{Sersic}$.  The red points in the comparison
   to \citet{Laesker2014} indicate their 'best'
   solution. \label{fig_rebulgecomp}}
\end{figure*}

Fig. \ref{fig_rebulgecomp}
compares our bulge circularized half-luminosity radii $R_e$ to 
literature values from \citet{Laurikainen2010,
  Sani2011,Beifiori2012,Vika2012,Rusli2013a,Laesker2014}. 
We use the fitted bulge axis ratios $(b/a)$
tabulated by \citet{Sani2011} to transform their major-axis bulge
half-luminosity radii $a_e$ into
$R_e=a_e\times(b/a)^{0.5}$. \citet{Laurikainen2010,Vika2012,Laesker2014}
did not publish their fitted $b/a$ values, therefore we adopt the axis
ratios $(b/a)_{25}$ given by Hyperleda and compute
$R_e=a_e\times(b/a)_{25}^{0.5}$. The observed spread is larger than
our estimated errors, with several1 galaxies where our bulge $R_e$
differs by more than 0.3 dex from the literature values.  We discuss
these objects in Appendix B, where we
justify our choices. Some
discrepancies stem from differences in the photometric band and
the assumed ellipticity. Often the differences in $R_e$ correlate with
the fitted value of $n_{Ser}$: they are small when $n_{Ser}\approx
3-5$, which matches approximately our assumed extrapolation. Our
$R_e$ are larger than the literature values when $n_{Ser}\le 3$ and
smaller when $n_{Ser}\ge 5$. However, the major contributors are the
fitting procedures adopted.

If we perform the same check for the 16 galaxies discussed in the
Appendix B, Tables \ref{tab_maxfits} to \ref{tab:7582decomp}, where we
perform multi-component fits, we find that the $R_e$ we give in Table
\ref{tab_data} (that are derived from the curve of growth of the
spherical densities projected along the line of sight) agree with the
ones from the fits to within 9\%.

\section{The error matrix}
\label{sec_errors}

\begin{figure}
  \begin{center}
    \includegraphics[trim=5cm 9cm 0cm 5cm,clip,width=11cm]{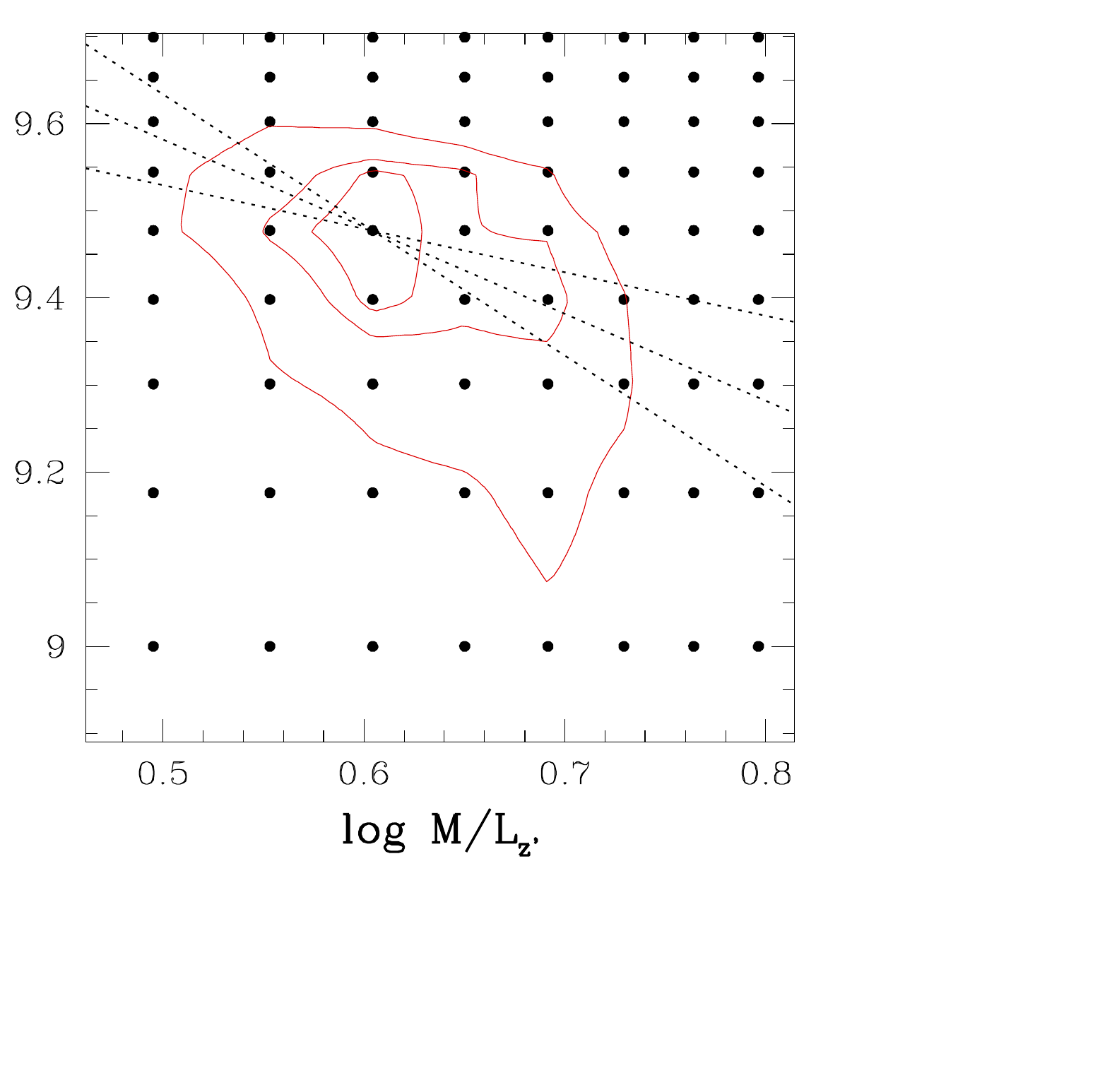}
 \end{center}
 \caption{The $\chi^2$ contours as a function of $\log M/L$ and $\log
   M_{BH}$ for NGC 3923.  The filled points show the grid of computed models.
The dotted lines show the lines of
   correlated errors $\Delta \log M_{BH}=a_{BH}\Delta \log M/L_{z'}$,
   for $a_{BH}=-0.5, -1, -1.5$.\label{fig_n3923chi}}
\end{figure}

In Sect. \ref{sec_data} we described the data sample. For each galaxy
we collected the distance $D$, the central black hole mass $M_{BH}$,
the average velocity dispersion $\sigma$, the bulge mass $M_{Bu}$
and mass-to-light ratio $M/L$, the bulge half-light radius $r_h$ and
the bulge average density $\rho_h$ within $r_h$. We also computed
errors on each quantity.  We now discuss how we estimated the total
error covariances on the decimal logarithms of the parameters
$\sigma$, $M_{BH}$, $M_{Bu}$, $r_h$, and $\rho_h$. All black hole
correlation analyses performed in the past have ignored covariances,
although some are obvious (e.g. black hole and bulge masses scale with
the distance). Here we attempt to quantify them systematically to
assess their importance (or lack thereof).  This is not always
possible in a rigorous way: some error correlations are hidden in the
modeling procedure (e.g, the anti-correlation between mass-to-light
ratio and black hole mass) and cannot be reconstructed from the
published material; what follows is sometimes simply our best guess.
The Tables \ref{tab_errormatrix} and  \ref{tab_errormatrixKormendy}
summarise our results for the two cases where the bulge masses are computed 
from mass-to-light ratios derived dynamically or from colors. In the first case
we have:

\begin{enumerate}
\item Since dynamically determined $M/L$ ratios scale as the square of 
velocities, we
  consider an additional error term on 
$(\delta\log M/L)^2$ of $4f(\delta \log \sigma)^2$,
  when we see that the quoted error on $M/L$ given in Table
  \ref{tab_ml} is too small given the kinematics available.  Here $f$
  is a fudge factor that can be either 0 or 1 and is listed in Table
  \ref{tab_errormatrix}. Therefore we use an effective error $(\delta
  M/L)_{eff}$, where we add in quadrature the two error contributions.
\item The total error on the black hole mass $\log M_{BH}$ comes from
  the fitting procedure and the error on the distance discussed in
  Sect. \ref{sec_data}. We add both terms in quadrature. Depending on
  the type of data and their spatial resolution, the errors on \mbh\
  and on the mass-to-light ratio $M/L$ can be anticorrelated
  \citep{Rusli2013a}: $\delta \log M_{BH}^{fit}=a_{BH}\delta \log M/L$
  with $a_{BH}\le 0$.  This dependency is important when computing the
  covariances $\delta \log M_{BH}\delta \log \sigma$ and $\delta \log
  M_{BH}\delta \log \rho_h$. We list the adopted values of $a_{BH}$
  (which can be 0, -1, or -2) in Table \ref{tab_errormatrix}: black
  hole masses not coming from stellar dynamical data (e.g., maser or
  gas dynamics measurements) must have $a_{BH}=0$. The values
  $a_{BH}=-1$ or $-2$ come from typical $\chi^2$ contour plots as a
  function of $\log M_{BH}$ and $\log M/L$
  \citep[see][]{Nowak2010,Rusli2011,Rusli2013a}. In
  Fig. \ref{fig_n3923chi} we show the case of NGC 3923, for which we
  assign $a_{BH}=-1$.
\item The total error on the bulge mass $\log M_{Bu}$ comes from the
  residual extrapolation of Eq. \ref{eq_LBu}, the error on the
  distance discussed above, and the error on the mass-to-light
  ratio. We add the three terms in quadrature. The external
  comparisons performed in the previous section show that this error
  estimate is probably too small. Therefore we also
  consider solutions where we add 0.15 dex in quadrature to $\delta
  \log M_{Bu}$.
\item The total error on the half-luminosity radius $r_h$ comes from
  the extrapolation in Eq. \ref{eq_LBu} and the error on the
  distance. We add both terms in quadrature, weighting the mass
  extrapolation term with the correlation coefficient $a_{rM}^2$.
\item The total error on the average density $\rho_h$ within the
  half-luminosity radius comes from the extrapolation in
  Eq. \ref{eq_LBu}, the error(s) on $M/L$ (see above) and the error on 
  the distance.  We add the four terms in quadrature, weighting the
  mass extrapolation term with the correlation coefficient $a_{\rho
    M}^2$ and the distance error by a factor $2^2$ (since $\rho\sim
  M/r^3\sim D^{-2}$). Following the reasoning applied to  $\delta
  \log M_{Bu}$, we also consider solutions where  we add 0.15 dex in quadrature to $\delta
  \log \rho_h$.
\item The errors on $M_{BH}$ and $\sigma$ can be correlated through the $M/L$
term, which scales as the square of velocity (see above).
\item The errors on $M_{BH}$ and $M_{Bu}$ are correlated through the
  distance and possibly anticorrelated through the 
$M/L$ term (since $a_{BH}$ is negative).
\item The errors on $M_{BH}$ and $r_h$ are correlated through the distance.
\item The errors on $M_{BH}$ and $\rho_h$ are anti-correlated through the
  distance and possibly through the $M/L$ term.
\item The errors on $M_{Bu}$ and $\sigma$ are correlated through the $M/L$
term, see above.
\item The errors on $M_{Bu}$ and $r_h$ are correlated through the mass
  extrapolation term  and the distance.
\item Errors on $M_{Bu}$ and $\rho_h$ are correlated through the mass
  extrapolation and the $M/L$ terms,  and
  anticorrelated through the distance. Moreover, when we consider solutions 
where  we add 0.15 dex in quadrature to $\delta \log M_{Bu}$ and $\delta  \log \rho$, 
we augment the covariance element
$\delta \log M_{Bu} \delta \log \rho_h$ by the same amount in quadrature.
\item Errors on $\log r_h$ and $\log \rho_h$ are correlated through the mass
  extrapolation term  and anticorrelated
  through the distance.
\item Errors on $\log \rho_h$ and $\log \sigma$ are correlated through the $M/L$
term.
\end{enumerate}

For some galaxies the black hole mass is determined independently from
the bulge $M/L$, for example from maser or gas rotation curves. In
this case $a_{BH}=0$. For some galaxies setting $a_{BH}=-2$ and $f=1$
produces covariance matrices with negative determinants. In these
cases we increase $a_{BH}$ to $-1$ or 0, and/or we set $f=0$.  Table
\ref{tab_errormatrix} reports the values of the terms averaged over
the sample. The most poorly determined parameter is $\rho_h$, followed
by $M_{BH}$. On average, the off-diagonal terms of the covariance
  matrix are smaller than the diagonal terms; therefore ignoring them,
  as done in the past, is not a bad approximation.

The Tables \ref{tab_data} and \ref{tab_erroffdiag1} list
the values of each term for each galaxy of the sample.

\begin{table*}
\caption{The error correlations between the parameters for mass-to-light ratios derived dynamically.}
\label{tab_errormatrix}
\begin{tabular}{lll}
\hline
\hline
N & Quantity & Mean value \\
\hline
0 & $(\delta \log \sigma)^2$ & 0.0006\\
1 & $(\delta \log M/L)^2_{eff}=(\delta \log M/L)^2+4f(\delta \log \sigma)^2$ & 0.0038\\
2 & $(\delta \log M_{BH})^2=(\delta \log M_{BH}^{fit})^2+(\delta \log D)^2$ & 0.023\\
3 & $(\delta \log M_{Bu})^2=(\delta \log M_{Bu}^{ext})^2+(\delta \log D)^2 +(\delta \log M/L)^2_{eff}$& 0.012\\
4 & $(\delta \log r_h)^2 = a^2_{rM}(\delta \log M_{Bu}^{ext})^2+(\delta \log D)^2$ & 0.015\\
5 & $(\delta \log \rho_h)^2 = a^2_{\rho M}(\delta \log M_{Bu}^{ext})^2 +(\delta \log M/L)^2_{eff}+4(\delta \log D)^2$ & 0.087\\ 
\hline
6 & $\delta \log M_{BH}\delta \log \sigma = 2fa_{BH}(\delta \log \sigma)^2$ & -0.0003\\
7 & $\delta \log M_{BH}\delta \log M_{Bu}=(\delta \log D)^2 +a_{BH}(\delta \log M/L)^2_{eff}$& 0.0023\\
8 & $\delta \log M_{BH}\delta \log r_h=(\delta \log D)^2$ & 0.0037\\
9 & $\delta \log M_{BH}\delta \log \rho_h=-2(\delta \log D)^2 +a_{BH}(\delta \log M/L)^2_{eff}$  & -0.009\\
10 & $\delta \log M_{Bu}\delta \log \sigma = 2f(\delta \log \sigma)^2$ & 0.0005\\
11 & $\delta \log M_{Bu}\delta \log r_h=a_{rM}(\delta\log M_{Bu}^{ext})^2+(\delta \log D)^2$ & 0.01 \\
12 & $\delta \log M_{Bu}\delta \log \rho_h=a_{\rho M}(\delta\log M_{Bu}^{ext})^2+(\delta \log M/L)^2_{eff}-2(\delta \log D)^2$ & -0.0185 \\
13 & $\delta \log r_h\delta \log \rho_h = a_{\rho M}a_{rM}(\delta \log M_{Bu}^{ext})^2-2(\log D)^2$ & -0.035 \\
14 & $\delta \log \rho_h \delta \log \sigma =  2f(\delta \log \sigma)^2$ & 0.0005\\
\hline
\end{tabular}
\end{table*}

In the case of mass-to-light ratios derived from colors, we get:

\begin{enumerate}
\item We assume that the error on the $\log M/L_C$ from colors is equal
  to the error on bulge masses quoted by \citet{KormendyHo2013}.
\item The total error on the black hole mass $\log M_{BH}$ is unchanged, 
and is as described above.
\item The total error squared on the bulge mass $\log M_{Bu,C}$ has an increased 
  dependence on the distance error, that now
  goes as $4(\delta \log D)^2$. This is because the mass-to-light ratio
  derived from colors is distance-independent, and therefore bulge masses scale as luminosities with
  the square of the distance.
\item The total error on $\log r_h$ is unchanged, and is as described above.
\item The total error squared on $\log \rho_{h,C}$ has a reduced dependence on the distance error, that now goes as  
$(\delta \log D)^2$. This stems from the increased dependence on distance of the bulge mass (see point 3).
\item Errors on $\log M_{BH}$ and $\log \sigma$ can be correlated through the $M/L$ as above and are unchanged.
\item The correlation between the errors on $\log M_{BH}$ and $\log M_{Bu,C}$ is only due to the distance and amounts to 
$2(\delta \log D)^2$ because of the dependence of bulge masses on distance (see above).
\item Errors on $\log M_{BH}$ and $\log r_h$ are correlated through the distance as before.
\item The anti-correlation between the errors on $\log M_{BH}$ and $\log \rho_{h,C}$ is only due to the distance and amounts to 
$-(\delta \log D)^2$ because of the dependence of bulge masses on distance (see above).
\item There is no correlation between the errors on  $\log M_{Bu,C}$ and on $\log \sigma$.
\item The errors on $\log M_{Bu,C}$ and $\log r_h$ are correlated
  through the mass extrapolation term and with a stronger dependence
  on the distance error squared (as $2(\delta \log D)^2$) from the
  $D^2$ scaling of bulges masses.
\item The errors on $\log M_{Bu,C}$ and $\log \rho_{h,C}$ are correlated
  through the mass extrapolation term and the $M/L_C$ dependency and anticorrelated through the distance, as in the previous case.
\item The errors on $\log r_h$ and $\log \rho_{h,C}$ are correlated through the mass
  extrapolation term  and anticorrelated (with the reduced dependency $-(\delta \log D)^2$)
  through the distance.
\item The errors on $\log \rho_{h,C}$ and $\log \sigma$ are not correlated.
\end{enumerate}

The average values of the covariance matrix elements listed in
  Tables \ref{tab_errormatrix} and \ref{tab_errormatrixKormendy} are
  similar, with $\delta \log M_{Bu,C}$ larger than $\delta \log
  M_{Bu}$.

The Tables \ref{tab_dataKormendy} and \ref{tab_erroffdiagKormendy}
list the values of each term for each galaxy of the sample of
\citet{KormendyHo2013} considered here.

\begin{table*}
\caption{The error correlations between the parameters for mass-to-light ratios derived from colors.}
\label{tab_errormatrixKormendy}
\begin{tabular}{lll}
\hline
\hline
N & Quantity & Mean value \\
\hline
0 & $(\delta \log \sigma)^2$ & 0.0005\\
1 & $(\delta \log M/L_C)^2= (\delta M_{Bu,C})^2$& 0.008 \\
2 & $(\delta \log M_{BH})^2=(\delta \log M_{BH}^{fit})^2+(\delta \log D)^2$ & 0.022\\
3 & $(\delta \log M_{Bu,C})^2=(\delta \log M_{Bu}^{ext})^2+4(\delta \log D)^2 +(\delta \log M/L_C)^2$& 0.019\\
4 & $(\delta \log r_h)^2 = a^2_{rM}(\delta \log M_{Bu}^{ext})^2+(\delta \log D)^2$ & 0.013\\
5 & $(\delta \log \rho_{h,C})^2 = a^2_{\rho M}(\delta \log M_{Bu}^{ext})^2 +(\delta \log M/L_C)^2+(\delta \log D)^2$ & 0.073\\ 
\hline
6 & $\delta \log M_{BH}\delta \log \sigma = 2fa_{BH}(\delta \log \sigma)^2$ & $-3\times10^{-6}$\\
7 & $\delta \log M_{BH}\delta \log M_{Bu,C}=2(\delta \log D)^2 $& 0.004\\
8 & $\delta \log M_{BH}\delta \log r_h=(\delta \log D)^2$ & 0.002\\
9 & $\delta \log M_{BH}\delta \log \rho_h=-2(\delta \log D)^2$ &-0.002   \\
10 & $\delta \log M_{Bu,C}\delta \log \sigma = 0$ & 0 \\
11 & $\delta \log M_{Bu,C}\delta \log r_h=a_{rM}(\delta\log M_{Bu}^{ext})^2+2(\delta \log D)^2$ & 0.01\\
12 & $\delta \log M_{Bu,C}\delta \log \rho_{h,C}=a_{\rho M}(\delta\log M_{Bu}^{ext})^2+(\delta M/L_C)^2-2(\delta \log D)^2$ & -0.01\\
13 & $\delta \log r_h\delta \log \rho_{h,C} = a_{\rho M}a_{rM}(\delta \log M_{Bu}^{ext})^2-2(\log D)^2$ & -0.03\\
14 & $\delta \log \rho_{h,C} \delta \log \sigma = 0$  & 0\\
\hline
\end{tabular}
\end{table*}

\begin{deluxetable}{llllllllllll}
\tablewidth{0pt}
\tabletypesize{\tiny}
\tablecaption{The values of the error covariance matrices: off-diagonal terms.
\label{tab_erroffdiag1}}
\tablehead{}
\startdata
Galaxy & $a_{BH}$ & $f$ & $\delta \log M_{BH}$ & $\delta \log M_{BH}$& $\delta \log M_{BH}$ & $\delta \log M_{BH}$& $\delta \log M_{Bu}$ & $\delta \log M_{Bu}$& $\delta \log M_{Bu}$ & $\delta \log r_h$ &  $\delta \log \rho_h$ \\ 
 &  & & $ \delta \log \sigma$ & $ \delta \log M_{Bu}$ & $ \delta \log r_h$ & $ \delta \log \rho_h$ & $ \delta \log \sigma $ & $ \delta \log r_h$ & $ \delta \log \rho_h$ & $ \delta \log \rho_h$ & $ \delta \log \sigma $ \\ 
\hline
          MW           &           0           &           0           &  0.00000000           &  0.00102400           &  0.00102400           & -0.00204800           &  0.00000000           &  0.00279996           & -0.00360376           & -0.00528875           &  0.00000000          \\
    Circinus           &           0           &           0           &  0.00000000           &  0.00518400           &  0.00518400           & -0.01036800           &  0.00000000           &  0.00518913           & -0.00772816           & -0.01038157           &  0.00000000          \\
       A1836           &           0           &           1           &  0.00000000           &  0.00057600           &  0.00057600           & -0.00115200           &  0.00089280           &  0.00267647           &  0.01019260           & -0.00828498           &  0.00089280          \\
      IC1459           &          -1           &           0           & -0.00000000           &  0.00117500           &  0.00313600           & -0.00823300           &  0.00000000           &  0.00415655           & -0.00672214           & -0.01005465           &  0.00000000          \\
      IC4296           &           0           &           1           &  0.00000000           &  0.00102400           &  0.00102400           & -0.00204800           &  0.00093300           &  0.01312244           & -0.02733637           & -0.05108090           &  0.00093300          \\
\enddata
\tablenotetext{\ }{Column 1: galaxy name; Column 2 and 3: $a_{BH}$ and $f$ 
factors, see \ref{sec_errors}; 
Column 4: error covariances between $M_{BH}$ and $\sigma$;
Column 5: error covariances between $M_{BH}$ and $M_{Bu}$; 
Column 6: error covariances between $M_{BH}$ and $r_h$; 
Column 7: error covariances between $M_{BH}$ and $\rho_h$; 
Column 8: error covariances between $M_{Bu}$ and $\sigma$;
Column 9: error covariances between $M_{Bu}$ and $r_h$; 
Column 10: error covariances between $M_{Bu}$ and $\rho_h$; 
Column 11: error covariances between $r_h$ and $\rho_h$;
Column 12: error covariances between $\rho_h$ and $\sigma$.}
\end{deluxetable}

\begin{deluxetable}{llllll}
\tablewidth{0pt}
\tabletypesize{\tiny}
\tablecaption{The values of the error covariance matrices: off-diagonal terms for the \citet{KormendyHo2013} sample.
\label{tab_erroffdiagKormendy}}
\tablehead{}
\startdata
Galaxy &  $\delta \log M_{BH}$   & $\delta \log M_{BH}$ & $\delta \log M_{Bu,C}$ & $\delta \log M_{Bu,C}$   & $\delta \log r_h$ \\
       &  $\delta \log M_{Bu,C}$ & $\delta \log \rho_{h,C}$ & $\delta \log r_h$ & $\delta \log \rho_{h,C}$ & $\delta \rho_{h,C}$\\
\hline
      IC1459           &    0.006272           &   -0.003136           &    0.007293           &  -0.0005831           &   -0.006919          \\
     NGC0221           &    0.000512           &   -0.000256           &    0.000598           &    0.009294           &  -0.0005159          \\
     NGC0224           &    0.000648           &   -0.000324           &   0.0006492           &     0.00745           &  -0.0003268          \\
     NGC0524           &      0.0032           &     -0.0016           &    0.004885           &    0.000887           &   -0.008082          \\
     NGC0821           &    0.002312           &   -0.001156           &     0.01613           &    -0.02763           &    -0.05871          \\
     NGC1023           &    0.002048           &   -0.001024           &    0.002049           &    0.006049           &   -0.001029          \\
\enddata
\tablenotetext{\ }{Column 1: galaxy name;
Column 2: error covariances between $M_{BH}$ and $M_{Bu,C}$; 
Column 3: error covariances between $M_{BH}$ and $\rho_{h,C}$; 
Column 4: error covariances between $M_{Bu,C}$ and $r_h$; 
Column 5: error covariances between $M_{Bu,C}$ and $\rho_{h,C}$; 
Column 6: error covariances between $r_h$ and $\rho_{h,C}$.}
\end{deluxetable}

\section{Exploring multivariate correlations}
\label{sec_correlations}

In the following we investigate the correlations
between all measured parameters. We assume that the $n$ measured
values of the dependent variable $\eta_i$ (with $i$ running from 1 to $n$), 
can be expressed as:
\begin{equation}
\eta_i=\vec{\alpha}^T \cdot \vec{\xi_i}+ZP+N(0,\epsilon),
\label{eq_multivariate}
\end{equation}  
where $\vec{\xi_i}$ are the measured values of the independent vector
of variables, $N(0,\epsilon)$ is a normal random variable with zero
mean and variance $\epsilon^2$, with $\epsilon$ representing the
intrinsic scatter in $\eta_i$, and $ZP$ and $\vec{\alpha}$ are the
zero-point and the multi-linear coefficients. We compute $ZP$,
$\vec{\alpha}$ and $\epsilon$ following \citet{Kelly2007} and making
use of his IDL routines. To this purpose we compute the covariance
error matrices:
\begin{equation}
C_i=\left(
\begin{array}{ll}
\delta^2 \eta_i & \delta \eta_i\delta \vec{\xi}_i\\
\delta \eta_i\delta \vec{\xi}_i & \delta \vec{\xi}_i\delta \vec{\xi}_i\\
\end{array}
 \right).
\label{eq_covariance}
\end{equation}
The square roots of their diagonal terms are given in the Tables
\ref{tab_data} and \ref{tab_dataKormendy}, the off-diagonal terms can
be found in the Tables \ref{tab_erroffdiag1} and
\ref{tab_erroffdiagKormendy}.

\begin{figure*}
  \begin{center}
    \includegraphics[trim=0cm 6cm 0 4cm,clip,width=16cm]{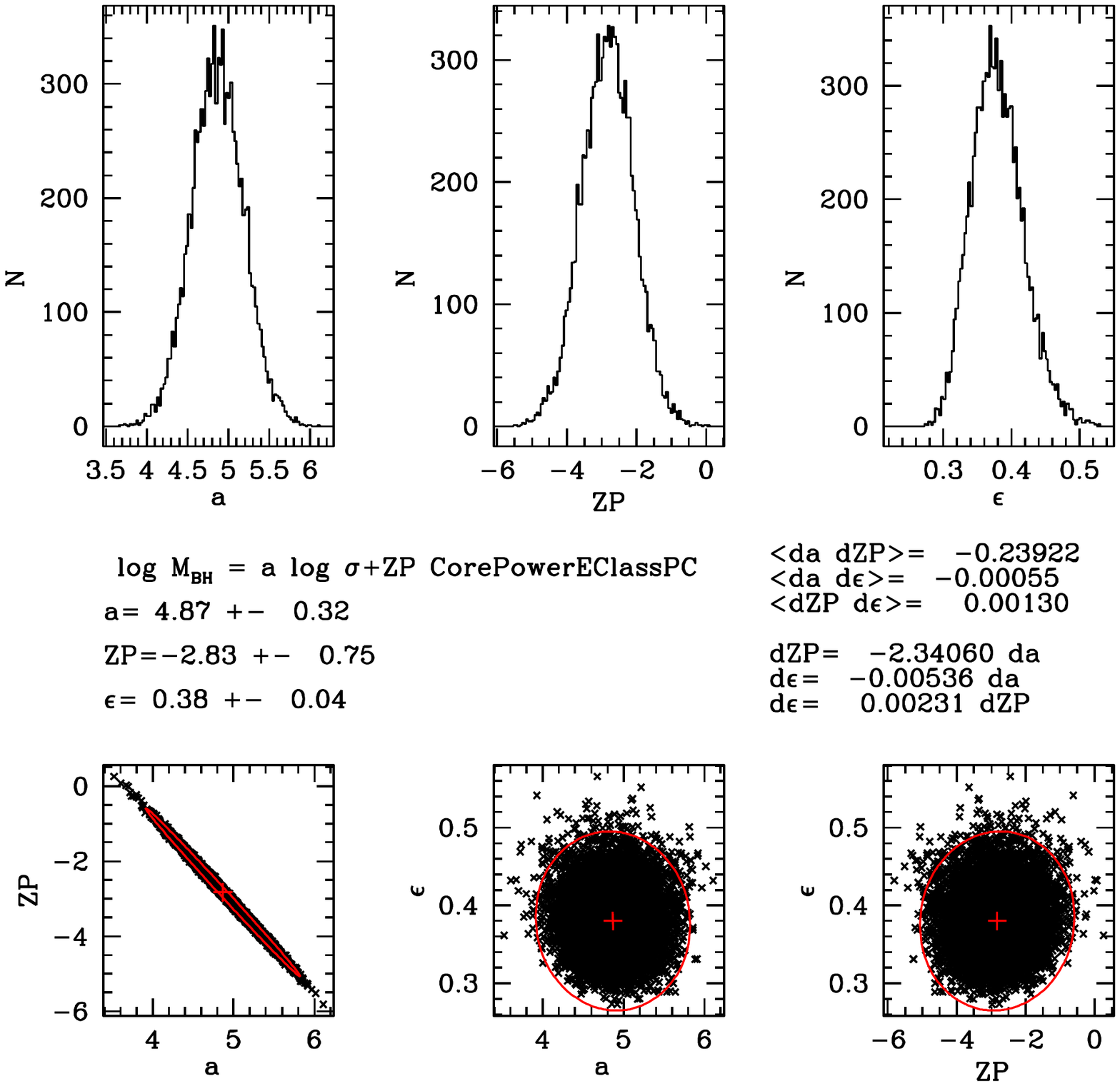}
 \end{center}
  \caption{Top row: the posterior probability distributions of the fitted 
parameters of the correlation between $\log M_{BH}$ and $\log \sigma$ for 
the subsample CorePowerEClassPC. Middle row: mean values of the parameters
 and errors, off-diagonal terms of the parameter variance matrix and resulting 
correlations between parameter errors.
Bottom row: the correlations between all possible pairs of
parameters. The red ellipse shows the
 $3\sigma$ contours.
   \label{fig_probMBHSig}}
\end{figure*}

\begin{figure*}
  \begin{center}
    \includegraphics[trim=0 6cm 0 4cm,clip,width=16cm]{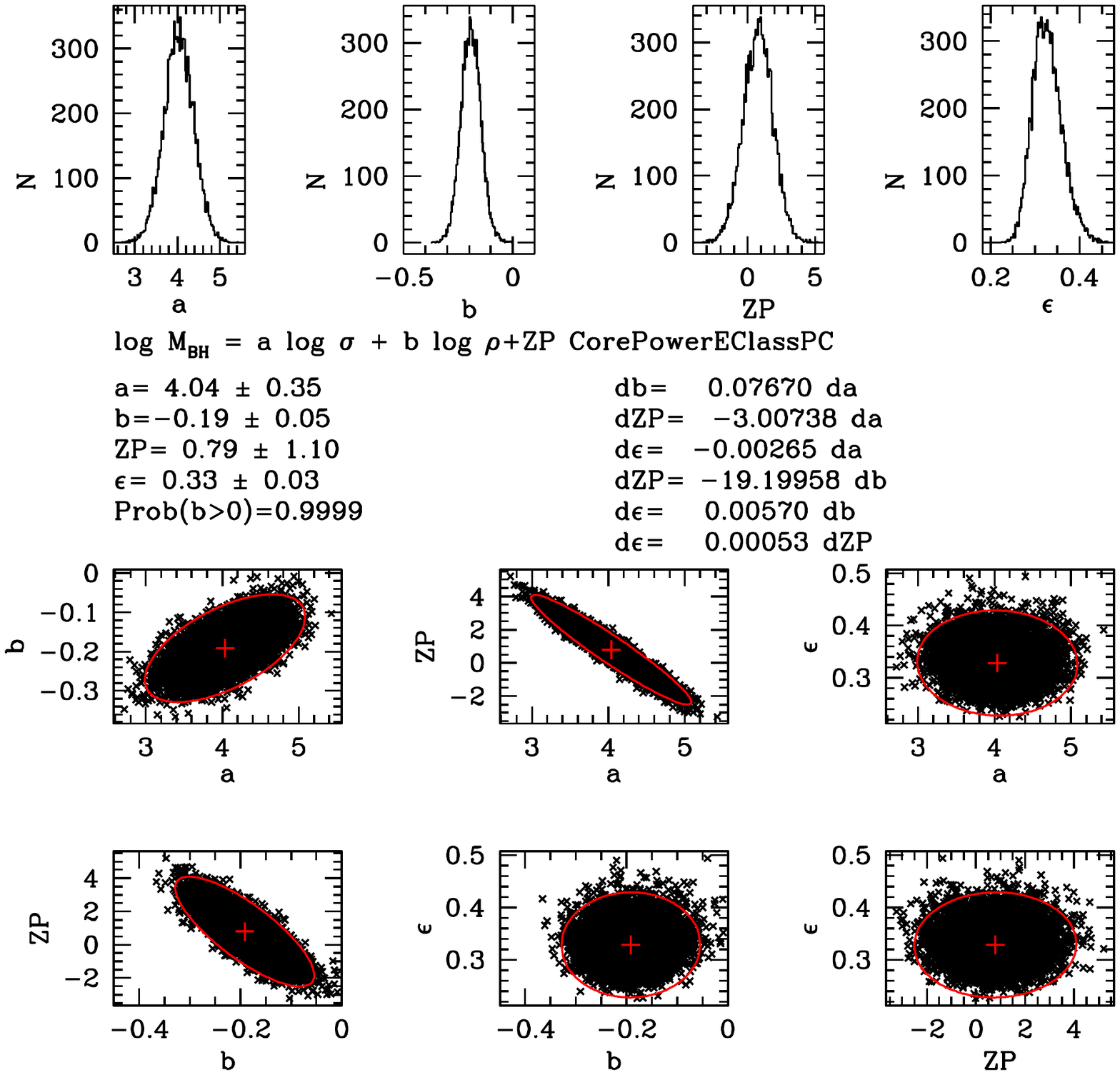}
 \end{center}
 \caption{Top row: the posterior probability distributions of the fit to the
   relation $\log M_{BH}=\alpha \sigma + \beta \rho+ZP$ for the subsample 
CorePowerEClassPC. Middle row: mean values of the parameters 
and errors, and correlations between parameter errors. The plots at
   the bottom  show the correlations between all possible pairs of
   parameters. The red ellipses show the
   $3\sigma$ contours. 
   \label{fig_ProbMBHSigrho}}
\end{figure*}

Kelly's routines provide the posterior probabilities 
  $P(\vec{\alpha},ZP,\epsilon)$ of the fitted parameters
$\vec{\alpha},ZP,\epsilon$. We quote as best-fit parameters the
averages of these distributions with errors given by the rms. We show
two examples of this procedure in Fig. \ref{fig_probMBHSig} and
\ref{fig_ProbMBHSigrho}.  Fig. \ref{fig_probMBHSig} shows the
one-dimensional case of the $\log M_{BH}=a \log \sigma +ZP$ relation
for the subsample CorePowerEClassPC (see below and Table
\ref{tab_subsamples}).  As expected, the errors in the slope and the
zero-point are highly correlated.

Fig. \ref{fig_ProbMBHSigrho} shows the two-dimensional case of the
$\log M_{BH}=a \log \sigma+b\log \rho_h+ZP$ relation discussed below,
again for the subsample CorePowerEClassPC. Strong error correlations
are present between the two fitted slopes and the zero-point. By
integrating the posterior probability for positive values of $b$
($P(b>0)=0.9999$) we show that we have detected this bivariate
correlation robustly.  Furthermore, to assess whether we are 
  overfitting the data by considering bivariate correlations involving
  black hole masses, we also compute the corrected Akaike information
  criterion \citep[][cAIC]{Akaike1973,HurvichTsai1989}.  It is defined
  as:
\begin{equation}
\label{cAIC}
cAIC(k) = AIC(k) +\frac{2k(k+1)}{n-k-1},
\end{equation}
where $n$ is the number of data points, $k$ is the number of free parameters 
($k=3$ for monovariate correlations and $k=4$ for bivariate correlations) and
$AIC$ is the Akaike information criterion:
\begin{equation}
\label{AIC}
AIC(k)= 2k-2ln P_{best}(k),
\end{equation}
where $P_{best}$ is the likelihood of the best-fitting solution with 
$k$ parameters. There is evidence for bivariate correlations if 
\begin{equation}
\label{eq_DcAIC}
\Delta cAIC=cAIC(k=4)-cAIC(k=3)<0.
\end{equation}
The relative probability of the two solutions
  is $RP=exp(\Delta cAIC/2)$, so the bivariate correlation is strongly
  preferred when $\Delta cAIC<-2$.  Equivalent conclusions are
  obtained by considering the Bayesian information criterion
  $BIC=-2\ln P_{best}+k\ln n$ \citep{Schwarz1978} and $\Delta
  BIC=BIC(k=4)-BIC(k=3)$.

  We implement this schema as follows.  We marginalize the posterior
  probability of the bivariate fits $P(a,b,ZP,\epsilon)$ over
  $(b,ZP,\epsilon)$ or $(a,ZP,\epsilon)$ to get the posterior
  distribution of $a$, $P_{k=4}(a)$ or $b$, $P_{k=4}(b)$,
  respectively. This is well approximated by a Gaussian,
  $P_{k=4}(a)=\frac{1}{\sqrt{2\pi}\delta
    a_{best}}exp[-\frac{(a-a_{best})^2}{2\delta a_{best}^2}]$ or
  $P_{k=4}(b)=\frac{1}{\sqrt{2\pi}\delta
    b_{best}}exp[-\frac{(b-b_{best})^2}{2\delta b_{best}^2}]$, see
  Fig. \ref{fig_ProbMBHSigrho}, where $a_{best}$ and $\delta
  a_{best}$, $b_{best}$ and $\delta b_{best}$ are given in the Tables
  \ref{tab_2dim}, \ref{tab_twokormendy} and \ref{tab_2dimstrict},
  where we omit the label $best$ for simplicity.  Therefore the
  probability of our best fitting bivariate solutions is
  $P_{best,k=4}=\frac{1}{\sqrt{2\pi}\delta a_{best}}$ or
  $P_{best,k=4}=\frac{1}{\sqrt{2\pi}\delta b_{best}}$. We compare this
  to the probability of one of the two possible best fitting
  monovariate (i.e., $k=3$) solutions for each combination of
  parameters we considered, having either $a=0$ or $b=0$.  We choose
  the one with the smallest intrinsic and measured scatter. In this
  way we are sure to get the most stringent test for the evidence of
  bivariate correlations.  For instance, in the case of the \mbus\
  correlation, this is \msig, which corresponds to the bivariate
  solution with $a=0$. In all other cases we consider the bivariate
  solution with $b=0$.

  In practice, we compute the marginalized posterior
  distribution $P_{k=4}(a,b)$, verify that the maximum of
  $P_{k=4}(a=0,b)$ is reached for $b=b_{best,k=3}$ for the \mbus\
  correlation and that the maximum of $P_{k=4}(a,b=0)$ is reached for
  $a=a_{best,k=3}$ for all the other correlations, where
  $a_{best,k=3}$ or $b_{best,k=3}$ are given in the Tables
  \ref{tab_1dim}, \ref{tab_kormendy} and  \ref{tab_1dimstrict},
  listing our monovariate (i.e., $k=3$) solutions 
(where again we dropped the label $best,k=3$ for simplicity). Then we set
  $P_{best,k=3}=P_{best,k=4}(a=0)=\frac{1}{\sqrt{2\pi}\delta
    a_{best}}exp[-(a_{best}/\delta a_{best})^2/2]$ for the \mbus\
  correlation or
  $P_{best,k=3}=P_{best,k=4}(b=0)=\frac{1}{\sqrt{2\pi}\delta
    b_{best}}exp[-(b_{best}/\delta b_{best})^2/2]$ for all the other
  correlations.

Finally we get:
\begin{equation}
\label{eq_lnPbestb}
2ln[P_{best}(k=4)/P_{best}(k=3)]=(b_{best}/\delta b_{best})^2
\end{equation}
or
\begin{equation}
\label{eq_lnPbesta}
2ln[P_{best}(k=4)/P_{best}(k=3)]=(a_{best}/\delta a_{best})^2
\end{equation}
for the the \mbus\ correlation or all the others, respectively.

Following the discussion given in the introduction, we consider 12
(sub-)samples. The first one (All) is made of the galaxies of
  Table \ref{tab_data}, with the exception of NGC 4486b. This galaxy
  deviates strongly from every correlation involving its black hole
  mass, despite the fact that our SINFONI BH mass (see Appendix B) is
  30\% smaller than the value adopted by \citet{KormendyHo2013}. As
  done by \citet{KormendyHo2013}, we discard it from all the subsamples and
  fits discussed below. The other subsamples are core ellipticals
  (CoreE); core and power-law ellipticals (CorePowerE); core and
  power-law ellipticals, plus classical bulges (CorePowerEClass); core
  and power-law ellipticals, classical bulges and classical bulge
  components of (composite) pseudo bulges (CorePowerEClassPC); the
  same for just SINFONI measurements (CorePowerEClassPCSINFONI); the
  same for just measurements from the literature
  (CorePowerEClassPCLit); the same without barred galaxies
  (CorePowerEClassnoBars); power-law ellipticals (PowerE); power-law
  ellipticals and classical bulges (PowerEClass); power-law
  ellipticals classical bulges and classical bulge components of
  pseudo (composite) bulges (PowerEClassPC); pseudo bulges
(Pseudo). Furthermore, we consider more or less stringent selection
criteria to include or exclude measurements of different quality.
Table \ref{tab_subsamples} describes how the subsamples are
constructed from Table \ref{tab_data} using the flags listed there.

With this set of subsamples we aim at assessing two questions. On the one
hand, we want to understand the influence of our SINFONI data set,
which provides almost 1/4 of the full database. On the other hand, we
want to explore the degree to which we can unify the different types
of galaxies in one common picture.  The Tables \ref{tab_onedimbulge}
to \ref{tab_2dim} report the results of the fits obtained for the
different families of objects.

\begin{deluxetable}{lll}
\tablecaption{The (sub-)samples considered in the correlation analysis. \label{tab_subsamples}}
\tabletypesize{\scriptsize}
\tablehead{
\colhead{Sample} & \colhead{Selection flags} & \colhead{Explanation}}
\startdata
All                      & $T\le 3$ and $b\le 1$ and $B\le 2$ & All galaxies without NGC 4486b\\
CoreE                    & $T=0$ & Core ellipticals\\       
CorePowerE               & $T\le 1$ & Core and power-law ellipticals without NGC 4486b\\
CorePowerEClass          & $T\le 2$ & Core and power-law ellipticals\\
                         &          & plus classical bulges without NGC 4486b\\
CorePowerEClassPC        & $T\le 2$ plus ($T=3$ and $B=2$)& Core and power-law ellipticals, \\
                         &                                & classical bulges \\
                         & & and classical component \\
                         & & of composite bulges without NGC 4486b\\
CorePowerEClassPCSINFONI & $T\le 2$ and $S=1$ plus ($T=3$ and $B=2$ and $S=1$)) & Core and power-law ellipticals,\\
                         & &  classical bulges and \\
                         & & classical component of \\
                         & & composite bulges, only \\
                         & & SINFONI measurements without NGC 4486b\\
CorePowerEClassPCLit     & $T\le 2$ and $S=0$ plus ($T=3$ and $B=2$ and $S=0$) & Core and power-law ellipticals, \\
                         & & classical bulges and\\
                         & & classical  component \\
                         & & of composite bulges,\\
                         & & only literature\\
                         & &  measurements\\
CorePowerEClassnoBars         & $T\le 2$ and $b\le 0.5$& Core and power-law ellipticals,\\
                         & &  classical bulges and\\
                         & &  classical component of \\
                         & & composite bulges but \\
                         & & without barred objects and without NGC 4486b\\
PowerE                   & $T=1$& Power-law ellipticals without NGC 4486b\\
PowerEClass              & $1\le T\le 2$& Power-law ellipticals\\
                         & &  and classical bulges without NGC 4486b\\
PowerEClassPC            & $1\le T\le 2$ plus ($T=3$ and $B=2$)& Power-law ellipticals\\
                         & & classical bulges and\\
                         & & classical component of \\
                         & & composite bulges  without NGC 4486b\\
Pseudo                   & $T=3$ and $B\ne 2$ & Pseudo bulges\\
\enddata
\tablenotetext{\ }{{\bf Flags} (see also Table \ref{tab_data}): 
T (type, 0 for core ellipticals, 1 for power-law ellipticals, 2 for classical 
bulges, 3 for pseudo bulges), b (0: no bar, 1: barred), 
B (0 for one-component galaxy, 1 for bulge plus disk galaxy, 2 for galaxy with 
composite (classical plus pseudo) bulge plus disk. }
\end{deluxetable}

\begin{figure*}
  \begin{center}
    \includegraphics[trim=0 4cm 0 4cm,clip,width=16cm]{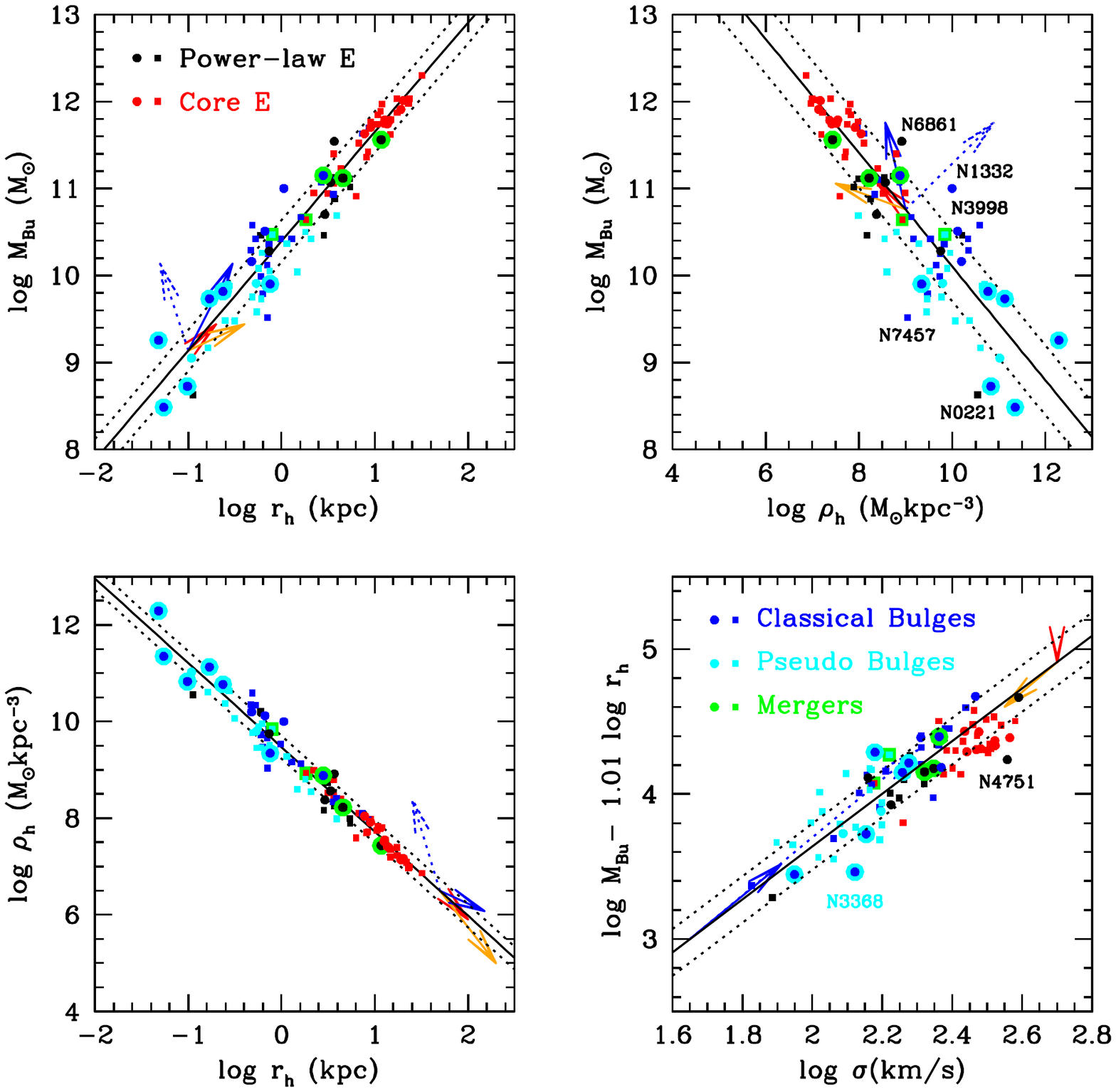}
 \end{center}
 \caption{The correlations between the bulge parameters $M_{Bu}$,
   $r_h$, $\rho_h$ and $\sigma$. Squares are data from the
   literature, circles are galaxies from our SINFONI survey.  We plot
   core galaxies in red, power-law ellipticals in black, classical
   bulges in blue, pseudo bulges in cyan. Mergers have a green
   annulus. Blue dots with a cyan annulus are the classical bulge
   components of composite bulges. The labels name particularly
   deviant galaxies.  The full lines indicate the best fit relations
   given in the Tables \ref{tab_onedimbulge} and \ref{tab_twodimbulge}
   for the CorePowerEClassPC sample. The dotted lines indicate the
   estimated intrinsic scatter. Arrows describe the effect of an equal
   mass dry merger (red), of a sequence of minor mergers doubling the
   bulge mass (orange), an equal-mass, gas-rich merger of two spiral
   galaxies with 20\% bulge mass and bulge-scales ratio $r_h^f/r_h^i$
   of 3 (blue) or 0.5 (dotted blue), see also Sect. \ref{discussion}
   and Table \ref{tab_scenarios}. As expected, the red arrow in the
   bottom right plot has zero length. \label{fig_phot}}
\end{figure*}

Fig. \ref{fig_phot} shows the correlations between the parameters
$M_{Bu}$, $r_h$ and $\rho_h$ for the galaxies of Table \ref{tab_data}
without NGC 4486b, see above.  The fourth plot presents
the virial relation between $M_{Bu}$, $r_h$ and velocity dispersions
$\sigma$ for the same sample.  There are no obvious outliers; the
galaxies NGC 1332, NGC 3998 and NGC 6861 have denser bulges than
expected given their bulge masses, and NGC 7457 and NGC 221 have less
dense bulges.

Bulge masses (see Table \ref{tab_onedimbulge}, sample All) scale as $M_{Bu}\sim
r_h^{1.3\pm 0.04}$ with estimated intrinsic scatter $\epsilon=0.24\pm0.02$ dex, 
or as $M_{Bu}\sim
\rho_h^{-0.7\pm 0.04}$ with $\epsilon=0.42\pm0.03$. Consistently,
bulge average densities scale as $\rho_h\sim r_h^{-1.72\pm
  0.04}$. These correlations hold within the errors for all subsamples
considered, with no appreciable differences between the literature and
the SINFONI sample. The
correlations for core ellipticals are tighter. 
Only pseudo bulges follow relations which are different at
the $2-3\sigma$ level ($M_{Bu}\sim r_h^{1.02\pm 0.16}$, $M_{Bu}\sim
\rho_h^{-0.4\pm 0.1}$, $\rho_h\sim r_h^{-1.96\pm 0.16}$). 

For pressure-supported, self-gravitating systems in virial equilibrium 
dynamical masses are expected to scale
as:
\begin{equation}
\label{eq_mdyn}
M=10^{6.064} (R_e/kpc) (\sigma/km s^{-1})^2M_\odot
\end{equation}
\citep{Cappellari2006}.  Using spherical half-luminosity radii we find
\begin{equation}
\label{eq_mdynbulges}
M_{Bu}=10^{6.67\pm0.38}r_h^{0.98\pm 0.04}\sigma^{1.65\pm0.17},
\end{equation}
with 40\% scatter (see Table \ref{tab_twodimbulge}, sample
CorePowerEClassPC).  The coefficients are similar within $1-2\sigma$
for all the subsamples which exclude pseudo bulges. In particular,
there is no significant difference if we use just the SINFONI or just
the literature sample.  For pseudo bulges the dynamical masses are
roughly factor 3 smaller for given $r_h$ and $\sigma$ with larger
scatter.  Core ellipticals follow a tighter relation with just 30\%
intrinsic scatter and a shallower dependence on velocity dispersion
($M_{Bu}=10^{7.96\pm0.99}r_h^{1.03\pm 0.13}\sigma^{1.09\pm0.44}$). The
slope difference is significant only at the $2\sigma$ level and could
point to the systematic uncertainties in the role of dark matter, as
discussed in \citet{Thomas2011}. Discussing in detail the origin
  of the differences between Eqs. 
  \ref{eq_mdyn} and \ref{eq_mdynbulges} goes beyond the scope of the paper.

When we augment the errors on the bulge mass and density by 0.15 dex
(see Sect. \ref{sec_errors}), the fitted parameters remain essentially
the same with somewhat larger errors, except for the estimated
intrinsic scatter, which is approximately reduced by 0.15 dex in
quadrature and therefore statistically compatible with zero. 

\begin{deluxetable}{lllllllllll}
\tablewidth{0pt}
\tabletypesize{\scriptsize}
\tablecaption{One-dimensional correlations within bulge parameters. \label{tab_onedimbulge}}
\tablehead{
\colhead{Fit} & \colhead{Sample} & \colhead{N} & \colhead{a} & \colhead{da} & \colhead{ZP} & \colhead{dZP} & \colhead{$\epsilon$} & \colhead{$d\epsilon$} & \colhead{rms}
}
\startdata
   \mbulger\           &         All           &          96           &       1.285           &       0.041           &       10.35           &       0.029           &       0.241           &        0.02           &       0.251          \\
                       &       CoreE           &          31           &       1.246           &       0.104           &       10.41           &       0.101           &       0.131           &       0.028           &       0.151          \\
                       &  CorePowerE           &          47           &        1.37           &       0.072           &       10.27           &       0.062           &       0.186           &       0.026           &         0.2          \\
                       & CorePowerEClass           &          71           &       1.245           &       0.055           &        10.4           &       0.041           &       0.228           &       0.022           &       0.232          \\
                       & CorePowerEClassPC           &          77           &       1.256           &       0.046           &       10.39           &       0.034           &       0.238           &       0.022           &       0.245          \\
                       & CorePowerEClassnoBars           &          61           &       1.261           &       0.065           &       10.39           &       0.052           &       0.231           &       0.025           &       0.228          \\
                       & CorePowerEClassPCSINFONI           &          22           &        1.24           &       0.094           &       10.43           &       0.078           &       0.268           &        0.05           &       0.232          \\
                       & CorePowerEClassPCLit           &          57           &       1.261           &       0.056           &       10.38           &        0.04           &       0.236           &       0.026           &       0.245          \\
                       &      PowerE           &          16           &       1.348           &       0.174           &       10.25           &       0.108           &       0.302           &       0.073           &        0.26          \\
                       & PowerEClass           &          40           &       1.204           &       0.108           &        10.4           &       0.053           &       0.287           &       0.037           &       0.278          \\
                       & PowerEClassPC           &          46           &       1.234           &       0.081           &       10.39           &       0.045           &       0.294           &       0.035           &       0.291          \\
                       &      Pseudo           &          19           &       1.018           &       0.156           &       10.16           &       0.069           &       0.234           &        0.05           &       0.209          \\
      \hline
 \mbulgerho\           &         All           &          96           &      -0.678           &        0.04           &       16.78           &       0.363           &       0.414           &       0.033           &       0.423          \\
                       &       CoreE           &          31           &       -0.63           &       0.101           &       16.53           &        0.79           &       0.219           &       0.044           &       0.245          \\
                       &  CorePowerE           &          47           &      -0.743           &       0.078           &       17.35           &       0.637           &       0.332           &       0.043           &       0.347          \\
                       & CorePowerEClass           &          71           &      -0.626           &       0.051           &       16.42           &       0.442           &       0.379           &       0.036           &       0.379          \\
                       & CorePowerEClassPC           &          77           &      -0.654           &       0.044           &       16.64           &       0.389           &       0.401           &       0.036           &       0.405          \\
                       & CorePowerEClassnoBars           &          61           &      -0.624           &       0.061           &       16.41           &       0.511           &       0.386           &        0.04           &       0.375          \\
                       & CorePowerEClassPCSINFONI           &          22           &      -0.643           &       0.087           &        16.6           &       0.751           &       0.443           &       0.082           &       0.386          \\
                       & CorePowerEClassPCLit           &          57           &      -0.655           &       0.053           &       16.63           &       0.466           &       0.401           &       0.043           &       0.404          \\
                       &      PowerE           &          16           &      -0.666           &       0.171           &       16.57           &        1.48           &       0.499           &        0.12           &       0.442          \\
                       & PowerEClass           &          40           &      -0.514           &       0.089           &       15.34           &       0.812           &        0.45           &       0.057           &       0.434          \\
                       & PowerEClassPC           &          46           &      -0.582           &       0.074           &       15.91           &       0.696           &       0.479           &       0.055           &       0.468          \\
                       &      Pseudo           &          19           &       -0.42           &       0.113           &          14           &       1.098           &       0.345           &       0.075           &       0.307          \\
      \hline
      \rhor\           &         All           &          96           &      -1.716           &       0.041           &       9.428           &       0.029           &       0.242           &        0.02           &       0.251          \\
                       &       CoreE           &          31           &      -1.756           &       0.105           &        9.49           &       0.104           &       0.133           &       0.027           &       0.151          \\
                       &  CorePowerE           &          47           &      -1.634           &       0.072           &       9.352           &       0.063           &       0.187           &       0.027           &       0.199          \\
                       & CorePowerEClass           &          71           &      -1.758           &       0.056           &        9.48           &       0.041           &       0.229           &       0.022           &       0.231          \\
                       & CorePowerEClassPC           &          77           &      -1.745           &       0.047           &       9.468           &       0.035           &       0.238           &       0.022           &       0.244          \\
                       & CorePowerEClassnoBars           &          61           &      -1.743           &       0.065           &       9.466           &       0.052           &       0.231           &       0.024           &       0.227          \\
                       & CorePowerEClassPCSINFONI           &          22           &      -1.758           &       0.093           &       9.503           &       0.077           &       0.264           &       0.049           &       0.231          \\
                       & CorePowerEClassPCLit           &          57           &      -1.737           &       0.055           &       9.457           &        0.04           &       0.236           &       0.026           &       0.245          \\
                       &      PowerE           &          16           &      -1.662           &       0.177           &       9.336           &       0.111           &       0.306           &       0.074           &       0.259          \\
                       & PowerEClass           &          40           &      -1.795           &       0.106           &       9.482           &       0.052           &       0.286           &       0.037           &       0.277          \\
                       & PowerEClassPC           &          46           &      -1.764           &       0.082           &       9.468           &       0.045           &       0.294           &       0.035           &        0.29          \\
                       &      Pseudo           &          19           &      -1.959           &       0.156           &       9.249           &       0.069           &       0.233           &       0.051           &        0.21          \\

\enddata
\tablenotetext{\ }{Column 1: Fit type; Column 2: Sample type, see
Table  \ref{tab_subsamples}; Column 3: number of data points; Column 4 and
  5: slope of the correlation and its error; Column 6 and 7:
  zero-point of the correlation and its errors; Column 8 and 9:
  intrinsic scatter and its errors; Column 10: measured scatter.}
\end{deluxetable}

\begin{deluxetable}{llllllllllll}
\tablewidth{0pt}
\tabletypesize{\scriptsize}
\tablecaption{The two-dimensional correlation $\log M_{Bu}=a \log r_h +b\log \sigma+ZP$ for different samples.\label{tab_twodimbulge}}
\tablehead{
\colhead{Fit} & \colhead{Sample} & \colhead{a} & \colhead{da} & \colhead{b} & \colhead{db} &\colhead{ZP} & \colhead{dZP} & \colhead{$\epsilon$} & \colhead{$d\epsilon$} &\colhead{rms} & \colhead{$P(b\neq 0)$}
}
\startdata
      \mburs           &         All           &       0.974           &       0.039           &       1.638           &       0.148           &       6.699           &        0.33           &       0.146           &       0.013           &       0.157           &      0.9999          \\
                       &       CoreE           &       1.033           &       0.133           &       1.091           &       0.443           &       7.957           &       0.993           &       0.117           &       0.022           &       0.119           &      0.9923          \\
                       &  CorePowerE           &       1.024           &       0.078           &       1.487           &       0.245           &       6.991           &        0.54           &       0.126           &       0.017           &       0.128           &      0.9999          \\
                       & CorePowerEClass           &        0.94           &       0.047           &       1.633           &       0.167           &       6.735           &       0.375           &       0.136           &       0.014           &       0.136           &      0.9999          \\
                       & CorePowerEClassPC           &       0.978           &       0.042           &        1.65           &        0.17           &       6.666           &       0.383                  &       0.146           &       0.014           &        0.15           &      0.9999          \\
                       & CorePowerEClassnoBars           &       0.911           &       0.054           &       1.721           &       0.177           &       6.546           &       0.395                  &       0.133           &       0.015           &       0.131           &      0.9999          \\
                       & CorePowerEClassPCSINFONI           &        0.93           &       0.084           &       1.814           &        0.35           &       6.271           &       0.803                  &       0.161           &       0.032           &       0.138           &      0.9999          \\
                       & CorePowerEClassPCLit           &       0.993           &       0.051           &       1.645           &       0.202           &       6.687           &       0.454                  &       0.147           &       0.017           &       0.152           &      0.9999          \\
                       &      PowerE           &       0.996           &        0.13           &       1.671           &       0.382           &       6.575           &       0.841           &        0.17           &       0.043           &       0.136           &      0.9995          \\
                       & PowerEClass           &       0.966           &       0.067           &        1.76           &       0.195           &       6.454           &       0.439           &        0.15           &        0.02           &       0.143           &      0.9999          \\
                       & PowerEClassPC           &       1.012           &       0.055           &       1.818           &         0.2           &       6.301           &        0.45           &       0.161           &        0.02           &        0.16           &      0.9999          \\
                       &      Pseudo           &       0.975           &       0.132           &       1.472           &       0.604           &       7.066           &       1.274           &       0.185           &       0.045           &        0.18           &       0.988          \\

\enddata
\tablenotetext{\ }{Column 1: Fit type; Column 2: Sample type, see
Table  \ref{tab_subsamples}; Column 3 and 4: first variable slope of the
  correlation and its error; Column 5 and 6: second variable slope of
  the correlation and its error; Column 7 and 8: zero-point of the
  correlation and its errors; Column 9 and 10: intrinsic scatter and
  its errors; Column 11: measured scatter; Column 12: probability of
  the bivariate correlation (see text).}
\end{deluxetable}

\section{BH Correlation analysis}
\label{correlations}

\begin{figure*}
  \begin{center}
    \includegraphics[trim=0 0cm 0 6cm,clip,width=16cm]{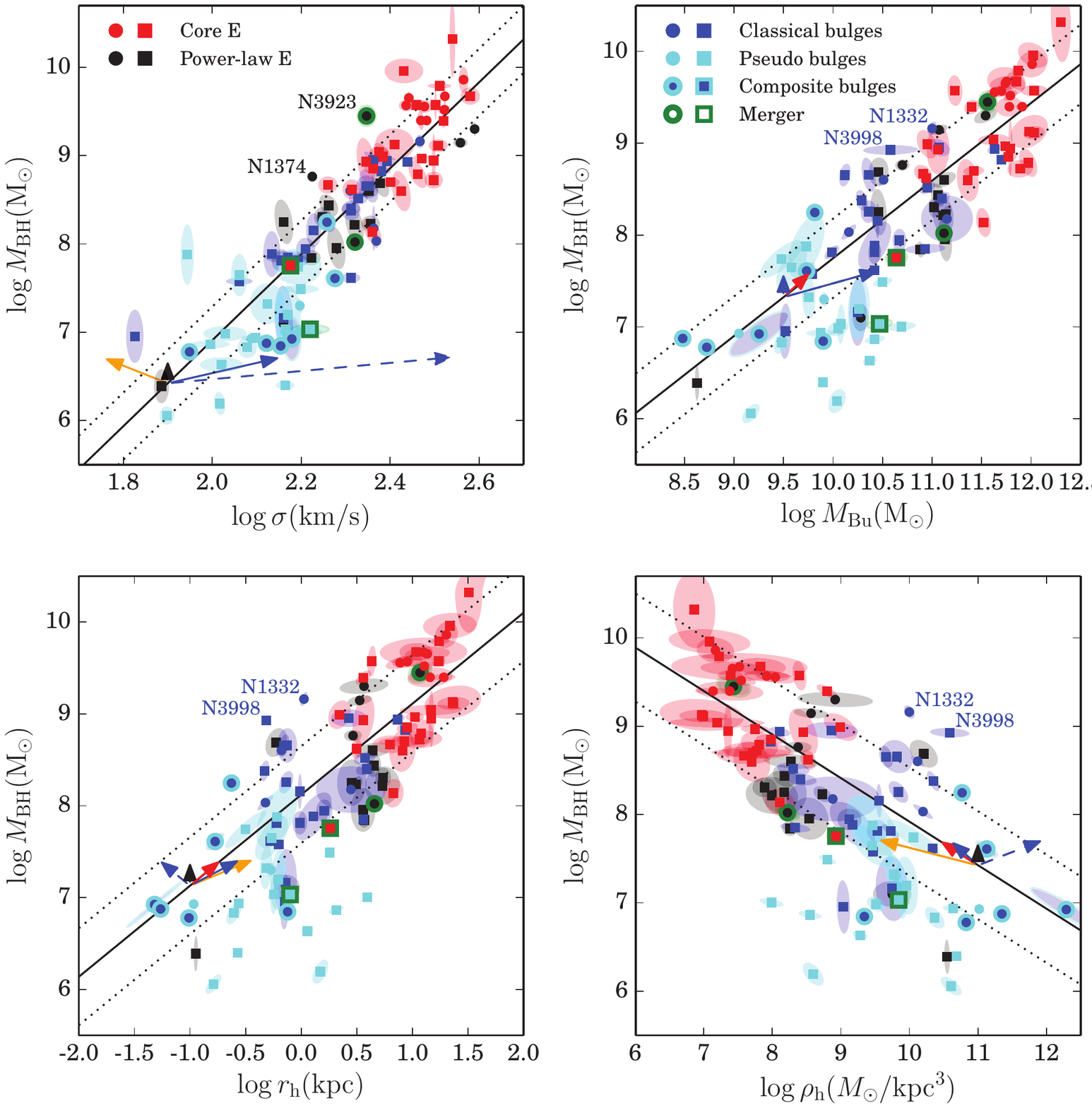}
  \end{center}
  \caption{ Top: the \msig\ (left) and \mbu\ (right)
    relations. Bottom: the \mr\ (left) and \mrho\ (right) relations.
   Symbols as in Fig. \ref{fig_phot}. The ellipses show the 1$\sigma$ errors. 
   The labels name particularly deviant galaxies.
    The full lines indicate the relations given in Tab. \ref{tab_1dim} for 
    the CorePowerEClassPC sample. The dotted lines indicate the 
estimated intrinsic
    scatter. Arrows describe the effect of an equal mass dry merger
    (red), of a sequence of minor mergers doubling the bulge mass
    (orange), an equal-mass, gas-rich merger of two spiral galaxies
    with 20\% bulge mass with bulge-scales ratio $r_h^f/r_h^i$ 3 (blue) or 0.5 
    (dotted blue), and doubling the BH mass through
    accretion or BH merging (black), see Sect. \ref{discussion} and 
    Table \ref{tab_scenarios}.  
    \label{fig_onedim}}
\end{figure*}

We now proceed to examine correlations involving the BH mass.
We start by investigating four one-dimensional
correlations of the type $y=ax+ZP+N(0,\epsilon)$:
\begin{equation}
\label{eq_MBHSig}
M_{BH}-\sigma: \log M_{BH}= a \log \sigma+ ZP,
\end{equation}
\begin{equation}
\label{eq_MBHMBu}
M_{BH}-M_{Bu}: \log M_{BH}= a \log M_{Bu}+ ZP,
\end{equation}
\begin{equation}
\label{eq_MBHRhalf}
M_{BH}-r_h: \log M_{BH}= a \log r_h+ ZP,
\end{equation}
\begin{equation}
\label{eq_MBHrho}
M_{BH}-\rho_h: \log M_{BH}= a \log \rho_h+ ZP.
\end{equation}
Table \ref{tab_1dim} summarizes the results, giving the number of
galaxies in each subsample, the best-fit values of $a$ and $ZP$ with
errors, the estimated intrinsic scatter $\epsilon$, and the measured
scatter $rms$. In addition, we list the Spearman coefficient $r_S$ and
the probability $P_{r_s}$ of its value being greater caused by chance.
Fig. \ref{fig_onedim} presents the correlation plots.

In agreement with the literature, we find that the strongest
correlations with the lowest measured and intrinsic scatter are with
$\sigma$.  The correlations with $M_{Bu}$ are strong, except for the
pseudo bulge subsample. Correlations with sizes or anti-correlations
with densities are generally weaker (and non-existent for pseudo
bulges), but still robust for several subclasses. The measured
slopes of the correlations of the Eqs. \ref{eq_MBHSig} to \ref{eq_MBHrho} are
steeper for the subsamples including core ellipticals. In contrast,
all correlations for the subsamples excluding core ellipticals and
pseudo bulges are similar within the errors. BH masses in pseudo
bulges correlate (weakly) only with $\sigma$, and this with a flatter 
slope than for the other samples.

As noticed before, when we augment the errors on the bulge mass and
density by 0.15 dex (see Sect. \ref{sec_errors}), the fitted
parameters of the \mbu\ and \mrho\ correlations remain essentially the
same with somewhat larger errors; but the estimated intrinsic scatter
is approximately reduced by 0.15 dex in quadrature.

This confirms the results reported in the literature. In particular:

\begin{itemize}

\item[a)] the slopes of the \msig\ and \mbu\ relations agree with published results 
within the  quoted errors;
\item[b)] core ellipticals have more massive BHs than other classical bulges, 
at a given $\sigma$ or bulge mass, when correlations derived using samples including
non-core galaxies are adopted; moreover, the smallest intrinsic and measured scatter  
of the \msig\ and \mbu\ relations are measured for the sample of core ellipticals; 
\item[c)]  power-law early type galaxies
and classical bulges follow similar \msig\ and \mbu\ relations; 
\item[d)] pseudo bulges have smaller BH masses than the rest of the
sample at a given $\sigma$ or $M_{Bu}$. 
\end{itemize}

Our sample indicates that the \msig\ relation is possibly a better
predictor (i.e. with lower intrinsic scatter) of \mbh\ than the \mbu\
relation; this remains true even when we augment the errors on the
bulge mass by 0.15 dex in quadrature. However, the difference is not
statistically significant.  \citet{KormendyHo2013} argue on the basis
of their KH45 sample  that both
relations are equivalent (see Table \ref{tab_kormendy} and discussion
below).

\citet{Graham2013} claim that ``Sersic galaxies'' follow a quadratic
\mbu\ relation.  We disagree with this interpretation: {\it classical}
bulges (all with a Sersic profile) and Core-Sersic galaxies follow the
same linear \mbu\ relations within the errors. The steepening of the
relation at low bulge masses for ``Sersic galaxies'' seen by
\citet{Graham2013} stems from their possibly uncertain bulge masses
(see Fig. \ref{fig_MbulgeScott}) and the fact that they do not
distinguish between pseudo and classical bulges. We also disagree with
their interpretation of the role of barred galaxies: considering
barred galaxies with classical bulges delivers the same \msig\ and
\mbu\ relations within the errors that we derive for non-barred
classical bulges and early-type galaxies. Again, it is the pseudo
bulges, not the barred classical bulges, that deviate from the \msig\
and \mbu\ relations.

The intrinsic and measured scatter of our  \msig\ and \mbu\ relations
are generally larger than the values quoted in previous studies,
although the coefficients of the relations are compatible within the
errors. This stems from the sample of objects that comes from our
SINFONI survey, where we deliberately observed objects with extreme
properties (i.e., objects with small or large velocity dispersions, 
particularly compact objects, or merger remnants).

\begin{deluxetable}{llllllllllll}
\tablewidth{0pt}
\tabletypesize{\tiny}
\tablecaption{One-dimensional correlations with black hole masses. \label{tab_1dim}}
\tablehead{
\colhead{Fit} & \colhead{Sample} & \colhead{N} & \colhead{a} & \colhead{da} & \colhead{ZP} & \colhead{dZP} & \colhead{$\epsilon$} & \colhead{$d\epsilon$} & \colhead{rms} & \colhead{$r_S$}& \colhead{$P(r_S)$}
}
\startdata
      \msig\           &         All           &          96           &       5.246           &       0.274           &       -3.77           &       0.631           &       0.417           &       0.037           &       0.459           &      0.9109           &    6.59e-38          \\
                       &       CoreE           &          31           &       4.772           &       0.794           &      -2.476           &       1.942           &       0.332           &        0.06           &       0.366           &      0.6936           &    1.52e-05          \\
                       &  CorePowerE           &          47           &       4.713           &       0.458           &      -2.393           &       1.099           &        0.38           &        0.05           &       0.399           &      0.8005           &    1.45e-11          \\
                       & CorePowerEClass           &          71           &       4.546           &        0.33           &       -2.03           &        0.78           &       0.348           &       0.037           &        0.38           &      0.8677           &    1.24e-22          \\
                       & CorePowerEClassPC           &          77           &       4.868           &        0.32           &      -2.827           &        0.75           &        0.38           &       0.038           &       0.409           &       0.883           &    2.42e-26          \\
                       & CorePowerEClassPCSINFONI           &          22           &       4.976           &        0.72           &      -3.025           &        1.72           &       0.502           &       0.093           &       0.446           &      0.7134           &    0.000194          \\
\enddata
\tablenotetext{\ }{Column 1: Fit type; Column 2: Sample type, see
Table  \ref{tab_subsamples}; Column 3: number of data points; Column 4 and
  5: slope of the correlation and its error; Column 6 and 7:
  zero-point of the correlation and its errors; Column 8 and 9:
  intrinsic scatter and its errors; Column 10: measured scatter;
  Column 11 and 12: Spearman coefficient and its probability.}
\end{deluxetable}

Finally, we explicitly show that galaxy sizes and densities correlate
with BH masses too, although with larger scatter. This and the
discrepant results about the existence of the ``BH Fundamental Plane''
quoted in Sect. \ref{intro} motivate our next step, which is to
investigate five two-parameter correlations of the type
$z=ax+by+ZP+N(0,\epsilon)$:

\begin{equation}
\label{eq_MBHSigrho}
M_{BH}-\sigma-\rho_h: \log M_{BH}= a \log \sigma+ b \log \rho_h+ZP,
\end{equation}
\begin{equation}
\label{eq_MBHMBurho}
M_{BH}-M_{Bu}-\rho_h:\log M_{BH}= a \log M_{Bu}+ b \log \rho_h + ZP,
\end{equation}
\begin{equation}
\label{eq_MBHSigRhalf}
M_{BH}-\sigma-r_h: \log M_{BH}= a \log \sigma + b \log r_h+ ZP,
\end{equation}
\begin{equation}
\label{eq_MBHMBuRhalf}
M_{BH}-M_{Bu}-r_h: \log M_{BH}= a \log M_{Bu} + b \log r_h+ ZP.
\end{equation}
\begin{equation}
\label{eq_MBHSigMBu}
M_{BH}-M_{Bu}-\sigma: \log M_{BH}= a \log M_{Bu}+b \log \sigma+ ZP,
\end{equation}

Table \ref{tab_2dim} gives the results for the 12 (sub-)samples
considered above: the values of the best-fit parameters $a$, $b$, and
$ZP$ with their errors of the correlations of the type $z=a x +b y
+ZP$, the intrinsic and measured scatter $\epsilon$ and $rms$, the
probability of the bivariate correlation $P(b\neq 0)$ of $b\le0$ (if
$b_{fit}>0$), or of $b\ge0$ (if $b_{fit}<0$),  the value of $\Delta
cAIC$ and of $RP=\exp(\Delta cAIC/2)$ (the relative probability of the
mono- and bivariate solutions), the Spearman
correlation coefficient of the residuals $z-ax$ with $y$, and the
probability of the latter's value being greater due to chance.

\begin{figure*}
  \begin{center}
\includegraphics[trim=0cm 6cm 0cm 3cm,clip,width=16cm]{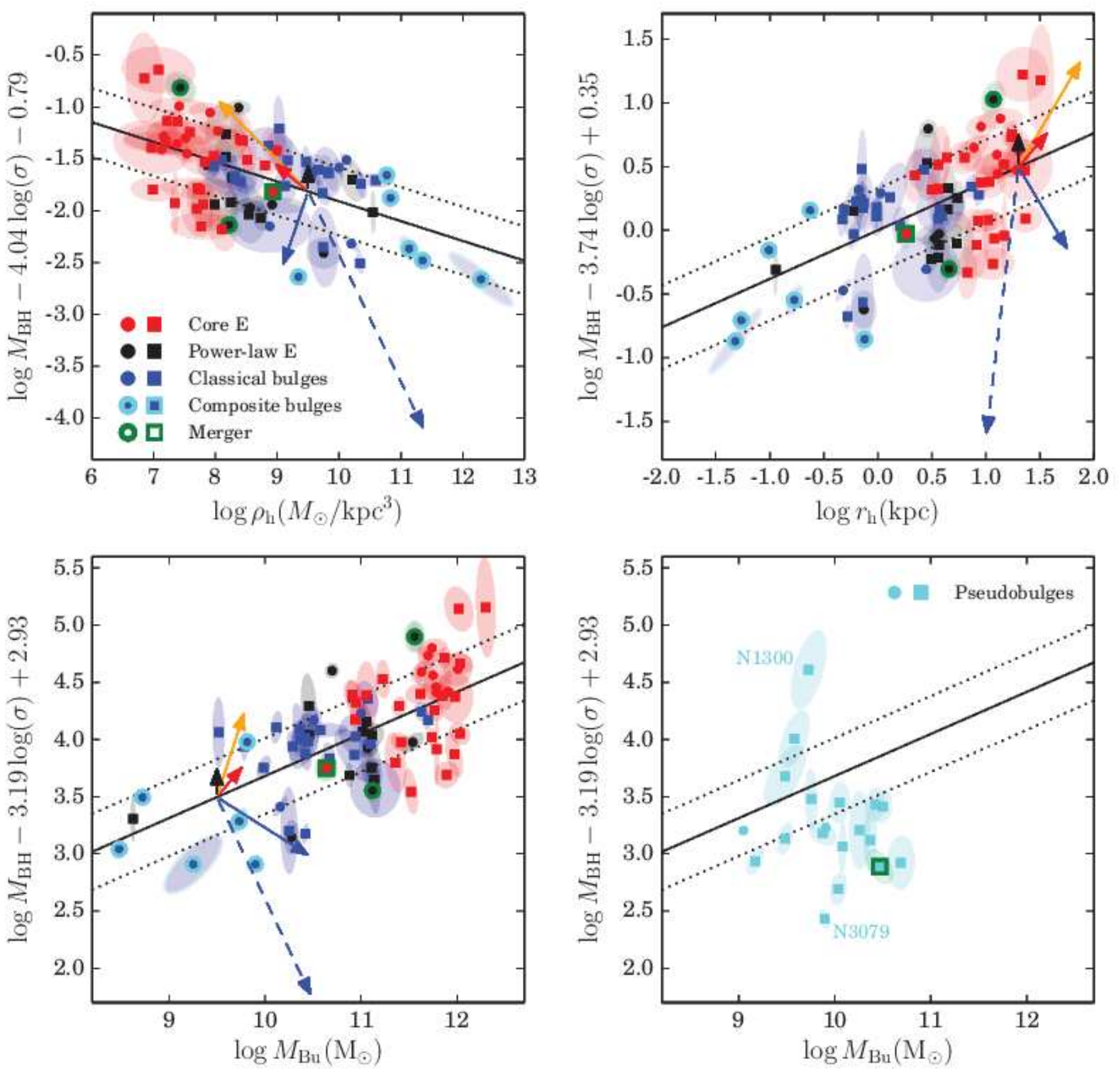}
  \end{center}
  \caption{Illustration of the bivariate correlations \msigrho\ (top
    left), \msigr (top right) and \mbus\ (bottom left and right).  In
    each panel we plot the corresponding best-fit relations for the
    CorePowerEClassPC subsample using full lines, with dotted lines showing
    the estimated intrinsic scatter. Galaxy data points from the various
    subsamples are plotted with different colors: core galaxies in
    red, power-law ellipticals in black, classical bulges in blue,
    pseudo bulges in cyan. Mergers have a green annulus.  Blue dots
    with a cyan annulus are the classical bulge components of
    (composite) pseudo bulges.  Squares indicate data from the
    literature and circles data from our SINFONI survey. The ellipses 
    show the 1$\sigma$ errors.  Arrows describe the
    effect of an equal mass dry merger (red), of a sequence of minor
    mergers which double the bulge mass (orange), an equal-mass,
    gas-rich merger of two spiral galaxies with 20\% bulge mass and
    bulge-scales ratio $r_h^f/r_h^i$ of 3 (blue) or 0.5 (dotted blue),
    and doubling the BH mass through accretion or BH merging (black),
    see also Sect. \ref{discussion} and Table
    \ref{tab_scenarios}. \label{fig_Sigma}}
\end{figure*}

The correlations involving $\sigma$ (\msigrho,\msigr,\mbus) are 
established with high confidence for all subsamples
except core ellipticals, power-law ellipticals, and pseudo bulges 
(each of these classes have $\approx 30$ or less objects). 
In particular, for the
sample CorePowerEClassPC (the 77 galaxies that are not
pseudo bulges) the relations:
\begin{equation}
\label{eq_MBHSigrhonoPseudo}
\begin{array}{ll}
\log M_{BH}=& (4.04\pm 0.35) \log \sigma-(0.19\pm 0.05)\log\rho_h \\
& + (0.79\pm 1.11), \\
\end{array}
\end{equation}
\begin{equation}
\label{eq_MBHSigRhalfnoPseudo}
\begin{array}{ll}
\log M_{BH}=&(3.74\pm 0.40) \log \sigma+(0.38\pm 0.09)\log r_h\\
           & - (0.35\pm 0.90)\\ 
\end{array}
\end{equation}
and 
\begin{equation}
\label{eq_MBHSigMbunoPseudo}
\begin{array}{ll}
\log M_{BH}=& (0.37\pm 0.09)M_{Bu}+(3.19\pm 0.52) \log \sigma \\
           &- (2.93\pm 0.66)\\ 
\end{array}
\end{equation}
have a measured/intrinsic scatter of just 0.36/0.33 dex, 0.05 dex less
than the respective one-dimensional solution \msig, a probability
$P(b\ne0)>0.99$, very negative $\Delta cAIC$ values ($<-13$, or 
relative probabilities of the mono- and bivariate solutions
less than 0.001) and a strong Spearman coefficient value ($|r_S|>0.5$
with $P(r_S)<10^{-6}$) of the residuals $z-ax$ with the $y$ variable.
We illustrate the three correlations in the top and bottom left panels
of Fig. \ref{fig_Sigma}.

\begin{figure*}
  \begin{center} 
\includegraphics[trim=0 4cm 0 4cm,clip,width=15cm]{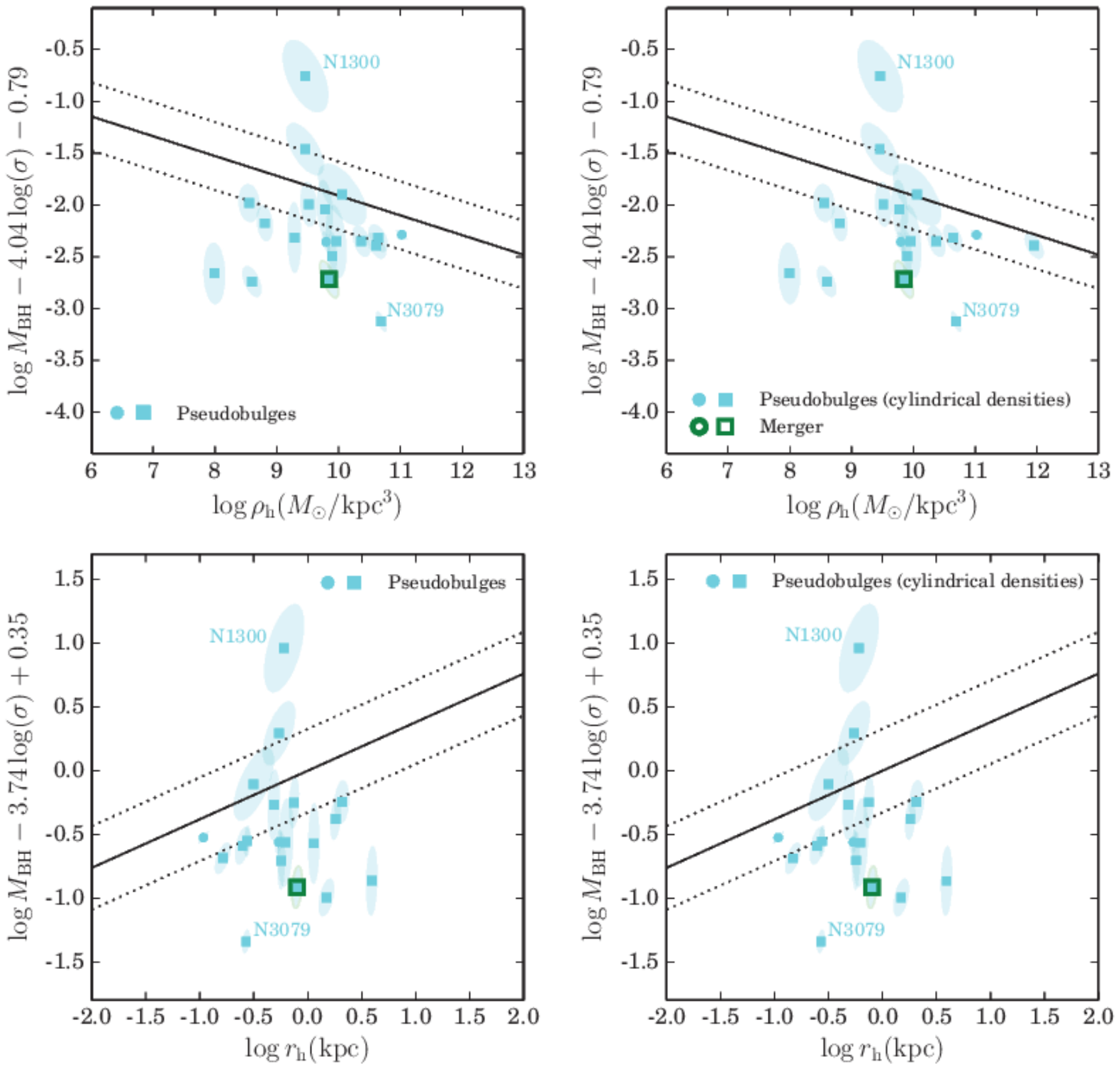}\\
  \end{center}
  \caption{Illustration of the bivariate correlations \msigrho\ (top) and
    \msigr\ (bottom) for
    pseudo bulges (cyan points). Squares indicate data from the
    literature and circles data from our SINFONI survey. The ellipses 
    show the 1$\sigma$ errors. Lines and arrows as in
    Fig. \ref{fig_Sigma}. The left plot is for spherical densities
    and radii (see Eq. \ref{eq_rho}), the plot to the right for cylindrical
    densities and radii with $h_z=0.2a_e/1.67$ 
    (see Eq. \ref{eq_pseudodensity}). 
\label{fig_Sigmapseudo}}
\end{figure*}

As in the single variable correlations presented in
Fig. \ref{fig_onedim}, the subsample of pseudo bulges (see
Fig. \ref{fig_Sigma}, bottom right, and Fig. \ref{fig_Sigmapseudo},
top and bottom left) has smaller black hole masses for a
given velocity dispersion. The smallest offsets are observed at the
largest densities and smallest scale-lengths, and pseudo bulges smaller
than 1 kpc, or denser than $10^{10} M_\odot/kpc^3$ roughly follow the
correlations defined by the other subsamples. As a consequence, it is
not possible to derive a tight bivariate correlation that
simultaneously describes the behavior of core, power-law ellipticals,
classical and pseudo bulges. If we fit the sample All, we get a
steeper $\sigma$ coefficient and a measured and intrinsic scatter
larger by 0.06-0.07 dex than for the subsample without pseudo
bulges. Nevertheless, the bivariate correlation remains highly
significant ($P(b\ne0)>0.99$) and the Spearman coefficient value of the
correlation of the residuals $z-ax$ with the $y$ variable is strong 
($|r_S|>0.49$, $P(r_S)<4\times 10^{-7}$).

The right-hand plots of Fig. \ref{eq_pseudodensity} illustrate what happens
if we use the cylindrical densities and radii for pseudo bulges that
we calculate from Eq. \ref{fig_Sigmapseudo} and
$h_z=0.2a_e/1.67$. Since densities are increased and scalelengths do
not change much, only the density threshold above which pseudo bulges
follow the scaling relations of the other subsamples changes, to about
$10^{11}M_\odot/kpc^3$.

The evidence for the correlations \mburho\ and \mbur\ is weaker but
still convincing.  Here the bivariate correlations derived for the
CorePowerEClassPC sample are significant at the 98\% level, the
$\Delta cAIC$ values are negative ($<-2.3$) with relative probability
of the mono- and bivariate solutions less than 0.32, with large
Spearman correlation coefficients ($|r_S|>0.45$) and low $P(r_S)$
probabilities ($P(r_S)<4\times 10^{-5}$ for the residuals $z-ay$, but
only a marginal reduction (by 0.01 dex) of the measured scatter
compared to the monovariate correlations is achieved. We derive the
following relations:
\begin{equation}
\label{eq_MBHMburhonoPseudo}
\begin{array}{ll}
\log M_{BH}=&(1.11\pm 0.14) \log M_{Bu}+(0.23\pm 0.11)\log\rho_h \\
           &- (5.52\pm 2.42),\\
\end{array}
\end{equation}
\begin{equation}
\label{eq_MBHMbuRhalfnoPseudo}
\begin{array}{ll}
\log M_{BH}=&(1.33\pm 0.24) \log M_{Bu}-(0.68\pm 0.32)\log r_h\\
           & - (5.74\pm 2.53).\\
\end{array}
\end{equation}
Stronger bivariate correlations are obtained when excluding Core
ellipticals (the subsample PowerEClassPC), see discussion below.

\begin{figure*}
  \begin{center}
    \includegraphics[trim=0 4cm 0 4cm,clip,width=16cm]{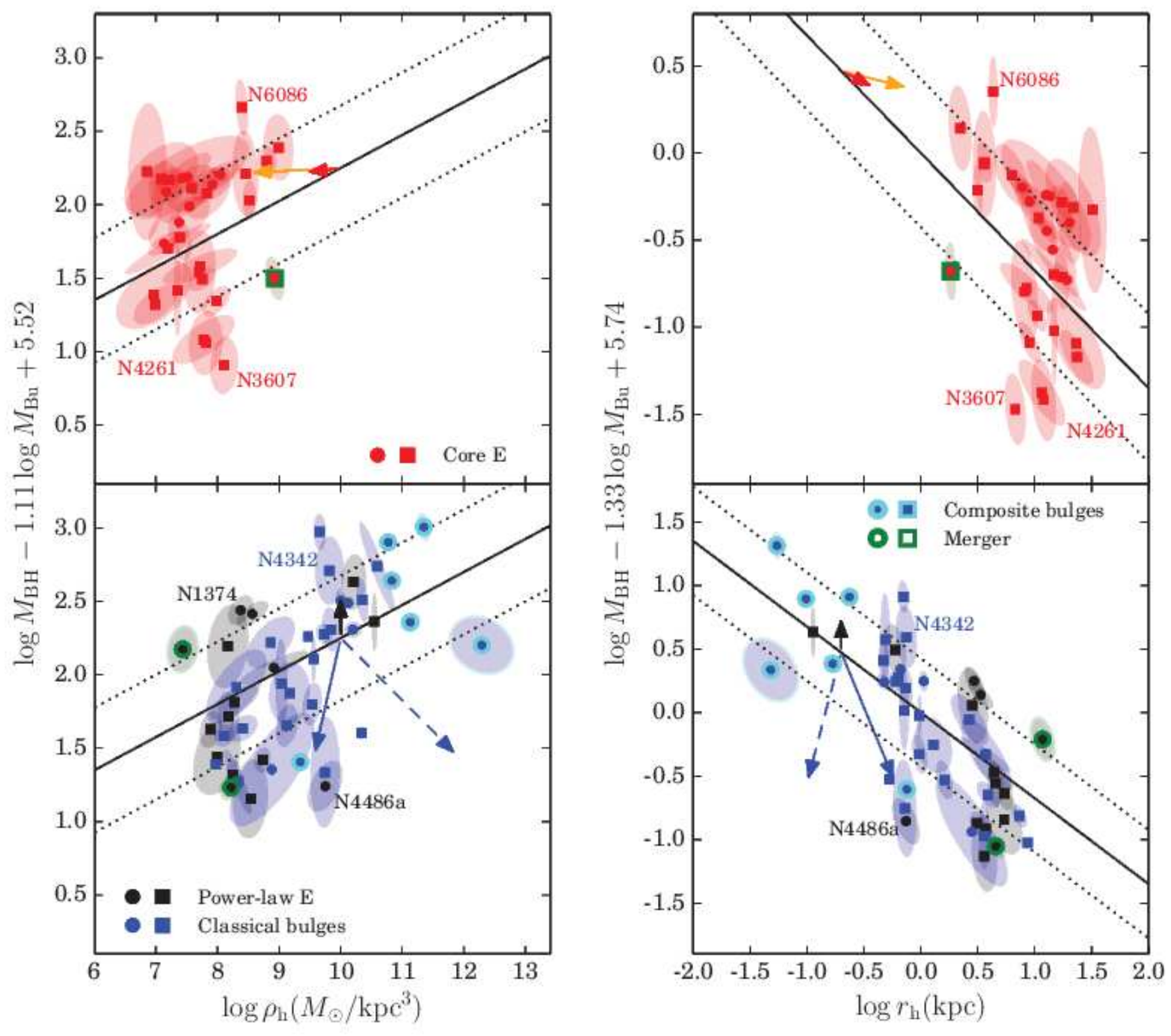}
  \end{center}
  \caption{As for Fig. \ref{fig_Sigma}, but showing how different subsamples 
relate to the \mburho\ (left panels) and the \mbur\ (right panels) bivariate 
correlations. In each set of panels we plot the corresponding best-fit 
relations for the CorePowerEClassPC sample using full lines, with the dotted 
lines showing estimated  intrinsic scatter. Individual panels show different 
subsamples: core galaxies (top panels), and PowerEClassPC (bottom panels), 
The labels name particularly deviant galaxies.
 The ellipses show the 1$\sigma$ errors. Colors, arrows and point types as
    in Fig. \ref{fig_Sigma}. 
    \label{fig_Mbu}}
\end{figure*}

This lets us present the correlations  \mburho\ and \mbur\  
in a slightly different fashion in Fig. \ref{fig_Mbu}. There we plot
the two relations separately for the subsamples CoreE,
PowerEClassPC and Pseudo.  First, we note that the subsample of
Core Es is offset to the left of the PoweEClassPC sequence at
lower densities and to the right at larger scale lengths. Indeed,
fitting the \mburho\ and \mbur\ relations to the PowerEClassPC
sample delivers steeper slopes for both the $\rho_h$ and $r_h$
dependence.

\begin{figure*}
  \begin{center}
    \includegraphics[trim=0 4cm 0 4cm,clip,width=16cm]{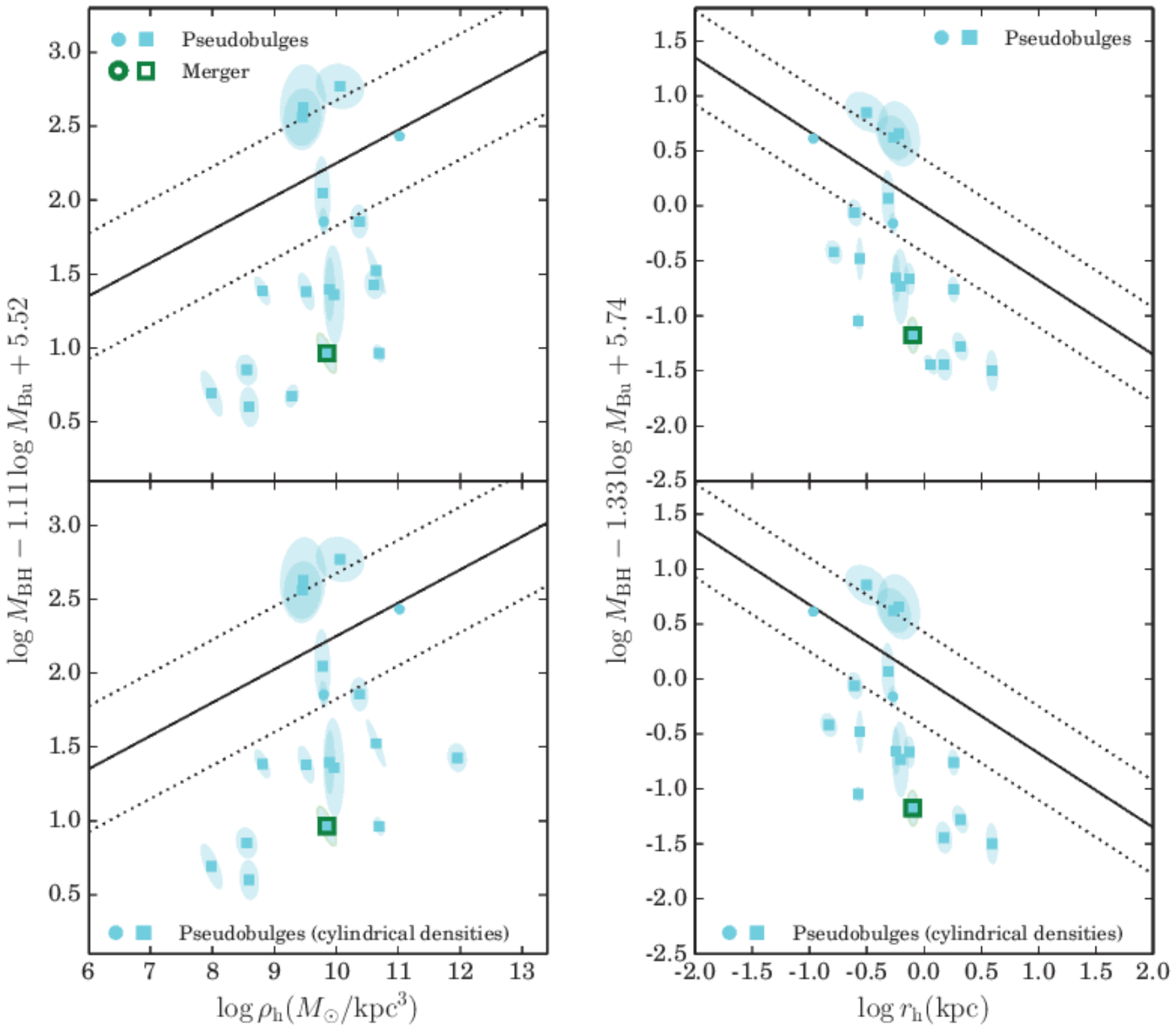}
  \end{center}
  \caption{As for Fig. \ref{fig_Mbu}, but showing pseudo bulges only.  
The plots at the top are for spherical densities
    and radii (see Eq. \ref{eq_rho}), the plots at the bottom for cylindrical
    densities and radii with $h_z=0.2a_e/1.67$ 
    (see Eq. \ref{eq_pseudodensity}). 
    \label{fig_MbuPseudo}}
\end{figure*}

Second, as found above, the pseudo bulges tend to have lower black hole masses
at any bulge mass. Fig. \ref{fig_MbuPseudo} shows that
pseudo bulges with large densities or small scalelengths tend to be
closer to the \mburho\ and \mbur\ relations defined by the other
subsamples.

\begin{deluxetable}{llllllllllllllll}
\tablewidth{0pt}
\rotate
\tabletypesize{\tiny}
\tablecaption{Two-dimensional correlations with black hole masses.\label{tab_2dim}}
\tablehead{
\colhead{Fit} & \colhead{Sample} & \colhead{a} & \colhead{da} & \colhead{b} & \colhead{db} &\colhead{ZP} & \colhead{dZP} & \colhead{$\epsilon$} & \colhead{$d\epsilon$} &\colhead{rms} & \colhead{$P(b\neq 0)$} &\colhead{$\Delta cAIC$} & \colhead{$RP$} & \colhead{$r_S$}& \colhead{$P(r_S)$} 
}
\startdata
   \msigrho\           &         All           &       4.511           &        0.32           &      -0.183           &       0.046           &      -0.449           &       1.038           &       0.379           &       0.035           &       0.409           &      0.9999           &      -13.41           &       0.001           &     -0.4934           &    3.27e-07          \\
                       &       CoreE           &       4.472           &       1.035           &      -0.103           &       0.181           &       -0.94           &       3.593           &       0.334           &       0.061           &       0.354           &      0.7242           &       2.328           &       3.203           &     -0.2972           &       0.104          \\
                       &  CorePowerE           &       3.887           &       0.533           &      -0.258           &       0.101           &       1.669           &       1.887           &       0.345           &       0.047           &       0.356           &      0.9929           &      -4.153           &       0.125           &     -0.4936           &    0.000422          \\
                       & CorePowerEClass           &       3.958           &       0.348           &      -0.176           &       0.052           &       0.859           &       1.116           &       0.314           &       0.035           &       0.339           &      0.9993           &      -9.336           &       0.009           &     -0.4546           &    6.83e-05          \\
                       & CorePowerEClassPC           &       4.037           &        0.35           &      -0.191           &       0.046           &       0.788           &       1.101           &       0.328           &       0.034           &       0.349           &      0.9999           &      -15.28           &           0           &     -0.5152           &    1.63e-06          \\
\enddata
\tablenotetext{\ }{Column 1: Fit type; Column 2: Sample type, see
  Table \ref{tab_subsamples}; Column 3 and 4: first variable slope of
  the correlation and its error; Column 5 and 6: second variable slope
  of the correlation and its error; Column 7 and 8: zero-point of the
  correlation and its errors; Column 9 and 10: intrinsic scatter and
  its errors; Column 11: measured scatter; Column 12: probability of
  the bivariate correlation; Column 13 and 14: $\Delta cAIC$ value and
  $RP=exp(\Delta cAIC/2)$ (the relative probability of the mono- and
  bivariate solutions); they are computed
  matching the bivariate solutions of this table to the monovariate
  solutions of Table \ref{tab_1dim} of the respective datasets. The
  pairings are: \msigrho\ with \msig, \mburho\ with \mbu, \msigr\ with
  \msig, \mbur with \mbu, \mbus\ with \msig, see
  Sect. \ref{sec_correlations}.  Column 15 and 16: Spearman
  coefficient and its probability.}
\end{deluxetable}

We now discuss the effects of considering different samples.  The
exclusion of barred objects weakens only slightly the significance of
all the bivariate correlations discussed above; we come back to this
issue at the end of this section.  If we just fit the 22 galaxies with
SINFONI black hole mass determinations, we find good evidence for the
bivariate correlations \msigrho, \msigr\ and \mbus\ ($P(\beta)\ne
0>0.97$, $\Delta cAIC<-0.9$), but weaker or no evidence ($P(\beta)\ne
0>0.95$ but positive $\Delta cAIC$) for the bivariate correlations
\mburho, \mbur.  Similar numbers are found if we fit only the 57
galaxies with black hole mass measurements from the literature.  The
results do not change if we minimize the correlated errors by setting
$a_{BH}=f=0$ for all galaxies in the equations of Table
\ref{tab_errormatrix}. Similar to what was noticed for the one-dimensional
correlations, we do not see any statistically significant change in
the fitted parameters of the \mburho, \mbus, and \msigrho\
correlations when we augment the errors on the bulge mass and density
by 0.15 dex (see Sect. \ref{sec_errors}). Only the estimated intrinsic
scatter is reduced by approximately 0.15 dex in quadrature.

We also repeat the one- and bivariate analysis by deleting from the
sample NGC 2974, NGC 3414, NGC 4552, NGC 4621, NGC 5813 and NGC 5846,
for which only uncertain BH mass determinations are available
\citep{Cappellari2008}, and NGC 3079 and NGC 4151, for which
\citet{KormendyHo2013} do not trust the black hole masses.  The
results are presented in Appendix D, Tables \ref{tab_1dimstrict} and
\ref{tab_2dimstrict}. The changes are not significant, therefore we
prefer the values derived including these galaxies (Tables
\ref{tab_1dim} and \ref{tab_2dim}) to maximize the size of the sample.

Finally, we repeated the \msig, \mbu\ and \mbus\ fits using the BH 
and bulge masses, plus velocity dispersions and errors of 
\citet{KormendyHo2013} for the KH45 sample to derive
  the solutions given in their Eqs. (5) and (10).  The results are
  given in Table \ref{tab_kormendy}, third and fourth row, and
  demonstrate that our fit methodology recovers the results of
  \citet{KormendyHo2013} for the \msig\ and \mbu\ correlations in
  terms of coefficients and intrinsic scatter within the errors.
  \footnote{We note in passing that the zero-point errors quoted by
    \citet{KormendyHo2013} are not marginalized, but given at the
    best-fit value of the slope.}  This remains true when we repeat
  the fits using our estimated covariance matrix (Tables
  \ref{tab_dataKormendy} and \ref{tab_erroffdiagKormendy}), see the
  results quoted in the fifth and sixth row of Table
  \ref{tab_kormendy}. Finally, using our bulge masses and errors
  (Table \ref{tab_data} and \ref{tab_errormatrix}) for the KH45
  sample, we get the results listed in the seventh and eight
  row of Table \ref{tab_kormendy}.  The coefficients agree at the
  1-$\sigma$ level, but the estimate of the measured and intrinsic
  scatter for the \mbu\ correlation is larger.

  Table \ref{tab_twokormendy} reports the results of the bivariate
  analysis on the same datasets.  The first row shows that the \mbus\
  relation is well established when using the data and errors of
  \citet{KormendyHo2013}. The intrinsic and measured scatter are just
  0.26 dex, the values of $P(\beta\ne 0)$, $\Delta cAIC$ and
  $RP=exp(\Delta cAIC/2)$ all demonstrate the existence of the bivariate
  correlation. When we consider the Tables \ref{tab_dataKormendy} and
  \ref{tab_erroffdiagKormendy}, we find results similar to the ones
  discussed above. The evidence for the \mbus\, \msigrho\ and \msigr\
  correlations is strong, while the \mburho\ and \mbur\ are only
  marginally detected (second to sixth row of Table
  \ref{tab_twokormendy}). The same is true when we consider the 
  bulge masses and errors (Table \ref{tab_data} and
  \ref{tab_errormatrix}) for the KH45 sample (seventh to
  eleventh row of Table \ref{tab_twokormendy}). Also in this case, the
  coefficients agree at the 1-$\sigma$ levels, but the estimates of
  the measured and intrinsic scatter for the correlations involving
  $M_{Bu}$ are larger. 

\begin{deluxetable}{lllllllll}
\tablecaption{One-dimensional correlations derived using the KH45 sample.\label{tab_kormendy}}
\tabletypesize{\scriptsize}
\tablehead{
\colhead{Fit} & \colhead{Sample} & \colhead{a} & \colhead{da} &\colhead{ZP} & \colhead{dZP} & \colhead{$\epsilon$} & \colhead{$d\epsilon$} &\colhead{rms}   
}
\startdata
\msig\ & Eq. (5)  \citet{KormendyHo2013} & 4.41 & 0.3  & -1.66 & 0.05 & 0.28 & - & -\\
\mbu\  & Eq. (10) \citet{KormendyHo2013} & 1.16 & 0.08 & -4.07 & 0.10 & 0.29 & - & -\\
\hline      \msig\           & \citet{KormendyHo2013}                &       4.342           &       0.323           &      -1.497           &       0.762           &       0.309           &       0.041           &       0.321          \\
       \mbu\           & \citet{KormendyHo2013}                &       1.153           &       0.088           &       -3.98           &       0.966           &       0.309           &       0.043           &       0.318          \\
      \hline
      \mhop\           & \citet{KormendyHo2013}                &        1.19           &       0.073           &      -3.431           &       0.745           &        0.25           &       0.035           &       0.263          \\
      \mfeo\           & \citet{KormendyHo2013}                &       0.796           &        0.05           &      -3.796           &       0.783           &       0.256           &       0.036           &       0.268          \\
      \hline
      \msig\           & Table \ref{tab_dataKormendy}  and \ref{tab_erroffdiagKormendy}           &       4.404           &       0.348           &      -1.658           &        0.82           &       0.323           &       0.043           &       0.337          \\
       \mbu\           & Table \ref{tab_dataKormendy}  and \ref{tab_erroffdiagKormendy}           &       1.172           &       0.089           &      -4.186           &        0.98           &       0.289           &       0.046           &       0.321          \\
      \hline
      \mhop\           & Table \ref{tab_dataKormendy}  and \ref{tab_erroffdiagKormendy}           &       1.204           &       0.073           &      -3.577           &       0.746           &       0.251           &       0.036           &        0.27          \\
      \mfeo\           & Table \ref{tab_dataKormendy}  and \ref{tab_erroffdiagKormendy}           &       0.803           &       0.051           &      -3.898           &       0.806           &       0.256           &       0.038           &       0.273          \\
      \hline
      \msig\           & Table \ref{tab_data}  and \ref{tab_errormatrix}           &       4.418           &       0.351           &      -1.693           &       0.826           &       0.325           &       0.045           &       0.337          \\
       \mbu\           & Table \ref{tab_data}  and \ref{tab_errormatrix}           &       0.952           &       0.095           &      -1.813           &       1.045           &       0.404           &       0.053           &       0.426          \\
      \hline
      \mhop\           & Table \ref{tab_data}  and \ref{tab_errormatrix}           &       1.115           &       0.079           &      -2.687           &       0.813           &       0.296           &       0.042           &       0.324          \\
      \mfeo\           & Table \ref{tab_data}  and \ref{tab_errormatrix}           &       0.715           &       0.056           &      -2.554           &       0.886           &       0.324           &       0.045           &       0.351          \\

\enddata 
\tablenotetext{\ }{Column 1: Fit type; Column 2: Sample type, see text;
Column 3 and 4: first variable slope of the correlation and its error; 
Column 5 and 6: zero-point of the correlation and its errors; 
Column 7 and 8: intrinsic scatter and its errors; 
Column 9: measured scatter.}
\end{deluxetable}

\begin{deluxetable}{llllllllllllll}
\tablecaption{ Two-dimensional correlations derived using the KH45 sample.\label{tab_twokormendy}}
\tabletypesize{\tiny}
\tablehead{
\colhead{Fit} & \colhead{Sample} & \colhead{a} & \colhead{da} &\colhead{b} & \colhead{db} & \colhead{ZP} & \colhead{dZP} & \colhead{$\epsilon$} & \colhead{$d\epsilon$} &\colhead{rms} &\colhead{$P(\beta\neq0)$} &\colhead{$\Delta cAIC$} & \colhead{$RP$}  
}
\startdata
      \mbus\           & \citet{KormendyHo2013}                &       0.621           &       0.164           &       2.275           &       0.614           &      -3.469           &       0.822           &       0.255           &       0.035           &       0.262           &      0.9997           &      -11.98           &       0.003          \\
      \hline
      \mbus\           & Table \ref{tab_dataKormendy}  and \ref{tab_erroffdiagKormendy}           &       0.701           &       0.188           &       2.027           &       0.716           &       -3.77           &       0.873           &       0.256           &       0.037           &       0.269           &      0.9995           &      -11.47           &       0.003          \\
   \msigrho\           & Table \ref{tab_dataKormendy}  and \ref{tab_erroffdiagKormendy}           &        4.02           &       0.338           &      -0.141           &       0.047           &       0.461           &       1.028           &       0.282           &       0.041           &       0.308           &      0.9965           &      -6.414           &        0.04          \\
    \mburho\           & Table \ref{tab_dataKormendy}  and \ref{tab_erroffdiagKormendy}           &       1.298           &       0.124           &       0.109           &       0.067           &      -6.508           &       1.807           &       0.294           &       0.046           &       0.312           &      0.9503           &      -0.251           &       0.882          \\
     \msigr\           & Table \ref{tab_dataKormendy}  and \ref{tab_erroffdiagKormendy}           &       3.621           &       0.385           &       0.385           &       0.113           &           0           &       0.874           &       0.272           &        0.04           &       0.295           &      0.9993           &      -9.198           &        0.01          \\
      \mbur\           & Table \ref{tab_dataKormendy}  and \ref{tab_erroffdiagKormendy}           &       1.396           &       0.177           &      -0.318           &       0.202           &       -6.48           &       1.858           &       0.297           &       0.047           &       0.311           &      0.9471           &      -0.062           &        0.97          \\
      \hline
      \mbus\           & Table \ref{tab_data}  and \ref{tab_errormatrix}           &       0.345           &       0.136           &       3.157           &       0.594           &      -2.534           &       0.813           &       0.293           &       0.042           &       0.311           &      0.9951           &      -4.073           &        0.13          \\
   \msigrho\           & Table \ref{tab_data}  and \ref{tab_errormatrix}           &        3.91           &       0.331           &      -0.194           &       0.054           &       1.191           &       1.067           &       0.267           &       0.039           &       0.291           &      0.9996           &      -10.64           &       0.005          \\
    \mburho\           & Table \ref{tab_data}  and \ref{tab_errormatrix}           &       1.159           &       0.162           &       0.203           &       0.119           &      -5.841           &       2.681           &       0.415           &       0.056           &        0.41           &       0.958           &      -0.512           &       0.774          \\
     \msigr\           & Table \ref{tab_data}  and \ref{tab_errormatrix}           &       3.616           &       0.386           &       0.386           &       0.116           &       0.011           &       0.876           &       0.275           &        0.04           &       0.295           &       0.999           &      -8.687           &       0.013          \\
      \mbur\           & Table \ref{tab_data}  and \ref{tab_errormatrix}           &       1.355           &       0.272           &      -0.599           &       0.362           &      -5.954           &        2.83           &       0.415           &       0.055           &        0.41           &      0.9546           &      -0.319           &       0.853          \\

\enddata 
\tablenotetext{\ }{Column 1: Fit type; Column 2: Sample type, text;
  Column 3 and 4: first variable slope of the correlation and its
  error; Column 5 and 6: second variable slope of the correlation and
  its error; Column 7 and 8: zero-point of the correlation and its
  errors; Column 9 and 10: intrinsic scatter and its errors; Column
  11: measured scatter; Column 12: probability of the bivariate
  correlation; Column 13 and 14: $\Delta cAIC$ value and $RP=exp(\Delta
  cAIC/2)$ (the relative probability of the mono- and bivariate
  solutions). They are computed matching the
    bivariate solutions of this table to the monovariate solutions of
    Table \ref{tab_kormendy} of the respective datasets. The pairings
    are: \mbus\ with \msig, \msigrho\ with \msig, \mburho\ with \mbu,
    \msigr\ with \msig, \mbur with \mbu, see
    Sect. \ref{sec_correlations}. }
\end{deluxetable}

These tests let us conclude that our large, combined
SINFONI-plus-literature database establishes the Equations
\ref{eq_MBHSigrhonoPseudo} to \ref{eq_MBHMbuRhalfnoPseudo}
convincingly  for the entire population of galaxies where
  dynamical BH masses have been measured. The ''best results''
  (i.e. lowest measured and intrinsic scatter) are obtained when the
  list of galaxies of \citet{KormendyHo2013} is considered
  and bulge masses derived from colour-based $M/L_C$ are used. 

To first order, these equations deliver a consistent, unifying
description of the relations between black holes on the one hand and
core and power-law ellipticals and classical bulges on the other.

The one- and two-dimensional correlations discussed above
have substantial intrinsic scatter ($ \ge 0.3$ dex), despite the
increased number of structural parameters investigated. This is true
even when we increase the errors on the bulge masses by 0.15 dex in
quadrature or when we use the bulge masses derived from colors of
  \citet{KormendyHo2013}.  The correlations derived for the sample of
core ellipticals tend to have the smallest intrinsic and measured
scatter, broadly in agreement with the averaging effect described by
\citet{Peng2007}.

We conclude by clarifying again the role of barred galaxies. 
In Table \ref{tab_1dim} and
\ref{tab_2dim} we quote the results of fitting the CorePowerEClassnoBars
sample, where we drop 16 barred galaxies contained in the CorePowerEClassPC
sample. The coefficients and scatter of the one-dimensional
correlations hardly change within the errors, in particular the ones
of Eqs. \ref{eq_Hopkins} or \ref{eq_Feoli}. The same is true for the
bivariate correlations, but their significance is slightly decreased. 
This effect is partly due to the reduced number of fitted
galaxies. We conclude that the BH FP is
not driven solely  by barred galaxies, contrary to the  suggestion of
\citet{Graham2008}.

\section {Discussion}
\label{discussion}

\subsection {Comparing   $M_{Bu}^{a_{best}}\sigma^{b_{best}}$, $M_{Bu}^{0.5}\sigma^2$,
$M_{Bu}\sigma^2$ and $M_{Bu}\sigma$}
\label{sec_modcomp}

In Sect. \ref{intro} we reviewed the different physical
 interpretations proposed to explain the bivariate correlations with
  $M_{BH}$.  \citet{Hopkins2007a,Hopkins2007b}
concluded that the bivariate correlations mirror the correlation
between $M_{BH}$ and $M_{Bu}^{0.5}\sigma^2$, which can be expected on
simple physical grounds (see below). \citet{Feoli2005}, \citet{Aller2007} and
\citet{Mancini2012} argued that the bulge's kinetic or gravitational energy
$M_{Bu}\sigma^2$ of a galaxy sets its black hole mass.
Finally, \citet{Soker2011} suggested the bulge's momentum $M_{Bu}\sigma$ as the
key physical quantity. 

We compare the three options $M_{Bu}^{0.5}\sigma^2$,
$M_{Bu}\sigma^2$ and $M_{Bu}\sigma$ to our best fit
solutions given in Eq. \ref{eq_MBHSigMbunoPseudo} and Table
\ref{tab_twokormendy} (row 2 and 7) using the model comparison
formalism. Similar to what is discussed in
Sect. \ref{sec_correlations}, we marginalize the posterior
  probability distribution $P(a,b,ZP,\epsilon)$ over the parameters
  $ZP$ and $\epsilon$ and derive the equivalent of
  Eq. \ref{eq_lnPbestb} as:
\begin{equation}
\label{eq_lnPbest2}
\begin{array}{l}
-2\ln [P(a_{best},b_{best})/P(a,b)]=\\
(a-a_{best},b-b_{best})V^{-1}\left(
\begin{array}{l}
a-a_{best}\\
b-b_{best}\\
\end{array}
\right) ,\\
\end{array}
\end{equation} 
where 
\begin{equation}
\label{eq_V}
V=\left(
\begin{array}{ll}
(\delta a_{best})^2        & <\delta a \delta b>_{best}\\
<\delta a \delta b>_{best} & (\delta b_{best})^2\\
\end{array}
\right),
\end{equation}
$a_{best}$, $\delta a_{best}$, $b_{best}$, $\delta b_{best}$ are given
in Tables \ref{tab_2dim} and \ref{tab_twokormendy}  (as usual
  dropping the label $best$), and $<\delta a \delta b>_{best}$ (see
Fig. \ref{fig_ProbMBHSigrho}) is listed in Table \ref{tab_modcomp}.
The slopes $a$ and $b$ in Eq. \ref{eq_lnPbest2} obey $a/b=0.25, 0.5,
1$ for the cases $M_{Bu}^{0.5}\sigma^2$, $M_{Bu}\sigma^2$ and
$M_{Bu}\sigma$, respectively. The values of $a$ and $b$ reported in
Table \ref{tab_modcomp} maximize Eq. \ref{eq_lnPbest2} and are
identical to the values quoted in Tables \ref{tab_2dim} and
\ref{tab_kormendy} for the $M_{Bu}^{0.5}\sigma^2$ and $M_{Bu}\sigma^2$
fits. Next, we compute $\Delta cAIC$ and $RP=exp(\Delta
  cAIC/2)$. What are the correct values of $k$ to be used in Eq.
  \ref{eq_DcAIC}? Since $a_{best}$ is different from zero, we can
  recast our best-fit solution $M_{BH}\sim (M_{Bu}^{a_{best}}
  \sigma^{b_{best}})$ as $M_{BH}\sim
  (M_{Bu}\sigma^{b_{best}/a_{best}})^{a_{best}}$ and argue that $k=3$
  for the four relations $M_{Bu}\sigma^{b_{best}/a_{best}}$,
  $M_{Bu}^{0.5}\sigma^2$, $M_{Bu}\sigma^2$ or $M_{Bu}\sigma$. The
  values of $RP^{(3)}$ given in Table \ref{tab_modcomp} are
  computed following this reasoning. They stem directly from
  Eq. \ref{eq_lnPbest2}, since the terms involving $k$ in Eq.
  \ref{eq_DcAIC} are all equal and cancel out. These relative
  probabilities assess which of the three models
  $M_{Bu}^{0.5}\sigma^2$, $M_{Bu}\sigma^2$ or $M_{Bu}\sigma$ is
  nearest to our best solution.

  The values of $RP^{(4)}$ given in Table \ref{tab_modcomp}
  assume instead that our best-fit solution $M_{BH}\sim
  (M_{Bu}^{a_{best}}\sigma^{b_{best}})$ has $k=4$, while the  $M_{Bu}^{0.5}\sigma^2$,
  $M_{Bu}\sigma^2$ or $M_{Bu}\sigma$ models have $k=3$. They assess whether
  our 'complex' best-fit solution $M_{BH}\sim (M_{Bu}^{a_{best}}
  \sigma^{b_{best}})$ is really needed to describe the data, or
  whether the 'simpler' $M_{Bu}^{0.5}\sigma^2$, $M_{Bu}\sigma^2$ or
  $M_{Bu}\sigma$ models should be preferred. 

If we use $RP^{(3)}$ to rank the $M_{Bu}^{0.5}\sigma^2$,
  $M_{Bu}\sigma^2$ or $M_{Bu}\sigma$ models with respect to
  $M_{Bu}\sigma^{b_{best}/a_{best}}$, we find that the
$M_{Bu}^{0.5}\sigma^2$ model performs the best (i.e., it has the
largest $RP^{(3)}$), when compared with our best fitting solution,
whatever sample is considered (either our reference CorePowerEClassPC
sample, or the KH45 sample with the color-based bulge masses, or the
same sample with dynamical bulge masses). The $M_{Bu}\sigma$ model
performs worst; when using dynamical bulge masses we get
$RP^{(3)}<10^{-4}$ and only for the KH45 sample with bulge masses
derived from $M/L_C$ $RP^{(3)}$ is not too low (0.33) . 

If we consider $RP^{(4)}$, we conclude that the model
  $M_{Bu}^{0.5}\sigma^2$ provides a description of the data as good as
  the one of our best fitting solution, with $RP^{(4)}=0.41$ for our
  reference CorePowerEClassPC sample and $RP^{(4)}=0.88$ for the KH45
  sample with dynamical bulge masses. The model $M_{Bu}^{0.5}\sigma^2$
  is statistically preferred with respect to our best fitting solution
  when the KH45 sample with color-based bulge masses is considered
  ($RP^{(4)}=2.89$).  For this sample the same applies to
  the $M_{Bu}\sigma^2$ model ($RP^{(4)}=2.82$), while the
  $M_{Bu}\sigma$ model is statistically indistinguishable
  ($RP^{(4)}=1.1$) from the bivariate best-fit model.  This
  conclusion is partially driven by the smaller size (45 versus 77
  galaxies) and larger errors (see Sect. \ref{sec_errors}) of the KP45
  sample, compared to our reference CorePowerEClassPC sample. But it
  could also reflect the underlying physical difference between
  color-based and dynamical bulge masses.

In the following we give a closer look to the models that compare best
to our best fitting bivariate solution, the $M_{Bu}^{0.5}\sigma^2$ and
$M_{Bu}\sigma^2$ model.

\begin{table*}
\caption{Comparison of the $M_{Bu}^{a_{best}}\sigma^{b_{best}}$, $M_{Bu}^{0.5}\sigma^2$, $M_{Bu}\sigma^2$ and $M_{Bu}\sigma$ models. \label{tab_modcomp}}
\scriptsize
\begin{tabular}{llllllll}
\hline
\hline
Model & a & b & Subsample & $<\delta a \delta b>_{best}$ & $RP^{(3)}$ & $\Delta cAIC$ & $RP^{(4)}$\\
\hline
$M_{Bu}^{0.5}\sigma^{2}$ & 0.547 & 2.186 & CorePowerEClassPC & -0.041 & 0.14 & -1.78 & 0.41\\ 
                      & 0.602 & 2.408 & KH45 sample from Tables \ref{tab_dataKormendy}  and \ref{tab_erroffdiagKormendy} & -0.123 & 0.86 & 2.12 & 2.89\\                  
                      & 0.558 & 2.230 & as above, but from Tables \ref{tab_data}  and \ref{tab_errormatrix} & -0.068 & 0.26 & -0.25 & 0.88\\                  
\hline
$M_{Bu}\sigma^2$ & 0.673 & 1.346 & CorePowerEClassPC & -0.041 & 0.002 & -10.6 &  0.005\\
                & 0.803 & 1.606 & KH45 sample from Tables \ref{tab_dataKormendy}  and \ref{tab_erroffdiagKormendy} & -0.123 & 0.84 & 2.07 & 2.82\\                  
                & 0.715 & 1.430 & as above, but from Tables \ref{tab_data}  and \ref{tab_errormatrix} & -0.068 & 0.013 & -6.31 & 0.04\\
\hline
$M_{Bu}\sigma$   & 0.757 & 0.757 & CorePowerEClassPC & -0.041 & $1.5\times 10^{-5}$ & -19.9 & $4.7\times 10^{-5}$\\
                & 0.965 & 0.965  & KH45 sample from Tables \ref{tab_dataKormendy}  and \ref{tab_erroffdiagKormendy} & -0.123 & 0.33 & 0.2  & 1.1\\                  
                & 0.825 & 0.825 & as above, but from Tables \ref{tab_data}  and \ref{tab_errormatrix} & -0.068 &  $4\times 10^{-4}$ & -13.2 & $1.4\times10^{-3}$\\
 \hline
\end{tabular}
\tablenotetext{\ }{Column 1: tested model; Column 2 and 3: best-fitting values of $a$ and $b$; Column 4: dataset used; Column 5: see Eq. \ref{eq_V}; Column 6: the probabilities relative to the best-fitting \mbus\ correlation, see Sect. \ref{sec_modcomp}; Column 7: values of $\Delta cAIC$ computed matching the models of this table to the  $M_{Bu}^{a_{best}}\sigma^{b_{best}}$ model of the corresponding dataset, see Tables \ref{tab_2dim} and \ref{tab_twokormendy}; Column 8: $RP^{(4)}=exp(\Delta cAIC/2)$ gives the probability of the models of this table relative to the  $M_{Bu}^{a_{best}}\sigma^{b_{best}}$ model of the corresponding dataset.}
\end{table*}

\begin{figure*}
  \begin{center}
\includegraphics[trim=0 0cm 0 16cm,clip,width=16cm]{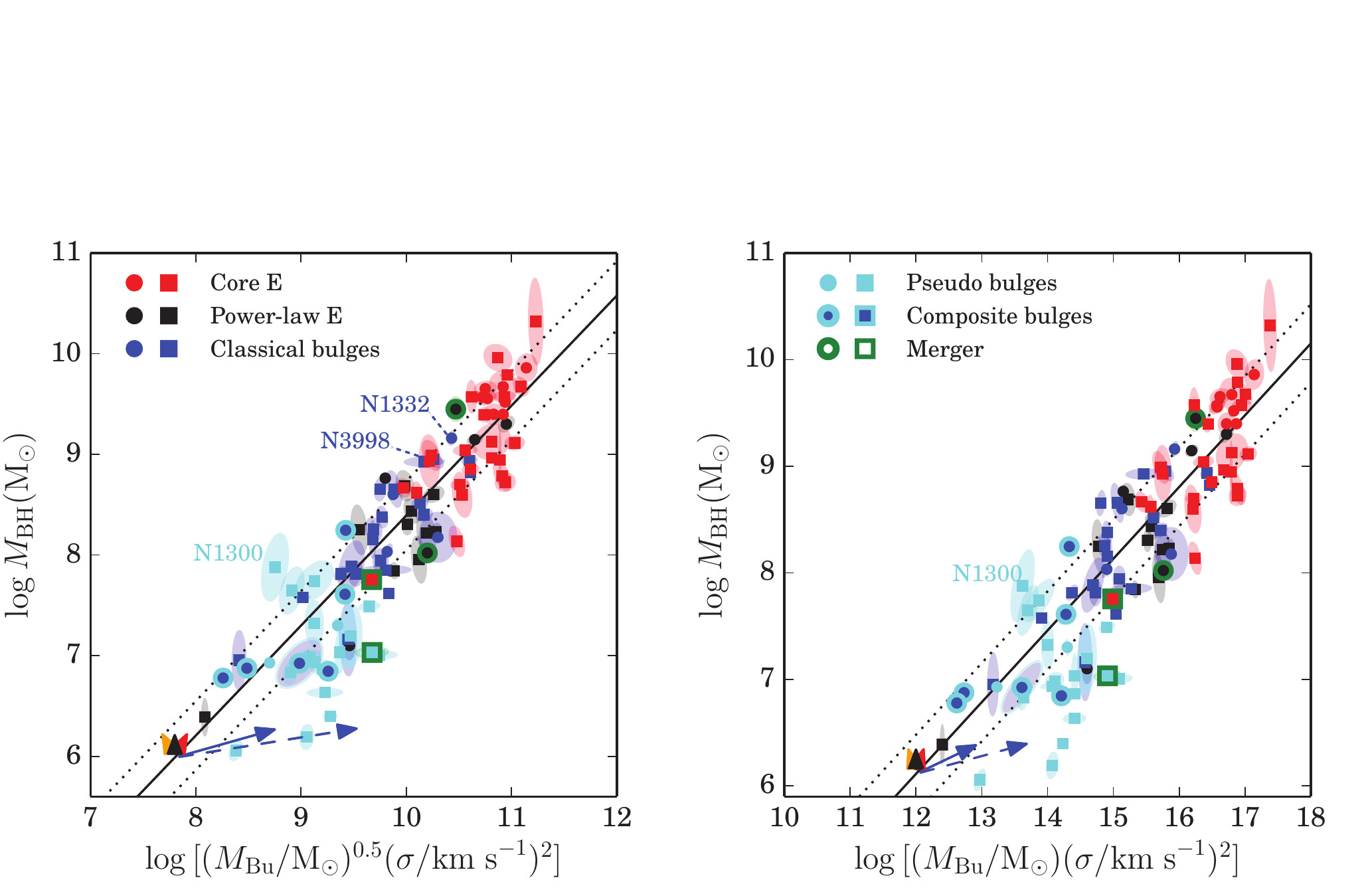}
  \end{center}
  \caption{ The correlation between $\log M_{BH}$ and $\log M_{Bu}^{0.5}\sigma^2$ (left) and between  $\log M_{BH}$
and $\log M_{Bu}\sigma^2$ (right). The corresponding best-fit 
relations for the CorePowerEClassPC sample are shown using full lines, with the dotted lines 
showing estimated  intrinsic scatter.  The ellipses show the 1$\sigma$ errors. Colors, arrows and point types as
    in Fig. \ref{fig_Sigma}.
\label{fig_Hopkins}}
\end{figure*}

 Table \ref{tab_1dim} presents
the results of fitting the relation $\log M_{BH}=a \log
(M_{Bu}^{0.5}\sigma^2)+ZP$. For the sample without pseudo bulges we
find:
\begin{equation}
\label{eq_Hopkins}
\log M_{BH}=(1.09\pm 0.06) \log (M_{Bu}^{0.5}\sigma^2) - (2.54\pm 0.64), 
\end{equation}
with intrinsic and measured scatter slightly lower than the \msig\
relation, and slightly larger than the ones given by
Eqs. \ref{eq_MBHSigrho}, \ref{eq_MBHMBurho} or \ref{eq_MBHSigRhalf}.
Fig. \ref{fig_Hopkins} (left) shows the relative correlation plot. We
recognize the features discussed above: core Es are slightly shifted
upwards, while pseudo bulges have systematically lower black hole
masses than predicted. This confirms that the picture presented by
\citet{Hopkins2007a,Hopkins2007b} is valid for ellipticals and
classical bulges, but breaks down for pseudo bulges.  The intrinsic
scatter is reduced to just 0.25 dex, when the KH45 sample with
color-based bulge masses is considered.

Fig. \ref{fig_Hopkins} (right) shows the
correlations between the black hole mass and the kinetic energy of
the bulge $M_{Bu} \sigma^2$ put forward by \citet{Feoli2005},
\citet{Aller2007}, \citet{Feoli2009}, and \citet{Mancini2012}. Table
\ref{tab_1dim} also presents the results of fitting the relation $\log
M_{BH}=a \log (M_{Bu}\sigma^2)+ZP$. For the sample without pseudo
bulges we find:
\begin{equation}
\label{eq_Feoli}
\log M_{BH}=(0.67\pm 0.04) \log (M_{Bu}\sigma^2) - (1.96\pm 0.65). 
\end{equation}
This is similar to the dependences found by \citet{Aller2007}, 
  \citet{Hopkins2007a} and \citet{Mancini2012}.  The correlation is
strong with intrinsic scatter slightly better than the one of the
\msig\ relation and slightly larger than the one given by
Eq. \ref{eq_Hopkins}.  In both cases the differences are not
  statistically significant.  When the KH45 sample with color-based
  bulge masses  is considered, the intrinsic scatter is
  reduced to just 0.256 dex.  Again, pseudo bulges are offset towards
lower black hole masses.

To conclude, our analysis shows that the model
  $M_{Bu}^{0.5}\sigma^2$ is slightly preferred to explain our
  bivariate correlations, especially when dynamically determined bulge
  masses are considered. However, the bulge kinetic energy
  $M_{Bu}\sigma^2$ produces correlations with black hole masses with
  statistically equivalent intrinsic and measured scatter. The
  bulge momentum $M_{Bu}\sigma$ is highly disfavoured in the case of
  dynamically determined bulge masses.

\subsection{A simple interpretation framework} 
\label{sec_interpretation}

As discussed above, Fig. \ref{fig_Hopkins} points to some sort of
gravitationally induced equilibrium configuration, coupled with a
feedback mechanism, which lies behind our empirical findings. 
Indeed, one could naively expect a correlation between  $M_{Bu}\sigma^2$  and
$M_{BH}$ as a consequence of an energy balance during the simultaneous
growth of bulges and black holes.  Why a correlation between
 $M_{Bu}^{0.5}\sigma^2$  and  $M_{BH}$ should exist is less obvious. 
\citet{Hopkins2007a,Hopkins2007b} argue that one needs to equate the
momentum of the galactic outflow $p\sim M_{Bu}\sigma$ to the injection
of momentum $\dot{p}\sim L_{Edd}/c\sim M_{BH}$ during the dynamical
time $t^{dyn}_{soi}\sim R_{soi}/\sigma$ near the black hole sphere of
influence with radius $R_{soi}\sim G M_{BH}/\sigma^2$. This gives
$\dot{p}\times t^{dyn}_{soi}\sim M^2_{BH}/\sigma^3\sim M_{Bu}\sigma$,
or $M_{BH}\sim M_{Bu}^{0.5}\sigma^2$.  Clearly this argument applies
only if the Eddington luminosity $L_{Edd}$ is high enough. This is
probably the case during 'wet' mergers of gas-rich galaxies, or 
during violent instabilities in high-density, high redshift disks, where
large amounts of gas could be funnelled to the center and accreted
quickly while simultaneously forming a classical bulge \citep{Bournaud2014,Menci2014}. 
However, this mechanism
should not work in the case of 'dry' mergers, when gas-poor galaxies
coalesce, or for pseudo bulges formed by secular evolution of disks,
where only small amounts of gas can flow towards the center at any
time (due to, for example, bar torques). Only pseudo bulges with
particularly high densities could be expected to have generated in the
past with accretion luminosities approching the Eddington limit.
 In the following we summarize these statements in a simple quantitative way.

 Dry mergers are mergers of (elliptical and S0) galaxies without cold
 gas and therefore are dissipationless. Equal mass dry mergers
 preserve the velocity dispersion $\sigma$ ($\sigma^f=\sigma^i$, 
where $i$ and $f$ indicate the initial and final state) of
 the resulting merged galaxy, double the size ($r_h^f/r_h^i=2$) and
 reduce the average density by 1/4 ($\rho^f_h/\rho^i_h=1/4$). They
 increase the central black hole masses through black hole merging by
 a factor 2 (if both merging galaxies obey the \mbu\ relation before
 the merging), see
 \citet{Naab2009}. These effects are indicated by the red arrows in
 the Figures \ref{fig_phot} to \ref{fig_Hopkins}. Dry mergers are also
 responsible for the formation of the cores of core ellipticals through
 the binary black hole scouring mechanism. The absence of gas implies
 that the central cusp cannot be reformed after the merging event.

 Several minor dry mergers (orange arrows in the Figures
 \ref{fig_phot} to \ref{fig_Hopkins}) can enlarge sizes and lower
 average densities even more (for a sequence of minor mergers doubling
 the mass, the size is increased by 4 and the density is decreased by
 1/32), while keeping $M_{BH}/M_{Bu}$ ratio constant and reducing
 $\sigma$ by a factor $1/2^{0.5}$ \citep{Naab2009}. These combined
 effects can qualitatively explain the position of the most massive,
 largest, and least dense core ellipticals in the Figures
 \ref{fig_phot}, \ref{fig_Sigma}, \ref{fig_Mbu} and \ref{fig_Hopkins}
 and the ``saturation effect'' discussed in
   \citet{KormendyBender2013}. These objects appear to have BH masses
 $\approx 0.3$ dex larger than the rest of the CorePowerEClassPC
 sample and could well be the result of dry merging of pre-existing
 power-law or core ellipticals. Moreover, the intrinsic and measured
 scatter of the correlations defined by core ellipticals alone tend to
 be the smallest, in line with the averaging effect expectations
 discussed by \citet{Peng2007}.

Wet mergers are mergers of (disk) galaxies with cold gas.  Measuring
\mbh\ in local galaxies known to be recent mergers of disk galaxies
allows one to sketch what could be the evolution with time of these
class of objects \citep{KormendyHo2013}.  If the merging progenitors
follow a black hole scaling relation and the black hole(s) do not
accrete substantial amounts of gas during the merging process, the
merger remnant would now fall below the relation by a factor
corresponding to the mean bulge-to-total ($B/T$) ratio of the
progenitors.  This is indicated by the blue arrows in the Figures
\ref{fig_phot} to \ref{fig_Hopkins} for an equal mass merger of
spirals with $B/T=0.2$ and a factor 3 increase in bulge size. The
velocity dispersion will change according to the Virial Theorem
$r_h^f\sigma^{f2}=2r_h^i\sigma^{i2}/(B/T)$, while the scale length of
the bulge will change by a factor $r_h^f/r_h^i=k$. \citet{Naab2006}
argue that $k$ can vary between 0.5 and 3 in disk plus bulge mergers
without gas. As a consequence, bulge masses and velocity dispersions
will generally increase, while scale radii or densities might either
increase or decrease. Therefore we could expect recent mergers to lie
generically below the black hole correlations.  This is indeed what we
see for NGC 1316, NGC 2960, NGC 5018, and NGC 5128 (plotted in green
in the Figures \ref{fig_phot} to \ref{fig_Hopkins}), all rather young,
gas-rich, merger remnants. NGC 3923 is likely to be (the late phase
of) a merger between an existing elliptical with a low velocity
dispersion dwarf galaxy. In general, the distribution of power-law
ellipticals and classical bulges in the Figures \ref{fig_phot} to
\ref{fig_Hopkins} is better explained by the full blue arrow, i.e.
when the scale length of the bulge resulting from the merger is larger
than the original value.

Finally, the BH can increase its mass by accretion. This process
has been also proposed for changing sizes and velocity dispersions of
galaxies through its feedback \citep{Fan2008}, but it fails to explain 
the redshift evolution of sizes of early-type galaxies
\citep{Saglia2010b}. For simplicity we therefore neglect
this possible effect on sizes and densities and just mark the importance 
of BH accretion in itself. The black arrows in the
Figs.  \ref{fig_onedim} to \ref{fig_Hopkins} indicate accretion that
doubles the mass of the BH. Table \ref{tab_scenarios} summarizes the
mechanisms discussed above. Ultimately the feedback
produced during the accretion event will set the galaxy back to the
black hole correlations as discussed above.

\begin{table*}
  \caption{The fractional changes in $M_{BH},M_{Bu},\sigma,r_h,\rho_h$ expected after
    a major or a minor 'dry' merger that double the galaxy and BH mass, a 'wet' merger 
    with initial bulge-to-total ratio $B/T$ and final-to-initial bulge-scales ratios $k$, and 
    gas accretion on the BH.}
\label{tab_scenarios}
\begin{tabular}{llllll}
\hline
\hline
Process & $M_{BH}^f/M_{BH}^i$ &  $M_{Bu}^f/M_{Bu}^i$ & $r_h^f/r_h^i$ & $\rho_h^f/\rho_h^i$ & $\sigma^f/\sigma_i$ \\
\hline
Major dry merger &  2 & 2                 & 2       & 1/4                   & 1 \\
Minor dry merger &  2 &         2         & 4       & 1/32                  & $1/2^{0.5}$\\
Wet merger       &  2 & $\frac{2}{(B/T)}$ & $\kappa$ & $\frac{2}{k^3(B/T)}$  & $\left[\frac{2}{k (B/T)}\right]^{0.5}$ \\
BH Accretion     &  2 & 1                 & 1       & 1                     & 1\\
\hline
\end{tabular}
\end{table*}

The distribution of power-law ellipticals and classical bulges in
Figures \ref{fig_onedim}, \ref{fig_Sigma} and \ref{fig_Mbu} 
can be seen as the
result of the combined effects of wet mergers, black hole accretion
and feedback mechanisms.  The exact steepness of the blue arrows
depends on $B/T$ and the relative amount of (dissipative) gas
available. Some gas will be accreted on the BH.   Some
gas will generate new stars, increasing the density of the newly
formed bulge and possibly forming a central power-law cusp.

As the average gas fraction of galaxies decreases with cosmic time, it
is plausible that galaxies that merged earlier had a better chance to
grow their black holes in lock-step (or even over-grow) with the
spheroid than is possible for present-day mergers. Thus, we would expect the
$M_{BH}/M_{Bu}$ ratio to increase with increasing redshift
\citep{Sijacki2015}. This is indeed observed in various samples of
quasars at $z>2$ (see review in \citealt{KormendyHo2013} and
references therein).

We can further speculate that those objects that were assembled early
and did not undergo late major mergers today harbour the most massive
black holes for a given bulge mass. These objects should have formed
from very gas-rich material that allowed their black holes to grow
efficiently. They also should have high stellar densities because
earlier formation implies higher dark matter and gas densities
\citep{Thomas2009}. As dry mergers decrease the mean stellar density,
a high density today also implies that the objects did not undergo
such events more recently. Examples for such objects could be very
compact bulges in old S0’s or early-type spirals, where the existence
of a substantial disk indicates the lack of late major mergers.

In fact, some of the most compact bulges known (we include NGC 1332,
NGC 3998, and NGC 4486b) do harbour unusually large black holes for
their bulge mass (see Fig. \ref{fig_onedim}). M32 (NGC 221) is also
extremely dense, but with a normal black hole for its bulge
mass. Except for NGC 4486b, these galaxies are not particularly
deviant when the \msig\ relation is considered, an indication that the
velocity dispersion (through the \msig\ relation) is a more robust
black hole mass predictor than the bulge mass (through the \mbu\
relation).

Pseudo bulges with spherical densities $\rho_h\approx 10^{10}
M_\odot/$kpc$^3$ such as NGC 4501 (where we detect molecular gas in
non-circular motions, see \citealt{Mazzalay2013,Mazzalay2014}) or NGC
3227 (where an active nucleus is present and a recent episode of star
formation took place, see \citealt{Davies2006}) have BHs with masses
similar to those predicted by the \msig\ or the \mbu\ relations of
ellipticals and classical bulges.  Pseudo bulges with lower densities
have smaller black holes than the classical-bulges
prediction. Therefore, the growth of black holes in galaxies that did
not undergo mergers, and therefore do not have a classical bulge,
follows a path decoupled from the rest of the galaxy and set by the
amount of gas that secular processes (such as bars) manage to funnel
towards the galaxy centers. The resulting black hole masses are much
smaller than the ones measured in early-type galaxies or classical
bulges. We speculate that only when the densities involved are above a
certain threshold are pseudo bulges able to fuel the black holes
efficiently enough to approach the feedback mechanism (see above) that
sets $M_{BH}/M_{Bu}$ in classical bulges. The exact value of the
density threshold is, however, uncertain by an order of magnitude, as
it depends on the unknown geometry of pseudo bulges.

Finally, we also find ‘composite systems’ where both a classical and
a pseudo bulge co-exist. In the Figures \ref{fig_onedim},
\ref{fig_Sigma} and \ref{fig_Mbu} we plot the position of 
NGC 1068, NGC 2787, NGC 3368,
NGC 3489, NGC 4371 and NGC 4699, all galaxies with composite bulges, using the
mass of their small classical bulges. These high-density components
form the high density, small size end of the power law and 
classical bulge bivariate correlations. We speculate that they formed together
with their BH at high redshifts. Possibly {\it every} pseudo bulge
has a small classical component at its center: \citet{Erwin2015a}
argue that it is not (yet) possible to present a clear case of a pure
pseudo bulge where the presence of a small classical component can
be excluded without doubt.

As noted in Section \ref{correlations}, all our one- and
two-dimensional correlations have sizeable intrinsic scatter.  The
  smallest intrinsic scatter (0.26 dex) is derived for the KH45 sample, using mass-to-light
  ratios derived from colors. On one hand, this could indicate that BH
  masses correlate best with the baryonic mass of classical
  bulges. On the other hand, having now explored the influence of all
galaxy structure parameters and their errors, we have to conclude that
the remaining factor $\approx 2$ uncertainty in the black hole mass
must stem from the unknown details of the accretion and feedback
mechanisms.

\section{Conclusions}
\label{conclusions}

We produced a merged SINFONI-plus-literature database of BH masses for
97 galaxies. For this sample we computed dynamical bulge mass
estimates $M_{Bu}$, and determined the bulge spherical half-luminosity
radius $r_h$ and averaged spherical density $\rho_h$ within $r_h$,
collecting bulge-plus-disk decompositions from the literature or
perfoming them ourselves (for 16 galaxies). We confirm that there is
an almost linear relation between BH mass and the mass of the
classical bulge of a galaxy. The quadratic relation suggested by
\citet{Graham2013} for ``Sersic galaxies'' is driven by the inclusion
of pseudo bulges and possibly uncertain bulge masses.

Densities and sizes of classical bulges turn out to be important. We
showed that BH masses correlate directly with both densities and
sizes, although with larger scatter than the more usual \msig\ or
\mbu\ relations. We established significant bivariate correlations
involving $\sigma$, $M_{BH}$-$\sigma$-$\rho_h$,
$M_{BH}$-$\sigma$-$r_h$, and $M_{BH}$-$\sigma$-$M_{Bu}$, valid for all
classical bulges (core and power-law ellipticals, classical bulges of
disk galaxies) with low intrinsic ($\le 0.34$ dex) and measured ($\le
0.37$ dex) scatter. Two further bivariate correlations involving
$M_{Bu}$ - \mburho\ and \mbur\ - are also robustly detected, but with
larger intrinsic ($\approx 0.43$ dex) and measured ($\approx 0.44$
dex) scatter and with core elliptical galaxies slightly offset.
  The five bivariate correlations are robustly detected also when the
  KH45 sample 
  is considered. For this sample, with bulge masses scaled from M/Ls
  derived from colors, the estimated intrinsic scatter is as low as
  0.26 dex. Contrary to the suggestion of \citet{Graham2008}, none of
these bivariate correlations are driven by the inclusion of barred
galaxies. The relations are point to a link between
black hole mass and $M_{Bu}^{0.5}\sigma^2$, as proposed by
\citet{Hopkins2007a,Hopkins2007b}, or the bulge kinetic energy $M_{Bu}\sigma^2$, 
as first suggested by \citet{Feoli2005}. In contrast, pseudo bulges have
systematically lower black hole masses, but approach the predictions
of all the above relations at (spherical) densities $\rho_h\ge 10^{10}
M_\odot/kpc^3$ or scale lengths $r_h\le 1$ kpc. These thresholds are
rather uncertain, because we do not know the true geometry of pseudo
bulges.

High densities and small sizes imply a large baryonic concentration
near the centre and make very efficient mass accretion onto the black
hole likely. Classical bulge densities/sizes, in turn, are set mainly
by two factors: (1) the formation redshift - earlier formation implies
higher halo and gas densities - and (2) the merging history - gas poor
mergers reduce the density and increase the size in each merger
generation. (1) could explain why compact classical bulges of S0
galaxies have the highest BH masses for their bulge mass; (2) implies
that slow-rotator, core ellipticals of a given mass have slightly more
massive black holes the lower their average density is or the larger
their size is. The averaging effect of a series of gas-free mergers
\citep{Peng2007} would also make plausible why the correlations derived for
the sample of core ellipticals only tend to have the smallest intrinsic
and measured scatter. Power-law, fast-rotator early-types and
classical bulges are the results of dissipational, gas-rich mergers of
disk-dominated progenitors. The feedback mechanism triggered by black
hole accretion, coupled with the gravitationally induced virial
equilibrium, creates the correlations between black hole mass and
galaxy structural parameters. In this case at a given $M_{Bu}$ or
$\sigma$ objects with larger average densities or smaller sizes have
larger $M_{BH}$. The bivariate correlations, however, are not as tight
as the mass Fundamental Plane of early-type galaxies
\citep[e.g.,][]{Hyde2009}. The tightest relation we derive is the
\mbus\ for the KH45 sample, where the estimated intrinsic scatter is 0.26
dex.  The unknown details of the black hole accretion physics and
feedback mechanisms are probably responsible for this sizeable
intrinsic and measured scatter.

Disk galaxies that do not experience major mergers might develop a
pseudo bulge through secular instabilities such as bars. These may
drive gas towards the center and feed the central black hole. However,
the lack of correlation between $M_{BH}$ and the structural parameters of
pseudo bulges shows that no efficient consistent feedback mechanism is at work
in these objects. Only pseudo bulges with extremely high
densities/small sizes manage to form black holes with masses
approaching those of classical-bulges.

\section*{Acknowledgements}

We thank the first referee of the paper, John Kormendy, for a careful
proof of Table 1; we thank the statistics editor of ApJ for his
valuable suggestions on model comparison; we thank the second
  anonymous referee for comments that helped us improving the
  presentation of the results. MO acknowledges support by the
Trans-regional Collaborative Research Centre TR22 ``The Dark Universe''
of the Deutsche Forschungsgemeinschaft (DFG) and the DFG Cluster of
Excellence ``Origin and Structure of the Universe''.  SPR was
supported by the DFG Cluster of Excellence ``Origin and Structure of
the Universe''. PE was supported by the DFG Priority Programme 1177
``Galaxy Evolution''.
Funding for the creation and distribution of the SDSS Archive has
  been provided by the Alfred P. Sloan Foundation, the Participating
  Institutions, the National Aeronautics and Space Administration, the
  National Science Foundation, the US Department of Energy, the
  Japanese Monbukagakusho and the Max Planck Society. The SDSS website
  is http://www.sdss.org/.  The SDSS is managed by the Astrophysical
  Research Consortium (ARC) for the Participating Institutions. The
  Participating Institu- tions are The University of Chicago,
  Fermilab, the Institute for Advanced Study, the Japan Participation
  Group, The Johns Hop- kins University, the Korean Scientist Group,
  Los Alamos National Laboratory, the Max-Planck-Institute for
  Astronomy (MPIA), the Max-Planck-Institute for Astrophysics (MPA),
  New Mexico State University, University of Pittsburgh, University of
  Portsmouth, Princeton University, the United States Naval
  Observatory and the University of Washington. This research made use
  of the of the NASA/IPAC Infrared Science Archive and the NASA/IPAC
  Extragalactic Database (NED) which are operated by the Jet
  Propulsion Lab- oratory, California Institute of Technology, under
  contract with the National Aeronautics and Space Administration. It
  also made use of the Lyon-Meudon Extragalactic Database (LEDA; part
  of HyperLeda at http://leda.univ-lyon1.fr/).

\bibliography{BHref}

\section*{Appendix A: The effective velocity dispersion for SINFONI galaxies}
\label{app_sigma}

We determined the effective velocity dispersion $\sigma$ for the
SINFONI sample using long-slit or integral field stellar kinematics
and total half-luminosity radii $R_{eT}$ taken from the Hyperleda or RC3 (see
Table \ref{tab_sigma}) for consistency with previous studies, in
combination with the photometry we used for the dynamical modeling.
These radii refer to the galaxy as a whole; the half-luminosity radii
of the bulge component of a galaxy can be much smaller, see
Sect. \ref{sec_data}. The $\sigma$ values given in Table
\ref{tab_data} were obtained by averaging the quantity:
\begin{equation}
v(R)=\sqrt{V(r)^2+\sigma(R)^2},
\label{eq_v}
\end{equation}
out to $R=R_{eT}$, where $V(R)$ and
$\sigma(R)$ are the stellar line-of-sight mean velocity and velocity
dispersion at a distance $R$ from the center:
\begin{equation}
\sigma=\sum_{R\le R_{eT}} v(R) w(R)/\sum_{R\le R_{eT}} w(R)
\label{eq_sigmae}
\end{equation}
We weighted each data point with its light contribution.  When
integral field data were available, this means we set $w(R)=I_c(R)$,
where $I_c(R)$ is the circularized surface brightness at the distance
$R$. When only long-slit data were available, we multiplied $I_c(R)$
by the corresponding circumference, $2\pi R$,
i.e. $w(R)=2\pi R I_c(R)$.  The errors given in Table
\ref{tab_data} for the SINFONI galaxies are the rms of the simple mean of 
the $v(R)$ ($\sigma^s$, obtained
by setting $w(R)=1$ in Eq. \ref{eq_sigmae}),
divided by the square root of the number of points.

In Table \ref{tab_sigma} we also list the following quantities:
\begin{description}
\item[$\sigma_{col}$:] the velocity dispersion derived by fitting 
the spectrum obtained by summing together
the spectra of the SINFONI datacube
\item[$\sigma_{SIN}$:] the velocity dispersion derived by applying 
Eq. \ref{eq_sigmae} to the SINFONI kinematics.
\item[$\sigma_{SIN}^s$:] the velocity dispersion derived by applying 
Eq. \ref{eq_sigmae} with $w(R)=1$ to the SINFONI kinematics.
\item[$\sigma_{e/2}$:] the velocity dispersion derived by applying 
Eq. \ref{eq_sigmae} out to $R_{eT}/2$.
\item[$\sigma_{e/2}^s$:] the velocity dispersion derived by applying 
Eq. \ref{eq_sigmae} with $w(R)=1$ out to $R_{eT}/2$.
\end{description}

\begin{figure*}
  \begin{center}
    \includegraphics[trim=0 4cm 0 2cm,clip,width=16cm]{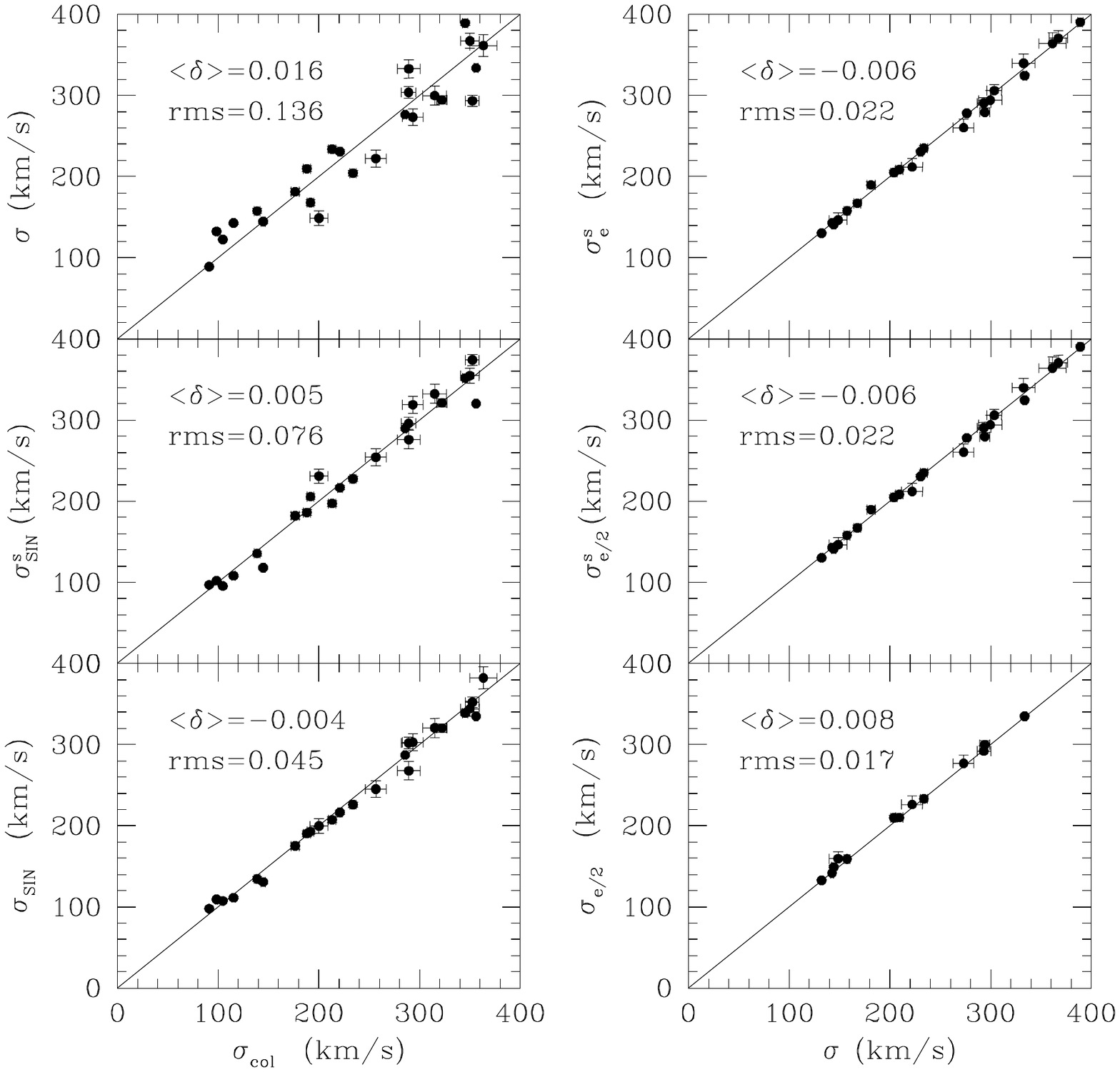}
  \end{center}
  \caption{The comparison between the different estimates of the 
average velocity dispersion of a galaxy and $\sigma$ for the SINFONI sample.
\label{fig_sigma}}
\end{figure*}

Fig. \ref{fig_sigma} compares the quantities listed in Tab.
\ref{tab_sigma} to $\sigma$. On one hand, the velocity dispersion
$\sigma_{col}$ derived from spectrum obtained by summing together the
spectra of the SINFONI datacube reproduces within 5\% $\sigma_{SIN}$
and within 8\% $\sigma_{SIN}^s$ with mean deviations less than 1\%.
So, averaging resolved kinematics according to Eq. \ref{eq_v} is
equivalent to deriving the velocity dispersion from an integrated
spectrum of a galaxy.  On the other hand, $\sigma$ matches the other
estimates $\sigma_{e}^s$, $\sigma_{e/2}^s$, $\sigma_{e/2}$ within less
than 3\% scatter and with mean deviations less than 0.8\%. So neither
the exact choice of the cutoff radius, nor the weighting scheme plays
a big role in the determination of the average velocity
dispersion. Moreover, the average fractional error is 0.02, which
matches the scatter well. Not surprisingly, the scatter obtained
comparing $\sigma_{col}$ to $\sigma$ is much larger (14\% with a mean
difference of 2\%), as $\sigma_{col}$ probes the inner regions of the
galaxies, where the presence of the supermassive black hole and/or the
influence of a compact classical bulge becomes dynamically important.

\citet{KormendyHo2013} give discrepant velocity dispersions for three
galaxies of our SINFONI sample. For NGC 1332 they have 328 km/s
\citep[from our previous determination in][]{Rusli2011}, that we now
revise to 293.1 km/s (see Appendix B). For NGC 4486a they quote 111
km/s while we get 144.5 km/s using the profiles of
\citet{Prugniel2011}.  For NGC 4486b their value is larger (185 km/s
compared to 148.6 km/s), but the combination of a steep surface brighness 
and velocity dispersion radial gradient makes the measurement difficult
(see Appendix B).

\begin{table*}
\caption{Velocity dispersions of the SINFONI sample.\label{tab_sigma}}
\begin{tabular}{lllllllll}
\hline
\hline
Galaxy & $R_{eT}$ & $R_{max}^{kin}/R_{eT}$ & $\sigma_{col}$ & $\sigma_{SIN}$ & $\sigma_{SIN}^s$ & $\sigma_{e/2}$ & $\sigma_{e/2}^s$ & $\sigma_e^s$\\
       & $('')$ & - & $(km s^{-1})$ &  $(km s^{-1})$ &   $(km s^{-1})$ &  $(km s^{-1})$ &  $(km s^{-1})$ &  $(km s^{-1})$ \\
\hline
     NGC0307           &       4.755           &       1.942           &       233.9           &       225.9           &       227.6           &       209.9           &       204.9           &       204.9          \\
     NGC1316           &       80.75           &      0.6128           &         221           &       216.4           &       216.5           &       230.6           &       230.4           &       230.4          \\
     NGC1332           &          28           &      0.7178           &       352.4           &       352.3           &       373.9           &         292           &       290.4           &       290.4          \\
     NGC1374           &       24.38           &      0.4825           &       191.7           &       192.9           &       205.6           &       167.8           &         167           &         167          \\
     NGC1398           &       52.13           &       0.693           &       213.1           &       207.3           &       197.1           &       233.2           &         235           &         235          \\
     NGC1407           &       70.33           &      0.5436           &       285.6           &       287.2           &       289.4           &       276.4           &       277.9           &       277.9          \\
     NGC1550           &       25.53           &       0.965           &       293.3           &       302.8           &       318.8           &       276.9           &       260.2           &       260.2          \\
     NGC3091           &       32.89           &      0.3684           &         315           &       320.5           &       332.3           &       299.7           &       294.1           &       294.1          \\
     NGC3368           &       73.64           &       1.058           &        98.5           &       109.2           &         102           &       132.8           &       130.2           &       130.2          \\
     NGC3489           &        20.3           &           -           &        91.1           &       97.72           &       96.85           &          -           &          -           &          -          \\
     NGC3627           &       67.16           &      0.4754           &       104.7           &       107.3           &       95.43           &       122.4           &       122.4           &       122.4          \\
     NGC3923           &       49.79           &      0.7626           &       256.6           &       245.2           &       254.1           &       226.3           &       211.8           &       211.8          \\
     NGC4371           &       23.29           &       1.099           &       115.4           &       111.3           &       108.1           &       141.5           &       142.8           &       142.8          \\
     NGC4472           &       225.5           &      0.3036           &       289.1           &       302.1           &       295.9           &       303.7           &       305.9           &       305.9          \\
    NGC4486a           &       5.459           &       1.952           &       144.8           &       130.7           &         118           &       149.1           &       141.1           &       141.1          \\
    NGC4486b           &       2.495           &         1.2           &       200.1           &       199.8           &       230.8           &       159.4           &       146.3           &       146.3          \\
     NGC4501           &       77.11           &      0.8222           &       138.7           &       134.5           &       135.5           &       159.1           &       157.8           &       157.8          \\
     NGC4699           &        30.7           &      0.5416           &       176.5           &         175           &         182           &       181.3           &       189.6           &       189.6          \\
     NGC4751           &       22.76           &      0.3563           &       363.3           &       382.3           &       416.7           &       361.4           &       363.9           &       363.9          \\
     NGC5018           &       22.76           &      0.9905           &       188.1           &       190.3           &         186           &       210.1           &       208.4           &       208.4          \\
     NGC5328           &       22.24           &        1.27           &       356.1           &         335           &       320.1           &         335           &       324.3           &       324.3          \\
     NGC5419           &       43.36           &      0.3201           &       349.9           &       344.1           &       354.6           &       367.2           &       370.4           &       370.4          \\
     NGC5516           &       22.09           &      0.3944           &       289.3           &       267.7           &       275.7           &       332.7           &       339.7           &       339.7          \\
     NGC6861           &       17.67           &      0.4878           &       345.3           &       338.8           &       351.7           &         389           &       390.2           &       390.2          \\
     NGC7619           &       36.91           &       1.837           &       322.2           &       320.3           &       321.1           &       299.9           &       279.3           &       279.3          \\

\hline
\multicolumn{9}{l}{Column 1: galaxy name; column 2: total half-luminosity radius $R_{e,T}$; column 3: ratio between 
$R_{max}^{kin}$ and $R_{e,T}$, where $R_{max}^{kin}$}\\
\multicolumn{9}{l}{is the distance from the center of the most distant
available stellar kinematic point; column 4 to 9: see text in Appendix A.}\\
\end{tabular}
\end{table*}

\section*{Appendix B: The  luminosity profiles and the bulge mass-to-light ratios.}
\label{app_ML}

Many of the profiles come from the ESO Key Programme described in
\citet[][hereafter KeyProg]{Scorza1998} or were derived from images
from the Sloan Digital Sky Survey \citep[][hereafter
SDSS]{York2000}. We corrected luminosity profiles and colors for
Galactic extinction (hereafter G.E.) following \citet{Schlegel1998};
however we quote the results of the fits in the Tables
\ref{tab_maxfits} to \ref{tab:7582decomp} without correction for
G.E.. The zero-points of the profiles derived from SDSS images and
corrected for G.E. are moved to the Johnson-Cousin bands when
necessary using the equations of \citet{Jordi2006} and extinction
corrected colors.  Most of the colors used in the conversions from one
band to the other come from the Hyperleda database
\citep{Paturel2003}.  For three galaxies we also use the colors of the
Simple Stellar Populations (SSP) of \citet{Maraston2005}.  We discuss
how we addressed these cases and how we obtained the bulge luminosity
profiles (which coincide with the total luminosity profile in the case
of elliptical galaxies) below.

For all the galaxies we managed either to collect the dynamically
determined bulge mass-to-light ratios from the literature or to
compute our own dynamical estimate by fitting available stellar
kinematics (see Appendix C). The ratios and their sources are quoted in Table \ref{tab_ml}. In
most of the cases the literature values do not include a correction
for G.E., which we apply here following
\citet{Schlegel1998}. Exceptions are \citet{Saglia1993,Saglia2000},
\citet{Moellenhoff1995},
\citet{Kronawitter2000}, \citet{Bower2001}, \citet{Cappellari2006},
\citet{DallaBonta2009}, \citet{Krajnovic2009}, \citet{Rusli2013b} who
already published $M/L$ values corrected for G.E. The
photometric band for which the $M/L$ values were computed does not
always coincide with the photometic band we used to derive the bulge
luminosity profile (see Table \ref{tab_ml}). Finally, we adjust distances
by adopting  the ones given by
\citet{KormendyHo2013}, or \citet{Sani2011} for the remaining objects. 

In general we scale the $M/L$ ratios to our
distances and bands using the equation
\begin{equation}
\begin{array}{ll}
M/L_{us}= & M/L_{lit}\times \frac{D_{lit}}{D_{us}}\times 10^{0.4(m_{us}-m_{lit})}\\
         & \times 10^{0.4(m_{\odot,lit}-m_{\odot,us})}.\\
\end{array}
\label{eq_MLtrans}
\end{equation}
The solar magnitudes used below are listed in Table \ref{tab_solmag}.
The original $M/L$ values and distances, together with colors and
G.E. values used to transform or compute the $M/L$s are
given in Table \ref{tab_MLorig}.

\begin{figure*}
  \begin{center}
    \includegraphics[trim=0 1cm 0 3cm,clip,width=15cm]{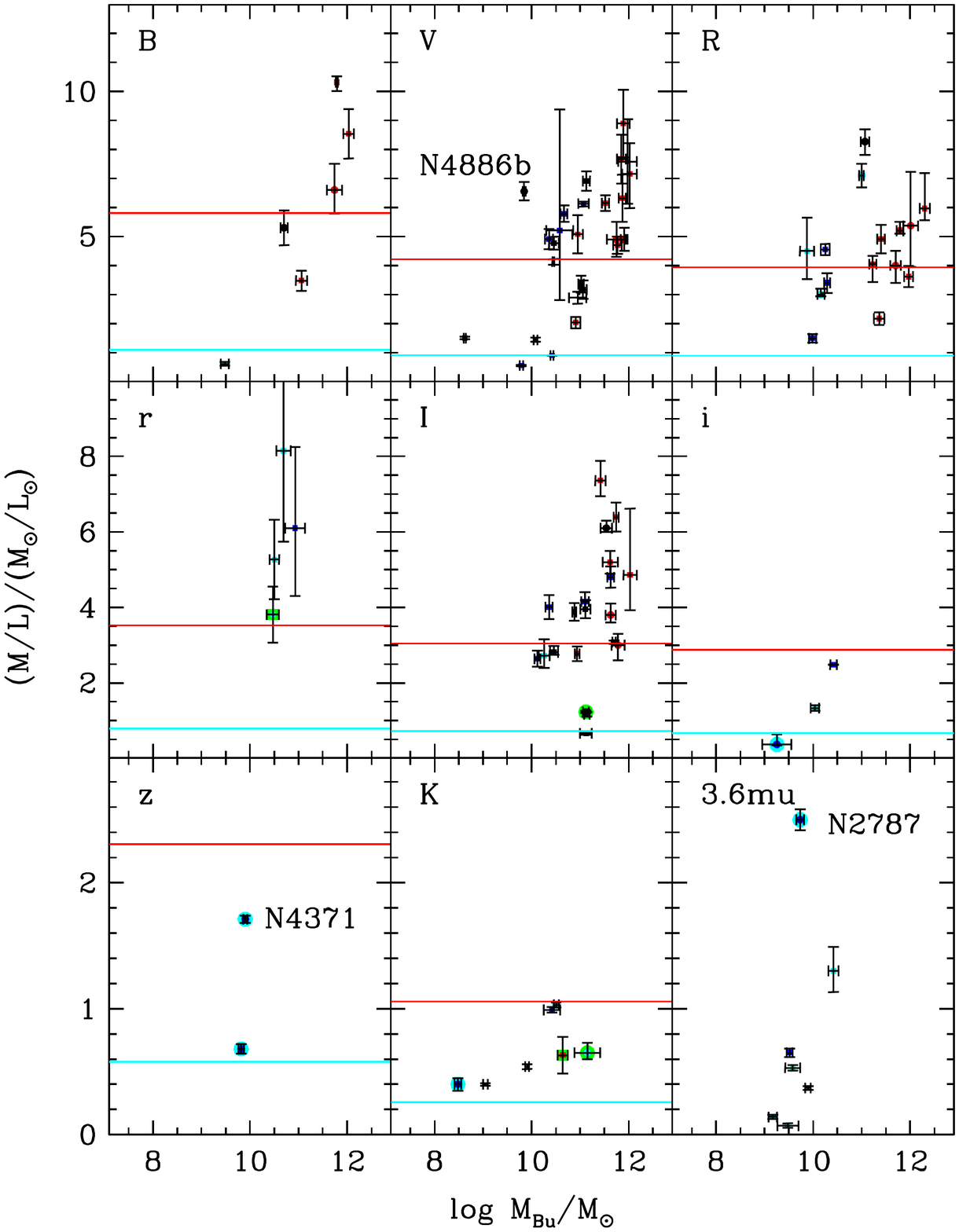}
  \end{center}
  \caption{The $M/L$ ratios in the 9 bands with more than one galaxy as
    a function of the dynamical bulge mass. Colors and point types as in
    Fig. \ref{fig_Sigma}. The lines show the values predicted by the
    simple stellar population models of \citet{Maraston2005} for a
    Kroupa IMF, solar metallicity and age of 2 (cyan) and 12 (red)
    Gyr.\label{fig_ML}}
\end{figure*}

Fig. \ref{fig_ML} shows the derived $M/L$ ratios in the 9 bands with
more than one galaxy as a function of the dynamical bulge mass. In general, they
increase with bulge mass; a comparison with the models of
\citet{Maraston2005} shows that simple stellar populations with a Kroupa
IMF, solar metallicity and ages ranging from 2 to 12 Gyr roughly
bracket the observed range.

\begin{table}
\caption{The solar magnitudes used in the conversion of the $M/L$.}
\label{tab_solmag}
\begin{tabular}{ll}
 \hline
\hline
$m_{\odot,B}$ & 5.48\\
$m_{\odot,V}$ & 4.83\\
$m_{\odot,g}$ & 5.36\\
$m_{\odot,R}$ & 4.42\\
$m_{\odot,r}$ & 4.67\\
$m_{\odot,I}$ & 4.08\\
$m_{\odot,i}$ & 4.57\\
$m_{\odot,z}$ & 4.52\\
$m_{\odot,zACS}$ & 3.98\\
$m_{\odot,H}$ & 3.32\\
$m_{\odot,K}$ & 3.28\\
$m_{\odot,3.6mu}$ & 3.24\\
\hline
\end{tabular}
\end{table}

\begin{deluxetable}{ll}
\tablecaption{The original $M/L$ and distances, with colors and galactic extinction (G.E.) 
values used in the $M/L$ conversions for the literature sample.\label{tab_MLorig}}
\tablewidth{0pt}
\tablehead{\colhead{Galaxy} & \colhead{Comment}}
\startdata
Circinus & $M/L_{3.6mu}$ derived fitting the stellar kinematics of \citet{Maiolino1998}, $A_{3.6mu}=0.107$\\
A1836	 & $M/L_I=5 M_\odot/L_\odot$, D=147.2 Mpc from \citet{DallaBonta2009}, G.E. corrected\\
         & $A_i=0.134,A_r=0.176,(r-i)=0.482$,\\
         & $(I-i)=-0.247\times[(r-i)_0+0.236]/1.007$\\
IC1459	 & $M/L_R=4.2 M_\odot/L_\odot$, D=29.2 Mpc from \citet{Haering2004}, not corrected for G.E.\\
 	 & $A_R=0.042,A_V=0.053, (V-R)=0.621$\\
IC2560	 & $M/L_{3.6mu}$ derived fitting the central $\sigma$ of \citet{Greene2010}, $A_{3.6mu}=0.015$\\
IC4296	 & $M/L_B=5.6 M_\odot/L_\odot$, D=75 Mpc from \citet{Saglia1993}, G.E. corrected\\
         & $A_B=0.265,A_I=0.119,(B-I)=2.41$\\
NGC0221	 & $M/L_V=2.16 M_\odot/L_\odot$, D=0.8 Mpc from \citet{Magorrian1998}, not corrected for G.E.\\
	 & $A_V=0.206,A_R=0.166, (V-R)=0.641$\\
NGC0224	 & $M/L_V=4.83 M_\odot/L_\odot$, D=0.8 Mpc from \citet{Magorrian1998}, \\
         & not corrected for G.E., $A_V=0.206$\\
NGC0524	 & $M/L_I=4.99 M_\odot/L_\odot$, D=23.3 Mpc from \citet{Cappellari2006}, G.E. corrected\\
NGC0821	 & $M/L_I=3.08 M_\odot/L_\odot$, D=23.44 Mpc from \citet{Cappellari2006}, G.E. corrected\\
	 & $A_R=0.294,A_I=0.213,(R-I)=0.6923$\\
NGC1023	 & $M/L_V=5.56 M_\odot/L_\odot$, D=10.2 Mpc from \citet{Bower2001}, G.E. corrected\\
	 & $A_V=0.201,A_{3.6mu}=0.009,(V-3.6mu)=3.695$\\
NGC1068	 & $M/L_K$ derived fitting the stellar kinematics of \citet{Emsellem2006}, $A_{K}=0.012$\\
NGC1194	 & $M/L_r$ derived fitting the stellar kinematics of \citet{Greene2010}, $A_{r}=0.21$\\
NGC1300	 & $M/L_{F606W}=2.29 M_\odot/L_\odot$, D=18.8 Mpc from \citet{Atkinson2005}, not corrected \\
         & for G.E., $A_V=0.1,A_R=0.081,(V-F606W)=0.36$\\
NGC1399	 & $M/L_B=10.2 M_\odot/L_\odot$, D=21.1 Mpc from \citet{Kronawitter2000}, G.E. corrected\\
NGC2273	 & $M/L_R$ derived fitting the stellar kinematics of \citet{Greene2010}, $A_{R}=0.189$\\
NGC2549	 & $M/L_R$ from \citet{Krajnovic2009}, G.E. corrected\\
         & $A_V=0.22, A_R=0.175, (V-R)_{SV}=0.567, (R_{SV}-r)=-0.267\times(V-R)_{SV}-0.088$\\
         & $(R-R_{SV})=0.27\times(R-V)_{SV}-0.22$\\
NGC2748	 & $M/L_{3.6mu}$ derived fitting the stellar kinematics of \citet{Batcheldor2005}, $A_{3.6mu}=0.004$\\
NGC2778	 & $M/L_V=8 M_\odot/L_\odot$, D=22.9 Mpc from \citet{Gebhardt2003}, not corrected for G.E.\\
         & $A_V=0.069, (V-R)=0.643, (B-V)=0.958$, \\
         & $(V-r)_0=-0.63(B-V)+1.646(V-R)+0.124-A_V$\\
NGC2787	 & $M/L_{3.6mu}$ derived fitting the stellar kinematics of \citet{Bertola1995}, $A_{3.6mu}=0.020$\\
NGC2960	 & $M/L_r$ derived fitting the stellar kinematics of \citet{Greene2010}, $A_{r}=0.123$\\		
NGC2974  & $M/L_I=4.52  M_\odot/L_\odot$, D=20.89 Mpc from \citet{Cappellari2006}, corrected for G.E.\\
         & $A_I=0.106 , A_{3.6mu}=0.008, (I-3.6mu)=2.374$\\	
NGC3031	 & $M/L_i$ derived fitting the stellar kinematics of \citet{Fabricius2012}, $A_i=0.167$\\						
NGC3079	 & $M/L_{3.6mu}$ derived fitting the stellar kinematics of \citet{Shaw1993}, $A_{3.6mu}=0.002$\\
NGC3115	 & $M/L_V=8.04 M_\odot/L_\odot$, D=8.4 Mpc from \citet{Magorrian1998}, not corrected for G.E., \\
         & $A_V=0.157$\\
NGC3227	 & $M/L_{Kp}$ from  \citet{Davies2006}, G.E. corrected, narrow band definition\\	
NGC3245	 & $M/L_R=3.7 M_\odot/L_\odot$, D=20.9 Mpc from \citet{Haering2004}, not corrected for G.E. \\
         & $A_B=0.108, A_V=0.083, A_R=0.067, A_i=0.052, (B-V)=0.8367$, \\
         & $(V-R)_{SSP}=0.508, (R-I)_{SSP}=0.48,$\\
         & $ (R-i)=1.007\times(R-I)_{SSP}-0.267\times(V-R)_{SSP}-0.236-0.088$\\	
NGC3377	 & $M/L_I=2.22 M_\odot/L_\odot$, D=10.91 Mpc from \citet{Cappellari2006}, G.E. corrected\\
         & $A_R=0.091, A_I=0.066, (R-I)=0.629$\\	
NGC3379	 & $M/L_I=2.8 M_\odot/L_\odot$, D=10.57 Mpc, from \citet{Cappellari2006}, G.E. corrected\\
NGC3384	 & $M/L_V=2.2 M_\odot/L_\odot$, D=11.7 Mpc from \citet{Schulze2011}, not corrected\\
         &  for G.E., $A_V=0.088,	A_R=0.071, A_I=	0.052, A_i=0.052, (R-I)=0.624$,\\ 
         & $(V-I)=1.18, (I-i)=-0.247\times(R-I)_0-0.329$\\
NGC3393	 & $M/L_I$ fitting the stellar velocity dispersion of \citet{Greene2010}, $A_I=0.146$\\
NGC3414	 & $M/L_I=4.26 M_\odot/L_\odot$, D=24.55 Mpc from \citet{Cappellari2006}, G.E. corrected\\
	 & $A_r=0.067,	A_i=0.051, (r-i)=0.419$, $(R-I)=[0.236+(r-i)_0]/1.007$,\\ 
         & $(i-I)=-0.247\times(R-I)-0.329$\\
NGC3585	 & $M/L_V=3.4 M_\odot/L_\odot$, D=21.2 Mpc from \citet{Gueltekin2009a}, not corrected for G.E.,\\
         & $A_V=0.212$\\
NGC3607	 & $M/L_V=7.5 M_\odot/L_\odot$, D=19.9 Mpc from \citet{Gueltekin2009a}, not corrected for G.E.\\
	 & $A_B=0.09, A_V=0.069, A_g=0.079, (B-V)=0.921$,\\ 
         & $(V-g)=-0.63\times(B-V)_0+0.124$\\
NGC3608	 & $M/L_V=3.1 M_\odot/L_\odot$, D=23 Mpc from \citet{Rusli2013b}, G.E. corrected\\
         & $A_V=0.069, A_I=0.041, (V-I)=1.24$\\ 
         & $A_R=0.056, A_I=0.041, A_i=0.044, (R-I)=0.608$,\\ 
         & $(I-i)=-0.247\times(R-I)_0-0.329$\\
NGC3842	 & $M/L_V$ from \citet{Rusli2013b}, G.E. corrected\\																		
NGC3998	 & $M/L_V=6.5 M_\odot/L_\odot$, D=17 Mpc from \citet{deFrancesco2006}, not corrected for G.E.\\
	 & $A_B=0.069,	A_V=0.053, A_I=	0.031, A_i=0.034, (B-V)=0.966$,\\ 
         & $(V-I)_{SSP}=1.135, (R-I)_{SSP}=0.555, (I-i)=-0.247\times(R-I)_{SSP}-0.329$\\					
NGC4026	 & $M/L_V=4.89 M_\odot/L_\odot$ from \citet{Gueltekin2009a}, not corrected for G.E.\\
         & $A_B=0.095, A_V=0.073, A_g=0.084, (B-V)=0.962$,\\
         & $(V-g)=-0.63\times(B-V)_0+0.124$\\
NGC4151	 & $M/L_R=1.4 M_\odot/L_\odot$, D=13.9 Mpc from \citet{Onken2007}, not corrected for G.E.,\\
         & $A_R=0.074$\\
NGC4258	 & $M/L_V=3.6 M_\odot/L_\odot$, D=7.28 Mpc from \citet{Haering2004}, not corrected for G.E. \\
         & $A_V=0.053,	A_{3.6mu}=0.002,	(V-3.6mu)=3.629$\\
NGC4261	 & $M/L_V=9.1 M_\odot/L_\odot$, D=31.6 Mpc from \citet{Rusli2013b}, G.E. corrected\\
NGC4291	 & $M/L_V=5.4 M_\odot/L_\odot$, D=25 Mpc from \citet{Rusli2013b}, G.E. corrected\\
NGC4342	 & $M/L_I=6.3 M_\odot/L_\odot$, D=15 Mpc from \citet{Cappellari2006}, G.E. corrected\\
         & $A_r=0.056, A_i=0.043, A_I=0.04, (r-i)=0.426, (R-I)=[(r-i)_0+0.236)]/1.007$,\\
         & $(I-i)=-0.247\times(R-I)-0.329$\\									
NGC4374	 & $M/L_I=4.36 M_\odot/L_\odot$, D=17.86 Mpc from \citet{Cappellari2006}, G.E. corrected\\
         & $A_V=0.134,	A_I=0.078, (V-I)=1.26$\\
NGC4388	 & $M/L_{3.6mu} M_\odot/L_\odot$ fitting the stellar velocity dispersion \citet{Greene2010},\\
         &  $A_{3.6mu}=0.005$\\
NGC4459	 & $M/L_I=2.51 M_\odot/L_\odot$, D=15.7 Mpc from \citet{Cappellari2006}, G.E. corrected\\
         & $A_V=0.153, A_I=0.09, (V-I)=1.306$\\
NGC4473	 & $M/L_V=6.8 M_\odot/L_\odot$, D=17 Mpc from \citet{Schulze2011}, not corrected for G.E.,\\
         & $A_V=0.094$\\	
NGC4486	 & $M/L_V=6.3 M_\odot/L_\odot$, D=17.9 Mpc from \citet{Gebhardt2009}, not corrected\\ 
         & $ for G.E., A_V=0.074$\\
NGC4526  & $M/L_I=2.65 M_\odot/L_\odot$, D=16.5 Mpc from \citet{Davis2013}, corrected for G.E.\\
         & $A_I=0.043, A_{3.6mu}=0.003, (I-3.6mu)=2.47$\\
NGC4552	 & $M/L_V=7.1 M_\odot/L_\odot$, D=15.85 Mpc from \citet{Rusli2013b}, G.E. corrected\\
NGC4564	 & $M/L_V$ fitting the stellar kinematics of \citet{BSG1994}, $A_V=0.116$\\
NGC4594	 & $M/L_I=3.4 M_\odot/L_\odot$, D=9.8 Mpc from \citet{Jardel2011}, not corrected for G.E.\\
         &  $A_I=0.099$\\
NGC4596	 & $M/L_K M_\odot/L_\odot$ fitting the stellar kinematics of \citet{Bettoni1997}, $A_K=0.008$\\
NGC4621	 & $M/L_I=3.03 M_\odot/L_\odot$, D=17.78 Mpc from \citet{Cappellari2006}, G.E. corrected\\
         & $A_I=0.064,	A_{3.6mu}=0.005,	(I-3.6mu)=2.515$\\
NGC4649	 & $M/L_V=7.3 M_\odot/L_\odot$,	D=17.3 Mpc from \citet{Rusli2013b}, G.E. corrected\\
NGC4697	 & $M/L_V=4.3 M_\odot/L_\odot$, D=12.4 Mpc from \citet{Schulze2011}, not corrected \\
         & for G.E., $A_V=0.101, A_R=0.081, (V-R)=0.59$\\
NGC4736	 & $M/L_B=1.8 M_\odot/L_\odot$, D=6.6 Mpc from \citet{Moellenhoff1995}, G.E. corrected\\
         & $A_B=0.076, A_V=0.059, A_R=0.047,A_I=0.034, A_z=0.026, (B-V)=0.9$, \\          
         & $(V-R)=0.84, (R-I)=0.74, (g-r)=1.646\times(V-R)_0-0.139$,\\
         & $(g-B)=-0.37\times(B-V)_0-0.124, (r-z)=1.586\times(R-I)_0-0.386$,\\ 
         & $(z-B)=(g-B)-(g-r)-(r-z)$\\
NGC4826	 & $M/L_i$ fitting the stellar kinematics of \citet{Heraudeau1998}, $A_i=0.086$\\
NGC4889	 & $M/L_r=6.025M_\odot/L_\odot$, D=103.2 Mpc from \citet{McConnell2012}, not corrected\\
         &  for G.E., $A_R=0.026, A_V=0.032, (V-R)=0.724$,\\
         &  $(r-R)=-0.267\times(V-R)_0-0.088$\\
NGC5077	 & $M/L_B=4 M_\odot/L_\odot$, D=56 Mpc from \citet{Pizzella1997}, not corrected for G.E.\\
         & $A_B=0.21,	A_V=0.161, (B-V)=1.04$\\
NGC5128	 & $M/L_K=0.65 M_\odot/L_\odot$, D=3.5 Mpc from \citet{Cappellari2009}, G.E. corrected \\
NGC5576	 & $M/L_V=3.7 M_\odot/L_\odot$, D=27.1 Mpc from \citet{Gueltekin2009a}, not corrected for G.E.\\
	 & $A_V=0.104, A_R=0.084, (V-R)=0.553$\\
NGC5813	 & $M/L_V=4.7 M_\odot/L_\odot$, D=32.2 Mpc from \citet{Rusli2013b}, G.E. corrected\\
NGC5845	 & $M/L_V=5.1  M_\odot/L_\odot$, D=28.7 Mpc from \citet{Schulze2011}, \\
         & not corrected for G.E., $A_V=0.177$\\
NGC5846	 & $M/L_I=5.2 M_\odot/L_\odot$, D=24.9 Mpc from \citet{Rusli2013b}, G.E. corrected\\
NGC6086	 & $M/L_R=4.2 M_\odot/L_\odot$,	D=133 Mpc from  \citet{Rusli2013b}, G.E. corrected\\
NGC6251	 & $M/L_R=6  M_\odot/L_\odot$,	D=106 Mpc from \citet{Haering2004}, not corrected for G.E.\\
         & $A_V=0.29, A_R=0.234, A_I=0.17, (R-I)=0.626, (R-I)_{SSP}=0.643,$\\
         & $(V-I)=1.408$\\
NGC6264	 & $M/L_r$ fitting \citet{Greene2010} minor axis velocity dispersion, $A_r=0.178$\\
NGC6323	 & $M/L_r$ fitting \citet{Greene2010} velocity dispersion, $A_r=0.047$\\
NGC7052	 & $M/L_R=3.5  M_\odot/L_\odot$, D= 58.7 Mpc from \citet{Haering2004}, not corrected for G.E.\\ 
         & $A_R=0.324$\\
NGC7457	 & $M/L_{3.6mu}$ fitting the stellar kinematics of \citet{Emsellem2004}, $A_{3.6mu}=0.008$\\
NGC7582	 & $M/L_{3.6mu}$ fitting the stellar kinematics of \citet{Oliva1995}, $A_{3.6mu}=0.002$\\
NGC7768	 & $M/L_V=7.8 M_\odot/L_\odot$, D= 112.8 Mpc from \citet{Rusli2013b}, G.E. corrected\\	
UGC3789	 & $M/L_H$ fitting the stellar velocity dispersion of \citet{Greene2010}, $A_H=0.037$\\

\enddata
\end{deluxetable}

In the following two
subsections we discuss in detail the galaxies with black hole masses
from the literature and from our SINFONI survey.

\section*{Literature Sample}

The surface brightness values reported in the Tables \ref{tab_maxfits} 
to \ref{tab:7582decomp} are in the Vega system, when Johnson and the $3.6mu$ filters are used, 
or in the AB system, when SDSS filters are used. 

\begin{figure*}
  \begin{center}
    \includegraphics[trim=0 3cm 0 3cm,clip,width=8cm]{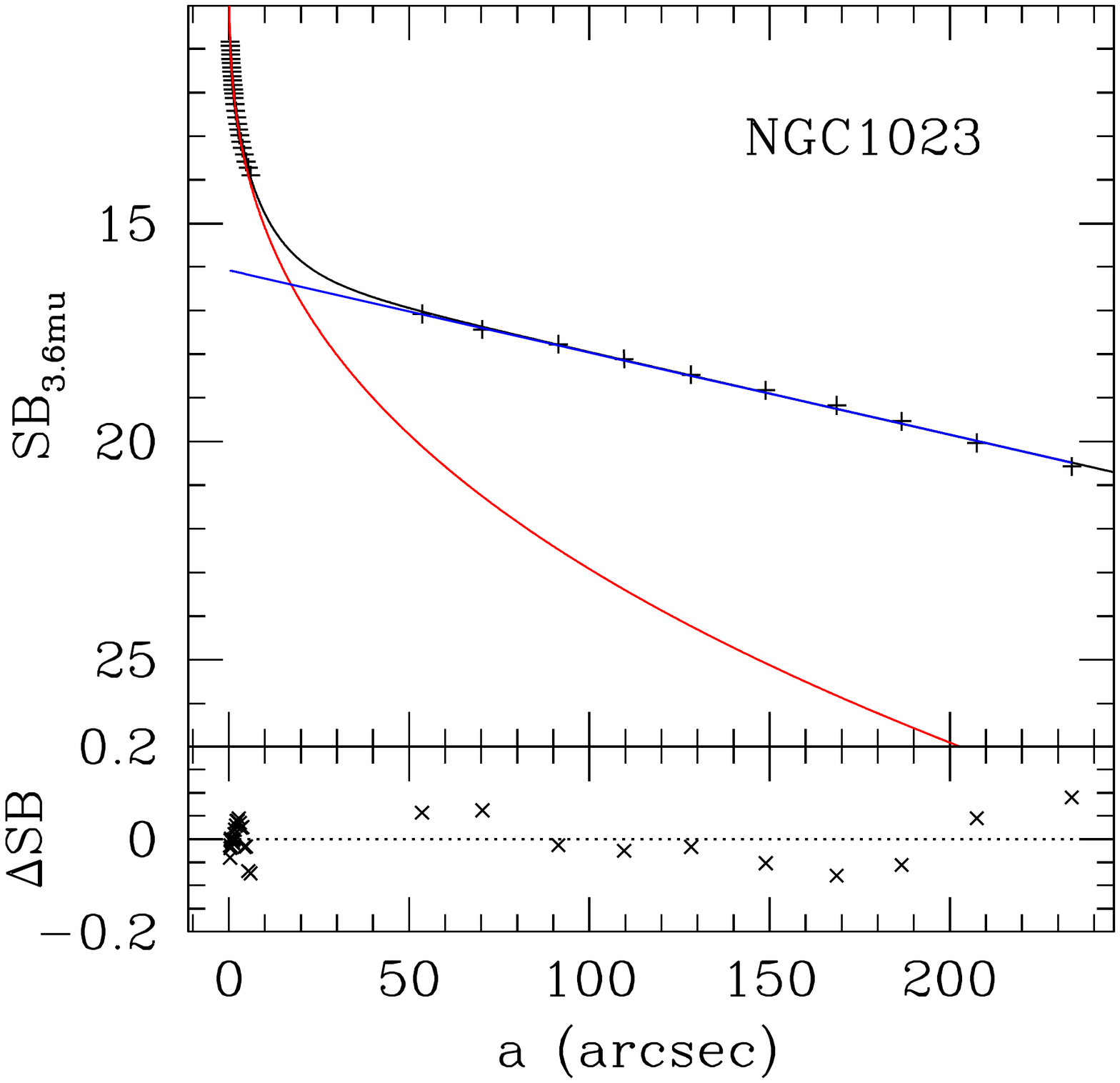}\\
    \includegraphics[trim=0 3cm 0 3cm,clip,width=8cm]{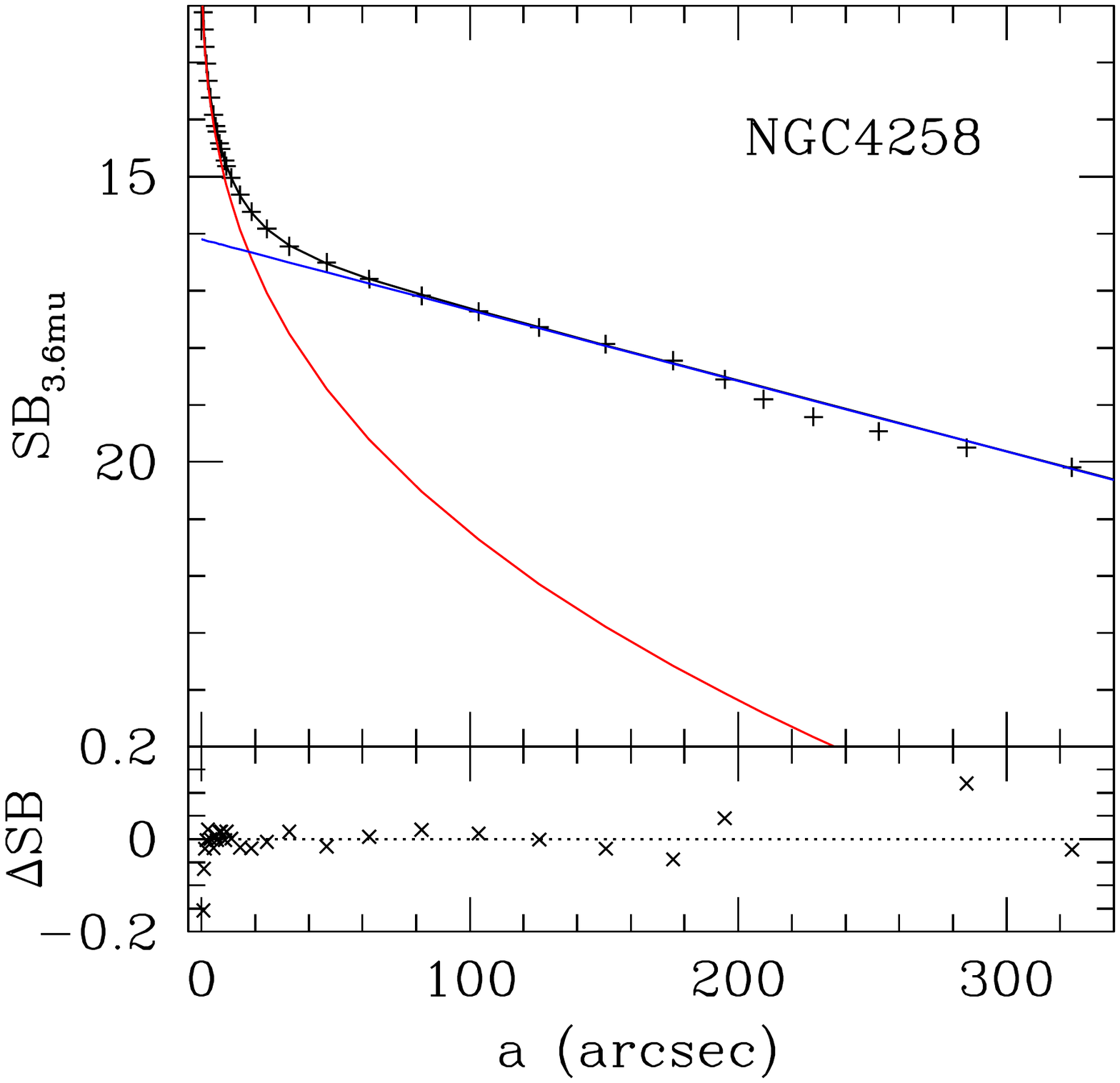}
    \includegraphics[trim=0 3cm 0 3cm,clip,width=8cm]{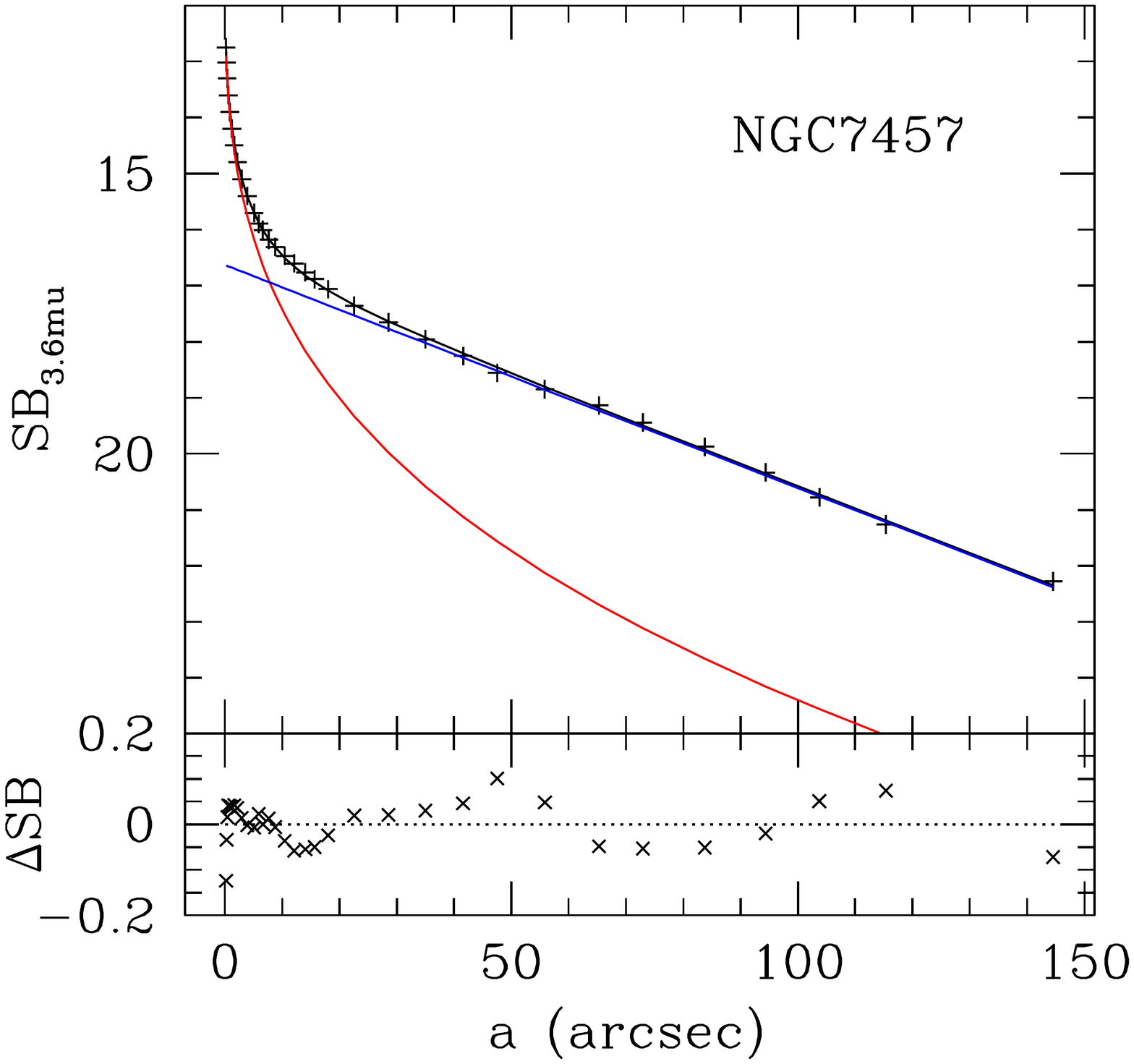}\\
 \end{center}
  \caption{The major-axis bulge plus disk fits to the  Spitzer $3.6mu$ profiles of NGC 1023, NGC 4258 and NGC 7457 (see text and Table \ref{tab_maxfits}).  
\label{fig_maxfits}}
\end{figure*}

\begin{description}
\item[Milky Way:] We adopt the axisymmetric bulge density profile
    of \citet{McMillan2011} scaled as follows. \citet{Portail2015}
    compute the total mass in the bulge volume $(\pm 2.2 \times \pm1.4
    \times \pm 1.2)$ kpc to be $1.84\times 10^{10} M_\odot$. A sphere
    with a 1.92 kpc radius has the same volume. Using the spherically
    averaged density profile of \citet{McMillan2011} we get a mass
    within this radius of $7.2\times 10^9 M_\odot$, so we scale up
    this profile by a factor 2.54.

\item[Circinus:] The bulge photometry in $3.6mu$-band is taken from
  the decomposition published by \citet{Sani2011}.  The
  $M/L$ is derived as in Appendix C and Fig. \ref{fig_kinprofile}
  to match the kinematic profiles of \citet{Maiolino1998}. 
We follow \citet{KormendyHo2013} in
  classifying the galaxy as barred.

\item[A1836:]  The $M/L$ is scaled to the distance of
    \citet{KormendyHo2013} and the $i-$band image is calibrated to
  the Cousins $I$ using the $(r-i)$ color and the equations in
  \citet{Jordi2006}. \citet{McConnellMa2013} and
  \citet{KormendyHo2013} consider the galaxy a core elliptical; we
  accept their classification, although strong nuclear dust makes it
  difficult to determine its core properties \citep[see discussion
  in][]{Rusli2013b}.

\item[IC 1459:] We correct the $M/L$ for G.E., scaled it from $R$ to
  $V-$band using the $(V-R)$ color from Hyperleda and adjust it to the
  distance of \citet{KormendyHo2013}. The galaxy is a core elliptical
  \citep{Rusli2013b}, despite the classification of
    \citet{KormendyHo2013}.  We use the profile of \citet{Rusli2013b}
  and derive $R_e=18.6''$, while their $n=7.6$ Core-Sersic fit profile
  has $R_e=45.4''$ (see Fig. \ref{fig_extrap}).  The difference is
  driven by the extrapolation.  The same applies to the result of
  \citet{Sani2011} who measure $R_e=53.3''$, and \citet{Laesker2014},
  who quote $R_e=38.4''$ for the 'classical' fit and $R_e=51.9''$ for
  the best fit.


\item[IC 4296:] \citet{McConnellMa2013} and \citet{KormendyHo2013}
  consider the galaxy a core elliptical; we accept their
  classification, although strong nuclear dust makes it difficult to
  determine its core properties \citep[see discussion
  in][]{Rusli2013b}. We measure the profile from $I-$band FORS images
  that we calibrate to Cousins $I$ using the photometry of
  \citet{Prugniel1998} and finally to the $B$ band using the mean
  $(B-I)=2.41$ color derived from \citet{Goudfrooij1994} and
  $A_B=0.265$. From this profile we measure $R_e=68.1''$, in agreement
  with \citet[][$R_e=70.9''$]{Laesker2014} 'best fit', but disagreeing
  with their 'classical fit' ($R_e=28.9''$) or the value reported by
  \citet[$32.3''$]{Sani2011}. The $M/L$ (from the model with dark
  matter and already corrected for G.E.) is scaled to the distance of
  \citet{KormendyHo2013}.

\item[NGC 221:] The $M/L$ is scaled from $V$ to $R-$band using the $(V-R)$ color
  from Hyperleda and to the distance of \citet{KormendyHo2013}.  

\item[NGC 224:] The $M/L$ is scaled to the distance of \citet{KormendyHo2013}. 
 We classify it as barred following \citet{Athanassoula2006}.

\item[NGC 524:] The $M/L$ is scaled to the distance of
  \citet{KormendyHo2013}.  We classify this galaxy with $T=2$ (a
    classical bulge). Although the galaxy is circular, significant
    stellar rotation is nonetheless seen
    \citep{Simien2000,Emsellem2004}; therefore the S0 classification is
    dubious and the galaxy could be a core elliptical
    \citep{McConnellMa2013}.  We deproject the multi-gaussian
  expansion of the photometry profile as given in
  \citet{Cappellari2006} and get $R_e=47.1''$.  We ignore the possible
  small disk detected by \citet{Sani2011} ($B/T\approx
  0.83$); \citet{KormendyHo2013} give $B/T=0.92$. Finally, we conclude that the bulge plus disk decomposition
  with $B/T=0.28$ and $R_e=8.9''$ of \citet{Laurikainen2010} does not
  describe the nature of the galaxy.

  \item[NGC 821:] \citet{KormendyHo2013} quote $B/T=0.95$, therefore
    we ignore the possible disk component and use the whole galaxy
    profile \citep{Graham2001}, getting
    $R_e=34.8''$. \citet{Beifiori2012} report $R_e=85.4''$ and
    \citet{Laesker2014} 'best fit' gives $3.5''$. The fit of
    \citet{Beifiori2012} overestimates the measured profile at radii
    larger than $70''$ by 0.4 mag. \citet{Laesker2014} consider a
    bulge, a (faint) disk and a halo and it is not clear that bulge
    and halo should be treated as separate components. The $M/L$ is
    scaled to the distance of \citet{KormendyHo2013} and from $I$ to
    $R-$band using the mean $(R-I)$ color from Hyperleda.

\item[NGC 1023:] The $M/L$ of \citet{Bower2001} is already corrected 
for G.E.; we scale it to the distance of \citet{KormendyHo2013}
  and calibrate it from the $V$-band to $3.6mu$ using the $V$-band aperture
  photometry from Hyperleda. The Spitzer images come from Program 69,
  PI Giovanni Fazio, Observer ID 2. We perform a Sersic bulge plus
  exponential disk decomposition along the major axis of the galaxy
  following \citet{Fisher2008} and masking the region where the bar is
  present (see Fig. \ref{fig_maxfits} and Table \ref{tab_maxfits}).
  This gives $R_e=7.5''$, which agrees with \citet{Fisher2010}. The
  2-dimensional fit (including a bar) of \citet{Sani2011} gives
  $R_e=18.8''$, but the peculiar fit residuals suggest that the bar
  has not been modeled correctly.

\item[NGC 1068:] We consider the small classical component of the
  composite bulge, getting $R_e=0.5''$ \citep{Erwin2015a}. This explains
  the large difference with \citet{Sani2011}, who derive $R_e=8.4''$
  for the pseudo bulge of the galaxy.  The $M/L$ is derived using the distance
  of \citet{KormendyHo2013} and fitting the major-axis kinematic
  profile $\sqrt{\sigma^2+V^2}$ of Fig. 5 of \citet{Emsellem2006},
  summing the bulge and disk contributions (see App. C
  and Fig. \ref{fig_kinprofile}). The value $M/L_K=0.7$
derived by \citet{Erwin2015a} for the stellar population of the classical bulge
is a factor 1.9 larger. We adopt this as the upper error estimate.  
 The lower error estimate comes from 
  the spherical mass profile derived from Fig. 13 of \citet{Emsellem2006}
  evaluated at $20''$ from the center. We classify
  the galaxy as ``barred'' following \citet{Erwin2004}.
  \citet{KormendyHo2013} call it a prototypical oval galaxy.

\item[NGC 1194:] The SDSS r-band bulge photometry is taken from the
  decomposition given in \cite{Greene2010}. The $M/L$ is derived
  following Appendix C to match the velocity
  dispersion within the effective aperture radius of $1.38''$, using the distance of \citet{KormendyHo2013} .


\item[NGC 1300:] The bulge model is taken from \citet{Fisher2008},
  who correctly discard the bar contribution. The huge difference
  between our $R_e$ ($4''$) and the value quoted by \citet{Sani2011},
  $45.98''$, stems from the fact that they do not separately fit the strong bar
  of the galaxy and thus include it in the bulge component.  The $M/L$ is
  derived by converting the average of the two values in the
  F160W band given in Table 4 of \citet{Atkinson2005}, using the color of
  Hyperleda. We adopt the distance of
  \citet{KormendyHo2013}.

\item[NGC 1399:] The galaxy is a core elliptical \citep{Rusli2013b};
  we use their profile and subtract the outer exponential component
  before deprojecting the galaxy, getting $R_e=103.7''$. The
  'classical' best fit of \citet{Laesker2014} gives $R_e=49.3''$, the
  'best' one $R_e=147''$. The $M/L$ is scaled to the distance of
  \citet{KormendyHo2013}.  Errors are taken from
  \citet{Kronawitter2000}.

\item[NGC 2273:] We performed a bulge plus disk decomposition on an
  $R-$band image from \citet{Erwin2003} using the \textsc{Imfit}
  software of \citet{Erwin2015c}. A Sersic bulge (function {\it
    Sersic} of {\it imfit}), a bar (function {\it Sersic\_GenEllipse}
  of \textsc{Imfit}), an exponential disk (function {\it Exponential}
  of \textsc{Imfit}) and an outer ring (function {\it
    GaussianRing2Side} of \textsc{Imfit}) were fit to the image; see
  \citet{Erwin2015c} for a definition of the function parameters. We
  set $\mu_e=-2.5 \log I_e$ and $\mu_0=-2.5 \log I_0$.  The parameter
  values of the fit are given in Table \ref{tab:2273decomp}.  Model
  images of the bar, disk and inner ring components (see
  Figs. \ref{fig_ngc2273} and \ref{fig_imfitiso}) were subtracted from
  the original image and the bulge photometry was derived on the
  residual image. The $M/L$ is computed as in Appendix C to match the
  velocity dispersion of \citet{Greene2010} within the effective
  aperture radius of $1.38''$, using the distance of
  \citet{KormendyHo2013}.

\item[NGC 2549:] The $M/L$ (already corrected for G.E.) is scaled to
  the distance of \citet{KormendyHo2013}.  The SDSS $r-$band image is
  calibrated to the $R-$band using the aperture photometry in the
  $R_{SV}$ and $V_{SV}$-band of \citet{Sandage1978}, using the color
  transformations of \citet{Prugniel1993} and \citet{Jordi2006}.  We
  performed an \textsc{Imfit} bulge (function {\it Sersic}) plus disk
  (function {\it EdgeOnDisk}) decomposition (see
  Figs. \ref{fig_ngc2549} and \ref{fig_imfitiso}).  The parameters of
  this fit are given in Table \ref{tab:2549decomp}.

\item[NGC 2748:] We performed an \textsc{Imfit} bulge (function {\it
    Sersic}) plus disk (function {\it Exponential}) plus double ring
  (functions {\it GaussianRing}) decomposition of the $3.6mu$ band
  Spitzer image (see Figs. \ref{fig_ngc2748} and
  \ref{fig_imfitiso}). The resulting values of the parameters of the
  fit are listed in Table \ref{tab:2748decomp}. The residual image
  shows that the galaxy has a low surface brightness polar ring, that
  we masked while fitting the main galaxy. The position angle of the
  bulge component (which contributes 15\% of the total light) is
  different from the rest of the galaxy; this probably stems from the
  modeling of the combination of a small bar (in which case the two
  rings make sense as an inner ring and an outer ring) and whatever
  small pseudobulge the galaxy may have. The Spitzer image does not
  really have the resolution to be sure about that.  The 'bulge' model
  of \citet{Sani2011} is 0.8 mag brighter and 3.5 times larger in size
  than our solution. The $M/L=0.5\pm 0.02 M_\odot/L_\odot$ in the
  $3.6mu$ band is derived as in Appendix C and Fig. \ref{fig_kinprofile}
  using the (circularized) surface brightness of the entire galaxy
  (i.e. with the polar ring component) to match the kinematic profile
  of \citet{Batcheldor2005}. The distance comes from 
  \citet{KormendyHo2013}. If we convert the best-fit value in the
  F160W-band given in the Table 4 of \citet{Atkinson2005} to the $3.6mu$
  band using 2MASS $J$, $H$, $K$ aperture magnitudes and the
  conversion equation given in \citet{Stephens2000}, we get
  $M/L_{3.6mu}=0.4\pm0.05 M_\odot/L_\odot$.  Since the galaxy is too
  dusty and edge-on to be sure about the presence or absence of a bar,
  we set $b=0.5$ in Table \ref{tab_data}.


\item[NGC 2787:] \citet{Erwin2003b} presented evidence for a composite
  bulge in this galaxy. Although their ``inner disk'' may perhaps be
  better understood as the projected box/peanut structure of the bar
  (rather than a ``disky pseudo bulge''), we use the parameters of
  their decomposition (their Table 5) giving $R_e=3.5''$, but shift
  the resulting total surface brightness to the $3.6mu$ band to match
  the total profile derived from Spitzer images (coming from Program
  30318, PI Giovanni Fazio, Observer ID 2;
  Fig. \ref{fig_maxfits}). \citet{Sani2011} derive $R_e=12.3''$, not
  distinguishing between the classical and the pseudo component of the
  bulge. The $M/L$ is computed by fitting the kinematics of
  \citet{Bertola1995} (see Appendix C and Fig. \ref{fig_kinprofile}),
  using the sum of the bulge and disk profiles derived
    above. The distance comes from \citet{KormendyHo2013}.

\item[NGC 2960:] We use the bulge plus disk decomposition of
  \citet{Greene2010} in the $r-$band, ignoring the E2 morphology
    of \citet{KormendyHo2013}, but accepting the merger appearance of
    the galaxy.  The $M/L$ is computed as in Appendix C to match the
  velocity dispersion of \citet{Greene2010} within an effective
  aperture radius of $1.6''$, using the distance of
  \citet{KormendyHo2013}.

\item[NGC 2974:] The $M/L$ is scaled to the distance of \citet{Sani2011}
  and from $I$ to $3.6mu-$band using the $(I-3.6mu)$ color from
  Hyperleda. The Spitzer images come from Program 30318, PI Giovanni
  Fazio, Observer ID 2. The BH mass comes
  from the uncertain determination of \citet{Cappellari2008}. We
  exclude this galaxy from the fits reported in Tables
  \ref{tab_1dimstrict} and \ref{tab_2dimstrict}.

\item[NGC 3031:] The SDSS $i-$band decomposed bulge profile comes from
  \citet{Beifiori2012}, which gives $R_e=41.4''$. \citet{Sani2011}
  quote $R_e=100.2''$, probably overestimating the size of the bulge
  by including the bar of the galaxy as part of the bulge in the fit.
  \citet{Fisher2010} are in better agreement with our adopted
  solution.  The $M/L$ is determined using the distance of
  \citet{KormendyHo2013} and by fitting the kinematics of
  \citet{Fabricius2012} (see Appendix C and
  Fig. \ref{fig_kinprofile}), summing the disk and the bulge
  profiles. We classify the galaxy as barred following \citet{Gutierrez2011}
  and \citet{Erwin2013}. Following \citet{Fabricius2012} we
    classify the galaxy bulge as classical, even if it is most likely
    a composite (pseudo plus classical) system.

\item[NGC 3079:] We model the Spitzer $3.6mu$-band image of this pseudo
  bulge galaxy with four components using \textsc{Imfit}: a pseudo
  bulge (function {\it Sersic}), a disk (function {\it Exponential}),
  a bar (a second {\it Sersic} function) and a ring (function {\it
    GaussianRing}). The results are shown in Fig. \ref{fig_ngc3079}
  and \ref{fig_imfitiso}.  The parameters of the fit are listed in
  Table \ref{tab:3079decomp}.  All four components are very flattened
  and have almost the same position angle. The small ($a_e=4.5''$)
  pseudo bulge component contributes 26\% of the total light of the
  galaxy. This is in contrast to the fit of \citet{Sani2011}, who
  claim the existence of an unrealistically large ($R_e=74''$) bulge with
  $B/T=0.87$.  The $M/L$ is computed as in Appendix C to match the
  stellar kinematic profile of \citet{Shaw1993} (see
  Fig. \ref{fig_kinprofile}) using the distance of
  \citet{Sani2011}. \citet{KormendyHo2013} judge  the
  black hole mass determination of \citet{Kondratko2005} unreliable, which
  however agrees with \citet{Yamauchi2004} and matches the value
  quoted by \citet{Sani2011} we used here.  We
  exclude this galaxy from the fits reported in Tables
  \ref{tab_1dimstrict} and \ref{tab_2dimstrict}.

\item[NGC 3115:] For this edge-on S0 we prefer the decomposition of
  \citet{Seifert1996} and \citet{Scorza1998}, which gives $R_e=42''$,
  to the one of \citet[][$R_e=15.7''$]{Sani2011} that shows systematic
  residuals. The $M/L$ is scaled to the distance of
  \citet{KormendyHo2013} and corrected for G.E..

\item[NGC 3227:] The bulge component is determined in
  \citet{Davies2006} from a high-resolution SINFONI image, giving
  $R_e=3.1''$. The 'classical fit' of \citet{Laesker2014} delivers
  $R_e=1.4''$, the 'best fit' $R_e=0.5''$, both results are probably
  affected by insufficient resolution (FWHM=$0.9''$). \citet{Sani2011}
  quote $R_e=58''$, which is definitely too large. The $K-$band $M/L$ 
\citep[scaled to the distance of][]{KormendyHo2013} is
  given in the unusual units quoted by \citet{Davies2006}.  We mark
  the 'peculiarity' of this definition by referring to the band as
  'Kp' in Table \ref{tab_ml} and by not plotting the galaxy in
  Fig. \ref{fig_ML}. \citet{KormendyHo2013} exclude the galaxy from their 
fits, arguing that the active nucleus makes the dynamical modeling challenging.
 
\item[NGC 3245:]  We use the decomposition of \citet{Beifiori2012} in the 
$i-$band, which
  gives $R_e=3.5''$  (with $B/T=0.27$) and agrees with the 'classical fit' of
  \citet{Laesker2014}. The 'best fit' of \citet{Laesker2014}, which
  includes a bar, gives $R_e=1.5''$. Since what \citet{Laesker2014}
  call a 'bar' could also be an 'oval' component, we stick to the
  decomposition of \citet{Beifiori2012}. We calibrate the fit 
  to the $R-$band from the original SDSS $i-$band using the $(R-I)$
  color estimated as follows and the $(i-R)$ conversion equation in
  \citet{Jordi2006}.  Since only the $(B-V)$ color was available in
  Hyperleda, we derived $(R-I)$ by linearly fitting the
  $(V-I)$vs.$(B-V)$ correlation of the simple
  stellar population models of \citet{Maraston2005}. The $M/L$ is corrected for 
  G.E. and scaled to the distance of \citet{KormendyHo2013}.

\item[NGC 3377:] For this E6 galaxy we use a single profile, which
  gives $R_e=39.1''$, roughly consistent with \citet{Arnold2014}.  The
  multi-component fit of \citet{Laesker2014} delivers $R_e=7''$, but
  their disk component is faint and the envelope could well be part of
  the bulge. The $M/L$ is scaled to the distance of
  \citet{KormendyHo2013} and from $I$ to $R-$band using the $(R-I)$
  color from Hyperleda.

\item[NGC 3379:] The galaxy is a core elliptical
  \citep{Rusli2013b}. We use their profile and $M/L$, which is scaled to the distance
  of \citet{KormendyHo2013}.

\item[NGC 3384:] The decomposed SDSS $i-$band bulge profile comes from
  \citet{Beifiori2012}. We calibrated it to the $I-$band using the
  equation of \citet{Jordi2006} and the $(R-I)$ color from
  Hyperleda. We get $R_e=7.6''$, while \citet{Laurikainen2010} quote
  $R_e=3.6''$ from a complex fit involving 2 bars. But
  \citet{Erwin2004} classified the object as a barred galaxy with an
  inner disk, rather than double-barred. Moreover, the residuals of
  the fit of \citet{Laurikainen2010} are not shown, so we keep our
  reasonable solution. We correct the $M/L$ of the model with dark
  matter of \citet{Schulze2011} for G.E., and scale it from $V$ to
  $I-$band using the $(V-I)$ color of Hyperleda, using the distance of
  \citet{KormendyHo2013}.

\item[NGC 3393:] We perfomed an \textsc{Imfit} bulge (function {\it
    Sersic}) plus bar (function {\it
    Sersic\_GenEllipse}) plus disk (function {\it
    Exponential}) decomposition (see
  Fig. \ref{fig_ngc3393} and \ref{fig_imfitiso}) on an $I-$band image
  observed with the 0.9 m CTIO telescope by \citet{Schmitt2000}. The
  parameters of the decomposition are given in Table
  \ref{tab:3393decomp}.  The $M/L$ is computed as in Appendix C
  to match the velocity dispersion of
  \citet{Greene2010} within an effective aperture of radius of
  $1.6''$, using the distance of \citet{KormendyHo2013}.

\item[NGC 3414:] We performed an \textsc{Imfit} bulge (function {\it
    Sersic}) plus bar (function {\it Sersic\_GenEllipse}) plus disk
  (function {\it Exponential}) decomposition on the SDSS $i-$band
  image of the galaxy (see Figs. \ref{fig_ngc3414} and
  \ref{fig_imfitiso}). The parameters are listed in Table
  \ref{tab:3414decomp}. We converted the $i-$band zeropoint into a
  Cousins $I-$band using two equations in \citet{Jordi2006} and the
  $(r-i)$ SDSS color.  The $M/L$ of \citet{Cappellari2006} is scaled to
  the distance of \citet{Sani2011}. Our decomposition gives $R_e=23''$
  and agrees with \citet{Sani2011}. \citet{Laurikainen2010} obtain
  $R_e=3.2''$ and classify the galaxy as 'spindle'.  Given the
  impossibility of verifying the quality of the fit of
  \citet{Laurikainen2010}, we stick to our solution but acknowledge
  that the galaxy is complex. For consistency with our fit that
  contains a bar we set $b=1$ in Table \ref{tab_data}.  The BH mass
  comes from the uncertain determination of \citet{Cappellari2008}. We
  exclude this galaxy from the fits reported in Tables
  \ref{tab_1dimstrict} and \ref{tab_2dimstrict}.

\item[NGC 3585:] For this edge-on S0 we prefer the decomposition of
  \citet{Scorza1998}, giving $R_e=27.4''$, to the one of
  \citet[][$R_e=11.5''$]{Sani2011}, which shows systematic residuals. The
  $M/L$ is corrected for G.E. and scaled to the distance
  of \citet{KormendyHo2013}.

\item[NGC 3607:] We calibrate the $g-$band SDSS image to the $V-$band
  using the $(B-V)$ color from Hyperleda and the $(g-V)$
  transformation of \citet{Jordi2006}.  \citet{McConnellMa2013} and
  \citet{KormendyHo2013} consider the galaxy a core elliptical; we
  accept their classification, although strong nuclear dust makes it
  difficult to determine its core properties \citep[see discussion
  in][]{Rusli2013b}. From the profile we measure $R_e=45.0''$, in fair
  agreement with \citet{Beifiori2012} who measure $R_e=56.34''$ from a
  Sersic fit to the SDSS $i-$band image. In contrast,
  \citet{Laurikainen2010} perform a bulge plus disk decomposition,
  deriving $R_e=6.1''$ with $B/T=0.32$. We do not think that this is a
  good description of the galaxy.  The $M/L$ is corrected for G.E. and
  scaled to the distance of \citet{KormendyHo2013}.

\item[NGC 3608:] The galaxy is a core elliptical
  \citep{Rusli2013b}. We calibrate the $i-$band SDSS image to the
  $I-$band using the $(R-I)$ color from Hyperleda and the
  tranformation of \citet{Jordi2006}. This gives a more extended
  profile than the one derived by \citet{Rusli2013b}, which comes from
  the $g-$band image.  We transform the $M/L_V$ of \citet{Rusli2013b}
  to the $I-$band using the $(V-I)$ color of Hyperleda and scaling it
  to the distance of \citet{KormendyHo2013}.  We measure $R_e=43''$,
  while \citet{Beifiori2012} obtain $R_e=161.9''$. This stems from the
  systematic overestimation (by 0.4 mag) of the light of the galaxy at
  radii larger than $60''$ in \citet{Beifiori2012}'s fit.

\item[NGC 3842:] The galaxy is a core elliptical
  \citep{Rusli2013b}. We use their profile and $M/L$, scaled 
  to the distance of \citet{KormendyHo2013}. 

\item[NGC 3998:] The SDSS $i-$band decomposed bulge profile comes
  from \citet{Beifiori2012}, is calibrated to the $I-$band using the
  equation of \citet{Jordi2006} and gives $R_e=5.2''$. This roughly
  agrees with the $R_e=3.4''$ of \citet{Laurikainen2010}. We prefer
  this solution to the multi-component 'best-fit' of
  \citet{Laesker2014}, which delivers $R_e=1.8''$, since the size of
  the fitted bar seems implausible. Since only the $(B-V)$ color was
  available in Hyperleda, we derived the $(V-I)$ and the $(R-I)$
  colors by linearly fitting the $(V-I)$vs.$(B-V)$ and
  $(R-I)$vs.$(B-V)$ correlation of the simple stellar population
  models of \citet{Maraston2005}.  We correct the $M/L$ for G.E. and
  scaled it to the distance of \citet{KormendyHo2013}; moreover, we
  adapt it to the $I-$band using the above colors. Errors on $M/L$ are
  half of the $2\sigma$ values.  We classify the galaxy as ``barred''
  following \citet{Gutierrez2011}.

\item[NGC 4026:] We correct the $M/L$ for G.E and scale it to the
  distance of \citet{KormendyHo2013}.  The SDSS $g-$band image was
  calibrated to the $V-$band using the transformation of
  \citet{Jordi2006} and the mean $(B-V)$ colors from Hyperleda. We
  performed an \textsc{Imfit} bulge (function {\it Sersic}) plus disk
  (function {\it EdgeOnDisk}) decomposition (see
  Figs. \ref{fig_ngc4026} and \ref{fig_imfitiso}). The best fit
  parameters are given in Table \ref{tab:4026decomp}. We set $b=0.5$
  in Table \ref{tab_data}, since the galaxy is too edge-on to be
  certain about the presence or the absence of a bar.

\item[NGC 4151:] We take the $R-$band bulge plus bar plus disk decomposition of
  \citet{Gadotti2008} and set $b=1$ in Table \ref{tab_data}
  \citep{Erwin2005}.  \citet{Sani2011} ignore the bar of the galaxy in
  their fit. We use the $R-$band $M/L$ value (and black hole
    mass) of \citet{Onken2007}, which we correct for G.E. We
    follow \citet{HoKim2014} and classify its bulge as classical, but
    \citet{Kormendy2013} disagrees. \citet{KormendyHo2013} exclude the
    galaxy from their fits, arguing that the active nucleus makes the
    dynamical modeling challenging. We exclude this galaxy from the
    fits reported in Tables \ref{tab_1dimstrict} and
    \ref{tab_2dimstrict}. 

\item[NGC 4258:] The Spitzer $3.6mu$ images come from Program 20801,
  PI Seppo Laine, Observer ID 14916. The Sersic bulge plus exponential
  disk decomposition is performed using the technique of
  \citet{Fisher2008}, fitting the major axis profile. The result is
  shown in Fig. \ref{fig_maxfits} and listed in Table
  \ref{tab_maxfits}. This gives $R_e=13.3''$. The fit of
  \citet{Sani2011} has $R_e=75.8''$ with strong residuals;
  \citet{Laesker2014} quote $R_e=118.9''$ for the 'classical'
  decomposition with strong residuals and $R_e=4.1''$ for the 6
  component fit; we prefer our simpler approach. The $M/L$ is scaled
  to the distance of \citet{KormendyHo2013} and from $V$ to
  $3.6mu$-band using the $(V-3.6mu)$ color using the $V$-band
  photometry of Hyperleda.

\item[NGC 4261:] The galaxy is a core elliptical
  \citep{Rusli2013b}. We use their profile and derive $R_e=54.2''$, in
  fair agreement with \citet{Beifiori2012}, who measure $R_e=45.82''$
  from a Sersic fit to the $i-$band SDSS image. In contrast,
  \citet{Sani2011} get $R_e=20''$ from an improbable bulge plus disk
  decomposition, given the galaxy type, and \citet{Vika2012} derive
  $R_e=21.6''$ from a fit with low $n_{Sersic}$.  The $M/L$ is scaled
  to the distance of \citet{KormendyHo2013}.

\item[NGC 4291:] The galaxy is a core elliptical
  \citep{Rusli2013b}. We use their profile and $M/L$ scaled to the distance of
  \citet{KormendyHo2013}. 

\item[NGC 4342:] We performed an \textsc{Imfit} bulge (function {\it
    Sersic}) plus disk (function {\it Exponential}) decomposition on a
  SDSS $i-$band image (see Figs. \ref{fig_ngc4342} and
  \ref{fig_imfitiso}), converting the zero-point to the Cousins $I-$band
  using the SDSS $(r-i)$ color and two equations from
  \citet{Jordi2006}. The parameters of the decomposition are given in
  Table \ref{tab:4342decomp}. Our fit gives $R_e=4.9''$, while
  \citet{Vika2012} quote $R_e=0.6''$ and \citet{Laesker2014} derive
  $R_e=1.1''$ for the classical fit and $R_e=0.7''$ for the 'best'
  fit. All fits have systematic residuals, but given that the
  resolution of the images considered by \citet{Vika2012} (FWHM=$0.5''$
  or 1.1 pixels) and \citet{Laesker2014} (FWHM=$0.6''$) is too near
  their quoted $R_e''$, we prefer our solution.  The $M/L$ is corrected
  for G.E. and scaled to the distance of \citet{KormendyHo2013}.
 
\item[NGC 4374:] The galaxy is a giant core elliptical
  \citep{Rusli2013b}.  We measure $R_e=84.0''$, while \citet{Vika2012}
  derive implausibly small bulge radii from one-component
  ($R_e=26.8''$) or bulge plus disk ($R_e=7.2''$) fits. The $M/L$ is
  scaled to the distance of \citet{KormendyHo2013} and from $I$ to
  $V-$band using the $(V-I)$ Hyperleda color.

 \item[NGC 4388:] Our best \textsc{Imfit} decomposition of this pseudo
  bulge galaxy is achieved by fitting a central point source (function
  {\it Gaussian}), a pseudo bulge (function {\it Sersic}), a disk
  (function {\it Exponential}) plus a ring (function {\it
    GaussianRing2Side}) (see Figs. \ref{fig_ngc4388} and
  \ref{fig_imfitiso}). The parameters of the fit are listed in Table
  \ref{tab:4388decomp}. The pseudo bulge component contributes 38\% of
  the total light of the galaxy; \citet{Greene2010} give $B/T=0.5$,
  while \citet{KormendyHo2013} quote $B/T=0.096$, which explains most
  of the discrepancy observed in Fig. \ref{fig_bulgecomp}.  Despite
  the use of four components, there are still significant residuals,
  stemming from the strong non-axisymmetric galaxy features.  The
  $M/L$ is computed as in Appendix C to match the velocity dispersion
  of \citet{Greene2010} within an effective aperture radius of
  $1.6''$, by considering the circularized surface brightness of the
  galaxy without the central point source and using the distance of
  \citet{KormendyHo2013}.  We classify the galaxy as barred following
  \citet{KormendyHo2013}. In our decomposition the bar is described by
  the ring component.

\item[NGC 4459:] For this E2 galaxy we measure $R_e=35.2''$ from the
  whole profile.  \citet{Beifiori2012} quote $140.5''$,
  \citet{Sani2011} give $9.4''$, and  \citet{Laurikainen2010} report
  $5.2''$. The large value of \citet{Beifiori2012} stems from a fit
  overestimating by 0.4 mag the light of the galaxy at radii larger
  than $100''$; the fit to a $K-$band image of the galaxy described by
  \citet{Beifiori2010} agrees with our value. The discrepancy with the
  $R_e$ of \citet{Sani2011} and \citet{Laurikainen2010} is driven by
  their (unplausible large) disk component, that we ignore following
  \citet{KormendyHo2013}. The $M/L$ is scaled to the distance of
  \citet{KormendyHo2013} and from $I$ to $V-$band using $(V-I)$ from
  Hyperleda.

\item[NGC 4473:] Contrary to the classification of
  \citet{McConnellMa2013}, the galaxy is not a core elliptical
  \citep[see discussion in][]{Rusli2013b}. We derive $R_e=35.7''$ using
  a single profile extending to $261.5''$. \citet{Vika2012} quote
  $R_e=16.3''$, fitting a single S\'ersic profile to just the inner
  $70''$. \citet{Beifiori2012} fit two components, which we do not
  believe to be real, getting $R_e=8''$. On the other hand, they derive
  $R_e=34''$, in agreement with our value, from the isophotal profile
  (their Table 3). We correct the $M/L$ for G.E. and
  scale it to the distance of \citet{KormendyHo2013}.

\item[NGC 4486:] The galaxy is a giant core elliptical \citep{Rusli2013b}.
  We derive $R_e=158.4''$, in disagreement with the 'classical fit' of
  \citet{Laesker2014} giving $R_e=59.1''$ and with the fit of
  \citet{Vika2012}, who quote an even smaller $R_e=34.6''$.  The $M/L$ is
  scaled to the distance of \citet{KormendyHo2013} and corrected for
  G.E..

\item[NGC 4526:] We considered a Spitzer $3.6mu$-band image that we
  calibrated to the $I-$band using the aperture photometry of
  Hyperleda. We performed an \textsc{Imfit} decomposition with 5
  structures: a classical bulge (function {\it Sersic}), an inner disk
  (function {\it Exponential}), a bar (function {\it
    Sersic\_GenEllipse}), an outer (edge-on and diamont-shaped) disk
  (function {\it Exponential\_GenEllipse}), and two off-center spurs
  (function {\it Sersic}), see Figs. \ref{fig_ngc4526} and
  \ref{fig_imfitiso}). The parameters of the decomposition are given
  in Table \ref{tab:4526decomp}. Our B/T (0.11) is much smaller than
  the value (0.65) quoted by \citet{KormendyHo2013}.  The $M/L$ is
  scaled to the distance of \citet{KormendyHo2013}.

\item[NGC 4552:] The galaxy is a core elliptical
  \citep{Rusli2013b}. We use their profile extending to 445.8'' and
  their $M/L$ scaled to the distance of \citet{Sani2011}. We measure
  $R_e=82.6''$.  \citet{Sani2011} and \citet{Vika2012} derive shorter
  scale lengths ($R_e=23.3''$ and $16.5''$ respectively) from a dubious
  bulge plus disk decomposition \citep{Sani2011}, given the galaxy type, 
or a fit to a too
  small ($\approx 100''$) image \citep{Vika2012}. The BH mass comes
  from the uncertain determination of \citet{Cappellari2008}. We
  exclude this galaxy from the fits reported in Tables
  \ref{tab_1dimstrict} and \ref{tab_2dimstrict}.

\item[NGC 4564:] We take the bulge profile as decomposed by
  \citet{Kormendy2009}. This gives $R_e= 15.32''$. We prefer it to the
  result of \citet[][$R_e=2.2''$]{Vika2012}, who fit a bulge and a bar
  (without a disk) to an image of average ($1.5''$) seeing. The $M/L$ is
  computed as in Appendix C to match the stellar
  kinematic profiles of \citet{BSG1994} (see
  Fig. \ref{fig_kinprofile}), using the distance of \citet{KormendyHo2013}.

\item[NGC 4594:] We use the Sersic bulge fit of \citet{Jardel2011}
  and correct their $M/L$ for G.E., scaling it to the distance of
  \citet{KormendyHo2013}.  Our $R_e=133.6''$ is much larger than the
  unreliable value of \citet[][$33.7''$]{Sani2011}.  We set $b=0.5$ in
  Table \ref{tab_data}, since the galaxy is too edge-on to be sure
  about the presence or absence of a bar.

\item[NGC 4596:] We use the bulge plus disk plus bar decomposition of
  \citet{Vika2012}, which gives $R_e=11.9''$ with small
  residuals. We prefer this to \citet{Laurikainen2010}, who quote
  $R_e=3.2''$ but do not show the fit. The fit of \citet{Beifiori2012}
  gives $R_e=44.9''$ without a bar component.  The $M/L$ is computed as
  in Appendix C to match the stellar kinematic profiles of
  \citet{Bettoni1997}, summing the disk, bulge and bar profiles (see
  Fig. \ref{fig_kinprofile}) and using the distance of \citet{KormendyHo2013}. 

\item[NGC 4621:] The $M/L$ is scaled to the distance of \citet{Sani2011}
  and calibrated from the $I-$band to $3.6mu$ using the $I-$band aperture
  photometry from Hyperleda. The Spitzer images come from Program
  13649, PI Patrick Cote, Observer ID 522.  The BH mass comes
  from the uncertain determination of \citet{Cappellari2008}. We
  exclude this galaxy from the fits reported in Tables
  \ref{tab_1dimstrict} and \ref{tab_2dimstrict}. 

\item[NGC 4649:] The galaxy is a core elliptical
  \citep{Rusli2013b}. We use their profile and get $R_e=99.5''$.
  \citet{Beifiori2012} quote $R_e=13.7''$, from a fit with two
  components, which we do not believe are real, given the galaxy type. 
The same applies to
  the decompositions put forward by \citet{Sani2011}, who quote
  $R_e=42.7''$, \citet{Vika2012}, who get $R_e=41.2''$, or the 'best
  fit' of \citet{Laesker2014}, who derive $R_e=46.8''$.  The $M/L$ of
  \citet{Rusli2013b} is scaled to the distance of
  \citet{KormendyHo2013}.

\item[NGC 4697:] The profile is measured on $R-$band images taken
  during the observations described in \citet{Erwin2008}. From this we
  derive $R_e=65.8''$, in rough agreement with
  \citet[][$R_e=81''$]{Sani2011}, \citet[][$R_e=107''$]{Beifiori2012}
  and the 'classical' fit of \citet[][$R_e=118''$]{Laesker2014}. The
  'best fit' of \citet{Laesker2014} has a bulge with a short
  scale-lenght ($R_e=4.8''$) but adds an envelope component. The $M/L$ is
  scaled to the distance of \citet{KormendyHo2013}, calibrated to the R
  band using $(V-R)=0.59$ from the aperture photometry in Hyperleda and
  corrected for G.E.

\item[NGC 4736:] An \textsc{Imfit} bulge (function {\it Sersic}) plus
  disk (function {\it Exponential}) plus outer ring (function {\it
    GaussianRing2Side}) decomposition was performed on a $z-$band image
  from SDSS (see Fig. \ref{fig_ngc4736} and \ref{fig_imfitiso}).  The
  parameters of the fit are given in Table \ref{tab:4736decomp}. A
  model image of the disk and the outer ring was created, which was
  then subtracted from the original image. On the residual image an
  ellipse fit was performed with \texttt{IRAF} to get the bulge
  profile. The $M/L$ of the bulge calculated by
  \citet{Moellenhoff1995} was converted to the $z-$band using the
  equations of \citet{Jordi2006} and the $B, V, R, I$ aperture
  magnitudes provided by Hyperleda, and scaled to the distance of
  \citet{KormendyHo2013}.  We classify the galaxy as ``barred''
  following \citet{Moellenhoff1995}.

\item[NGC 4826:] We performed an \textsc{Imfit} bulge (function {\it
    Sersic}) plus disk (function {\it Exponential}) decomposition on a SDSS $i-$band image (see
  Fig. \ref{fig_ngc4826}) after masking the very strong dust
  lane. The parameters for the fit are given in Table
  \ref{tab:4826decomp}. We subtracted a model image of the disk and
  performed an ellipse fit on the residual image for the bulge
  profile.  From this we derive $R_e=31.3''$, while
  \citet{Beifiori2012} get $R_e=4''$. The fit of \citet{Beifiori2012}
  fails to reproduce the photometry at large radii, where the disk
  dominates. \citet{Beifiori2010} discusses an alternative fit to a
  $K-$band image of the galaxy that agrees with us. The $M/L$ is computed
  as in Appendix C to match the stellar kinematic
  profiles of \citet{Heraudeau1998}, summing the disk and the bulge
  profiles (see Fig. \ref{fig_kinprofile}) and using the distance of \citet{KormendyHo2013}.

\item[NGC 4889:] The Gunn $r-$band photometry of \citet{Jorgensen1994}
  was calibrated to Cousins $R$ using the transformation of
  \citet{Jordi2006} and the $(V-R)$ colors from Hyperleda. The galaxy
  is a core elliptical \citep{Rusli2013b}. We derive $R_e=47.1''$,
  while \citet{Rusli2013b} quote $R_e=169.2''$ fitting a $n=9.8$
  Core-Sersic profile. The difference is driven by the extrapolation
  (see Fig. \ref{fig_extrap}). The $M/L_R$ value \cite[scaled to the
  distance of][]{KormendyHo2013} is the average of the results for
  four quadrants given by \citet{McConnell2012} corrected for
  G.E. \citet{Rusli2013b} get $M/L_R=5.8$ correcting for G.E., which is the
  value quoted in the Conclusions of \citet{McConnell2012}.

\item[NGC 5077:] The surface brightness profile comes from the
  observations described in \citet{Scorza1998}. The $M/L$ is scaled to
  the distance of \citet{KormendyHo2013}, corrected for G.E.  and
  converted from the $B$ to the $V-$band using the $(B-V)$ color from
  Hyperleda.  We follow \citet{McConnellMa2013} and
  \citet{KormendyHo2013} and consider the galaxy a core elliptical,
  even though the classification is uncertain \citep[see discussion
  in][]{Rusli2013b}.

\item[NGC 5128 (Cen A):] We deprojected the multi-gaussian expansion
  of the photometry profile as given in \citet{Cappellari2009} and
  their $M/L_K$ scaled to the distance of
  \citet{KormendyHo2013}. Despite the strong dust lane and the
  decomposition proposed by \citet{Sani2011}, CenA does not have a
  strong stellar disk.  \citet{McConnellMa2013} and
  \citet{KormendyHo2013} consider the galaxy a core elliptical; we
  accept their classification, although the strong nuclear dust makes
  it difficult to determine its core properties \citep[see discussion
  in][]{Rusli2013b}.

\item[NGC 5576:] We used an $r-$band SDSS image calibrated to the
  $R-$band. For this E6 galaxy we derive $R_e=26.1''$ from a profile
  extending to $200''$, while \citet{Beifiori2012} quote $64''$ fitting
  a Sersic model to an image extending to $120''$. Their isophotal
  profile (Table 3) gives $R_e=49''$. Our result agrees with
  \citet{Trujillo2004}. Contrary to the classification of
  \citet{McConnellMa2013} and \citet{KormendyHo2013}, the galaxy is
  not a core elliptical \citep[see discussion in][]{Rusli2013b}. The $M/L$ is
  corrected for G.E., scaled from the $V$ to the $R-$band
  using the Hyperleda $(V-R)$ color and using the distance of
  \citet{KormendyHo2013}.

\item[NGC 5813:] The galaxy is a core elliptical \citep{Rusli2013b},
  we use their profile and $M/L$, scaled to the distance of
  \citet{Sani2011}. The large discrepancy between our $R_e=42.43''$
  and the values reported by \citet[][$R_e=98.33''$]{Sani2011}  and
  \citet[][$R_e=106.8''$]{Vika2012} stem from the outer component of
  the galaxy \citep{Rusli2013b} \footnote{We note that Table 3 of
    \citet{Rusli2013b} reports wrong values for $n$, $r_e$ and
    $\mu_e$. The parameters corresponding to the fits presented in
    their Fig. 2 there are $n=2.07$, $r_e=55.98''$ and $\mu_e=21.57$
    mag/arcsec$^2$. }.  The BH mass comes from the uncertain
  determination of \citet{Cappellari2008}. We exclude this galaxy from
  the fits reported in Tables \ref{tab_1dimstrict} and
  \ref{tab_2dimstrict}.

\item[NGC 5845:] The surface brightness profile comes from the
  observations described in \citet{Scorza1998}.  The $M/L$ is
  corrected for G.E. and scaled to the distance of
  \citet{KormendyHo2013}.

\item[NGC 5846:] The galaxy is a core elliptical
  \citep{Rusli2013b}. We use their $M/L_i$ and profile extending to
  $214.5''$ from the center and measure $R_e=89.4''$. \citet{Sani2011}
  derive $R_e=36.5''$ by fitting (with systematic residuals) an image
  extending to $\approx 160''$ from the center.
  \citet{Laurikainen2010} perform an unrealistic bulge plus disk
  decomposition with $B/T=0.46$ which delivers $R_e=15.6''$.  The BH
  mass comes from the uncertain determination of
  \citet{Cappellari2008}. We exclude this galaxy from the fits
  reported in Tables \ref{tab_1dimstrict} and \ref{tab_2dimstrict}.

\item[NGC 6086:] The galaxy is a core elliptical
  \citep{Rusli2013b}. We use their profile and $M/L_R$, scaled to the
  distance of \citet{KormendyHo2013}, after subtracting the outer
  halo component fitted there.

\item[NGC 6251:] We confirm the core elliptical classification of
    the galaxy given by \citet{KormendyHo2013}.  We derive
  $R_e=16.5''$, while \citet{Sani2011} quote $R_e=38.9''$ from a
  Sersic fit (with $n_{Ser}=7$) and a small central point source. The
  difference is driven by the different amounts of extrapolation. The
  $M/L$ is corrected for G.E. and scaled from $R$ to $I-$band using
  $(R-I)=0.63$, using the distance of \citet{KormendyHo2013}.  There
  are no $(R-I)$ colors measured for this galaxy, but Hyperleda
  provides $(V-I)$. So we searched for elliptical galaxies with
  velocity dispersions and $(V-I)$ colors similar to NGC 6251 and
  adopted their mean $(R-I)$ value. This matches the predictions of
  the models of \citet{Maraston2005} for old and metal rich simple
  stellar populations.

\item[NGC 6264:] We took the bulge plus disk decomposition of Kormendy (in
  prep.), based on the $r-$band profile of \citet{Greene2010}. The $M/L$
  is computed by fitting the Jeans equations as in Appendix C to match
  the average velocity dispersion measured along the minor axis
  (118$\pm$10 km/s) within an effective aperture radius of $1.6''$,
  with a 20\% error and using the distance of \citet{KormendyHo2013}.
  The velocity dispersion along the major axis is much higher (166
  km/s) and would give a $M/L$ almost a factor 2 larger.

\item[NGC 6323:] We use the bulge plus bar plus disk decomposition of
  \citet{Greene2010}.  The $M/L$ is derived as described in 
  Appendix C, by matching the velocity dispersion of
  \citet{Greene2010} within an effective aperture radius of $1.38''$
  and using the distance of \citet{KormendyHo2013}. In this process we
  summed the density contributions of the small bulge and the
  prominent bar. The resulting $M/L$ (8.15 $M_\odot/L_\odot$ in the
  $r$-band) is suspiciously high, which might explain the discrepancy
  observed in Fig. \ref{fig_bulgecomp}. We classify the galaxy as
  barred following \citet{Graham2013}.

\item[NGC 7052:] \citet{McConnellMa2013} and \citet{KormendyHo2013} 
consider the galaxy a core
  elliptical, we accept their classification, although strong nuclear
  dust makes it difficult to determine its core properties \citep[see
  discussion in][]{Rusli2013b}. We derive $R_e=17.4''$ in agreement
  with \citet[][$R_e=18.6''$]{Laesker2014} and the one-component fit of
  \citet[][$R_e=15.7''$]{Vika2012}, and compatible with
  \citet[][$R_e=27.5''$]{Sani2011}, but different from the
  two-component fit of \citet[][$R_e=3''$]{Vika2012}. The $M/L$ is scaled
  to the distance of \citet{KormendyHo2013} and corrected for
  G.E.

\item[NGC 7457:] The Spitzer images come from Program 30318, PI
  Giovanni Fazio, Observer ID 2.  We perform a Sersic bulge plus
  exponential disk decomposition along the major axis, following
  \citet{Fisher2008}; the results are given in Fig. \ref{fig_maxfits}
  and Table \ref{fig_maxfits}. This gives $R_e=8.7''$, which agrees
  with \citet{Erwin2015a}, \citet{Laurikainen2010} and
  \citet{Sani2011}. The 'classical fit' of \citet{Laesker2014} is clearly too
  large a bulge ($R_e=39.4''$), while their multicomponent 'best fit'
  giving $R_e=2.3''$ is possibly affected by insufficient resolution
  ($FWHM=1.1''$). The $M/L$ is computed as in Appendix C to match the
  stellar kinematic profiles along the major axis of
  \citet{Emsellem2004}, summing the bulge and the disk profiles (see
  Fig. \ref{fig_kinprofile}) and using the distance of \citet{KormendyHo2013}.

\item[NGC 7582:] We perform a (pseudo) bulge plus bar plus disk
  decomposition using \textsc{Imfit} and the $3.6mu$ Spitzer image. A
  Sersic bulge (function {\it Sersic}), a bar (function {\it
    Sersic\_GenEllipse}), and an exponential disk (function {\it
    Exponential}) were fit to the image (see Figs. \ref{fig_ngc7582}
  and \ref{fig_imfitiso}). The parameters of the fits are given in
  Table \ref{tab:7582decomp}. The inclusion of a central point source
  does not improve the fit. The pseudo bulge contributes 29\% of the
  total light of the galaxy. The large discrepancy with the bulge
  luminosity and size fitted by \citet{Sani2011} stems from their
  inclusion in the bulge of the peanut bar present there.  The $M/L$
  is computed as in Appendix C to match the velocity dispersion of
  \citet{Oliva1995}, measured inside an equivalent $2.48''$ aperture
  radius, and considering the sum of the bulge, the disk and the bar
  profiles. The distance is taken from \citet{KormendyHo2013}.

\item[NGC 7768:] The galaxy is a core elliptical
  \citep{Rusli2013b}. We use their profile and $M/L$, scaled to the
  distance of \citet{KormendyHo2013}.

\item[UGC 3789:] We use the bulge-disk decomposition of Kormendy (in
  prep.), based on the $H-$band profile of \citet{Peletier1999}. The
  $M/L$ is computed as in Appendix C to match the average velocity
  dispersion measured by \citet{Greene2010} within a radius of $1.4''$
  (107$\pm$ 12 km/s) and using the distance of \citet{KormendyHo2013}.
  We classify the galaxy as barred following \citet{Graham2013};
  \citet{KormendyHo2013} disagree.
\end{description}

\begin{table*}
\caption{Bulge plus disk fits to the galaxies of Fig. \ref{fig_maxfits}.}
\label{tab_maxfits}
\tiny
\begin{tabular}{llllllllll}
\hline
\hline
Galaxy   & b/a  & $n_{Ser}^B$      & $a_e^B$        & $\mu_{a_e}$         & $m^B_{tot}$       & $h^D$           & $\mu_0^D$         & $m_{tot}^D$  & Bu/T \\
         &      &                &  (arcsec)     & (mag arcsec$^{-2})$ & (mag)           & (arcsec)        &  (mag arcsec$^{-2})$ & (mag)  & \\
\hline
NGC1023  & 0.75 & 2.71 $\pm$ 0.02 & 8.61 $\pm$ 0.19 & 14.8 $\pm$ 0.03  & 7.249 $\pm$ 0.04 & 57.6 $\pm$ 0.8 & 16.08 $\pm$ 0.04 & 5.59 $\pm$ 0.04 &0.18\\
NGC4258  &  0.6  & 2.91  $\pm$ 0.01 & 17.12 $\pm$ 0.21 & 16.27 $\pm$ 0.02 & 7.43  $\pm$ 0.03 & 87.6  $\pm$ 1.05 & 16.10  $\pm$ 0.02 & 4.95 $\pm$ 0.02 &0.09\\
NGC7457 & 0.75 & 3.14 $\pm$ 0.02  & 10.1 $\pm$ 0.4 & 17.43 $\pm$ 0.05 & 9.47 $\pm$ 0.07 & 27.3 $\pm$ 0.04 & 16.63 $\pm$ 0.04 & 7.77 $\pm$ 0.04 &0.17\\
\hline
\multicolumn{10}{l}{We list the galaxy name, Column 1; the flattening of the system, Column 2; 
the parameters of the bulge (major axis 
half-light radius $a_e$, Column 4; surface brightness at $a_e$,}\\
\multicolumn{10}{l}{Column 5; bulge magnitude, Column 6; exponential scale length of the disk, Column 7; disk central surface brightness, Column 8; disk magnitude, Column 9)}\\
\multicolumn{10}{l}{and bulge-to-total ratio, Column 10.  Surface brightnesses and magnitudes are given in the $3.6mu$ band.} \\
\end{tabular}
\end{table*}

\begin{table*}
\caption[NGC 2273: bulge plus disk decomposition]{Parameters of the bulge plus disk decomposition of NGC 2273.}
\label{tab:2273decomp}
\tiny
 \begin{tabular}{l c c c c c c c c}
 \hline
\hline
  Bulge  & PA$[^\circ]$        & ell                 &  & n                   & $\mu_e$[mag/arcsec$^2$, $R$] & $a_e$[arcsec] & $m_{Bu}$ [$R$ mag] & Bu/T\\ 
         &  $34.50 \pm 0.03$ & $0.3616 \pm 0.0003$ &  & $0.7937 \pm 0.0008$ & $17.2699 \pm 0.0008$      & $1.7801 \pm 0.0007$ & 13.91 & 0.08\\ 
 \hline 
  Bar   & PA$[^\circ]$         & ell                  & $c_0$                & n & $\mu_e$[mag/arcsec$^2$, $R$] & $a_e$[arcsec]  & $m_{Bar}$ [$R$ mag] & Bar/T\\
        &  $105.08 \pm 0.02$ & $0.4022 \pm 0.0003$  & $-0.305 \pm 0.002$ & $0.927 \pm 0.001$ & $20.2888\pm 0.0009$ & $14.67 \pm 0.01$ & 12.43 & 0.31\\
 \hline
  Disk & PA$[^\circ]$ & ell &&  &$\mu_0$ [mag/arcsec$^2$, $R$] & h [arcsec]  & $m_{Di}$ [$R$ mag] & Di/T\\
      &  $63.76 \pm 0.03$   &$0.3919 \pm 0.0004$ & & & $ 20.868 \pm 0.002$  &  $ 33.94\pm 0.03$ & 11.76 & 0.57\\
 \hline
  Inner Ring & PA$[^\circ]$ & ell &  & & $\mu_0$ [mag/arcsec$^2$, $R$] & $r_{ring}$[arcsec]  & $m_{Ri}$ [$R$ mag] & Ri/T\\
             & $64.65 \pm 0.05$ & $0.3069 \pm 0.0004$ & & & $21.884 \pm 0.002$ & $ 20.45 \pm 0.02 $ & 14.65 & 0.04\\
             & $\sigma_{in}$ [arcsec] & $\sigma_{out}$ [arcsec]& & & \\
             & $2.14 \pm 0.02 $ & $4.31 \pm 0.01$ \\
\hline
\end{tabular} 
\end{table*}

\begin{figure*}
 \begin{center}
    \includegraphics[width=16cm]{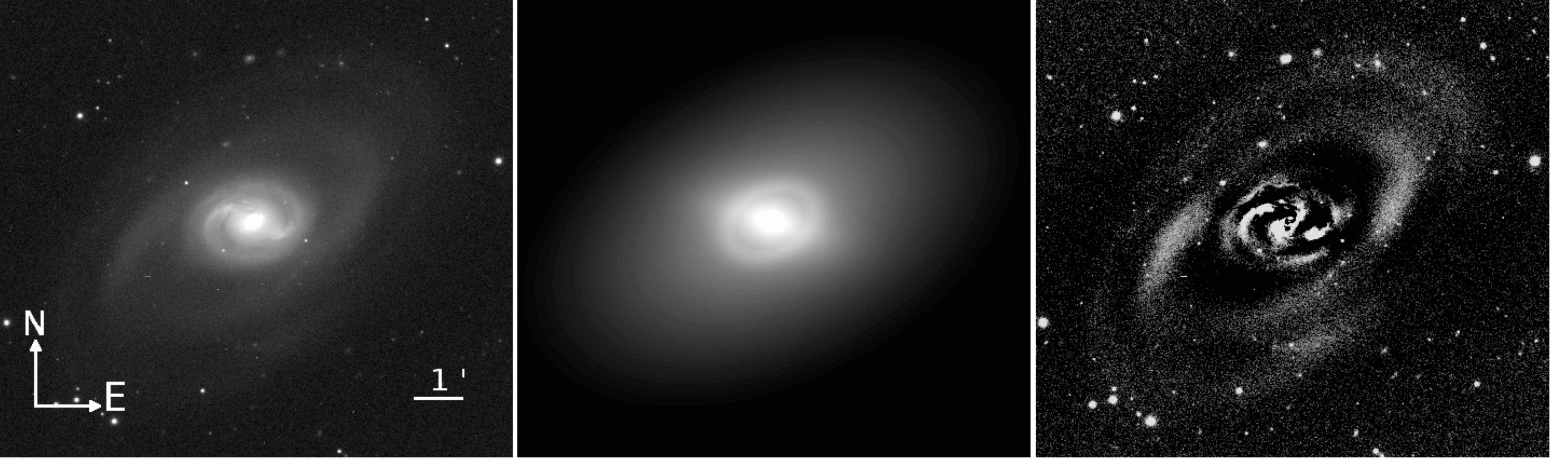}
  \end{center}
  \caption{The \textsc{Imfit} model for NGC 2273: the galaxy $R-$band image (left), the model (middle) and the residuals (right); 
see also Fig. \ref{fig_imfitiso}.
\label{fig_ngc2273}}
\end{figure*}

\begin{table*}
\caption[NGC 2549: bulge plus disk decomposition]{Parameters of the bulge plus disk decomposition of NGC 2549.}
\label{tab:2549decomp}
 \scriptsize
 \begin{tabular}{l c c c c c c c}
\hline
 \hline
  Bulge  & PA$[^\circ]$ & ell &   n &$\mu_e$[mag/arcsec$^2$, $R$] & $a_e$[arcsec] & $m_{Bu}$ [$R$ mag] & Bu/T\\ 
        &  $179.73 \pm 0.02$ & $0.5055 \pm 0.0002$&  $3.467 \pm 0.004$ & $19.23 \pm 0.003$ & $13.09 \pm 0.003$ & 11.1 & 0.77\\ 
 \hline 
  Edge-On-Disk & PA$[^\circ]$ & $\mu_0$ [mag/arcsec$^2$, $R$] & h [arcsec]  &  $\alpha=2/n$ & $z_0$ [arcsec] & $m_{Di}$ [$R$ mag] & Di/T\\
      &  $179.31 \pm 0.01$  & $ 20.14\pm 0.002$  & $26.43 \pm 0.03$ & $2.38 \pm 0.04$ &  $5.07 \pm 0.03$ & 12.39 & 0.23\\
 \hline
\end{tabular}
\end{table*}

\begin{figure*}
 \begin{center}
    \includegraphics[width=16cm]{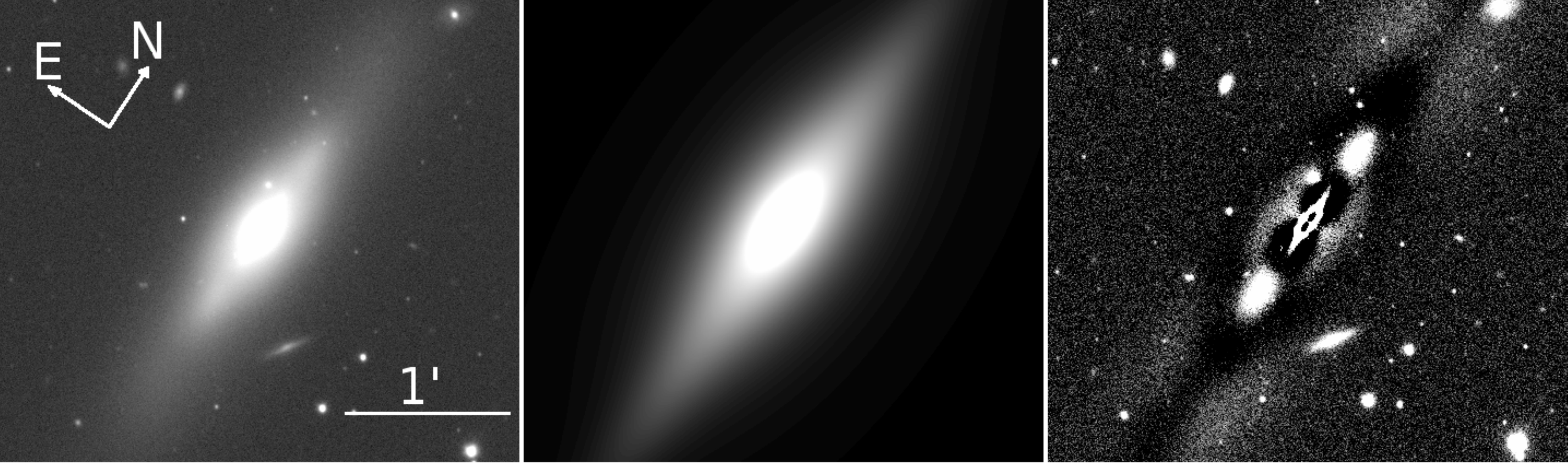}
  \end{center}
  \caption{The \textsc{Imfit} model for NGC 2549: the galaxy $r-$band image (left), the model (middle) and the residuals (right), see also Fig. \ref{fig_imfitiso}.
\label{fig_ngc2549}}
\end{figure*}

\begin{table*}
\caption[NGC 2748: Bulge-bar-disk decomposition]{Parameters of the bulge-bar-disk decomposition of NGC 2748.}
\label{tab:2748decomp}
\tiny
 \begin{tabular}{l c c c c c c c}
 \hline
\hline
  Bulge & PA$[^\circ]$       & ell                &  n                 & $\mu_e$[mag/arcsec$^2$, $3.6mu$] & $a_e$[arcsec] & $m_{Bu}$ [$3.6mu$ mag] & Bu/T\\ 
        & 13.68 $\pm$ 0.07  & 0.386  $\pm$ 0.001 &  1.419 $\pm$ 0.003 & 15.99  $\pm$    0.01      &  4.447  $\pm$    0.007 & 10.42 & 0.15\\
 \hline 
  Disk & PA$[^\circ]$       & ell                 &                    & $\mu_0$ [mag/arcsec$^2$, $3.6mu$] & h [arcsec] & $m_{Di}$ [$3.6mu$ mag] & Di/T   \\
       & 41.27 $\pm$ 0.01  & 0.719  $\pm$ 0.001  &                    & 15.50  $\pm$   0.01        & 16.75  $\pm$   0.01 & 8.35 & 0.69\\
 \hline
Inner Ring & PA$[^\circ]$   & ell                 &                    & $\mu_0$ [mag/arcsec$^2$, $3.6mu$] & $r_{ring}$ [arcsec] & $m_{Ri}$  [$3.6mu$ mag] & Ri/T \\
       & 44.28 $\pm$ 0.04  & 0.589 $\pm$  0.001  &                    & 17.30  $\pm$  0.01           & 10.73  $\pm$  0.01& 11.87 & 0.04\\
  & $\sigma_{ring}$ [arcsec] \\
  &     2.126  $\pm$    0.009 \\
 \hline
  Outer Ring   & PA$[^\circ]$       & ell                 &             & $\mu_0$ [mag/arcsec$^2$, $3.6mu$]   & $r_{ring}$ [arcsec] & $m_{Ro}$  [$3.6mu$ mag] & Ro/T   \\
               & 43.3 $\pm$ 0.005  & 0.822  $\pm$  0.001 &             &  17.49   $\pm$  0.01             &  28.80   $\pm$   0.01 & 9.09 & 0.12\\ 
  & $\sigma_{ring}$ [arcsec] \\
  &     6.90  $\pm$    0.01\\
\hline
\end{tabular}
\end{table*}

\begin{figure*}
 \begin{center}
    \includegraphics[width=5.3cm]{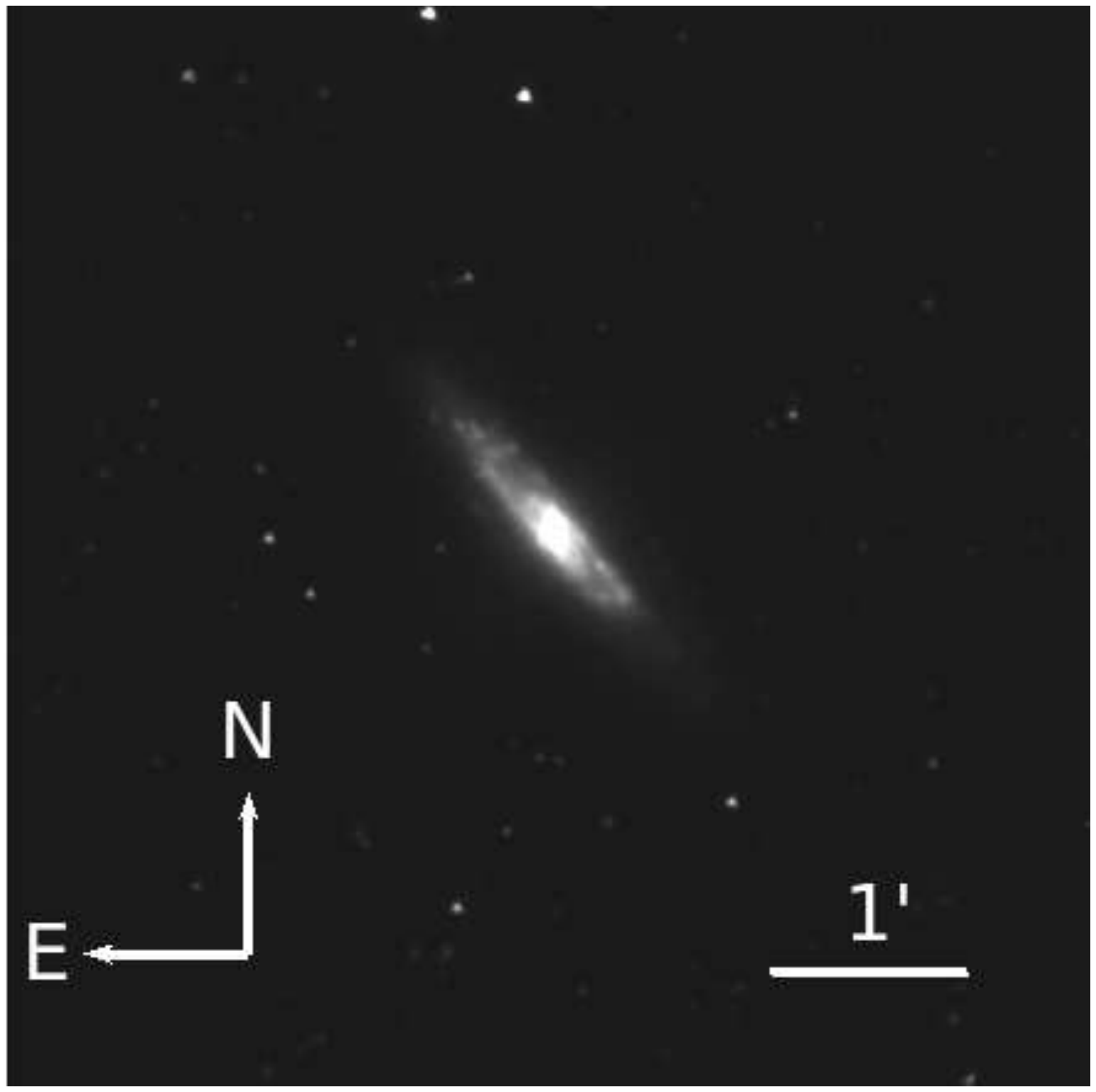}
    \includegraphics[width=5.3cm]{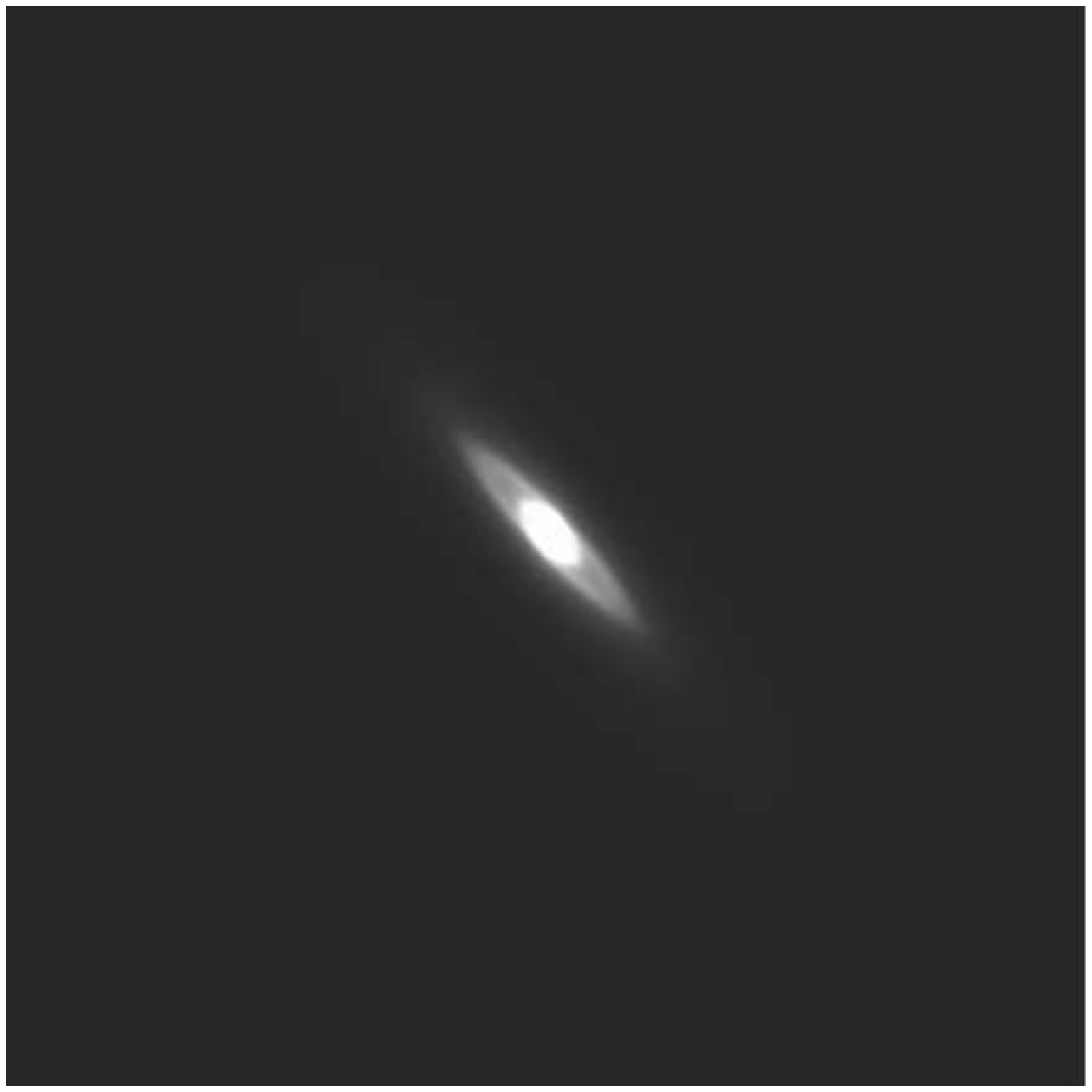}
    \includegraphics[width=5.3cm]{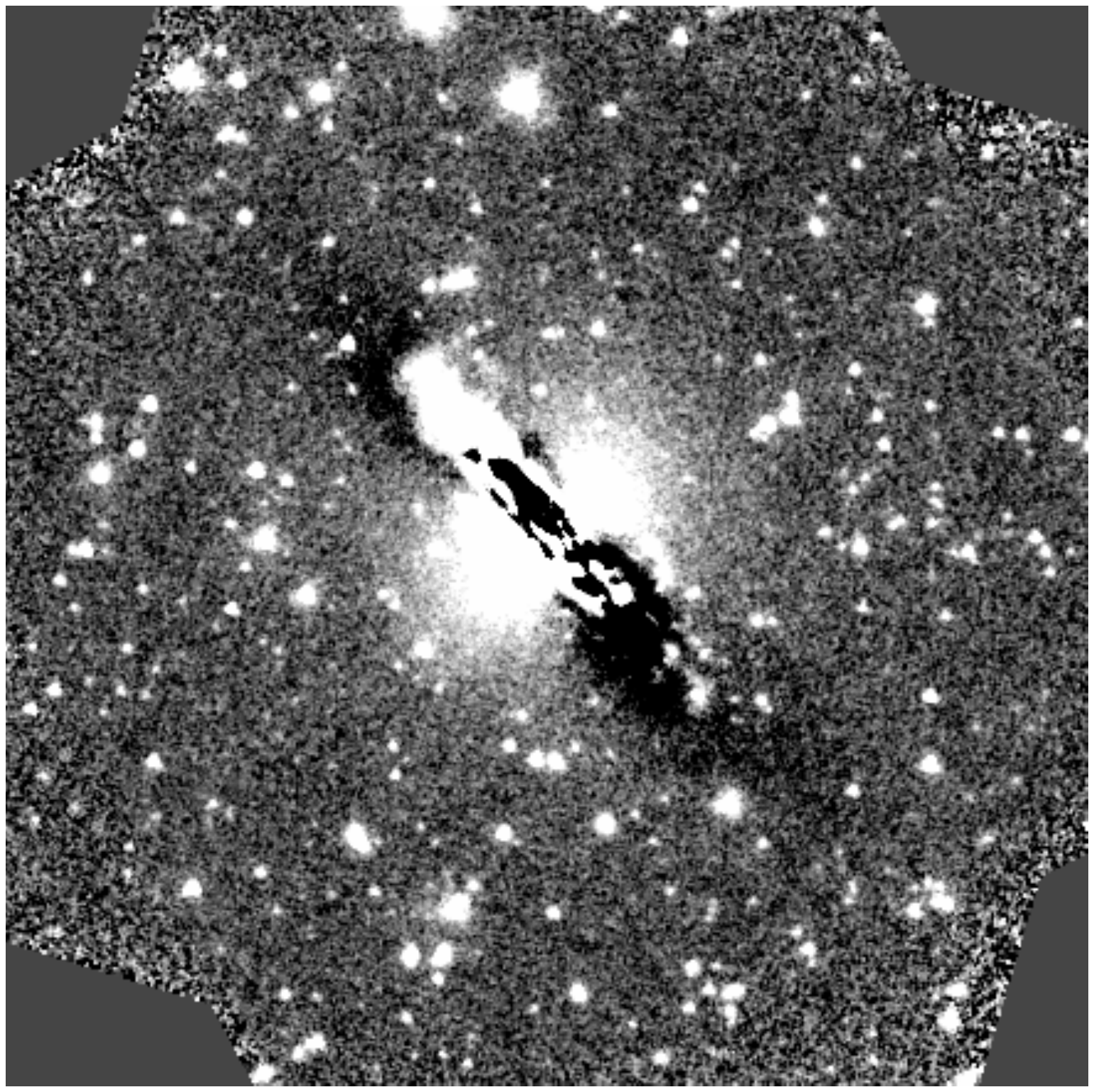}
  \end{center}
  \caption{The \textsc{Imfit} model for NGC 2748: the galaxy $3.6mu-$band image (left), the model (middle) and the residuals (right); 
see also Fig. \ref{fig_imfitiso}.
\label{fig_ngc2748}}
\end{figure*}

\begin{table*}
\caption[NGC 3079: Pseudo Bulge-Bar-Disk-Ring decomposition]{Parameters of the bulge-bar-disk-ring decomposition of NGC 3079.}
\label{tab:3079decomp}
\scriptsize
 \begin{tabular}{l c c c c c c c}
 \hline
 \hline
  Bulge & PA$[^\circ]$     & ell                &  n                 & $\mu_e$[mag/arcsec$^2$, $3.6mu$] & $a_e$[arcsec] &  $m_{Bu}$ [$3.6mu$ mag] & Bu/T\\ 
        & 169.4 $\pm$ 0.1 & 0.79  $\pm$ 0.01   &  2.724 $\pm$ 0.002 & 13.00 $\pm$ 0.011         & 4.50 $\pm$ 0.01& 8.24 & 0.26\\
\hline
  Bar   & PA$[^\circ]$     & ell                &  n               & $\mu_e$[mag/arcsec$^2$, $3.6mu$] &$a_e$[arcsec]& $m_{Bar}$ [$3.6mu$ mag] & Bar/T \\ 
        & 171.6 $\pm$ 0.1 & 0.69 $\pm$ 0.01    &  0.37 $\pm$ 0.01 &  17.43  $\pm$   0.01      &  40.77  $\pm$    0.01& 8.35 & 0.24\\
 \hline 
  Disk  & PA$[^\circ]$      & ell             & & $\mu_0$ [mag/arcsec$^2$,  $3.6mu$] & h [arcsec]   &   $m_{Di}$ [$3.6mu$ mag] & Di/T \\
        & 166.9 $\pm$ 0.1  & 0.86 $\pm$ 0.01 & & 16.35 $\pm$ 0.01           & 52.87 $\pm$   0.01 & 7.83 & 0.38\\
 \hline
  Ring   & PA$[^\circ]$       & ell              &  $\mu$ [mag/arcsec$^2$, $3.6mu$]   & $r_{ring}$ [arcsec]   & $\sigma_{ring}$ [arcsec] &   $m_{Ri}$ [$3.6mu$ mag] & Ri/T \\
         & 169.2  $\pm$  0.01& 0.90  $\pm$ 0.01 & 16.66  $\pm$  0.01              &  52.74   $\pm$  0.01 & 13.11   $\pm$  0.01 & 9.09 & 0.12 \\
\hline
\end{tabular}
\end{table*}

\begin{figure*}
 \begin{center}
    \includegraphics[width=5.3cm]{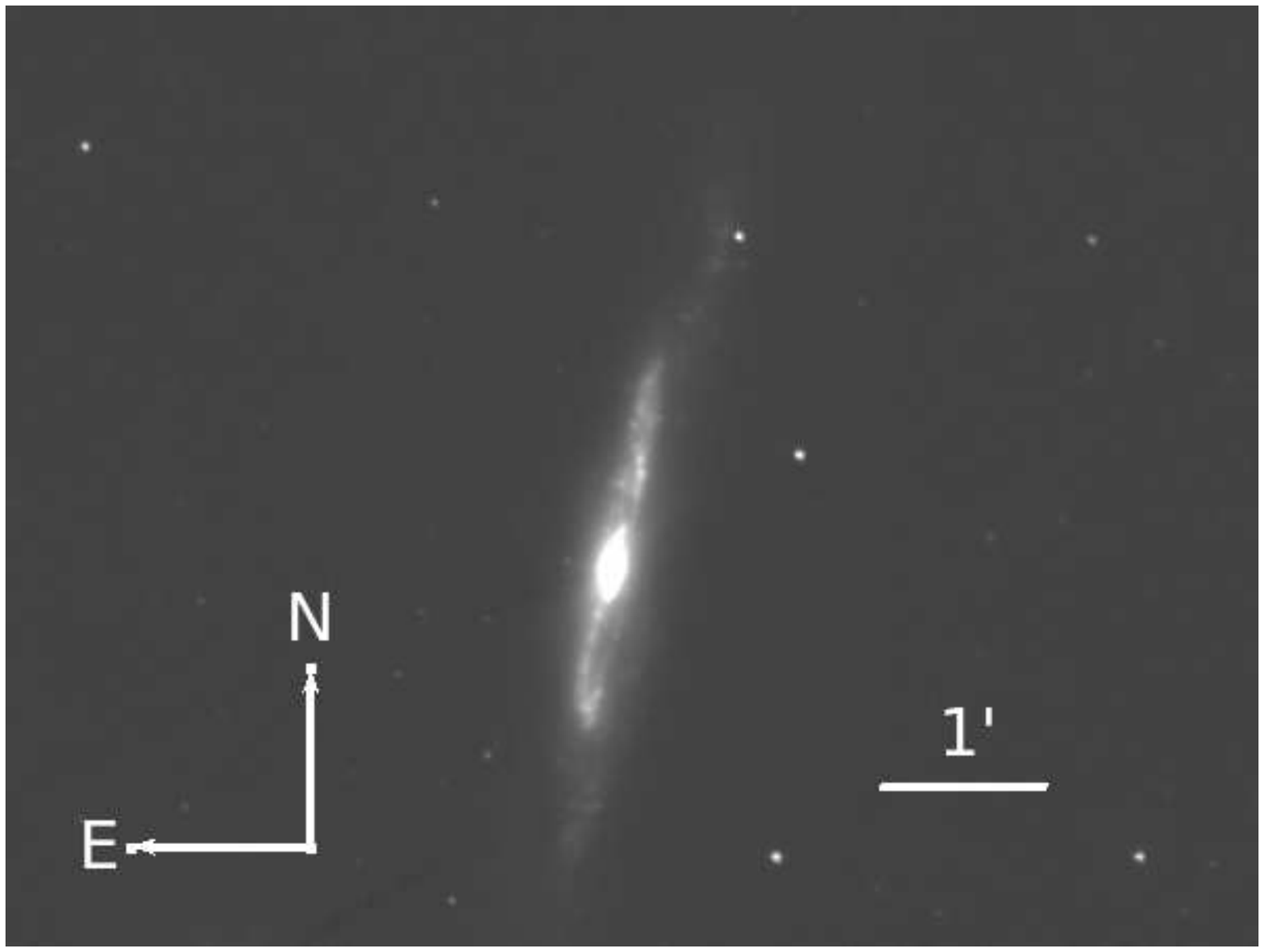}
    \includegraphics[width=5.3cm]{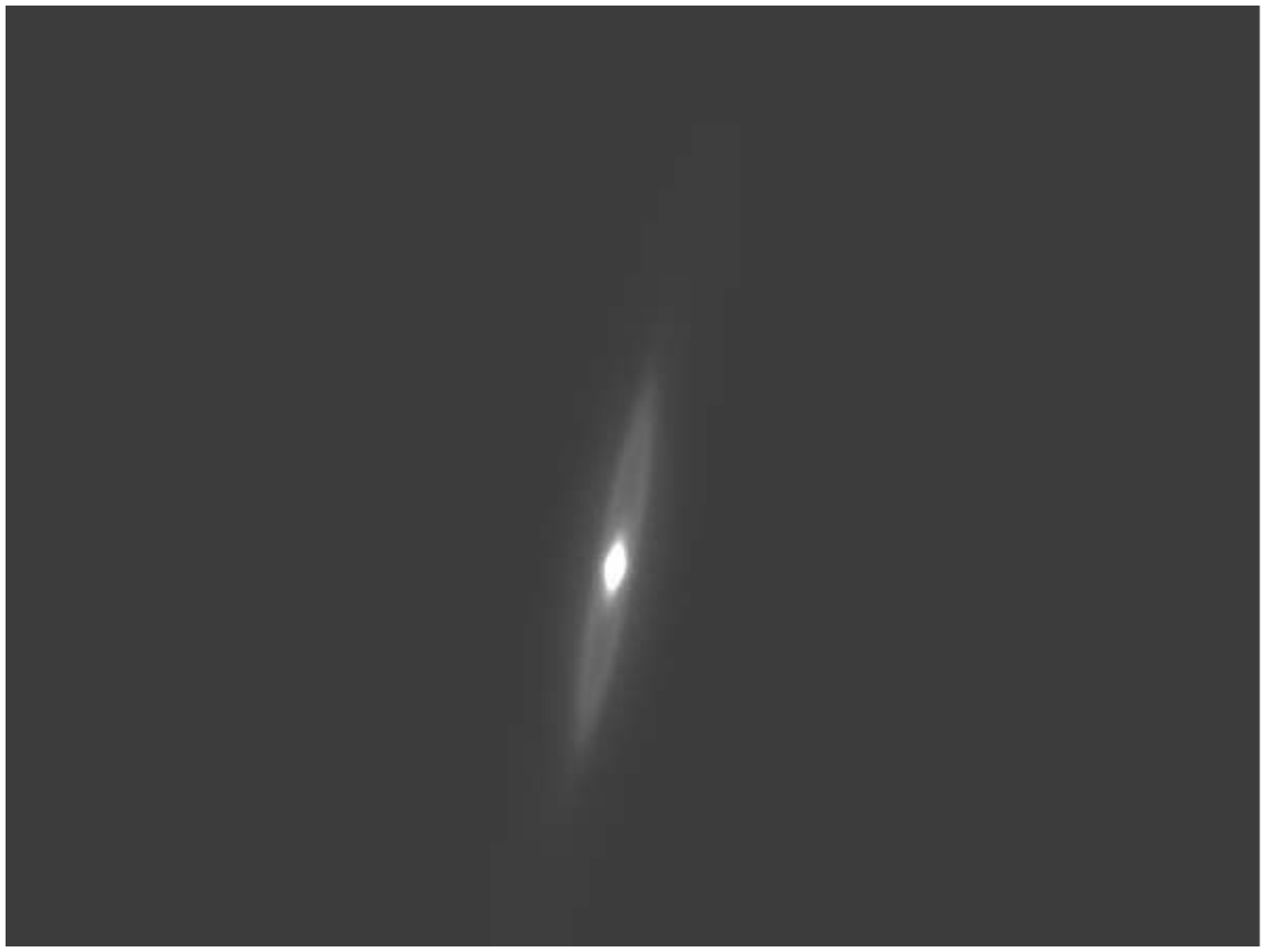}
    \includegraphics[width=5.3cm]{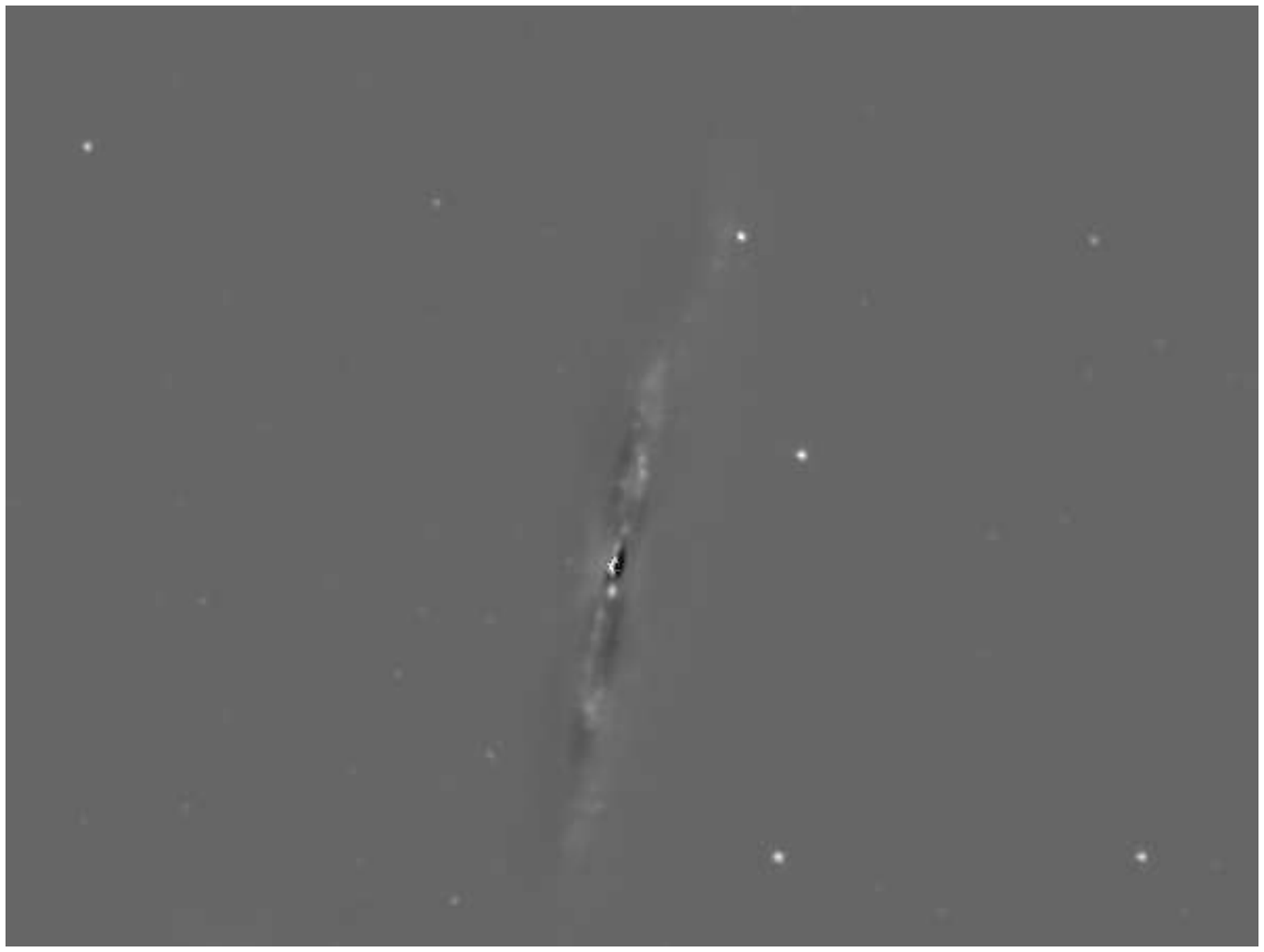}
  \end{center}
  \caption{The \textsc{Imfit} model for NGC 3079: the galaxy $3.6mu-$band image (left), the model (middle) and the residuals (right); 
see also Fig. \ref{fig_imfitiso}.
\label{fig_ngc3079}}
\end{figure*}

\begin{table*}
\caption[NGC 3393: bulge plus disk decomposition]{Parameters of the bulge plus disk decomposition of NGC 3393.}
\label{tab:3393decomp}
\scriptsize
 \begin{tabular}{l c c c c c c c c}
 \hline
 \hline
  Bulge  & PA$[^\circ]$ & ell &  &  n &$\mu_e$[mag/arcsec$^2$, $I$] & $a_e$[arcsec] &  $m_{Bu}$ [$I$ mag] & Bu/T \\ 
        &  $142.1 \pm 0.5$ & $0.107 \pm 0.002$&  & $1.45 \pm 0.01$ & $17.27 \pm 0.01$ & $1.91 \pm 0.01$ & 13.12 & 0.17\\ 
 \hline 
  Bar  & PA$[^\circ]$ & ell & $c_0$ & n & $\mu_e$[mag/arcsec$^2$, $I$] & $a_e$[arcsec] & $m_{Bar}$ [$I$ mag] & Bar/T \\ 
        &  $160.5 \pm 0.1$ & $0.429 \pm 0.002$  & $-0.35 \pm 0.01$ & $0.352 \pm 0.004$ & $19.632\pm 0.007$ & $9.98 \pm 0.03$ & 13.08 & 0.18\\ 
 \hline
  Disk & PA$[^\circ]$ & ell & & & $\mu_0$ [mag/arcsec$^2$, $I$] & h [arcsec] & $m_{Di}$ [$I$ mag] & Di/T  \\
      &  $29 \pm 4$  & $0.022 \pm 0.003$ &    &   &$ 20.022 \pm 0.005$  &  $18.98 \pm 0.06$ & 11.66 & 0.65\\
 \hline
\end{tabular}
\end{table*}

\begin{figure*}
 \begin{center}
    \includegraphics[width=16cm]{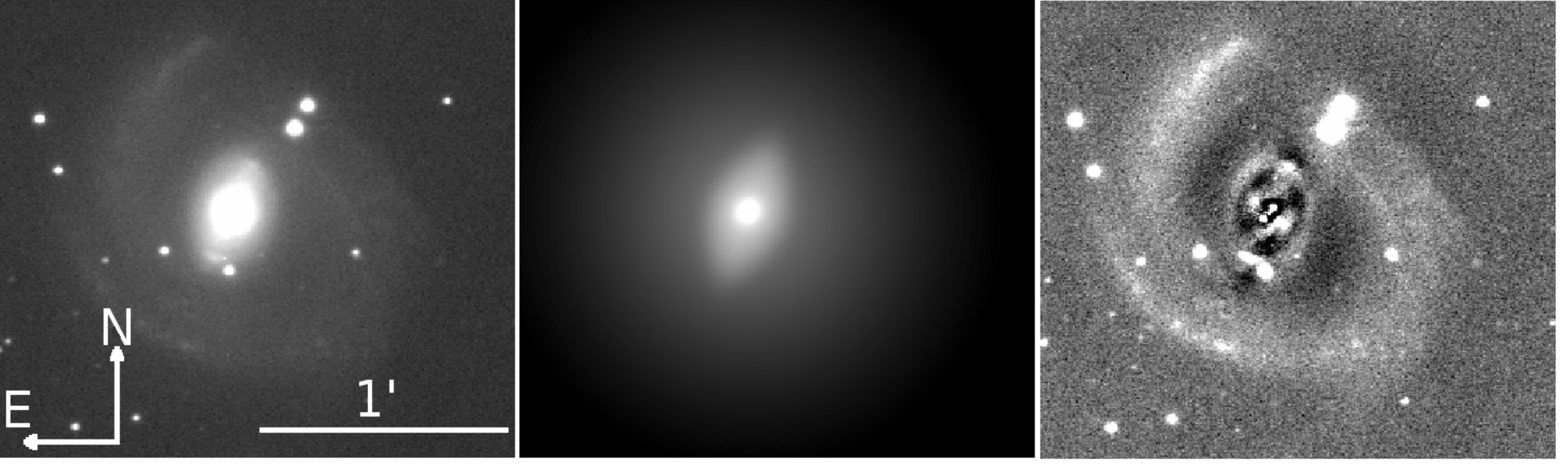}
  \end{center}
  \caption{The \textsc{Imfit} model for NGC 3393: the galaxy $I-$band image (left), the model (middle) and the residuals (right); see also Fig. \ref{fig_imfitiso}.
\label{fig_ngc3393}}
\end{figure*}

\begin{table*}
\caption[NGC 3414: bulge plus disk decomposition]{Parameters of the bulge plus disk decomposition of NGC 3414.}
\label{tab:3414decomp}
 \scriptsize \begin{tabular}{l c c c c c c c c}
 \hline
 \hline
  Bulge  & PA$[^\circ]$ & ell &  &  n &$\mu_e$[mag/arcsec$^2$, $I$] & $a_e$[arcsec] &  $m_{Bu}$ [$I$ mag] & Bu/T \\ 
        &  $176.01 \pm 0.01$ & $0.1918 \pm 0.0005$&  & $5.13 \pm 0.01$ & $20.827 \pm 0.008$ & $28.04 \pm 0.14$ & 9.9 & 0.79\\ 
 \hline 
  Bar  & PA$[^\circ]$ & ell & $c_0$ & n & $\mu_e$[mag/arcsec$^2$, $I$] & $a_e$[arcsec] & $m_{Bar}$ [$I$ mag] & Bar/T  \\ 
        &  $199.48 \pm 0.04$ & $0.8527 \pm 0.0006$  & $0.478 \pm 0.006$ &  $4.7 \pm 0.3$ & $22.177 \pm 0.006$ & $40.59 \pm 0.14$ & 13.2 & 0.04\\ 
 \hline
  Disk & PA$[^\circ]$ & ell & & & $\mu_0$ [mag/arcsec$^2$, $I$] & h [arcsec]  &    $m_{Di}$ [$I$ mag] & Di/T     \\
      &  $35.7 \pm 0.3$  & $0.307 \pm 0.003$ &    &   &$ 21.152 \pm 0.007$  &  $33.2 \pm 0.1$ & 11.54 & 0.17\\
 \hline
\end{tabular}
\end{table*}

\begin{figure*}
 \begin{center}
    \includegraphics[width=16cm]{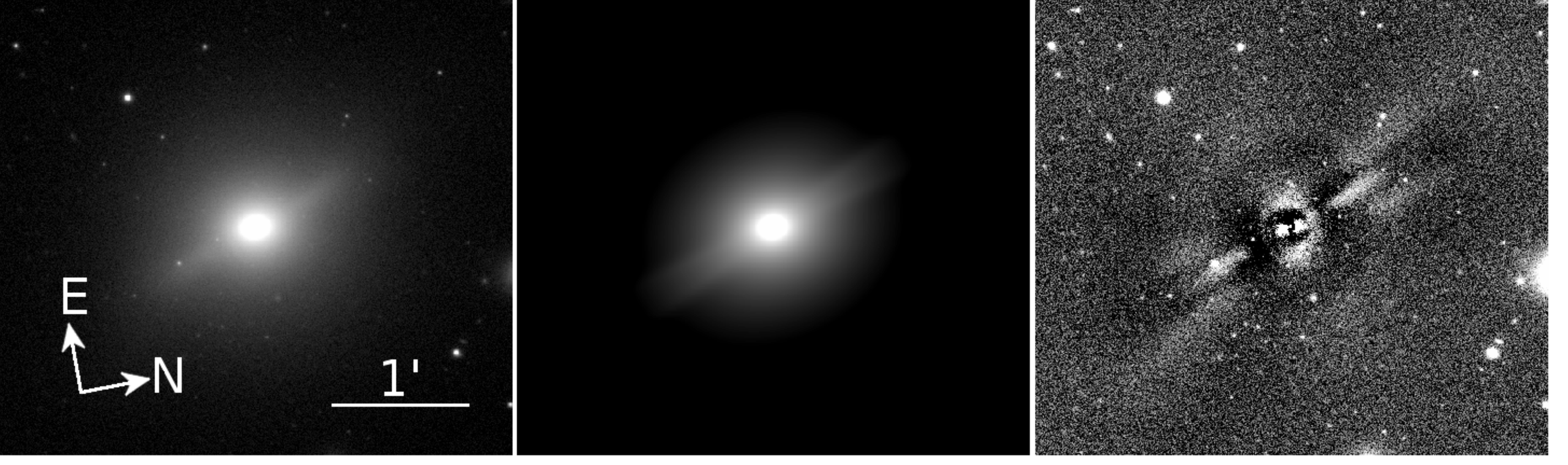}
  \end{center}
  \caption{The \textsc{Imfit} model for NGC 3414: the galaxy $I-$band image (left), the model (middle) and the residuals (right); 
see also Fig. \ref{fig_imfitiso}.
\label{fig_ngc3414}}
\end{figure*}

 \begin{table*}
\caption[NGC 4026: bulge plus disk decomposition]{Parameters of the bulge plus disk decomposition of NGC 4026.}
\label{tab:4026decomp}
\scriptsize
 \begin{tabular}{l c c c c c c c}
 \hline
 \hline
  Bulge  & PA$[^\circ]$ & ell &   n &$\mu_e$[mag/arcsec$^2$, $V$] & $a_e$[arcsec] &  $m_{Bu}$ [$V$ mag] & Bu/T  \\ 
        &  $180.9 \pm 0.1$ & $0.402457 \pm 0.0004$&  $3.242 \pm 0.004$ & $19.222 \pm 0.003$ & $10.67 \pm 0.02$ & 11.36 & 0.59\\ 
 \hline 
  Edge-On-Disk & PA$[^\circ]$ & $\mu_0$ [mag/arcsec$^2$, $V$] & h [arcsec]  &  $\alpha=2/n$ & $z_0$ [arcsec] &  $m_{Di}$ [$V$ mag] & Di/T \\
      &  $177.5 \pm 0.1$  & $ 19.893\pm 0.003$  & $34.34 \pm 0.03$ & $1.50 \pm 0.01$ &  $4.83 \pm 0.02$ & 11.76 & 0.41\\
 \hline
\end{tabular}
\end{table*}
\begin{figure*}
 \begin{center}
    \includegraphics[width=16cm]{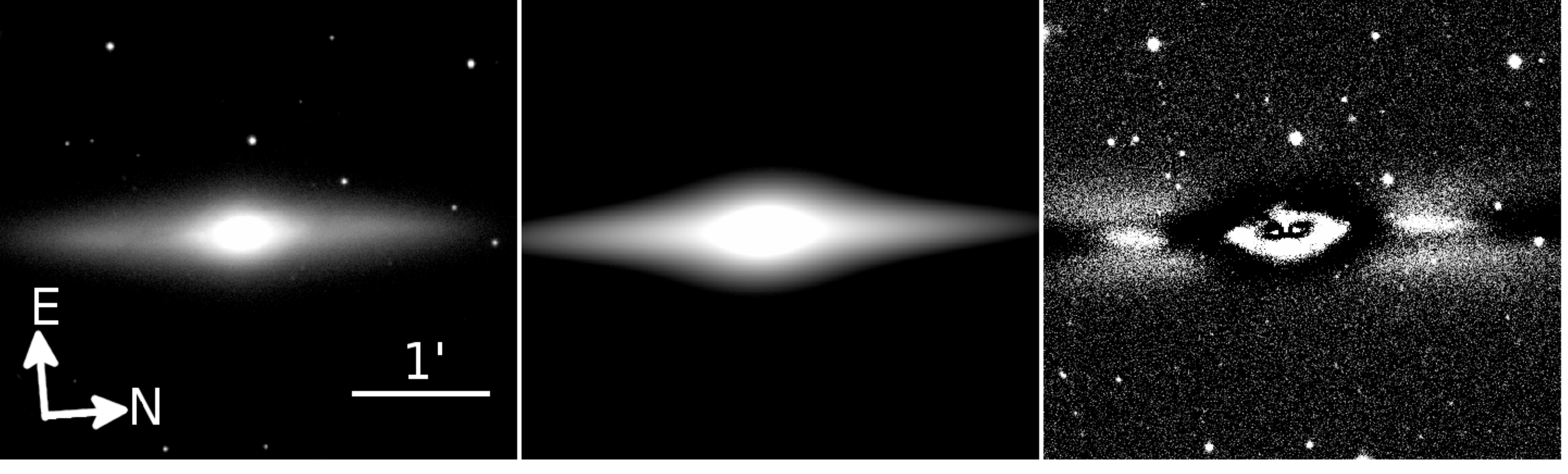}
  \end{center}
  \caption{The \textsc{Imfit} model for NGC 4026: the galaxy $V-$band image (left), the model (middle) and the residuals (right); 
see also Fig. \ref{fig_imfitiso}.
\label{fig_ngc4026}}
\end{figure*}

\begin{table*}
\caption[NGC 4342: bulge plus disk decomposition]{Parameters of the bulge plus disk decomposition of NGC 4342.}
\label{tab:4342decomp}
\scriptsize
 \begin{tabular}{l c c c c c c c}
 \hline
 \hline
  Bulge  & PA$[^\circ]$ & ell & n & $\mu_e$[mag/arcsec$^2$, $I$] & $a_e$[arcsec] &  $m_{Bu}$ [$I$ mag] & Bu/T\\ 
        &  $164.99 \pm 0.07$ & $0.3420 \pm 0.0009$  & $7.51 \pm 0.02$ & $18.50 \pm 0.008$ & $5.79 \pm 0.02$ & 11.42 & 0.65\\ 
 \hline 
  Disk & PA$[^\circ]$ & ell & & $\mu_0$ [mag/arcsec$^2$, $I$] & h [arcsec] &  $m_{Di}$ [$I$ mag] & Di/T \\
      &  $167.46 \pm 0.01$  & $0.7680 \pm 0.0003$ &      &$ 16.22 \pm 0.002$  &  $5.581 \pm 0.005$ & 12.07 & 0.35 \\
 \hline
\end{tabular}
\end{table*}

\begin{figure*}
 \begin{center}
    \includegraphics[width=16cm]{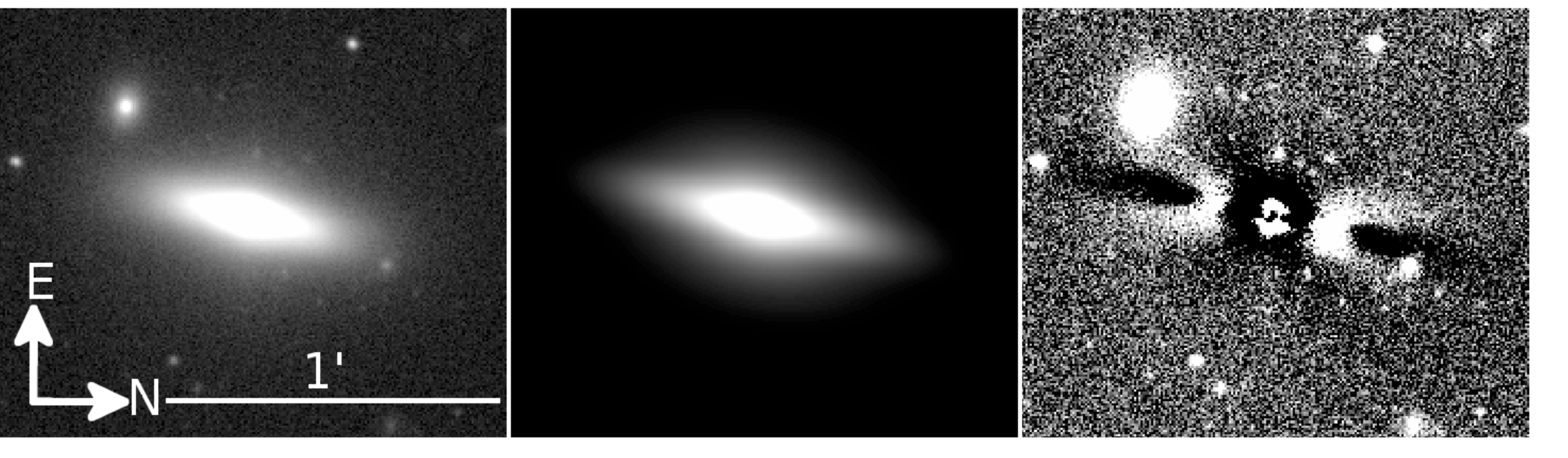}
  \end{center}
  \caption{The \textsc{Imfit} model for NGC 4342: the galaxy $I-$band image (left), the model (middle) and the residuals (right); 
see also Fig. \ref{fig_imfitiso}.
\label{fig_ngc4342}}
\end{figure*}

\clearpage
\begin{table*}
\caption[NGC 4388: bulge plus disk decomposition]{Parameters of the bulge--ring--disk plus point source decomposition of NGC 4388.}
\label{tab:4388decomp}
\tiny
 \begin{tabular}{l c c c c c c c}
 \hline
 \hline
  Bulge  & PA$[^\circ]$       & ell                & n               & $\mu_e$[mag/arcsec$^2$, $3.6mu$] & $a_e$[arcsec]  &  $m_{Bu}$ [$3.6mu$ mag] & Bu/T\\ 
         & 93.3 $\pm$ 0.1    & 0.494 $\pm$ 0.001  & 2.886  $\pm$ 0.002  & 18.23 $\pm$ 0.01          & 27.82 $\pm$ 0.01 & 8.53 & 0.38\\ 
 \hline 
  Disk   & PA$[^\circ]$       & ell               &      & $\mu_0$ [mag/arcsec$^2$, $3.6mu$] & h [arcsec] & $m_{Di}$ [$3.6mu$ mag] & Di/T \\
         & 88.75 $\pm$ 0.01  & 0.830 $\pm$ 0.001 &      & 17.29 $\pm$  0.01              & 48.64  $\pm$   0.01 & 8.78 & 0.31 \\

\hline
  Ring   & PA$[^\circ]$       & ell           &      & $\mu_0$ [mag/arcsec$^2$, $3.6mu$]   & $r_{ring}$ [arcsec] &  $m_{Ri}$ [$3.6mu$ mag] & Ri/T \\ 
         & 90.97  $\pm$ 0.01 & 0.82 $\pm$ 0.01 &    & 16.78 $\pm$   0.01 &      28.99 $\pm$     0.01 & 9.4 & 0.17\\
         & $\sigma_{in}$ [arcsec]       & $\sigma_{out}$ [arcsec]&      &                    &                               \\
         & 7.39  $\pm$   0.01  & 12.28  $\pm$   0.01 \\
\hline
Point Source &  PA$[^\circ]$       & ell                  &      & $\mu_0$ [mag/arcsec$^2$, $3.6mu$] & $\sigma$ [arcsec] &  $m_{PS}$ [$3.6mu$ mag] & PS/T\\ 
             &  63.55  $\pm$ 0.19 & 0.649  $\pm$   0.002 &      & 8.392  $\pm$   0.008 & 0.38  $\pm$  0.01 & 9.63 & 0.14\\
 \hline
\end{tabular}
\end{table*}

\begin{figure*}
 \begin{center}
    \includegraphics[width=5.3cm]{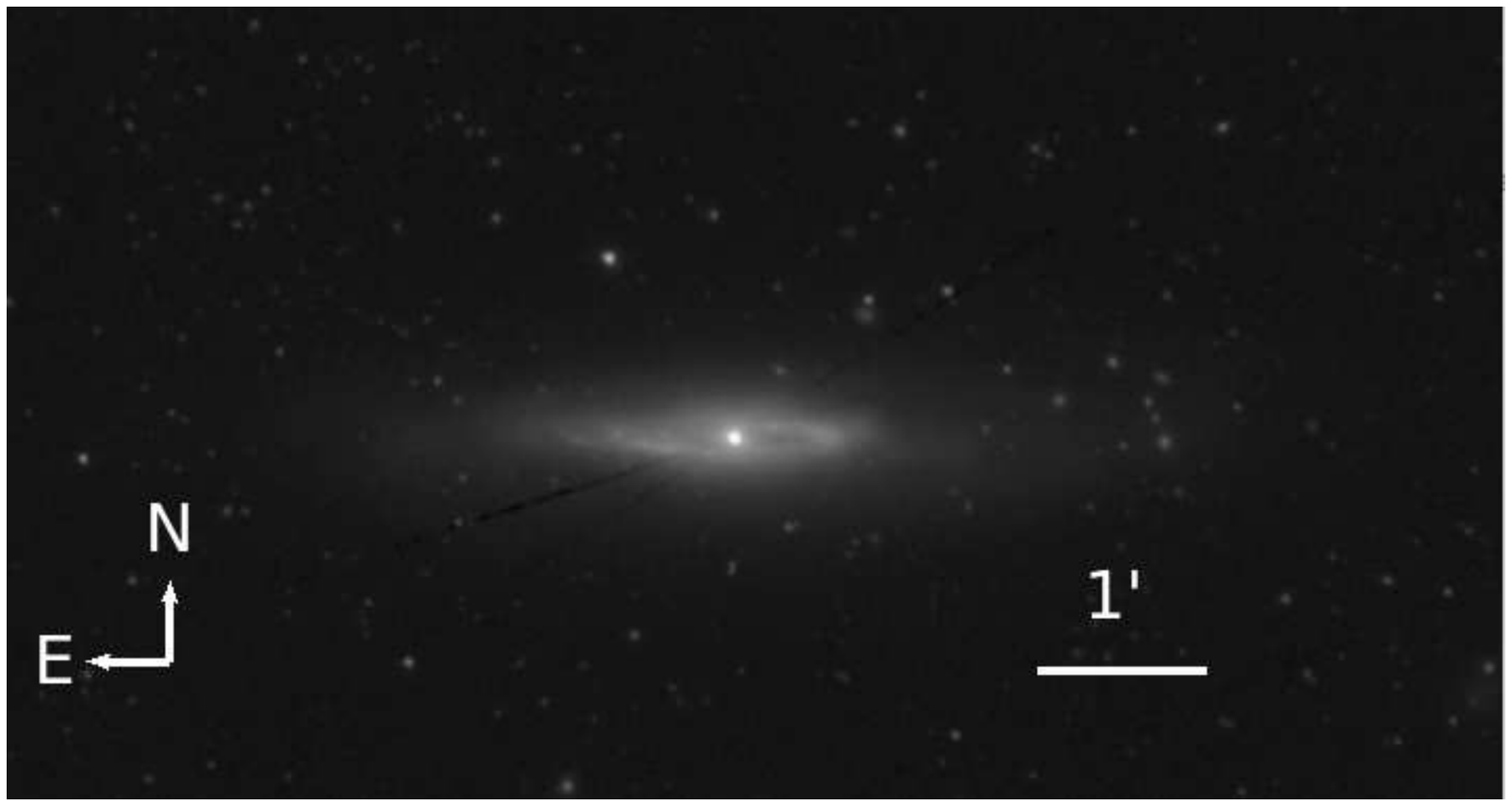}
    \includegraphics[width=5.3cm]{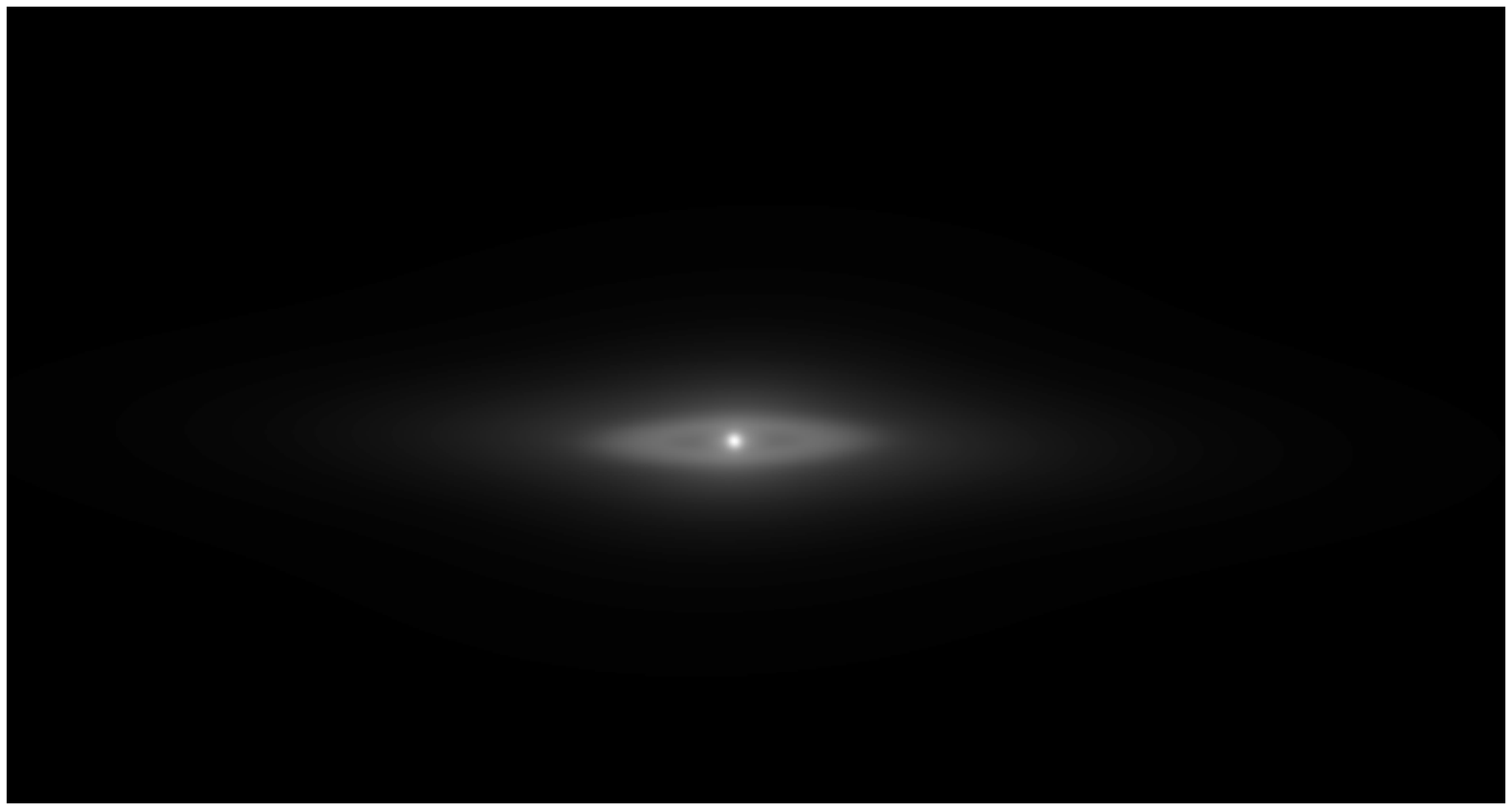}
    \includegraphics[width=5.3cm]{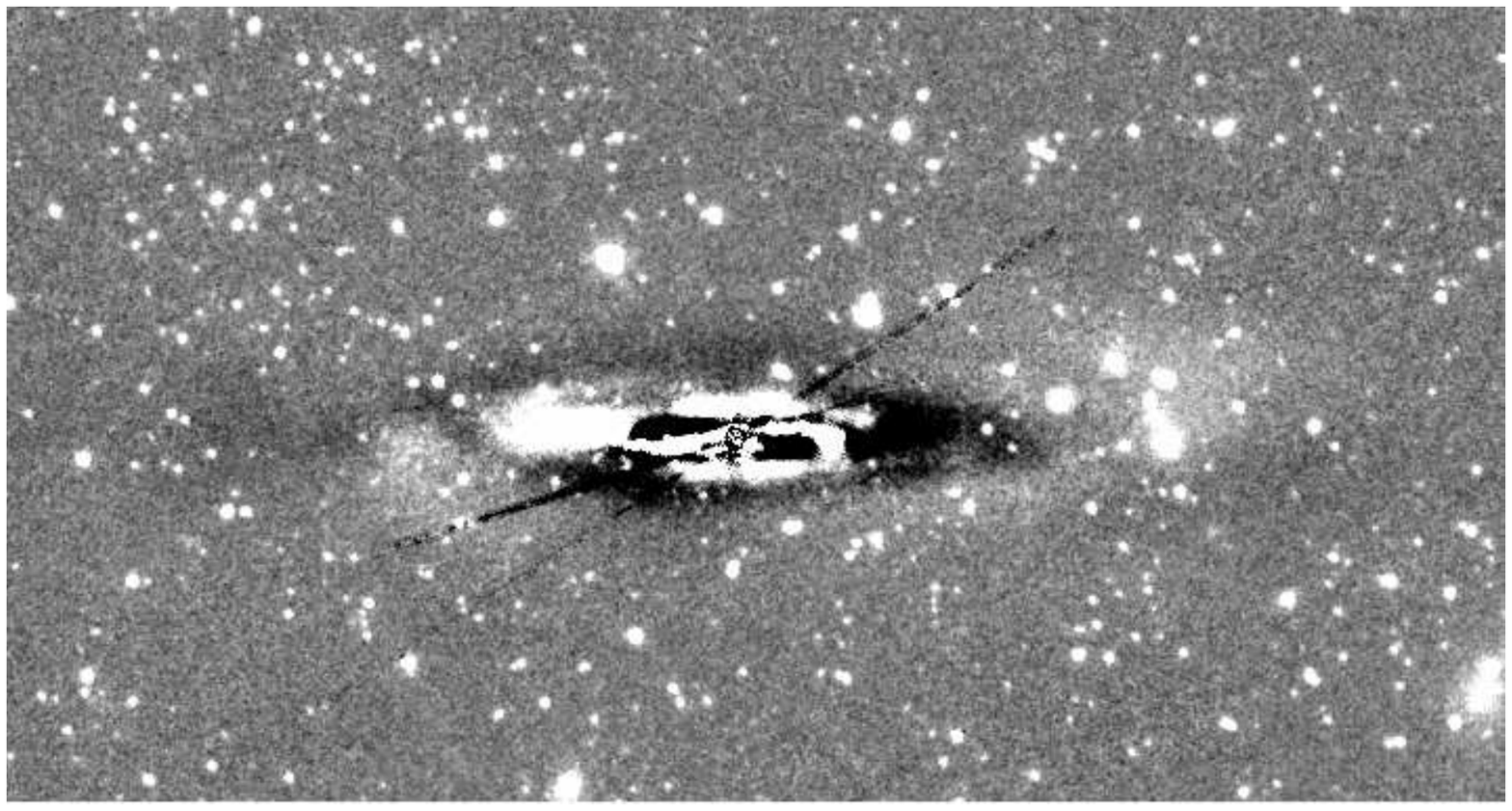}
  \end{center}
  \caption{The \textsc{Imfit} model for NGC 4388: the galaxy $3.6mu-$band image (left), the model (middle) and the residuals (right); 
see also Fig. \ref{fig_imfitiso}.
\label{fig_ngc4388}}
\end{figure*}

\begin{table*}
\caption[NGC 4526: bulge plus disk decomposition]{Parameters of the bulge plus disk decomposition of NGC 4526.}
\label{tab:4526decomp}
\tiny
 \begin{tabular}{l c c c c c c c c}
 \hline
 \hline
  Bulge  & PA$[^\circ]$    & ell     & n               & $\mu_e$[mag/arcsec$^2$, $3.6mu$] & $a_e$[arcsec] &  $m_{Bu}$ [$3.6mu$ mag] & Bu/T \\ 

         & $10.7 \pm 1.4$ & $0.15 \pm 0.01$ & $2.74 \pm 0.03$ &               $15.84 \pm   0.04$ & $7.3 \pm 0.2$ & 8.51 & 0.11\\
 \hline 
Inner  Disk & PA$[^\circ]$       & ell &  n  & $\mu_0$ [mag/arcsec$^2$, $3.6mu$] & h [arcsec]      &  $m_{Di}$ [$3.6mu$ mag] & Di/T \\
            & $21.58 \pm 0.05$ & $0.69 \pm 0.02$ & $12.23 \pm 0.01$             & $3.78 \pm 0.01$ &  8.63                  & 0.10 \\
 \hline
  Bar  & PA$[^\circ]$      & ell               & n                 & $\mu_e$[mag/arcsec$^2$, $3.6mu$] & $a_e$[arcsec] & $m_{Bar}$ [$3.6mu$ mag] & Bar/T \\ 
       & $28.28 \pm 0.07$ & $0.368 \pm 0.001$ &  $0.738 \pm 0.005$ & $16.33\pm 0.01$             &  $19.54 \pm 0.04$ & 7.66                  & 0.25  \\
     &  $c_0$            & &&&&&\\
     & $1.55 \pm 0.03$ &&&&&&\\
\hline
Outer  Disk & PA$[^\circ]$       & ell             & n & $\mu_0$ [mag/arcsec$^2$, $3.6mu$] & h [arcsec]      &  $m_{Do}$ [$3.6mu$ mag] & Do/T \\
            & $27.97 \pm 0.09$ & $0.710 \pm 0.003$ & 1 & $19.02\pm 0.02$                  & $154.9 \pm 2.6$ &  7.83                  & 0.21 \\
    &  $c_0$            & &&&&&\\
    &  $-0.92 \pm 0.01$ & &&&&&\\
\hline
Spur1       & PA$[^\circ]$       & ell                & n                 & $\mu_e$[mag/arcsec$^2$, $3.6mu$] & $a_e$[arcsec]    &  $m_{Sp1}$ [$3.6mu$ mag] & Sp1/T \\
            & $16.08 \pm 0.06$   & $0.711 \pm  0.001$ & $0.971 \pm 0.004$ & $18.99 \pm 0.01$                 & $83.62 \pm 0.35$ &   8.05                   & 0.18 \\       
            & RA(Spur1)-RA(Center) & DEC(Spur1)-DEC(Center) & & & & & & \\
            & [arcsec]             & [arcsec]               & & & & & & \\      
            & $31.2 \pm 0.1$                & $-18.1 \pm 0.2$                 & & & & & & \\ 
\hline
Spur2       & PA$[^\circ]$       & ell                & n                 & $\mu_e$[mag/arcsec$^2$, $3.6mu$] & $a_e$[arcsec]    &  $m_{Sp2}$ [$3.6mu$ mag] & Sp2/T \\
            & $17.37 \pm 0.06$   & $0.710 \pm  0.001$ & $0.756 \pm 0.003$ & $18.65 \pm 0.01$                 & $67.8 \pm 0.2$ &   8.27                   & 0.14 \\       
            & RA(Spur2)-RA(Center) & DEC(Spur2)-DEC(Center) & & & & & & \\
            & [arcsec]             & [arcsec]               & & & & & & \\
            & $-29.7 \pm 0.1$                & $-17.3 \pm 0.2$                 & & & & & & \\ 
\hline
\end{tabular}
\end{table*}

\begin{figure*}
 \begin{center}
   \includegraphics[width=5.3cm]{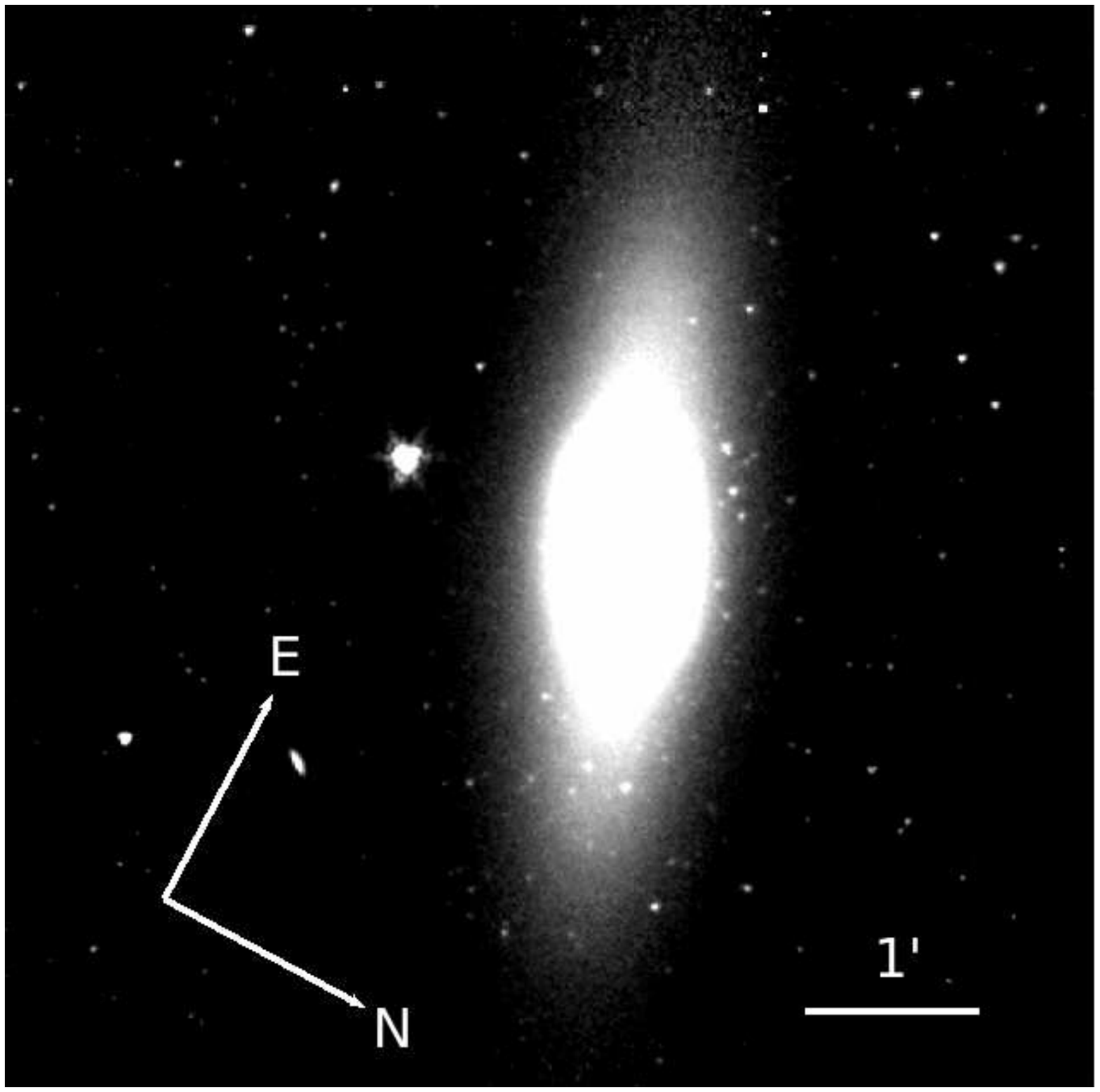}
   \includegraphics[width=5.3cm]{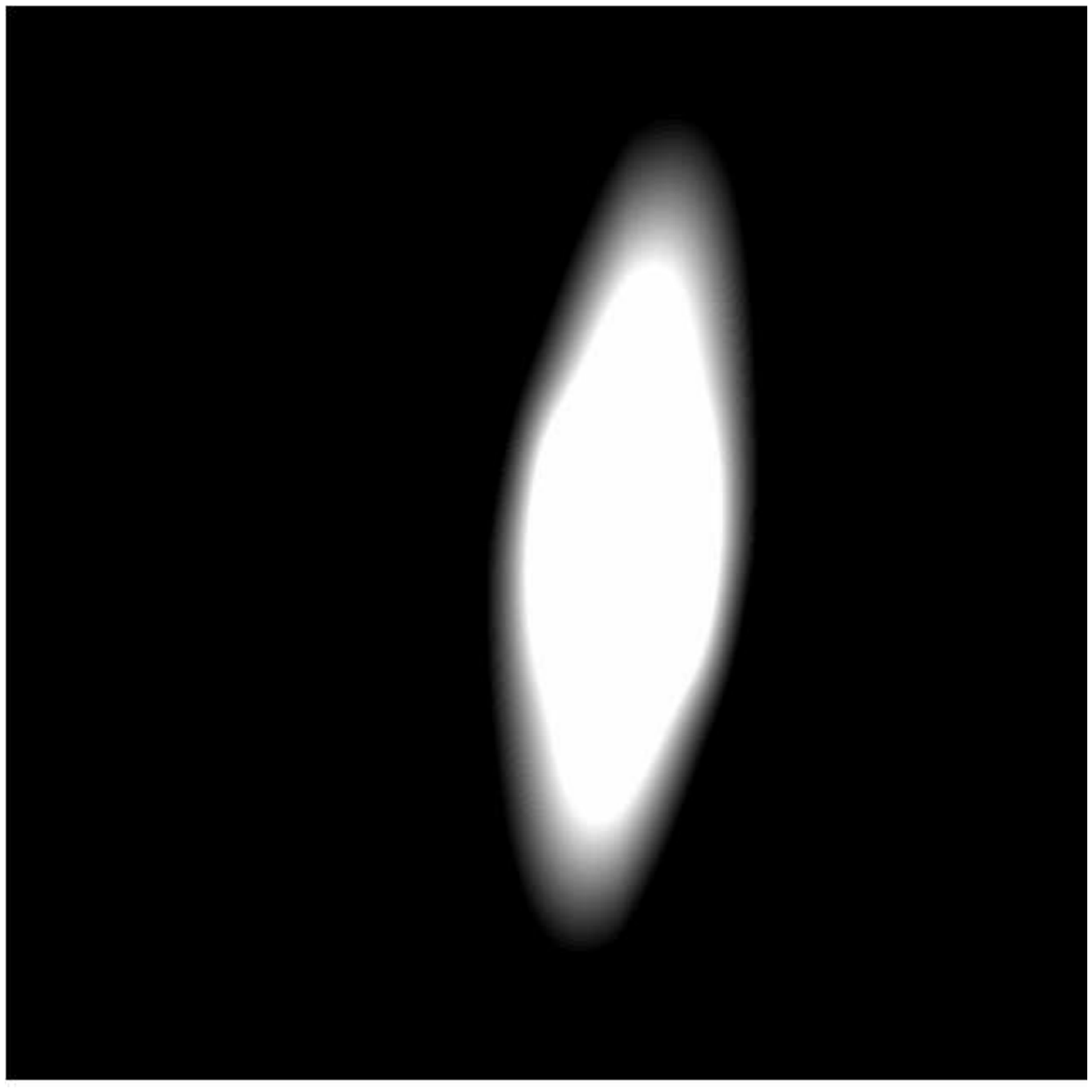}
   \includegraphics[width=5.3cm]{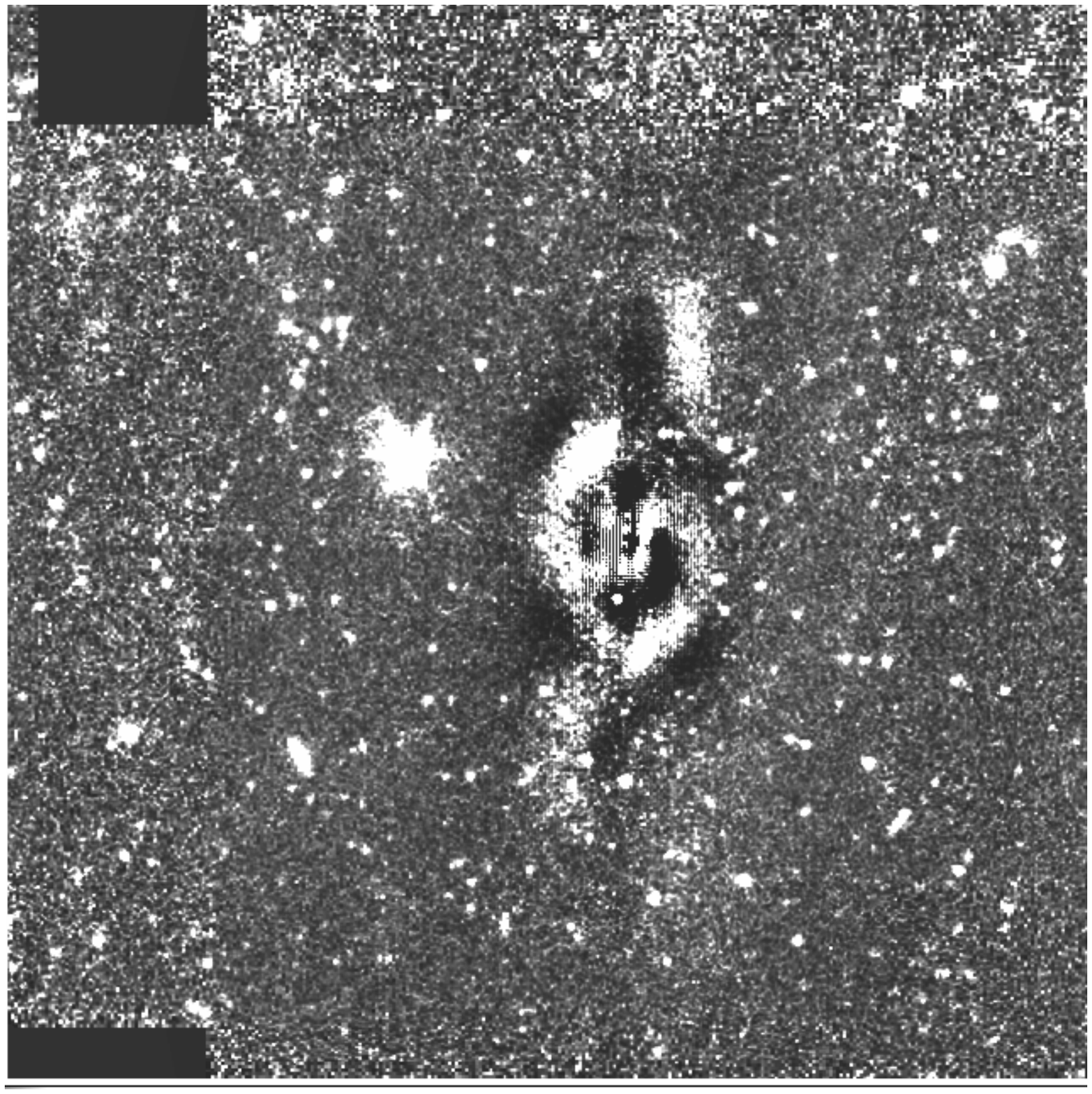}
  \end{center}
  \caption{The \textsc{Imfit} model for NGC 4526: the galaxy $3.6mu-$band image (left), the model (middle) and the residuals (right); see also Fig. \ref{fig_imfitiso}.
\label{fig_ngc4526}}
\end{figure*}

\begin{table*}
\caption[NGC 4736: bulge plus disk decomposition]{Parameters of the bulge plus disk decomposition of NGC 4736.}
\label{tab:4736decomp}
\scriptsize
 \begin{tabular}{l c c c c c c c}
 \hline
 \hline
  Bulge  & PA$[^\circ]$ & ell & n & $\mu_e$[mag/arcsec$^2$, $z$] & $a_e$[arcsec]  &  $m_{Bu}$ [$z$ mag] & Bu/T \\ 
        &  $25.80 \pm 0.06$ & $0.1582 \pm 0.0003$  & $1.405 \pm 0.001$ & $16.004 \pm 0.001$ & $8.071 \pm 0.007$ & 8.8 & 0.26\\ 
 \hline 
  Disk & PA$[^\circ]$ & ell &  &$\mu_0$ [mag/arcsec$^2$, $z$] & h [arcsec]  &  $m_{Di}$ [$z$ mag] & Di/T\\
      &  $106.14 \pm 0.04$  &  $0.1850 \pm 0.0003$ & &  $ 16.674 \pm 0.002$  &  $ 23.80\pm 0.02$ & 8.02 & 0.54\\
 \hline
  Outer Ring & PA$[^\circ]$ & ell &  & $\mu_0$ [mag/arcsec$^2$, $z$] & $r_{ring}$[arcsec] &  $m_{Ri}$ [$z$ mag] & Ri/T\\
             & $94.44 \pm 0.06$ & $0.2497 \pm 0.0005$ & & $21.011 \pm 0.001$ & $ 110.31 \pm 0.09 $ & 9.1 & 0.2\\
             & $\sigma_{in}$ [arcsec] & $\sigma_{out}$ [arcsec] & & &\\
             & $23.2 \pm 0.1 $ & $51.01 \pm 0.07$ \\
\hline
\end{tabular} 
\end{table*}

\begin{figure*}
 \begin{center}
    \includegraphics[width=16cm]{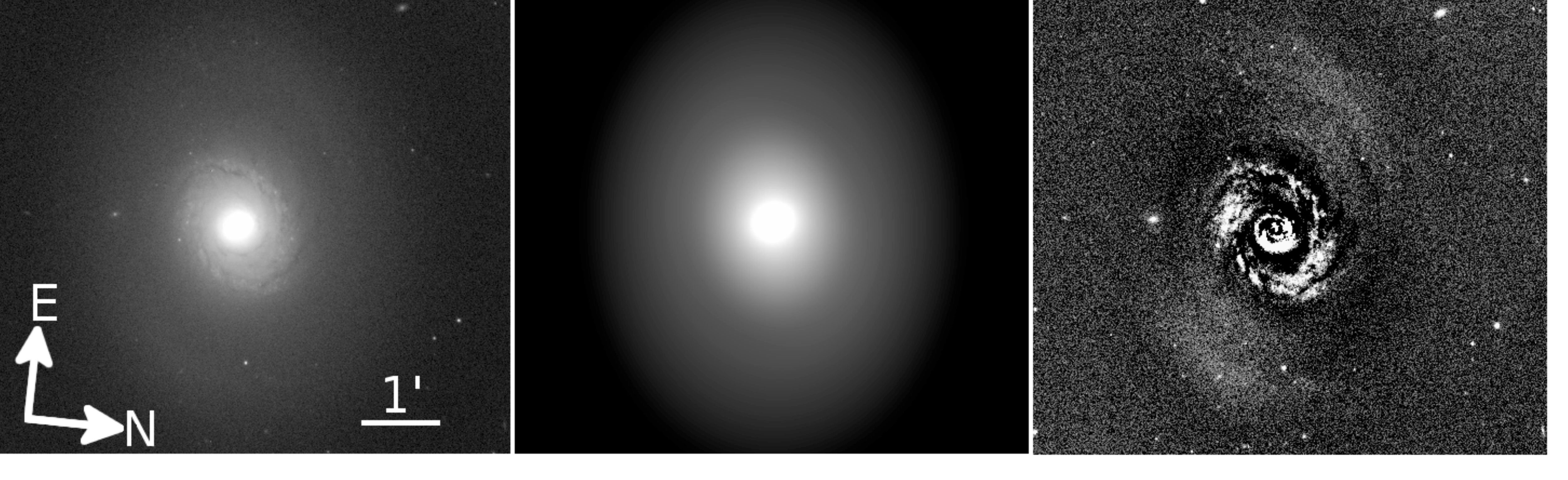}
  \end{center}
  \caption{The \textsc{Imfit} model for NGC 4736: the galaxy $z-$band image (left), the model (middle) and the residuals (right); 
see also Fig. \ref{fig_imfitiso}.
\label{fig_ngc4736}}
\end{figure*}


\begin{table*}
\caption[NGC 4826: bulge plus disk decomposition]{Parameters of the bulge plus disk decomposition of NGC 4826.}
\label{tab:4826decomp}
\scriptsize
 \begin{tabular}{l c c c c c c c}
 \hline
 \hline
  Bulge  & PA$[^\circ]$ & ell & n & $\mu_e$[mag/arcsec$^2$, $i$] & $a_e$[arcsec] &  $m_{Bu}$ [$i$ mag] & Bu/T \\ 
        &  $105.21 \pm 0.04$ & $0.2190 \pm 0.0003$  & $4.295 \pm 0.005$ & $20.106 \pm 0.004$ & $36.8 \pm 0.1$ & 9.12 & 0.28\\ 
 \hline 
  Disk & PA$[^\circ]$ & ell & & $\mu_0$ [mag/arcsec$^2$, $i$] & h [arcsec]  &  $m_{Di}$ [$i$ mag] & Di/T \\
      &  $111.399 \pm 0.006$  & $0.4556 \pm 0.0001$ &      &$ 18.1850 \pm 0.0004$  &  $ 56.561\pm 0.009$ & 8.08 & 0.72 \\
 \hline
\end{tabular}
\end{table*}

\begin{figure*}
 \begin{center}
    \includegraphics[width=16cm]{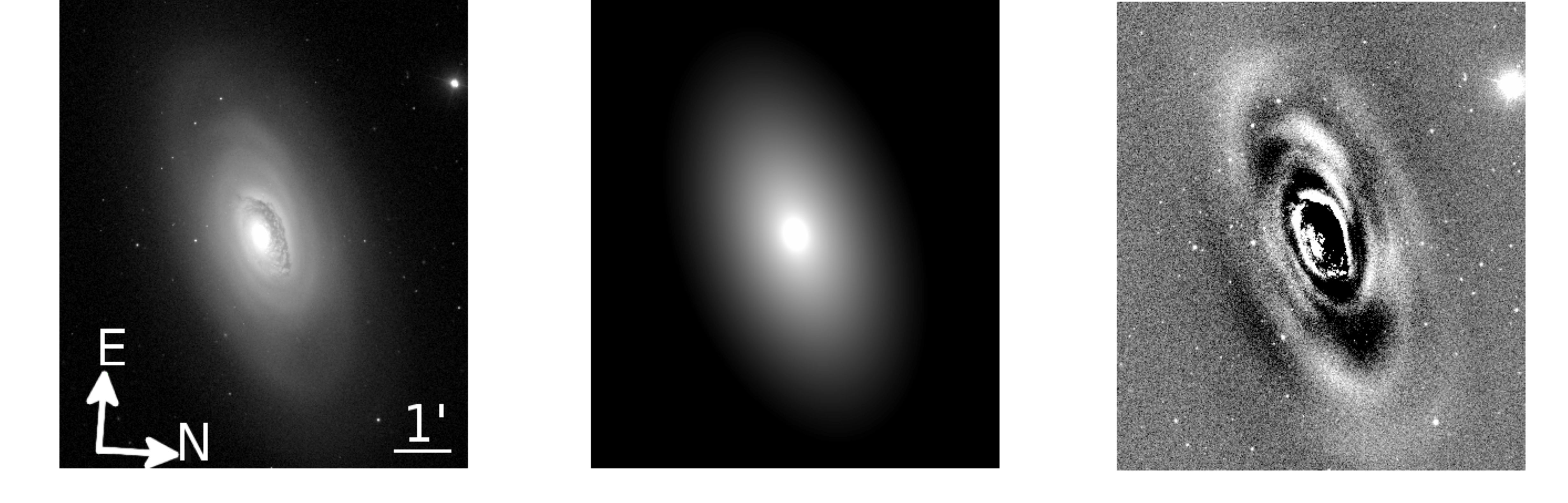}
  \end{center}
  \caption{The \textsc{Imfit} model for NGC 4826: the galaxy $i-$band image (left), the model (middle) and the residuals (right); see also Fig. \ref{fig_imfitiso}.
\label{fig_ngc4826}}
\end{figure*}

\begin{table*}
\caption[NGC 7582: bulge plus bar plus disk decomposition]{Parameters of the bulge plus bar plus disk decomposition of NGC 7582.}
\label{tab:7582decomp}
\tiny
 \begin{tabular}{l c c c c c c c c}
 \hline
 \hline
  Bulge  & PA$[^\circ]$       & ell        &        & n               & $\mu_e$[mag/arcsec$^2$, $3.6mu$] & $a_e$[arcsec] &  $m_{Bu}$ [$3.6mu$ mag] & Bu/T \\ 
         & $134.6 \pm 0.06$  & $ 0.419 \pm 0.001$ & & $2.59 \pm 0.01$ & $ 12.71\pm 0.01$          & $2.38 \pm 0.01$ & 8.27 & 0.29\\ 
\hline 
  Bar    & PA$[^\circ]$        & ell                & $c_0$             & n                 & $\mu_e$[mag/arcsec$^2$, $3.6mu$] & $a_e$[arcsec]  &  $m_{Bar}$ [$3.6mu$ mag] & Bar/T\\
         &  $140 \pm 0.01$    & $0.869 \pm 0.001$  & $1.74 \pm 0.02$   & $0.211 \pm 0.001$ & $17.20 \pm 0.01$                & $51.76 \pm 0.04$ & 8.58 & 0.22\\
 \hline 
  Disk   & PA$[^\circ]$ & ell & & & $\mu_0$ [mag/arcsec$^2$, $3.6mu$] & h [arcsec] & $m_{Di}$ [$3.6mu$ mag] & Di/T \\
         &  $141.4 \pm 0.03$  & $0.602 \pm 0.001$ &   &   &$ 16.54 \pm 0.01$  &  $ 37.3\pm 0.04$ & 7.69 & 0.49 \\
 \hline
\end{tabular}
\end{table*}

\begin{figure*}
 \begin{center}
    \includegraphics[width=5.3cm]{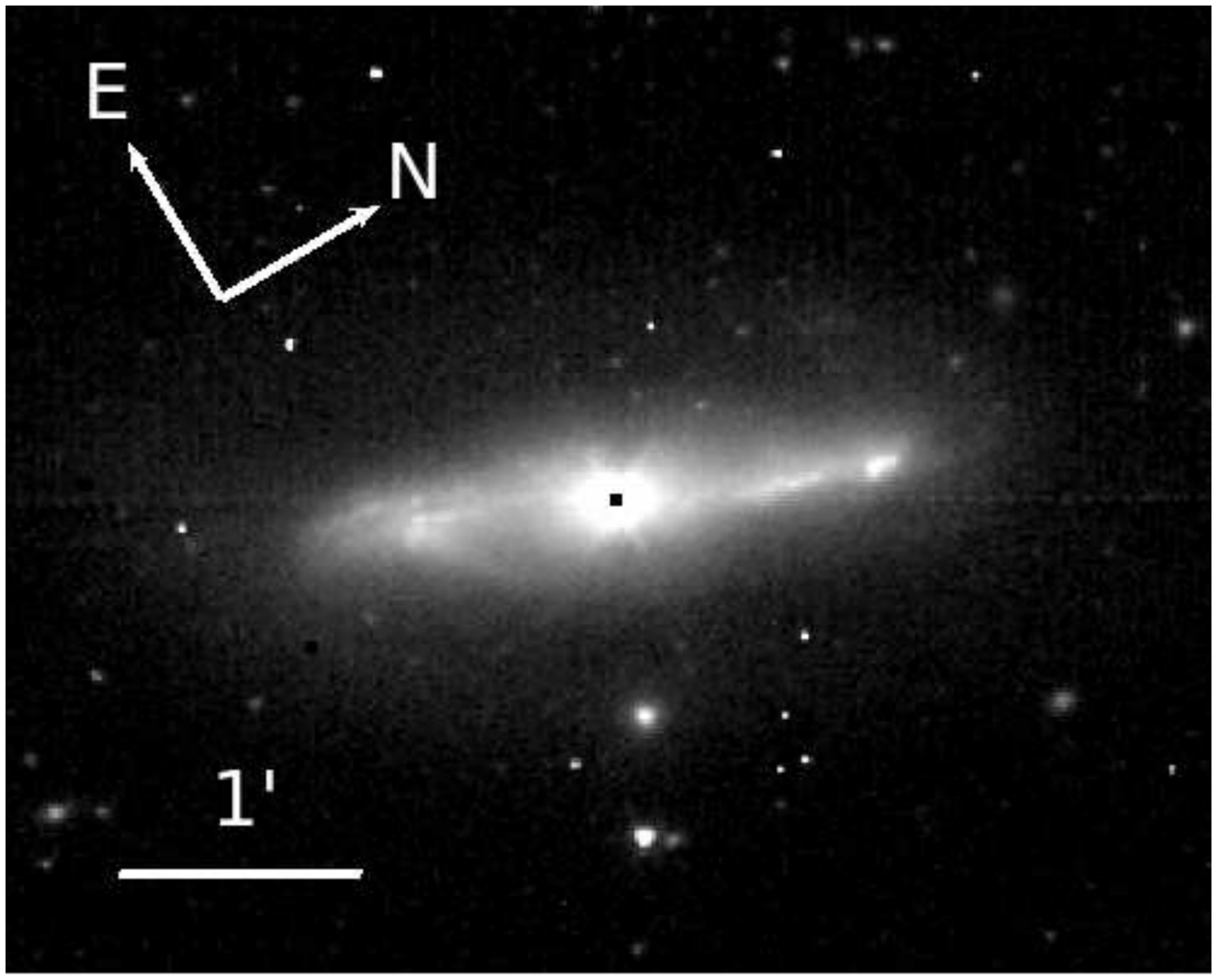}
    \includegraphics[trim=0 -7.5cm 0 0cm,clip,width=5.5cm]{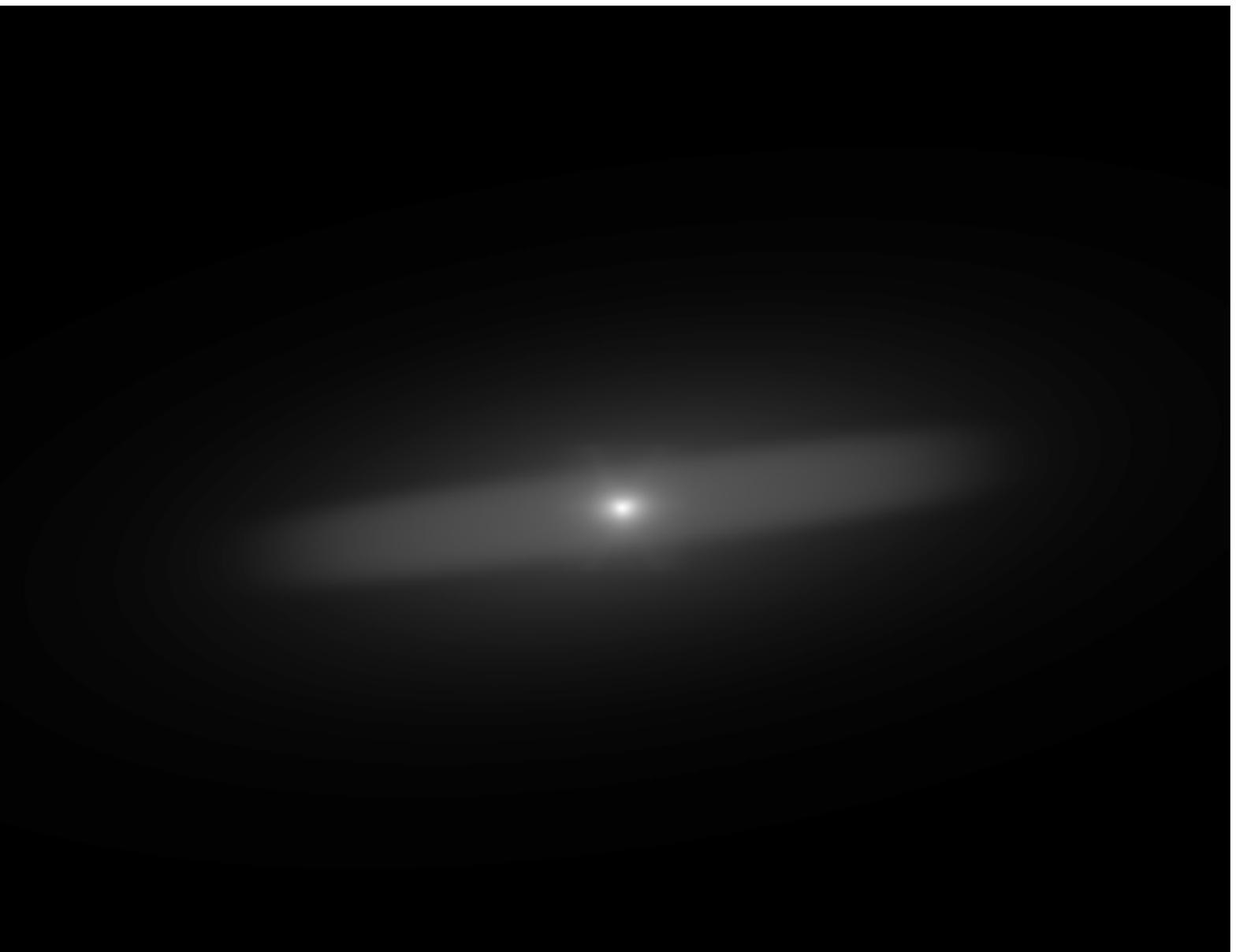}
    \includegraphics[width=5.3cm]{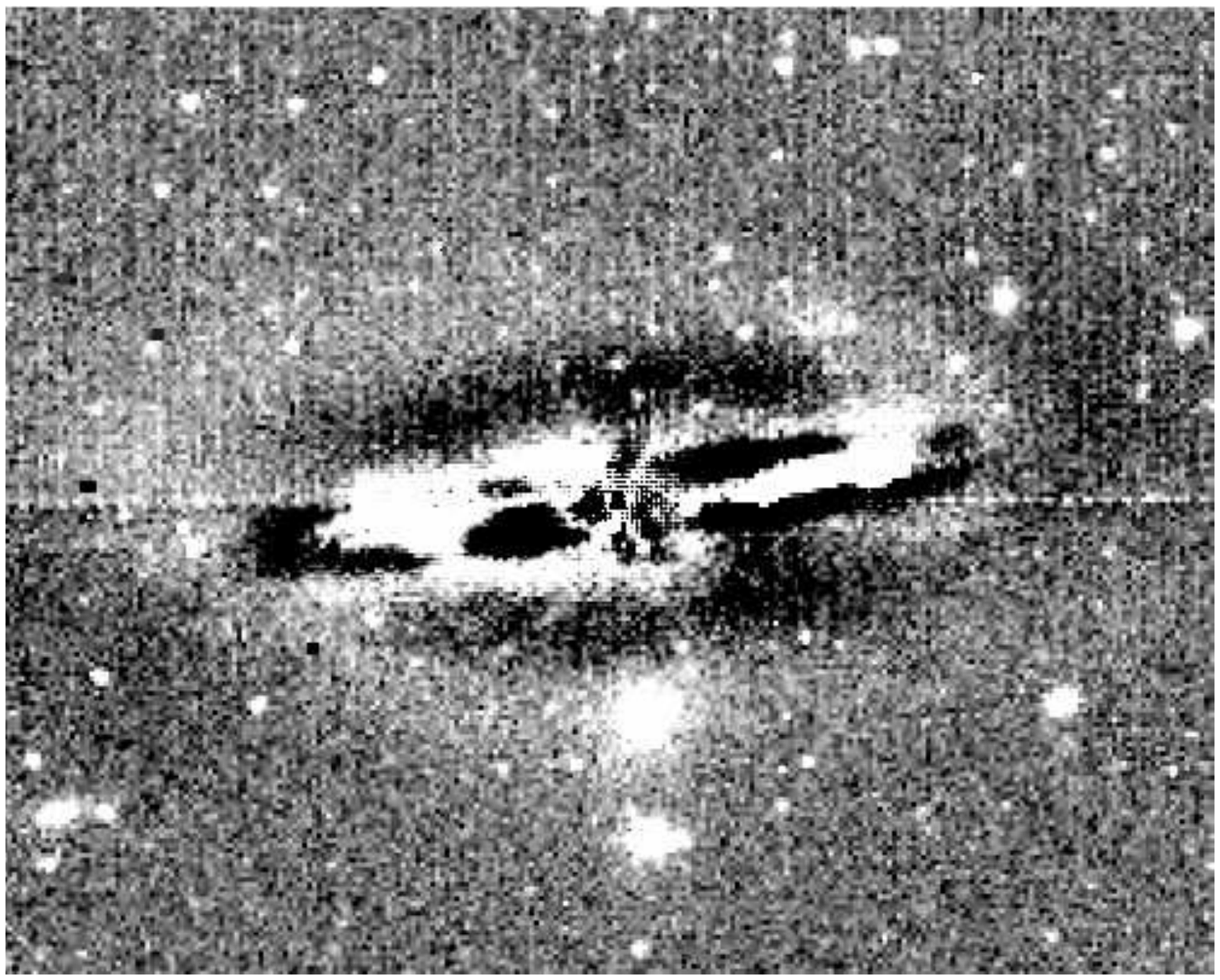}
  \end{center}
  \caption{The \textsc{Imfit} model for NGC 7582: the galaxy $3.6mu-$band image (left), the model (middle) and the residuals (right); 
see also Fig. \ref{fig_imfitiso}.
\label{fig_ngc7582}}
\end{figure*}

\begin{figure*}
 \begin{center}
    \includegraphics[trim=0 3cm 0cm 3cm,clip,width=8cm]{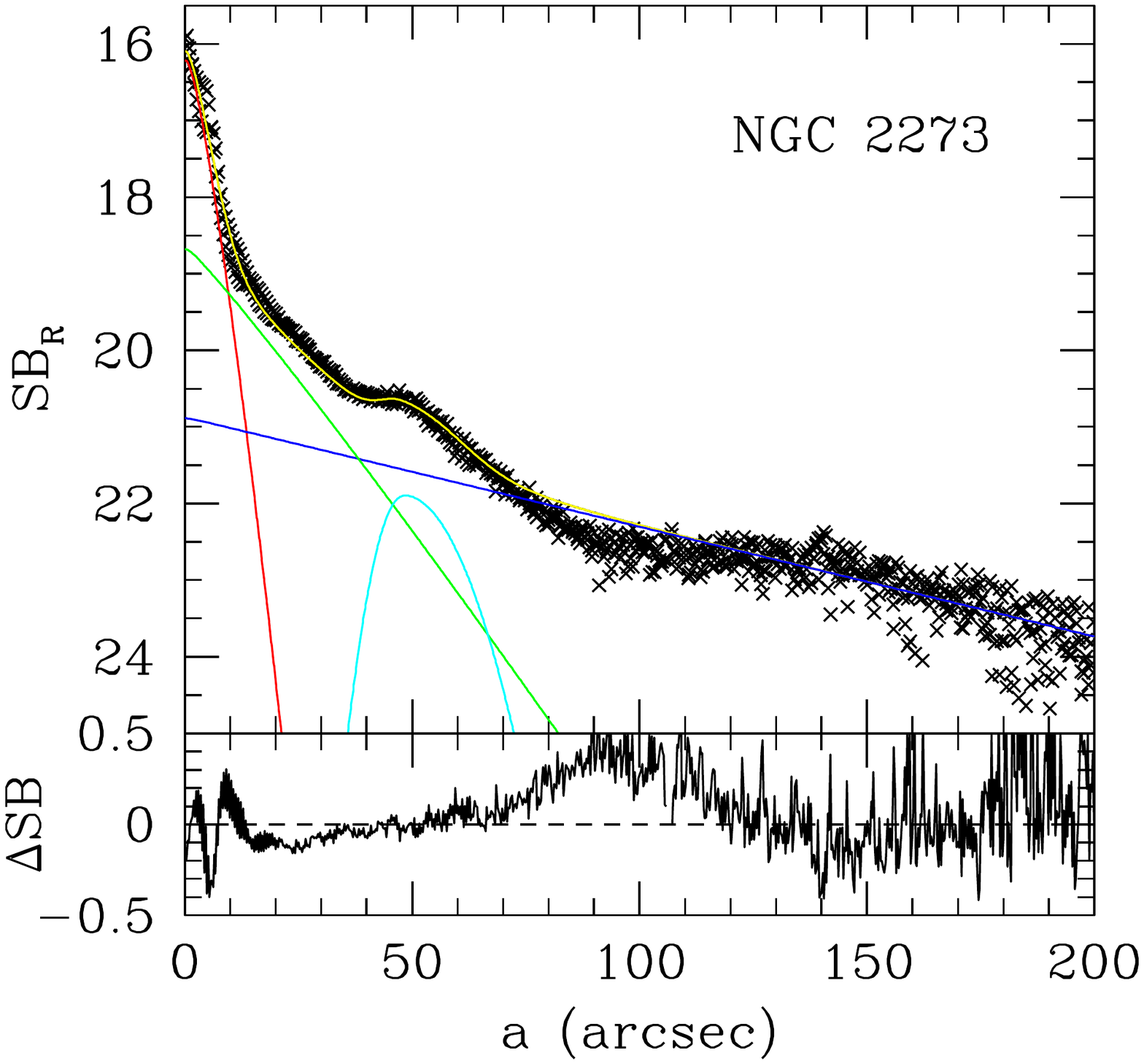}
    \includegraphics[trim=0 3cm 0cm 3cm,clip,width=8cm]{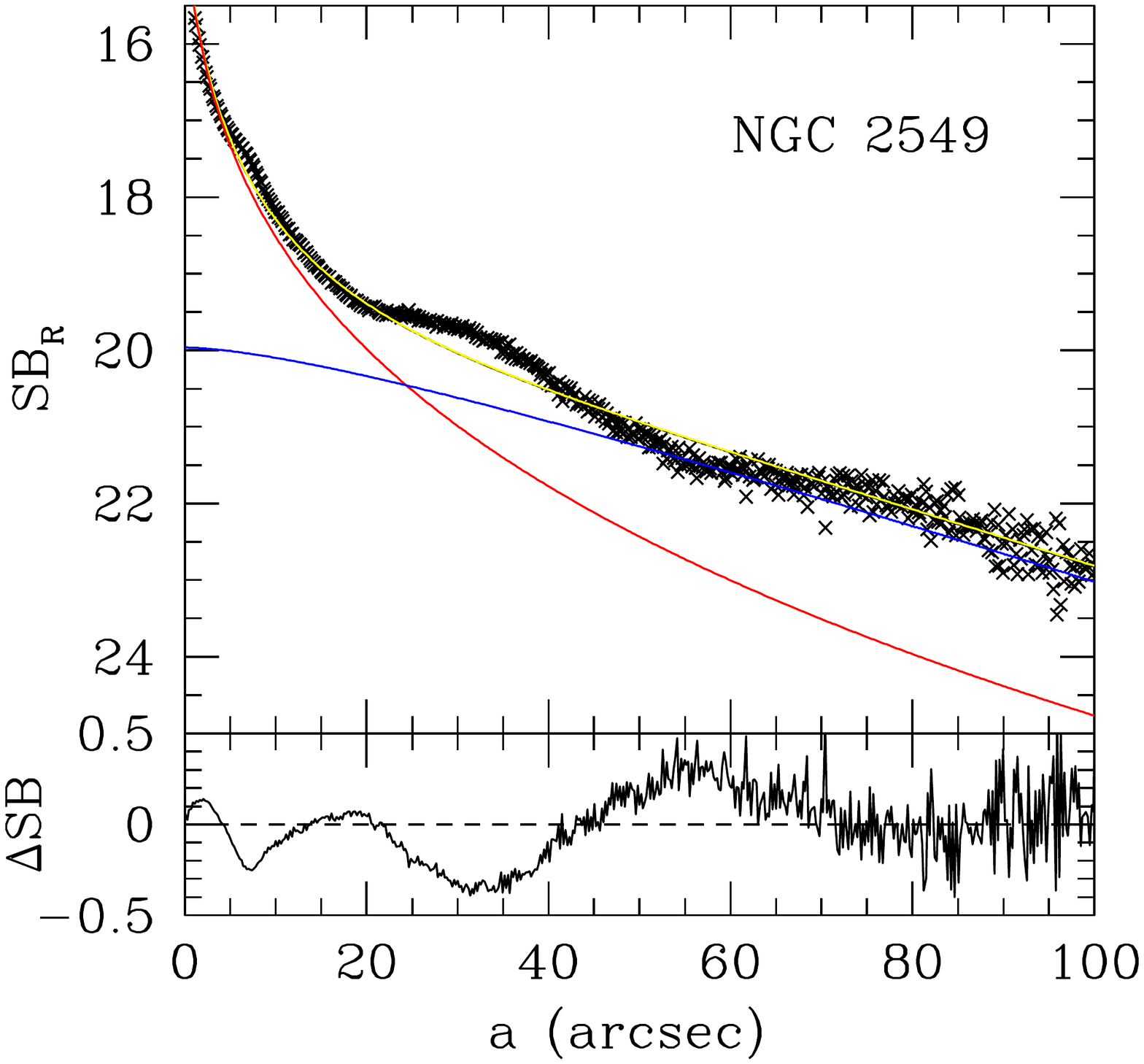}\\
    \includegraphics[trim=0 3cm 0cm 3cm,clip,width=8cm]{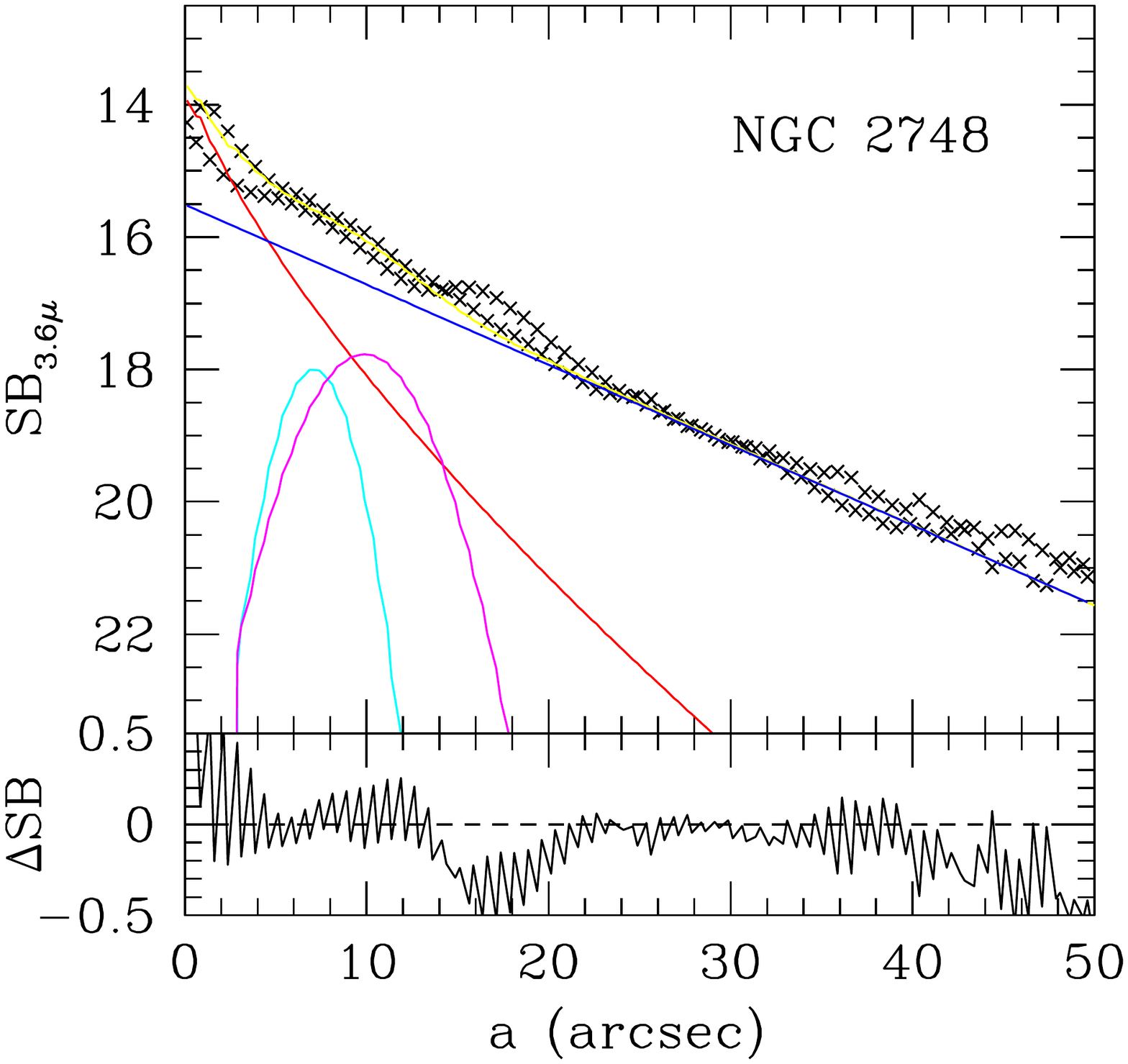}
    \includegraphics[trim=0 3cm 0cm 3cm,clip,width=8cm]{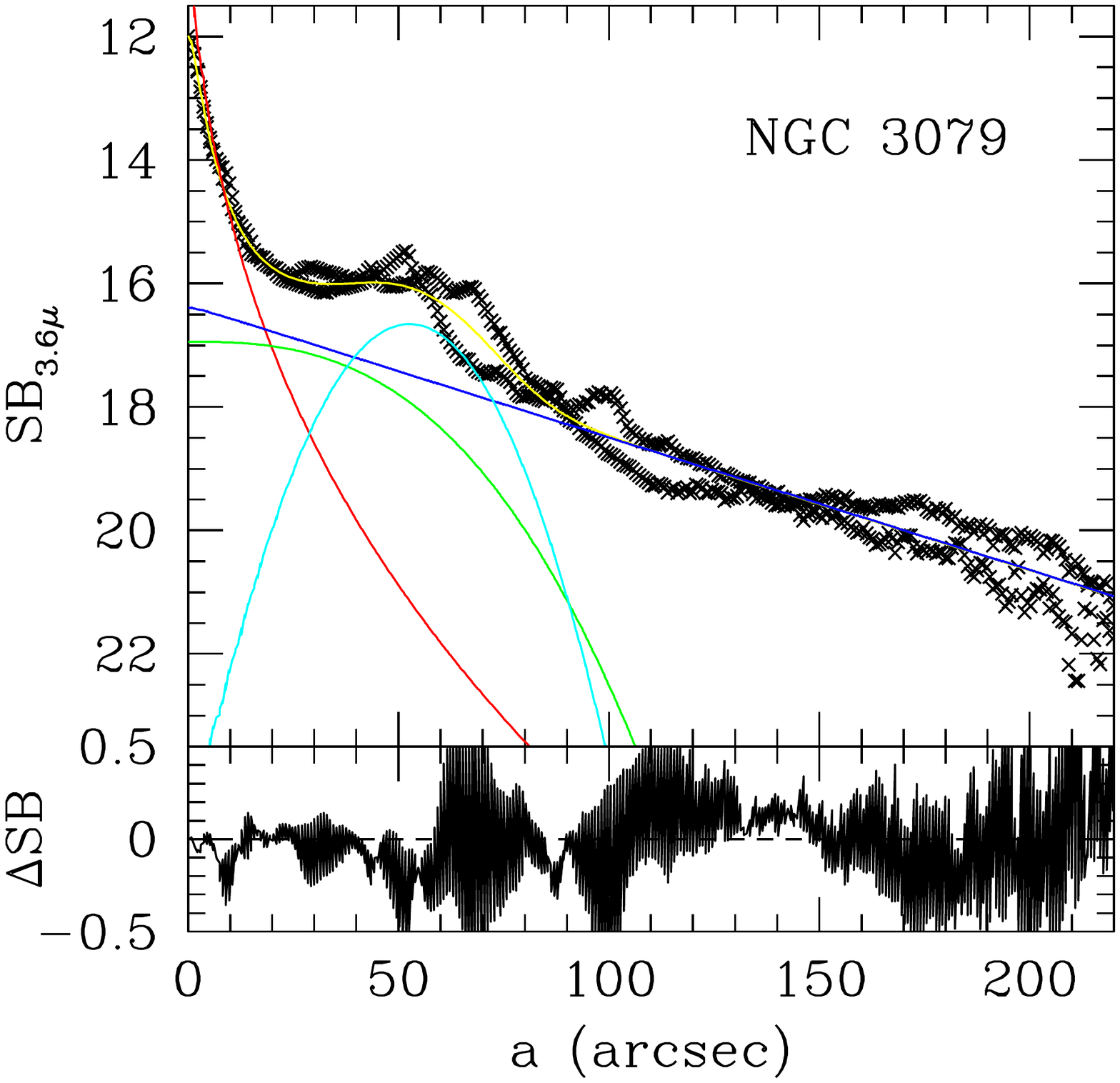}\\
  \end{center}
  \caption{Cuts along the bulge major axis for the galaxies where we performed
 \textsc{Imfit} decompositions. For each galaxy there are two plots. At the 
top we show the surface brightness along the bulge major axis  (crosses), 
the resulting point spread function- (PSF-) convolved fitted profile (yellow line) and the unconvolved 
fitted components (red line: bulge; blue 
line: disk; green line: bar; cyan line: ring or spurs, magenta line: point 
source), as a function of the distance
 from the center. At the bottom we show the difference in surface brightness 
between measured and PSF-convolved fitted profiles. 
\label{fig_imfitiso}}
\end{figure*}
\addtocounter{figure}{-1}
\begin{figure*}
 \begin{center}
   \includegraphics[trim=0 3cm 0cm 3cm,clip,width=8cm]{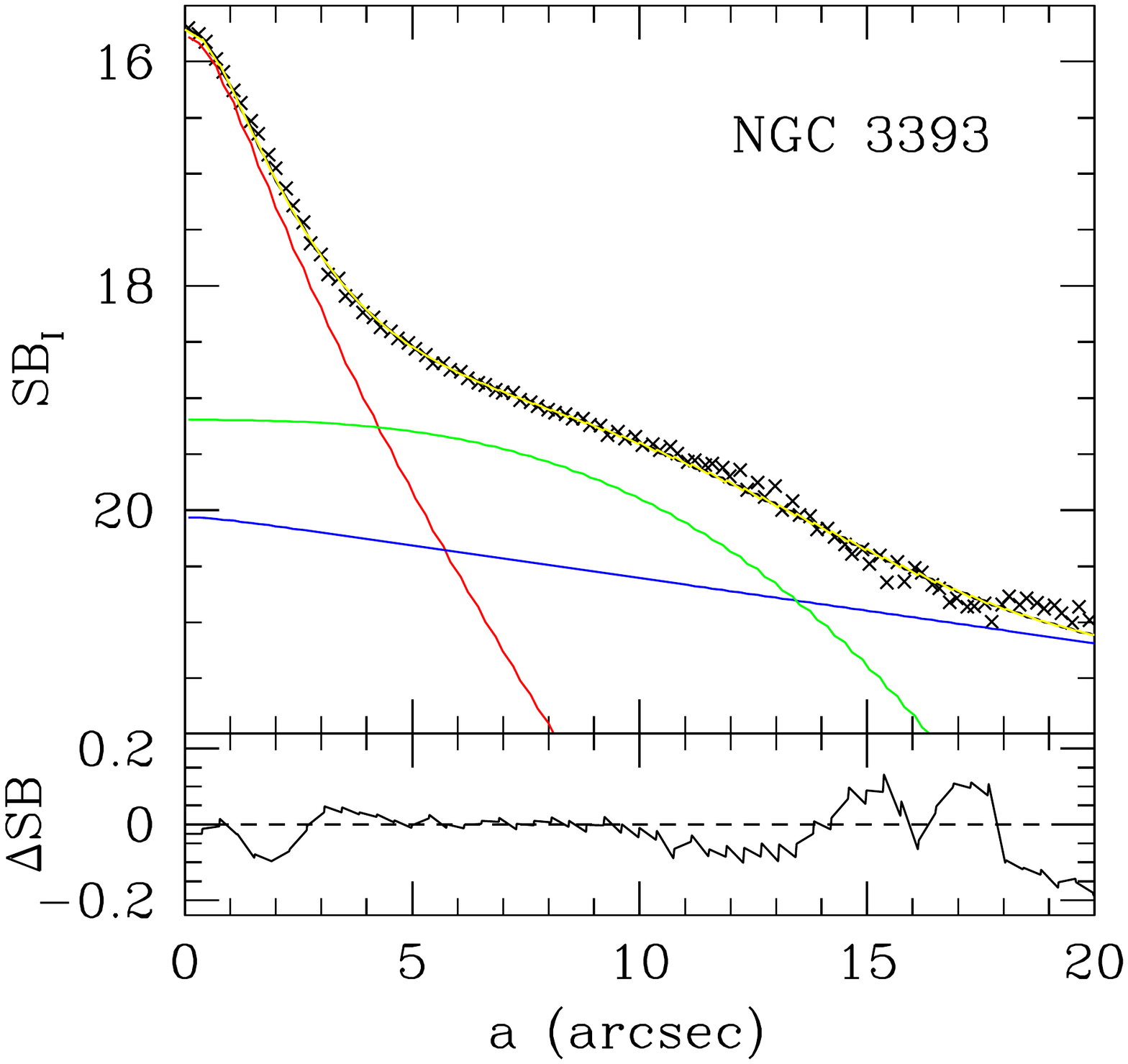}
   \includegraphics[trim=0 3cm 0cm 3cm,clip,width=8cm]{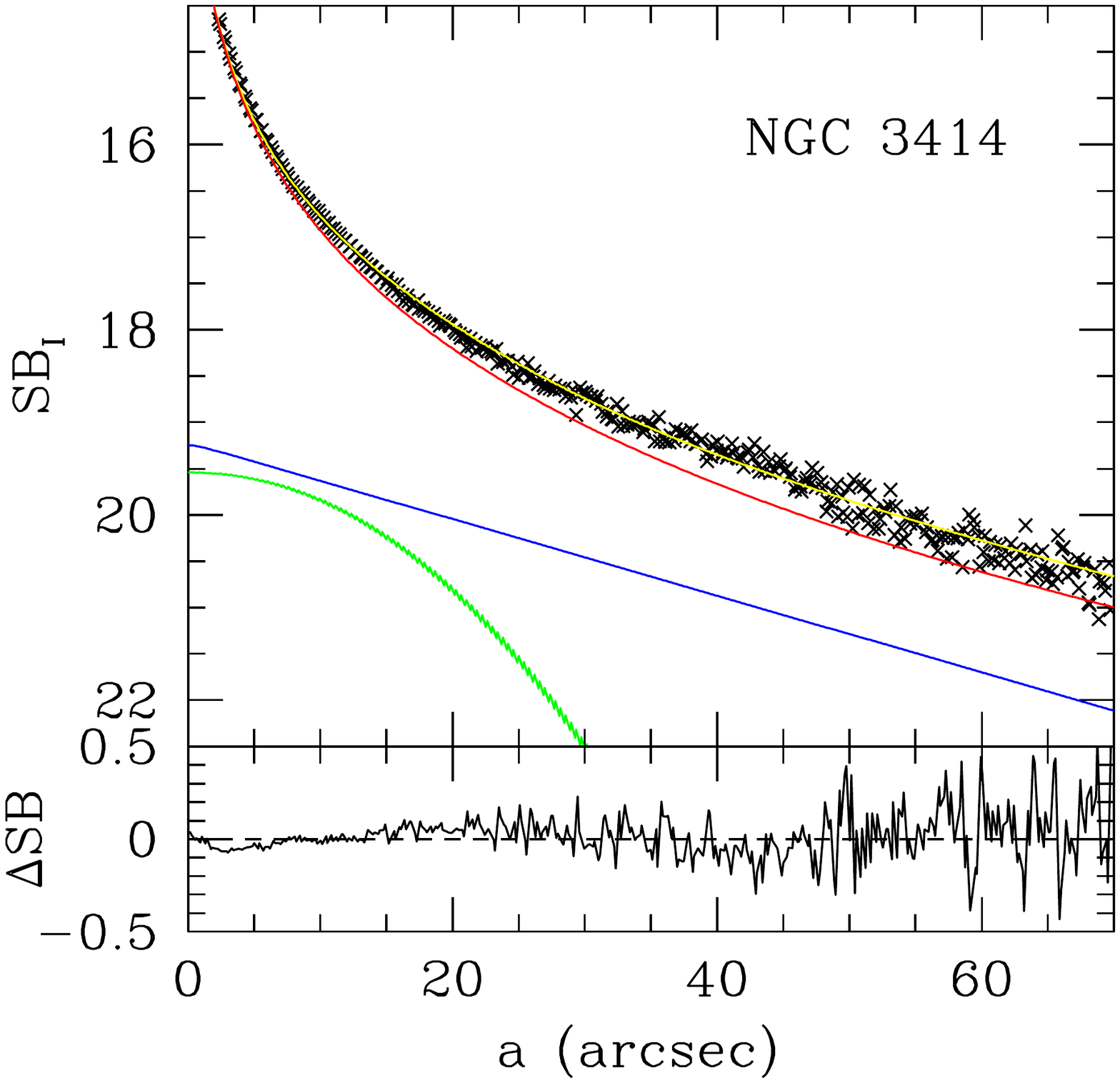}\\
    \includegraphics[trim=0 3cm 0cm 3cm,clip,width=8cm]{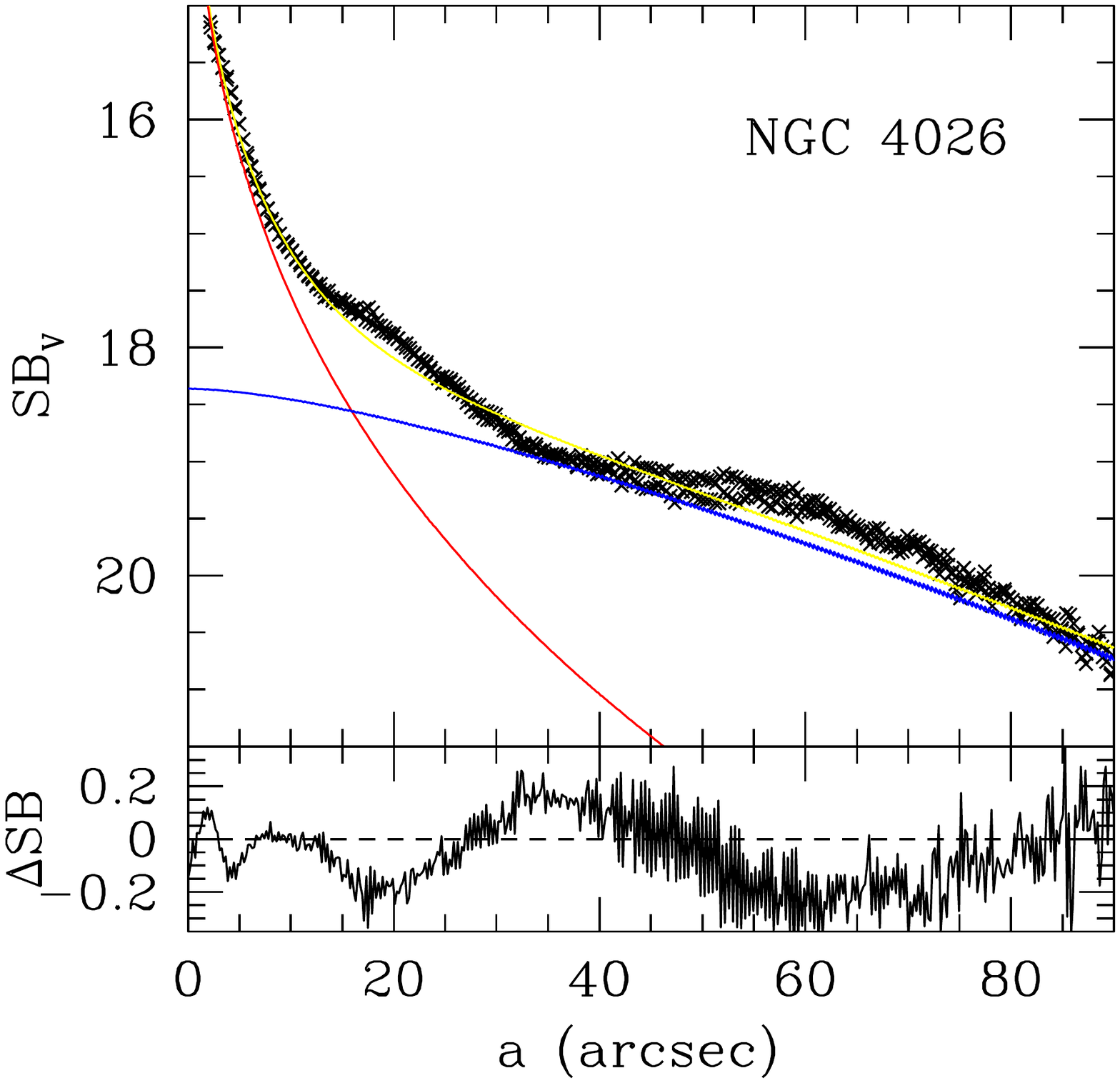}
    \includegraphics[trim=0 3cm 0cm 3cm,clip,width=8cm]{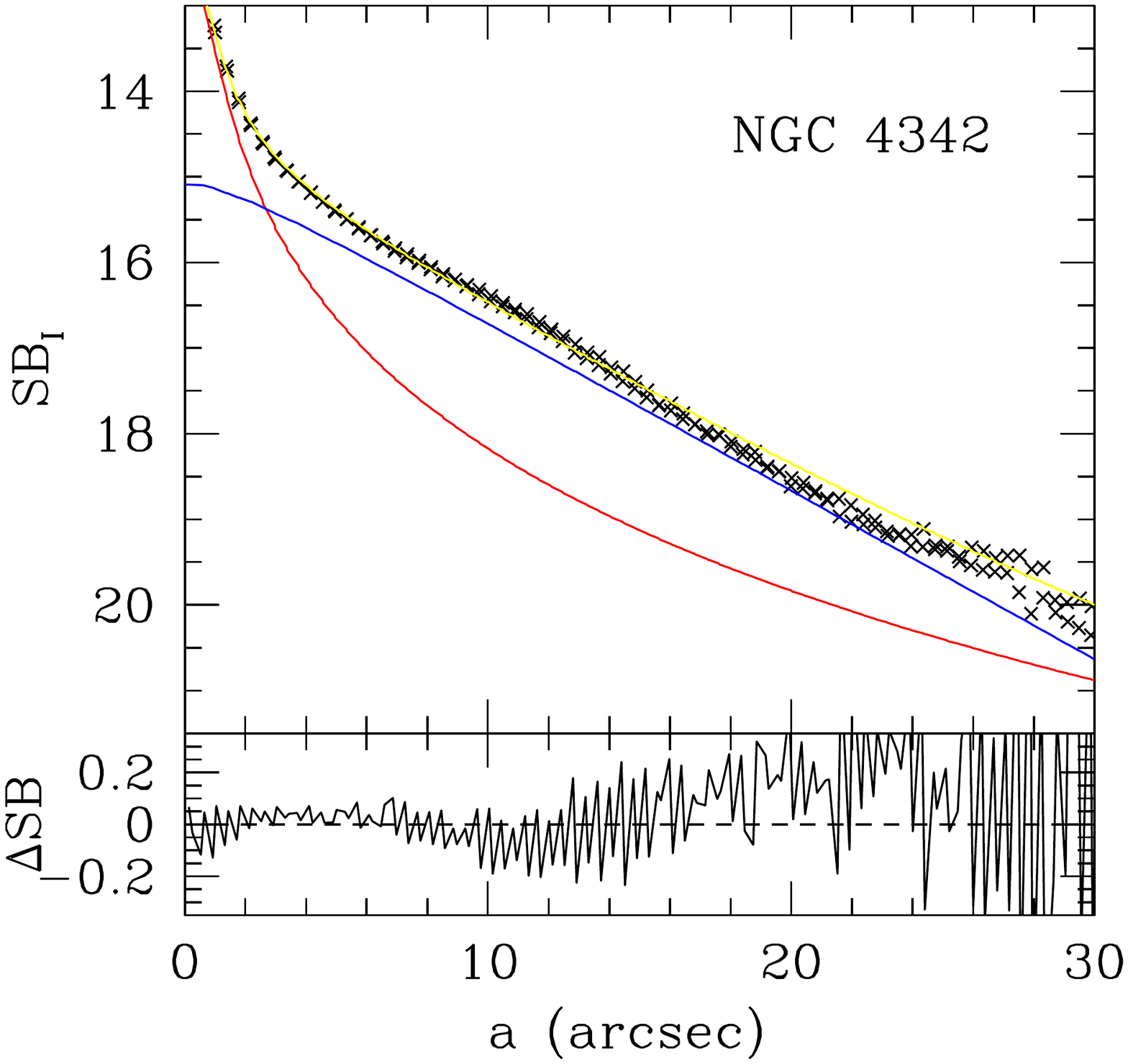}\\
  \end{center}
  \caption{Continued.}
\end{figure*}
\addtocounter{figure}{-1}
\begin{figure*}
 \begin{center}
    \includegraphics[trim=0 3cm 0cm 3cm,clip,width=8cm]{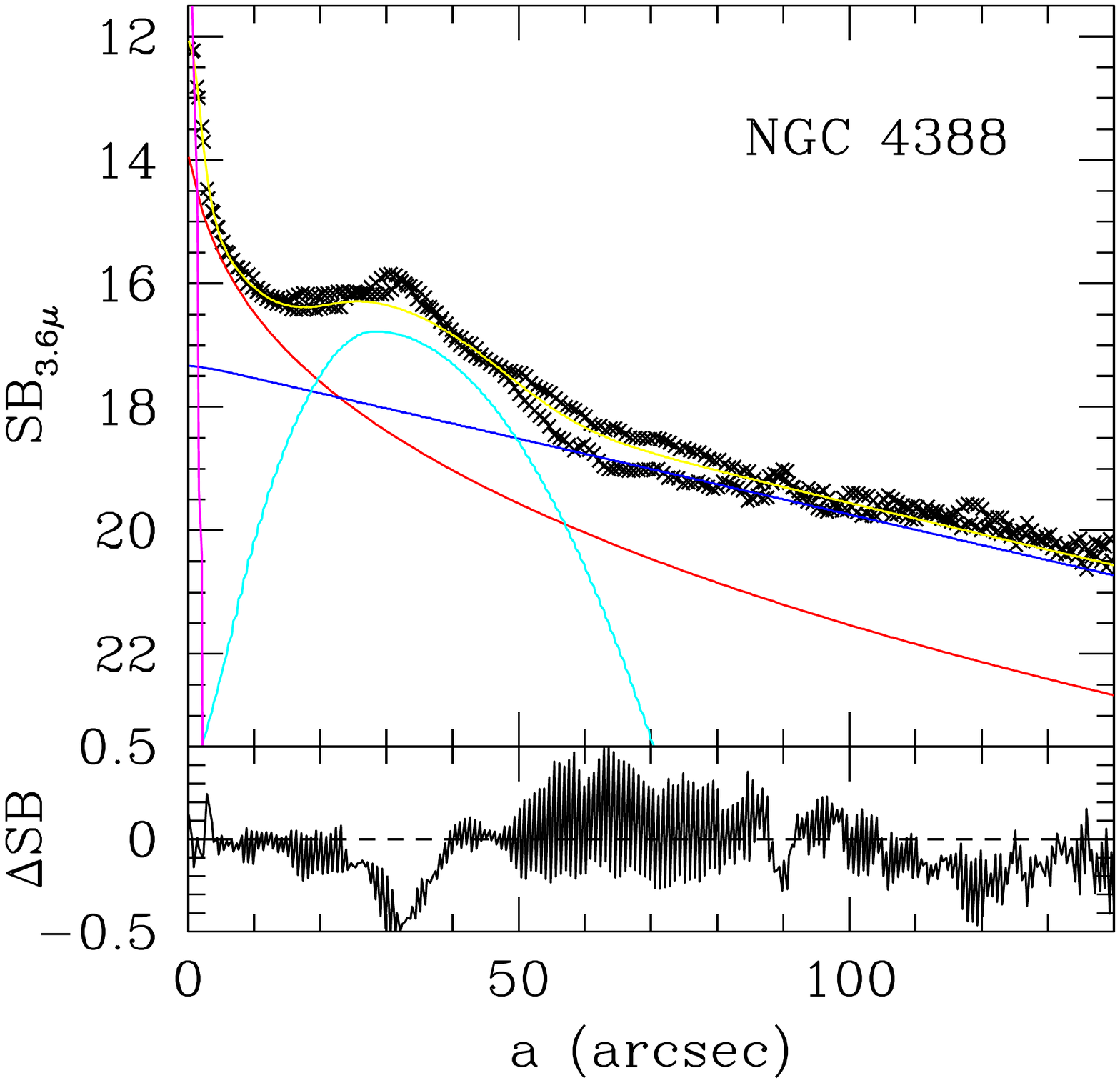}
    \includegraphics[trim=0 3cm 0cm 3cm,clip,width=8cm]{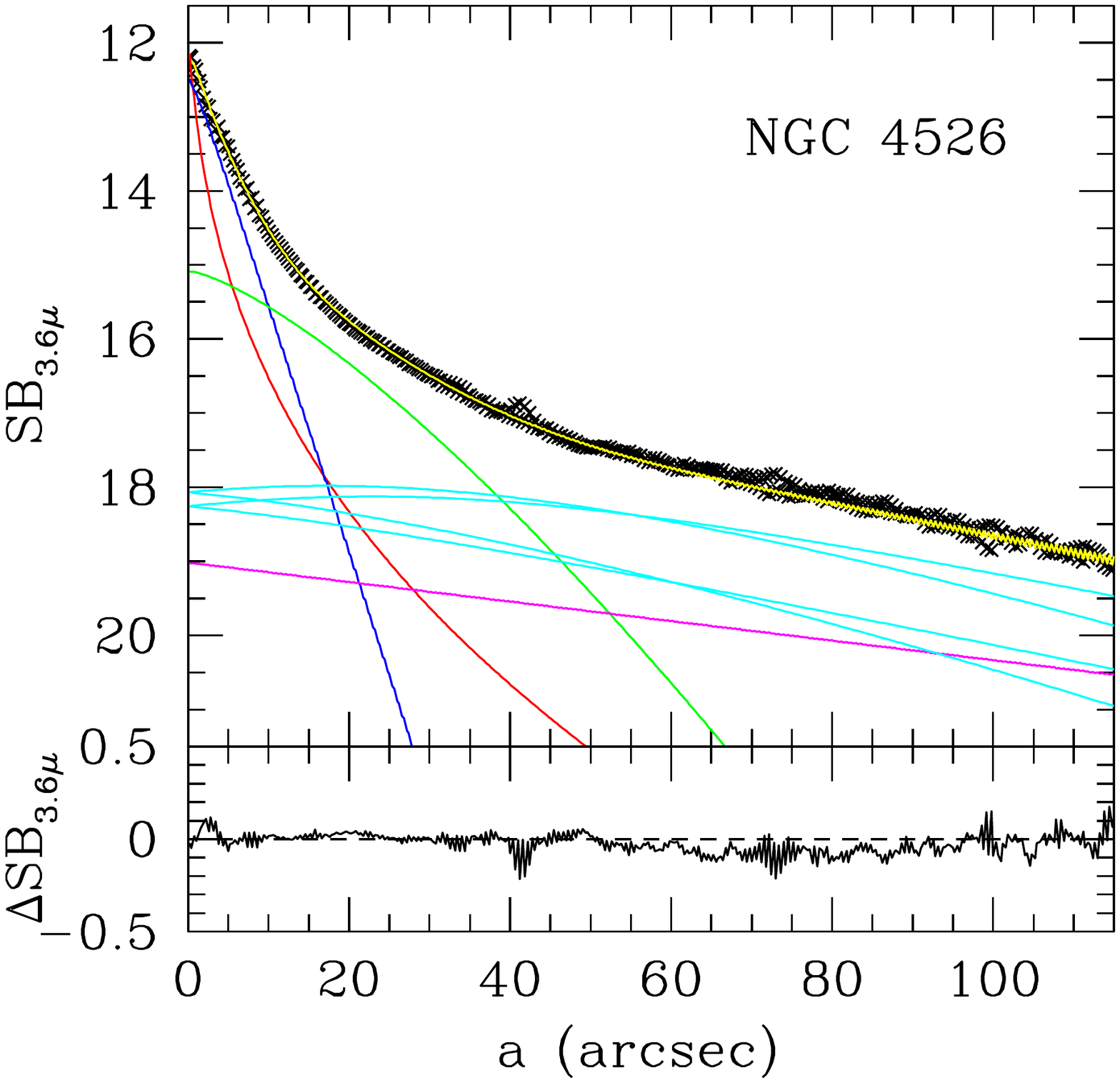}\\
    \includegraphics[trim=0 3cm 0cm 3cm,clip,width=8cm]{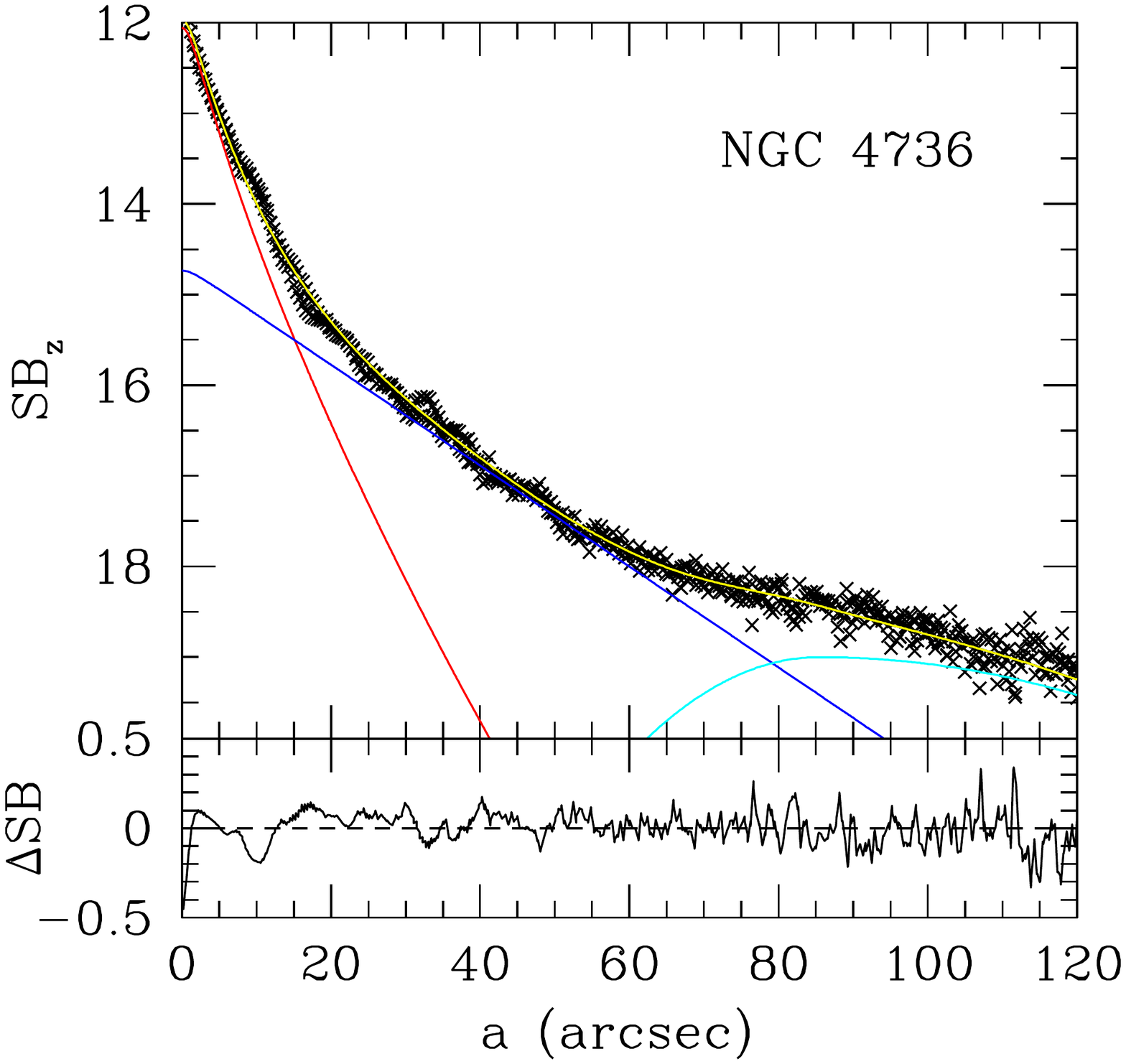}
    \includegraphics[trim=0 3cm 0cm 3cm,clip,width=8cm]{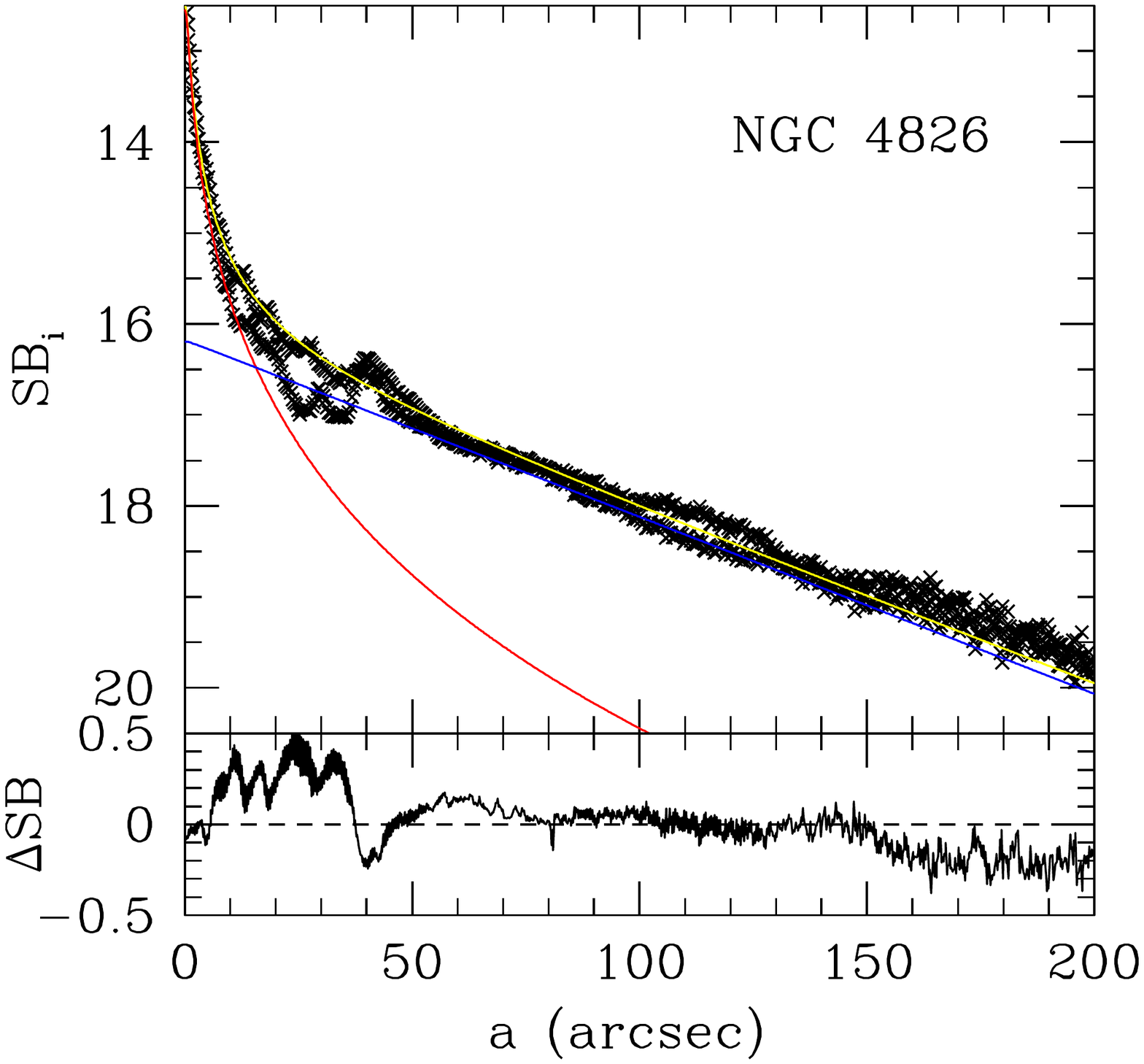}\\
  \end{center}
  \caption{Continued.}
\end{figure*}

\addtocounter{figure}{-1}
\begin{figure*}
 \begin{center}
    \includegraphics[trim=0 3cm 0cm 3cm,clip,width=8cm]{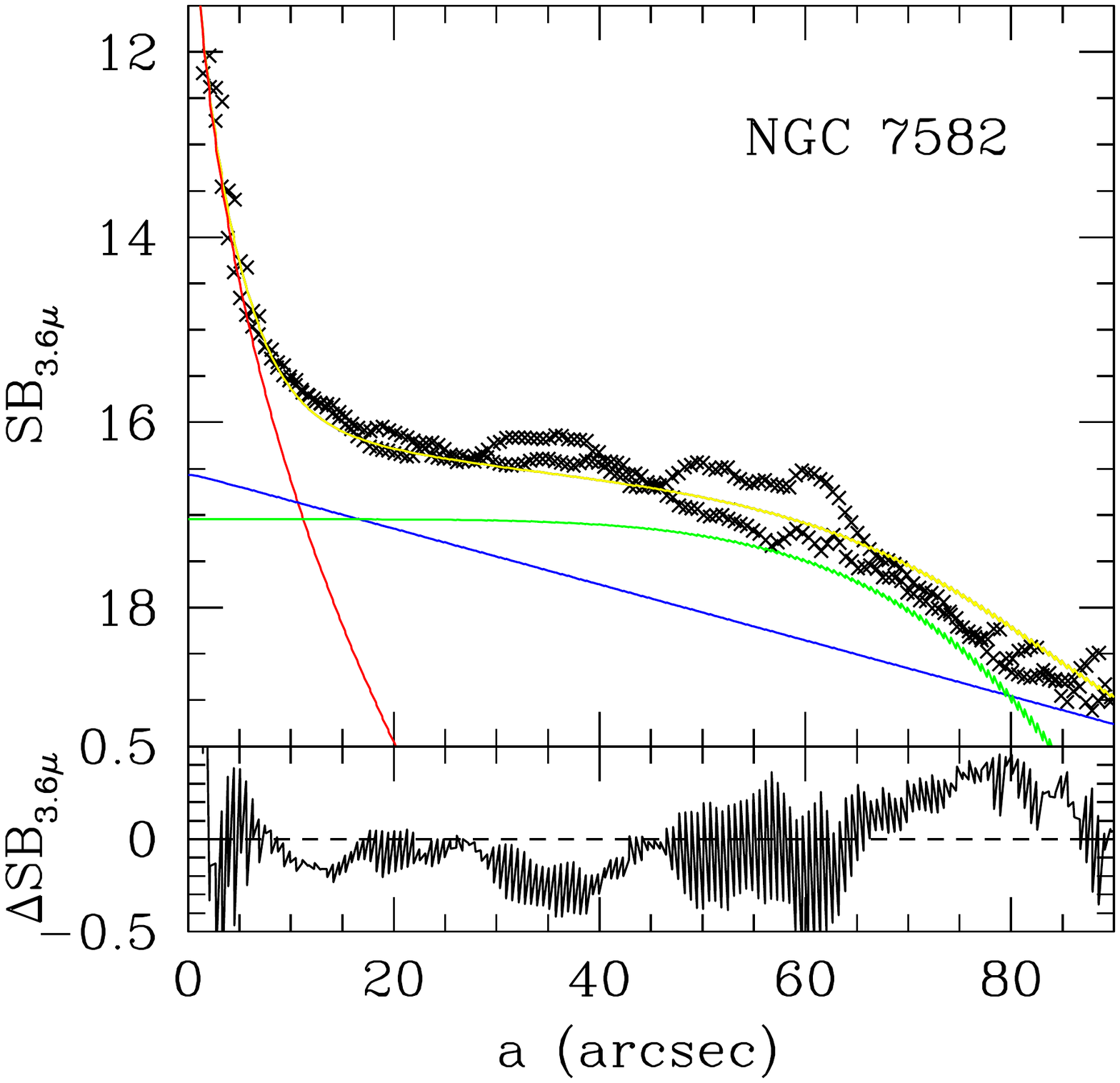}\\
  \end{center}
  \caption{Continued.}
\end{figure*}

\begin{deluxetable}{llllll}
  \tablecaption{The dynamical $M/L$ ratios of the literature and SINFONI samples.\label{tab_ml}} 
\tabletypesize{\footnotesize}
  \tablewidth{0pt} 
\tablehead{ 
\colhead{Galaxy} & \colhead{$M/L$} & \colhead{Band of $M/L$} &
\colhead{$(d\log M/L)^2$} & \colhead{Band of image} & \colhead{Reference}
}
\startdata
            MW           &       1.00            &           -           &    0.001886           &           -           &           {          -}          \\     
    Circinus           &       0.14            &       $3.6mu$           &    0.002651           &       $3.6mu$           &     \citet{   Sani2011}          \\     
       A1836           &       4.86            &           $I$           &     0.01443           &           $I$           &           {       SDSS}          \\     
      IC1459           &       4.91            &           $V$           &    0.001961           &           $V$           &     \citet{ Rusli2013b}          \\     
      IC4296           &       8.54            &           $B$           &    0.001961           &           $I$           &     \citet{ Erwin2015b}          \\     
      NGC221           &       1.50            &           $R$           &   0.0001532           &           $R$           &     \citet{ Peletier1993}          \\   
      NGC224           &       4.13            &           $V$           &   0.0001061           &           $V$           &     \citet{ Kormendy1999}          \\   
      NGC524           &       4.80            &           $I$           &   0.0006592           &           $I$           &     \citet{ Cappellari2006}          \\ 
      NGC821           &       3.95            &           $R$           &   0.0006942           &           $R$           &     \citet{ Graham2001}          \\     
     NGC1023           &       0.90            &       $3.6mu$           &   2.615e-06           &       $3.6mu$           &           {    Spitzer}          \\     
     NGC1068           &       0.37            &           $K$           &     0.02909           &           $K$           &     \citet{ Erwin2015a}          \\     
     NGC1194           &       6.10            &           $r$           &     0.01975           &           $r$           &           {       SDSS}          \\     
     NGC1300           &       2.13            &           $V$           &   1.451e-05           &           $V$           &     \citet{ Fisher2008}          \\     
     NGC1399           &      10.32            &           $B$           &   0.0001133           &           $B$           &     \citet{ Saglia2000}          \\     
     NGC2273           &       4.50            &           $R$           &     0.01044           &           $R$           &     \citet{  Erwin2003}          \\     
     NGC2549           &       4.55            &           $R$           &   0.0003426           &           $r$           &           {       SDSS}          \\     
     NGC2748           &       0.53            &       $3.6mu$           &   0.0003018           &       $3.6mu$           &           {    Spitzer}          \\     
     NGC2787           &       2.50            &       $3.6mu$           &   0.0002153           &       $3.6mu$           &           {    Spitzer}          \\     
     NGC2960           &       3.81            &           $r$           &    0.007132           &           $r$           &           {       SDSS}          \\     
     NGC2974           &       1.17            &       $3.6mu$           &   0.0005821           &       $3.6mu$           &           {    Spitzer}          \\     
     NGC3031           &       2.48            &           $i$           &   9.922e-07           &           $i$           &     \citet{ Beifiori2012}          \\   
     NGC3079           &       0.37            &       $3.6mu$           &   0.0003067           &       $3.6mu$           &           {    Spitzer}          \\     
     NGC3115           &       6.13            &           $V$           &   3.643e-05           &           $V$           &     \citet{ Scorza1998}          \\     
     NGC3227           &      19.68            &           $K$           &    0.001961           &           $K$           &     \citet{ Davies2006}          \\     
     NGC3245           &       3.40            &           $R$           &    0.001961           &           $i$           &     \citet{ Beifiori2012}          \\   
     NGC3377           &       2.81            &           $R$           &   0.0003275           &           $R$           &     \citet{ Graham2001}          \\     
     NGC3379           &       2.77            &           $I$           &   0.0009665           &           $I$           &     \citet{ Rusli2013b}          \\     
     NGC3384           &       1.44            &           $I$           &   0.0004682           &           $i$           &           {       SDSS}          \\     
     NGC3393           &       2.72            &           $I$           &    0.003697           &           $I$           &     \citet{ Schmitt2000}          \\    
     NGC3414           &       4.15            &           $I$           &   0.0005544           &           $i$           &           {       SDSS}          \\     
     NGC3585           &       2.89            &           $V$           &    0.000959           &           $V$           &     \citet{ Scorza1998}          \\     
     NGC3607           &       6.15            &           $V$           &    0.000346           &           $g$           &           {       SDSS}          \\    
     NGC3608           &       2.04            &           $I$           &    0.001753           &           $i$           &           {       SDSS}          \\     
     NGC3842           &       7.15            &           $V$           &    0.004632           &           $V$           &     \citet{ Rusli2013b}          \\     
     NGC3998           &       5.21            &           $I$           &     0.07487           &           $i$           &           {       SDSS}          \\     
     NGC4026           &       4.91            &           $V$           &   0.0009635           &           $g$           &           {       SDSS}          \\     
     NGC4151           &       1.49            &           $R$           &    0.001961           &           $R$           &     \citet{ Gadotti2008}          \\    
     NGC4258           &       0.55            &       $3.6mu$           &   0.0002412           &       $3.6mu$           &           {    Spitzer}          \\     
     NGC4261           &       8.89            &           $V$           &    0.003279           &           $V$           &     \citet{ Rusli2013b}          \\     
     NGC4291           &       5.08            &           $V$           &    0.003171           &           $V$           &     \citet{ Rusli2013b}          \\     
     NGC4342           &       4.01            &           $I$           &     0.00116           &           $i$           &           {       SDSS}          \\     
     NGC4374           &       6.39            &           $V$           &    0.000687           &           $V$           &     \citet{ Kormendy2009}          \\   
     NGC4388           &       1.30            &       $3.6mu$           &    0.003625           &       $3.6mu$           &           {    Spitzer}          \\     
     NGC4459           &       3.88            &           $V$           &   0.0006867           &           $V$           &     \citet{ Kormendy2009}          \\   
     NGC4473           &       6.91            &           $V$           &   0.0004416           &           $V$           &     \citet{ Kormendy2009}          \\   
     NGC4486           &       6.32            &           $V$           &    0.003142           &           $V$           &     \citet{ Kormendy2009}          \\   
     NGC4526           &       2.64            &           $I$           &    0.001193           &       $3.6mu$           &           {    Spitzer}          \\     
     NGC4552           &       7.36            &           $V$           &   0.0007571           &           $V$           &     \citet{ Kormendy2009}          \\   
     NGC4564           &       5.78            &           $V$           &   0.0004514           &           $V$           &     \citet{ Kormendy2009}          \\   
     NGC4594           &       3.08            &           $I$           &   3.265e-05           &           $I$           &     \citet{ Jardel2011}          \\     
     NGC4596           &       0.99            &           $K$           &   9.111e-05           &           $K$           &     \citet{   Vika2012}          \\     
     NGC4621           &       0.66            &       $3.6mu$           &   0.0006928           &       $3.6mu$           &           {    Spitzer}          \\     
     NGC4649           &       7.67            &           $V$           &    0.002266           &           $V$           &     \citet{ Rusli2013b}          \\     
     NGC4697           &       3.36            &           $R$           &    0.001467           &           $R$           &     \citet{  Erwin2008}          \\     
     NGC4736           &       0.61            &           $z$           &    0.001961           &           $z$           &           {       SDSS}          \\     
     NGC4826           &       1.33            &           $i$           &   0.0005272           &           $i$           &           {       SDSS}          \\     
     NGC4889           &       5.97            &           $R$           &    0.003468           &           $r$           &     \citet{ Jorgensen1994}          \\  
     NGC5077           &       3.48            &           $V$           &    0.001961           &           $V$           &           {    KeyProg}          \\     
     NGC5128           &       0.63            &           $K$           &     0.01004           &           $K$           &     \citet{ Cappellari2009}          \\ 
     NGC5576           &       3.17            &           $R$           &    0.001889           &           $r$           &           {       SDSS}          \\     
     NGC5813           &       4.70            &           $V$           &   0.0007684           &           $V$           &     \citet{ Rusli2013b}          \\     
     NGC5845           &       4.77            &           $V$           &   0.0004076           &           $V$           &           {    KeyProg}          \\     
     NGC5846           &       5.20            &           $I$           &   0.0006278           &           $i$           &     \citet{ Rusli2013b}          \\     
     NGC6086           &       4.05            &           $R$           &    0.002363           &           $R$           &     \citet{ Rusli2013b}          \\     
     NGC6251           &       3.62            &           $I$           &    0.001961           &           $I$           &     \citet{ Graham2001}          \\     
     NGC6264           &       5.27            &           $r$           &    0.007546           &           $r$           &     \citet{ Greene2010}          \\     
     NGC6323           &       8.15            &           $r$           &     0.01908           &           $r$           &     \citet{ Greene2010}          \\     
     NGC7052           &       2.17            &           $R$           &    0.001961           &           $R$           &     \citet{ Graham2001}          \\     
     NGC7457           &       0.65            &       $3.6mu$           &   0.0005046           &       $3.6mu$           &           {    Spitzer}          \\     
     NGC7582           &       0.07            &       $3.6mu$           &     0.01245           &       $3.6mu$           &           {    Spitzer}          \\     
     NGC7768           &       7.58            &           $V$           &    0.006975           &           $V$           &     \citet{ Rusli2013b}          \\     
       U3789           &       0.50            &           $H$           &    0.008882           &           $H$           &     \citet{ Peletier1999}          \\   
      NGC307           &       1.03            &           $K$           &   6.695e-05           &           $K$           &     \citet{ Erwin2015b}          \\     
     NGC1316           &       0.65            &           $K$           &    0.001886           &           $K$           &     \citet{  Nowak2008}          \\     
     NGC1332           &       7.10            &           $R$           &   0.0005986           &           $R$           &     \citet{  Rusli2011}          \\     
     NGC1374           &       5.30            &           $B$           &    0.002417           &           $B$           &     \citet{ Rusli2013a}          \\     
     NGC1398           &       3.00            &           $R$           &   0.0003303           &           $R$           &     \citet{ Erwin2015b}          \\     
     NGC1407           &       6.60            &           $B$           &    0.003128           &           $B$           &     \citet{ Rusli2013a}          \\     
     NGC1550           &       4.00            &           $R$           &    0.003566           &           $R$           &     \citet{ Rusli2013a}          \\     
     NGC3091           &       3.80            &           $I$           &   0.0008164           &           $I$           &     \citet{ Rusli2013a}          \\     
     NGC3368           &       0.40            &           $K$           &    0.002947           &           $K$           &     \citet{  Nowak2010}          \\     
     NGC3489           &       0.44            &           $H$           &   0.0003723           &           $H$           &     \citet{  Nowak2010}          \\     
     NGC3627           &       0.40            &           $K$           &   0.0001181           &           $K$           &     \citet{ Erwin2015b}          \\     
     NGC3923           &       4.22            &           $z$           &    0.003155           &           $z$           &     \citet{ Bender2015}          \\     
     NGC4371           &       1.71            &           $z$           &   5.221e-05           &           $z$           &     \citet{ Erwin2015b}          \\     
     NGC4472           &       4.90            &           $V$           &    0.001257           &           $V$           &     \citet{ Rusli2013a}          \\     
    NGC4486a           &       4.00            &           $zACS$        &   0.0007368           &           $zACS$        &     \citet{  Nowak2007}          \\     
    NGC4486b           &       6.56            &           $V$           &   0.0004456           &           $V$           &     \citet{ Bender2015}          \\     
     NGC4501           &       0.54            &           $K$           &   0.0002495           &           $K$           &     \citet{ Erwin2015b}          \\     
     NGC4699           &       0.68            &           $z$           &   0.0005982           &           $z$           &     \citet{ Erwin2015b}          \\     
     NGC4751           &       8.27            &           $R$           &   0.0005339           &           $R$           &     \citet{ Rusli2013a}          \\     
     NGC5018           &       1.23            &           $I$           &   0.0001083           &           $I$           &     \citet{ Bender2015}          \\     
     NGC5328           &       4.90            &           $V$           &    0.002828           &           $V$           &     \citet{ Rusli2013a}          \\     
     NGC5419           &       5.37            &           $R$           &     0.01759           &           $R$           &     \citet{ Mazzalay2015}          \\   
     NGC5516           &       5.20            &           $R$           &    0.000279           &           $R$           &     \citet{ Rusli2013a}          \\     
     NGC6861           &       6.10            &           $I$           &    0.000114           &           $I$           &     \citet{ Rusli2013a}          \\     
     NGC7619           &       3.00            &           $I$           &    0.002567           &           $I$           &     \citet{ Rusli2013a}          \\     

\enddata
\tablenotetext{\ }{Column 1, the object
    name (both literature and SINFONI samples); Column 2 to 4: the dynamical $M/L$, its band, and its logarithmic error squared; 
Column 5: the band 
    of the related image; Column 6: the references of the used profiles. 
    When no errors are
    available, we set the errors to the average value of all the
    available errors.}
\end{deluxetable}

\section*{SINFONI Sample}
\begin{description}
\item[NGC 307:] The galaxy has a classical bulge
  \citep{Thomas2014,Erwin2015b}.  We set $b=0.5$ in Table
  \ref{tab_data} since the galaxy is too edge-on to be sure about the
  presence or absence of a bar.  We derive its BH mass in
  \citet{Erwin2015b}, where we model the stellar kinematics allowing for
    different $M/L$ for the bulge and the disk components, and no dark
    matter halo, see Fig. \ref{fig_MBHML}. We adopt the distance
  derived from the Hyperleda radial velocity corrected for Local Group
  infall onto Virgo and $H_0=75$.

\item[NGC1316:] We consider the galaxy as a merger remnant and
  power-law elliptical \citep{Nowak2008}.  We derive its BH mass in 
\citet{Nowak2008}.

\item[NGC1332:] The galaxy has a prototypical classical bulge
  \citep{Erwin2015a}. We revised our previous velocity dispersion
  determination of 328 km/s in \citet{Rusli2011} to 293.1 km/s. This stems
  from the larger half-luminosity radius for the whole
  galaxy ($28''$) that we use now versus the bulge-only radius ($8.4''$) 
quoted in \citet{Rusli2011}. We derive its BH mass in \citet{Rusli2011}.

\item[NGC1374:] The galaxy is a power-law elliptical
  \citep{Rusli2013b}. We derive its BH mass in \citet{Rusli2013a}.

\item[NGC1398:] The galaxy has a classical bulge
  \citep{Erwin2015b}. We derive its BH mass in \citet{Erwin2015b},
 where we model the stellar kinematics allowing for
    different $M/L$ for the bulge and the disk components, and no dark
    matter halo, see Fig. \ref{fig_MBHML}.  The distance comes from \citet{Tully2009}.

\item[NGC1407:] The galaxy is a core elliptical \citep{Rusli2013b}. We
  derive its BH mass in \citet{Rusli2013a}.

\item[NGC1550:] The galaxy is a core elliptical \citep{Rusli2013b}. We
  derive its BH mass in \citet{Rusli2013a}.

\item[NGC3091:] The galaxy is a core elliptical
  \citep{Rusli2013b}. We derive its BH mass in \citet{Rusli2013a}.  We
  measure $R_e=22.4''$ using the profile of \citet{Rusli2013b}, who get
  $R_e=90''$ fitting a $n=9.3$ Core-Sersic profile. The difference is
  driven by the extrapolation (see Fig. \ref{fig_extrap}).

\item[NGC3368:] The galaxy has a composite (classical plus pseudo) bulge
  \citep{Nowak2010,Erwin2015a}. We derive its BH mass in
  \citet{Nowak2010}.


\item[NGC3489:] Following \citet{Nowak2010}, the galaxy has a
  composite (classical plus pseudo) bulge. We use that decomposition,
  although it is a bit uncertain \citep{Erwin2015a}. We derive its BH
  mass and velocity dispersion in \citet{Nowak2010} and do not compute
  $\sigma_{e/2},\sigma_{e/2}^S,\sigma_e^S$ in Table \ref{tab_sigma}
  \citep[see discussion in][]{Nowak2010}.

\item[NGC3627:] The galaxy has a pseudo bulge \citep{Erwin2015b}, that
  we fit without the box-peanut component.  We derive its BH mass in
  \citet{Erwin2015b}, where we model the stellar kinematics allowing for
    different $M/L$ for the bulge and the disk components, and no dark
    matter halo, see Fig. \ref{fig_MBHML}.  We analyze its gas emission and kinematics in
  \citet{Mazzalay2013,Mazzalay2014}, from which we take the distance.

\item[NGC3923:] The galaxy is a merger remnant \citep{Bender2015}. We
  derive the BH mass by fitting the stellar kinematics without dark
  matter halo, see Fig. \ref{fig_MBHML}.  We use the mean of the
  redshift-independent measurements given by the Nasa Extragalactic
  Database (NED).

\item[NGC4371:] The galaxy has a composite (classical plus pseudo) bulge
  \citep{Erwin2015a}. We derive its BH mass in \citet{Erwin2015b}, 
where we model the stellar kinematics allowing for
    different $M/L$ for the bulge and the disk components, and no dark
    matter halo, see Fig. \ref{fig_MBHML}.

\item[NGC4472:] The galaxy is a core elliptical \citep{Rusli2013b}. We
  derive its BH mass in \citet{Rusli2013a}.

\item[NGC4486a:] The galaxy is a power-law elliptical
  \citep{Nowak2007}. We derive its BH mass in \citet{Nowak2007}.  We
  derive the velocity dispersions quoted in Table \ref{tab_data} and
  \ref{tab_sigma} using the profiles of \citet{Prugniel2011}, see also
  Appendix A.

\item[NGC4486b:] The galaxy is a compact elliptical
  \citep{Bender2015}.  We derive the black hole mass by fitting
    the stellar kinematics without dark matter halo
    \citep{Bender2015}, see Fig. \ref{fig_MBHML}.  Our SINFONI black
    hole mass is 30\% smaller than the value used by
    \citet{KormendyHo2013}. We do not use it in the fits, since it is
    the largest outlier in the correlations involving BH masses. We
  derive the velocity dispersions quoted in Table \ref{tab_data} and
  \ref{tab_sigma} using the profiles of \citet{Kormendy1997}
  determined at distances betwen 1.5 and 3 arcsec from the center.
  This leads to a value of $\sigma$ different from the one used by
  \citet{KormendyHo2013}, see above.  We use the average of the three
  surface brightness fluctuation distances reported by NED.

\item[NGC4501:] The galaxy has a pseudo bulge \citep{Erwin2015b}. We
  derive its BH mass in \citet{Erwin2015b}, where we model the
    stellar kinematics allowing for different $M/L$ for the bulge and the
    disk components, and no dark matter halo, see
    Fig. \ref{fig_MBHML}.  We analyze its gas emission and
  kinematics in \citet{Mazzalay2013,Mazzalay2014}.

\item[NGC4699:] The galaxy has a composite (classical plus pseudo) bulge
  \citep{Erwin2015a}. We derive its BH mass in \citet{Erwin2015b},
 where we model the
    stellar kinematics allowing for different $M/L$ for the bulge and the
    disk components, and no dark matter halo, see
    Fig. \ref{fig_MBHML}.

\item[NGC4751:] The galaxy is a power-law elliptical
  \citep{Rusli2013a}.  We derive its BH mass in \citet{Rusli2013a},
  but correct the $M/L$, having discovered that \citet{Rusli2013a} incorrectly
  used the magnitude of the sun in the $V$ band instead of the $R$ band.


\item[NGC5018:] The galaxy is a power-law elliptical, see
  \citet{Thomas2015a} where we measure $M_{BH}$
allowing for a dark matter halo as in \citet{Rusli2013a}, see
    Fig. \ref{fig_MBHML}.  We take the
  Tully-Fisher mean distance of \citet{Theureau2007}.


\item[NGC5328:] The galaxy is a core elliptical
  \citep{Rusli2013b}. We derive its BH mass in \citet{Rusli2013a}.  We
  find $R_e=29.4''$ using the profile of \citet{Rusli2013b}, who get
  $R_e=76.8''$ fitting a $n=11.1$ Core-Sersic profile
\footnote{There is a mistake in their Table 2, where $\mu_b$ should read 16.73 instead of 17.07.}. 
However, the difference is compatible with our estimated errors due to extrapolation.

\item[NGC5419:] The galaxy is a core elliptical
  \citep[][where we measure its BH mass]{Mazzalay2015}. We derive the 
distance from the radial
  velocity corrected for Local Group infall onto Virgo from Hyperleda
  using $H_0=72$.

\item[NGC5516:] The galaxy is a core elliptical \citep{Rusli2013b}. 
\citet{KormendyHo2013} use a black hole mass slightly different 
from our value published in \citet{Rusli2013a}.

\item[NGC6861:] The galaxy is a power-law elliptical
  \citep{Rusli2013a}.  We derive its BH mass in \citet{Rusli2013a}.

\item[NGC7619:] The galaxy is a core elliptical \citep{Rusli2013b}. 
  We derive $R_e=42''$ using the profile of \citet{Rusli2013b}, who get
  $R_e=100''$ fitting a $n=9.3$ Core-Sersic profile. The difference is
  driven by the extrapolation (see
  Fig. \ref{fig_extrap}). \citet{KormendyHo2013} use a black hole mass
  slightly different from our value published in \citet{Rusli2013a}.

\end{description}

\begin{figure*}
  \begin{center}
    \includegraphics[trim=1cm 15cm 0cm 3cm,clip,width=16cm]{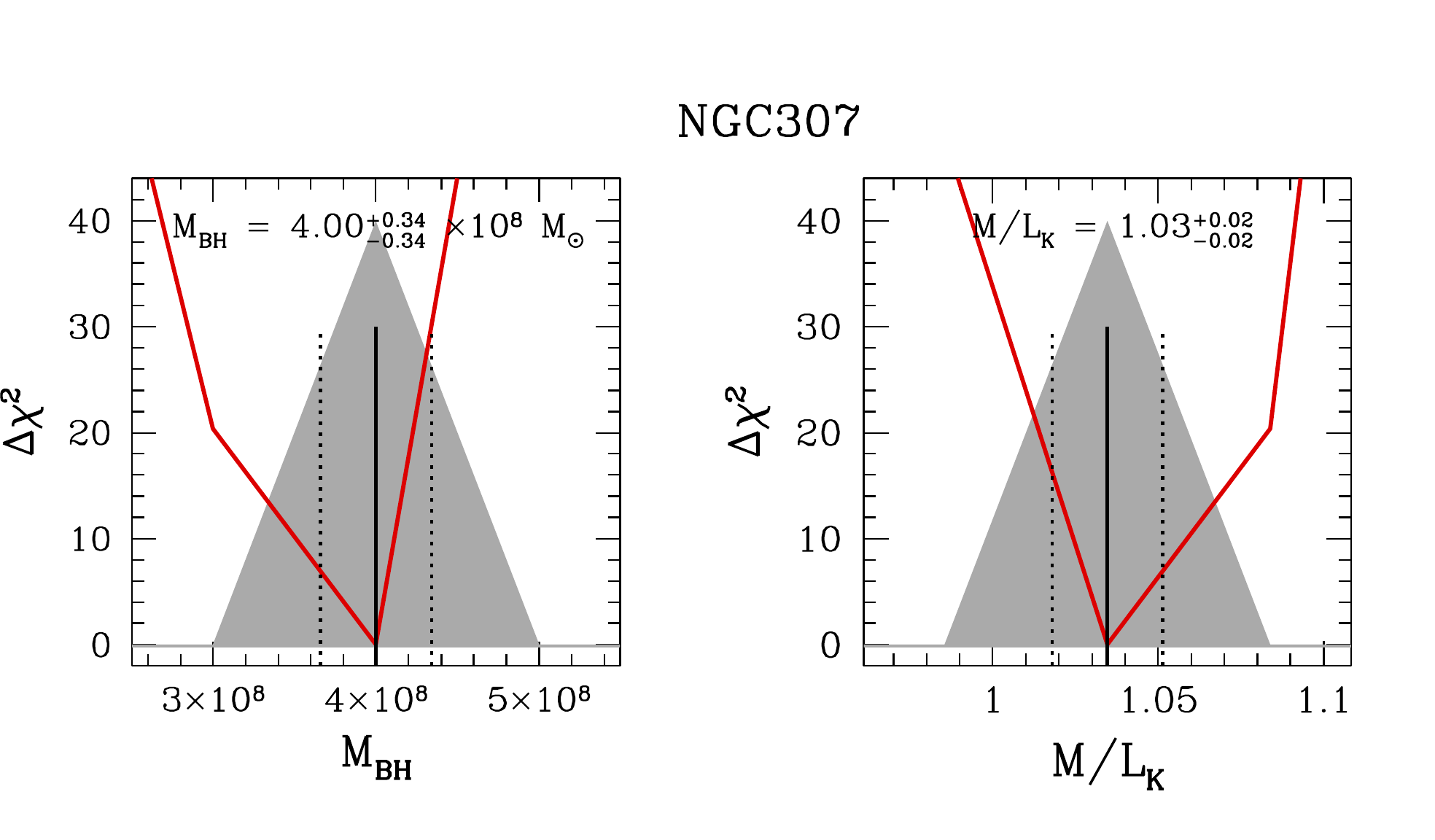}
    \includegraphics[trim=1cm 15cm 0cm 3cm,clip,width=16cm]{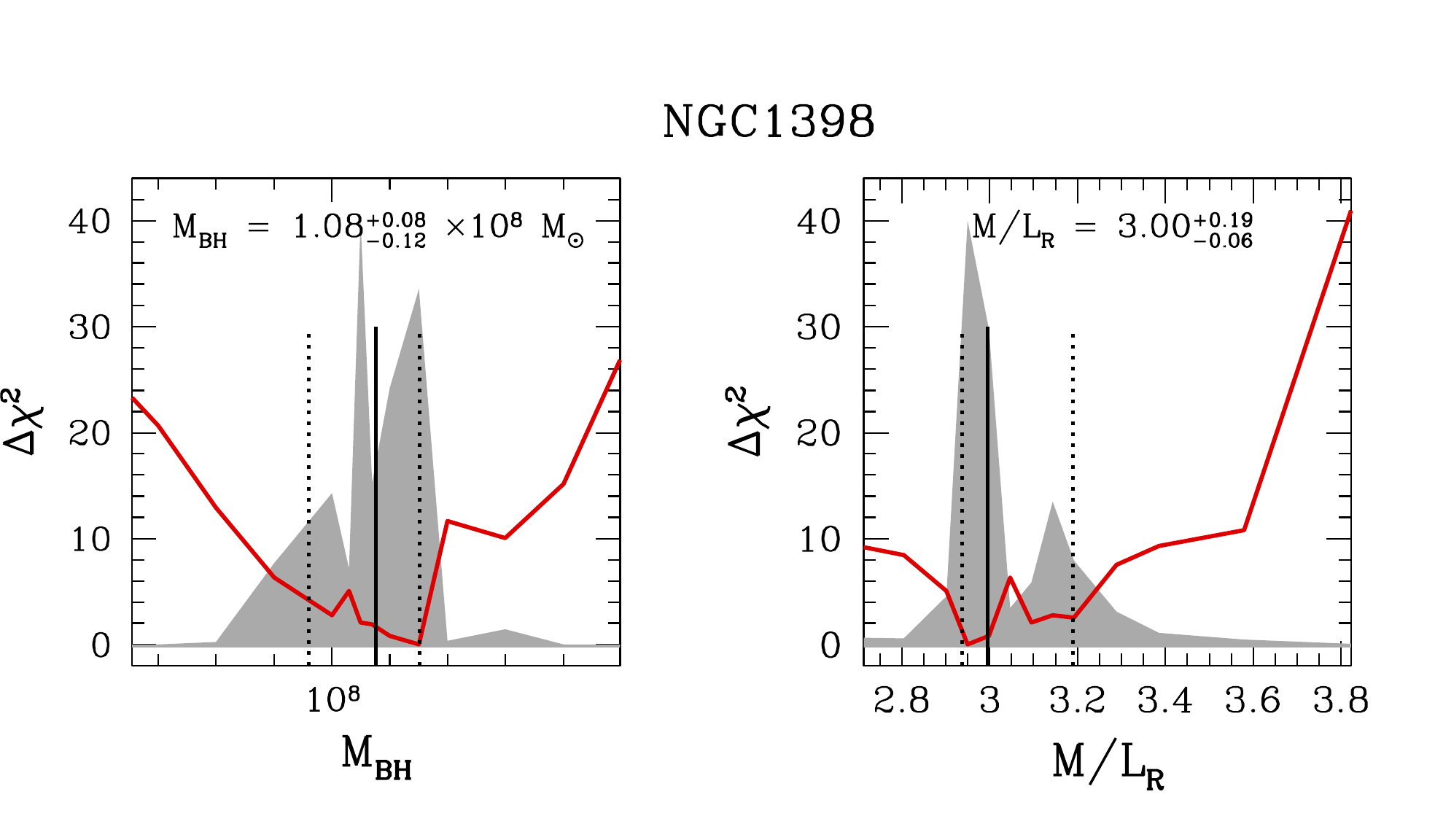}
  \end{center}
  \caption{We show two plots for each SINFONI galaxy where we have not
    yet published a complete analysis of our data. The left plot shows
    the marginalized posterior probability $P$, scaled arbitrarily to
    a maximum value of 40 \citep[shaded, see Eq. 4 of][]{Rusli2013a}
    and $\Delta\chi^2$ (red lines) vs. $M_{BH}$; the plot to the right
    shows $P$ and $\Delta \chi^2$ as a function of $M/L$. The vertical
    solid lines show the derived values; the vertical dashed lines
    show the $1\sigma$ errors. \label{fig_MBHML}}
\end{figure*}
\addtocounter{figure}{-1}
\begin{figure*}
  \begin{center}
    \includegraphics[trim=1cm 15cm 0cm 4cm,clip,width=16cm]{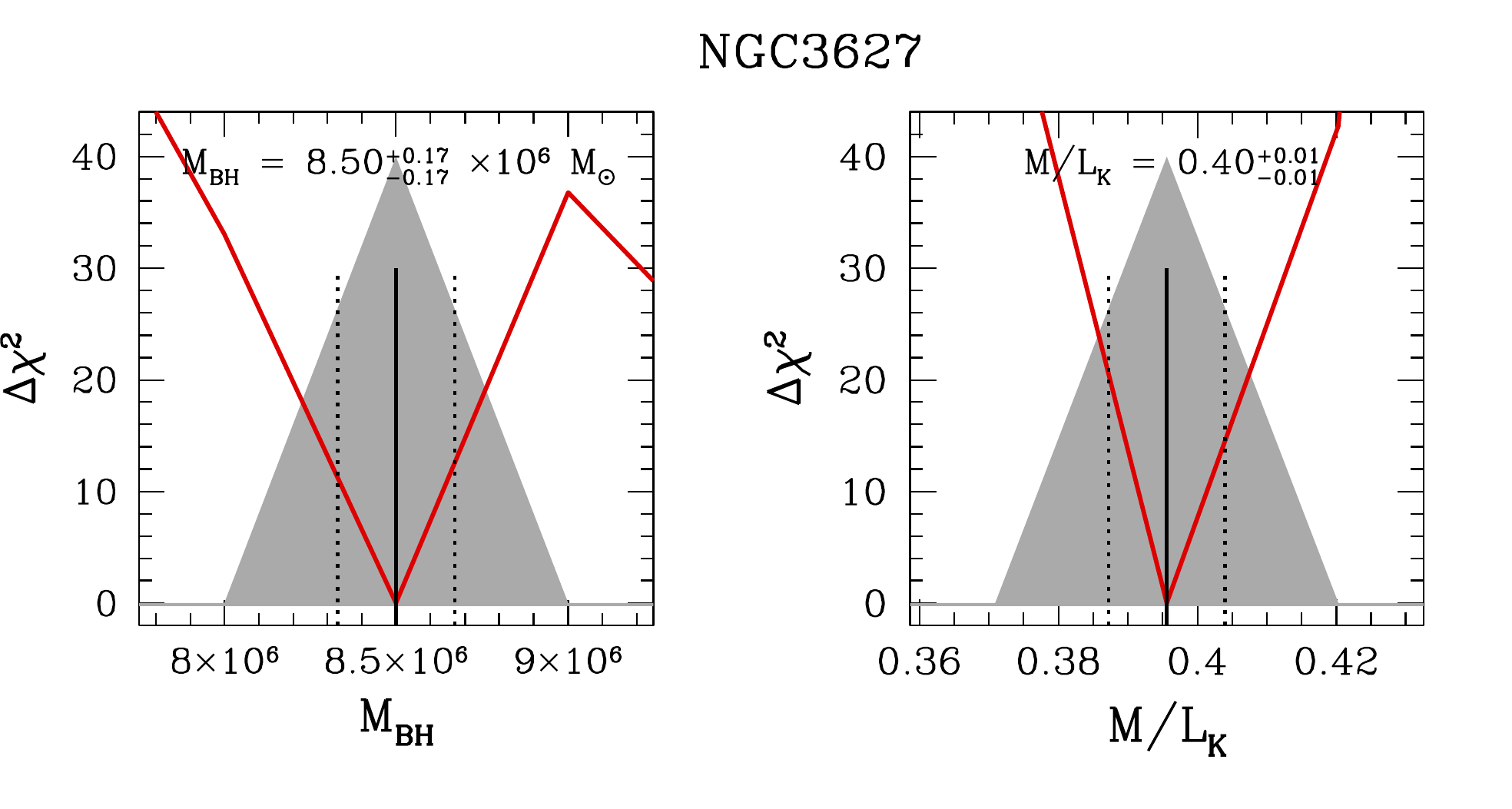}
    \includegraphics[trim=1cm 15cm 0cm 4cm,clip,width=16cm]{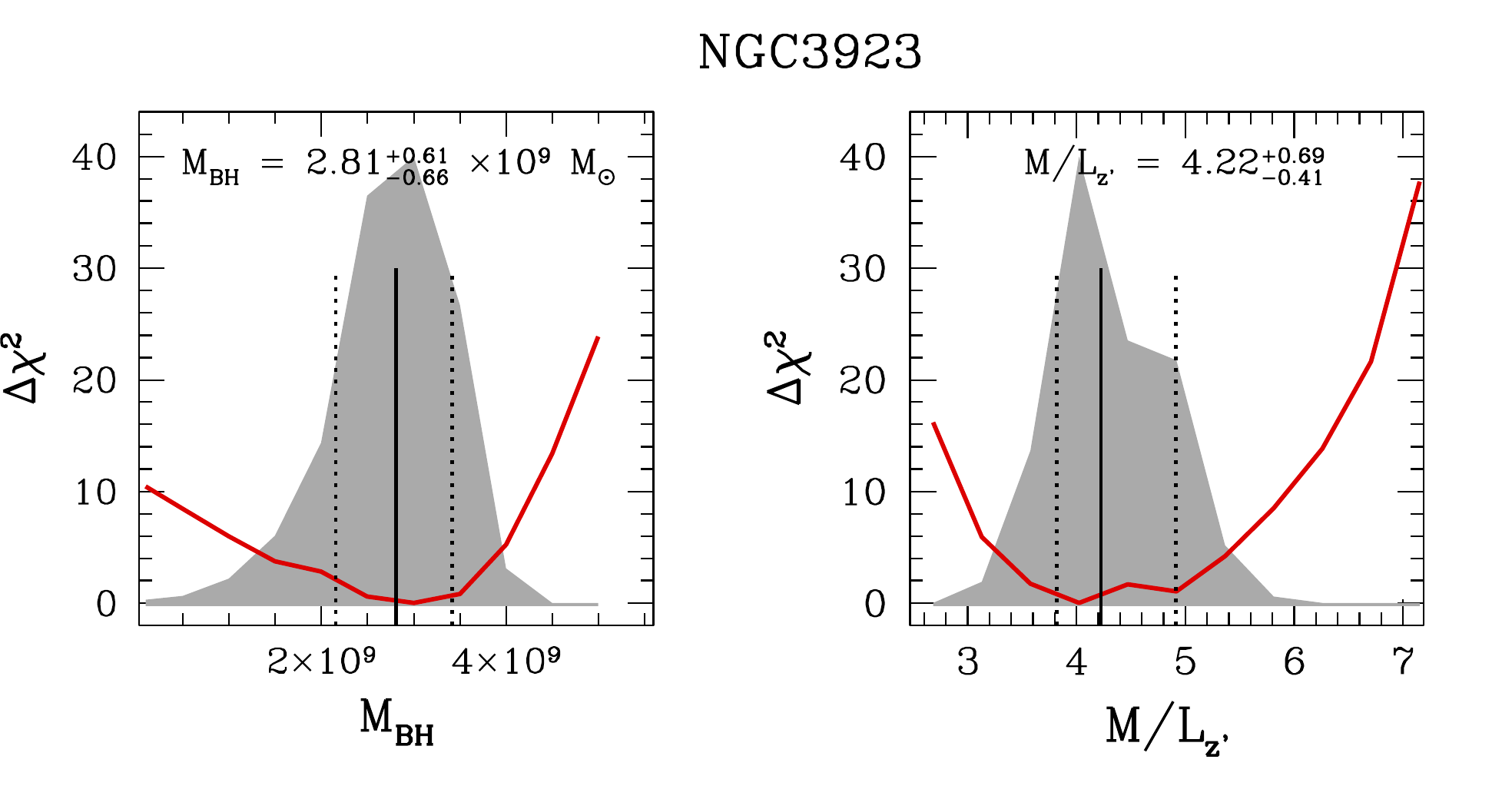}
\end{center}
\caption{Continued}
\end{figure*}
\addtocounter{figure}{-1}
\begin{figure*}
  \begin{center}
    \includegraphics[trim=1cm 15cm 0cm 4cm,clip,width=16cm]{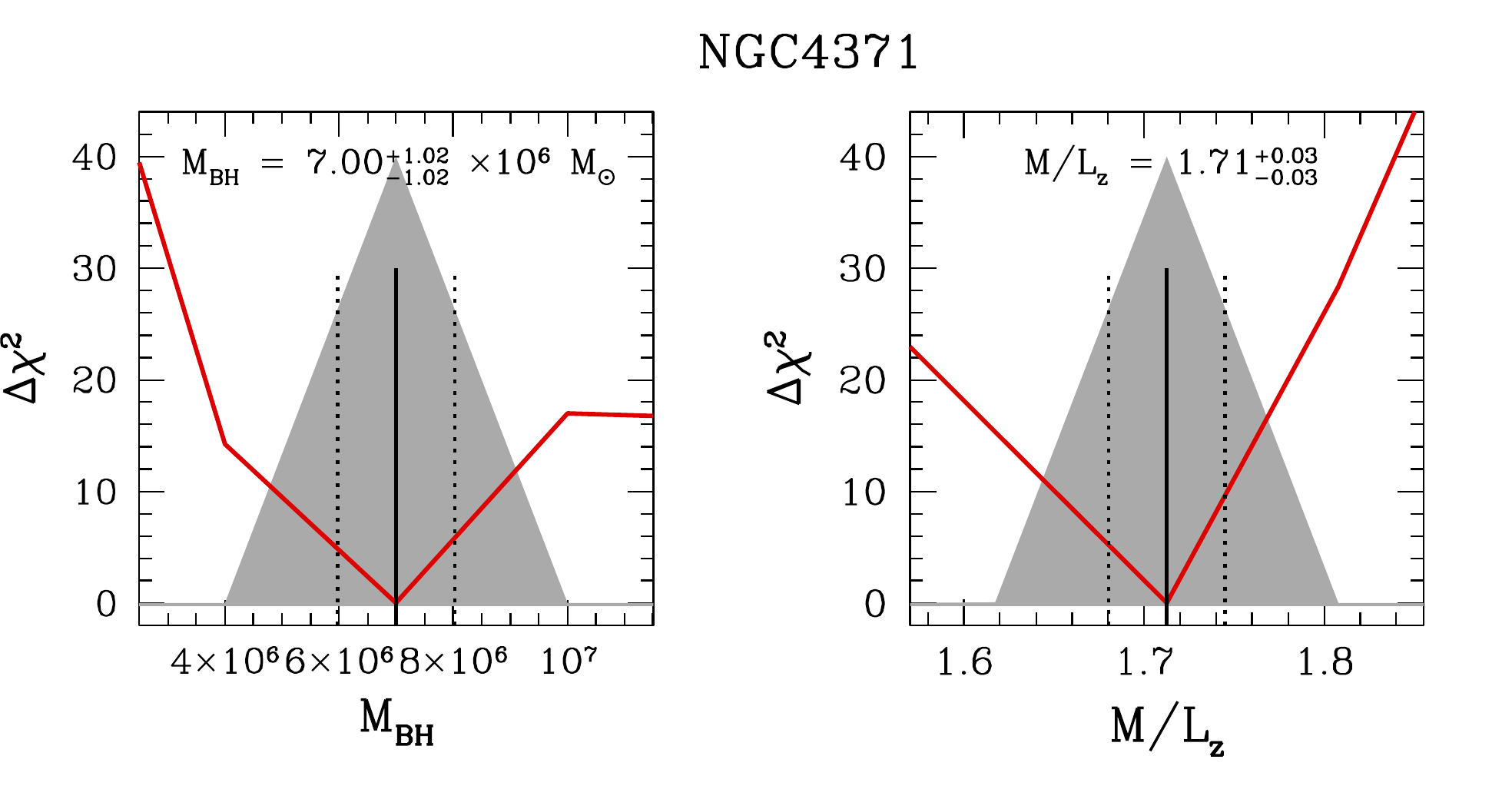}
    \includegraphics[trim=1cm 15cm 0cm 4cm,clip,width=16cm]{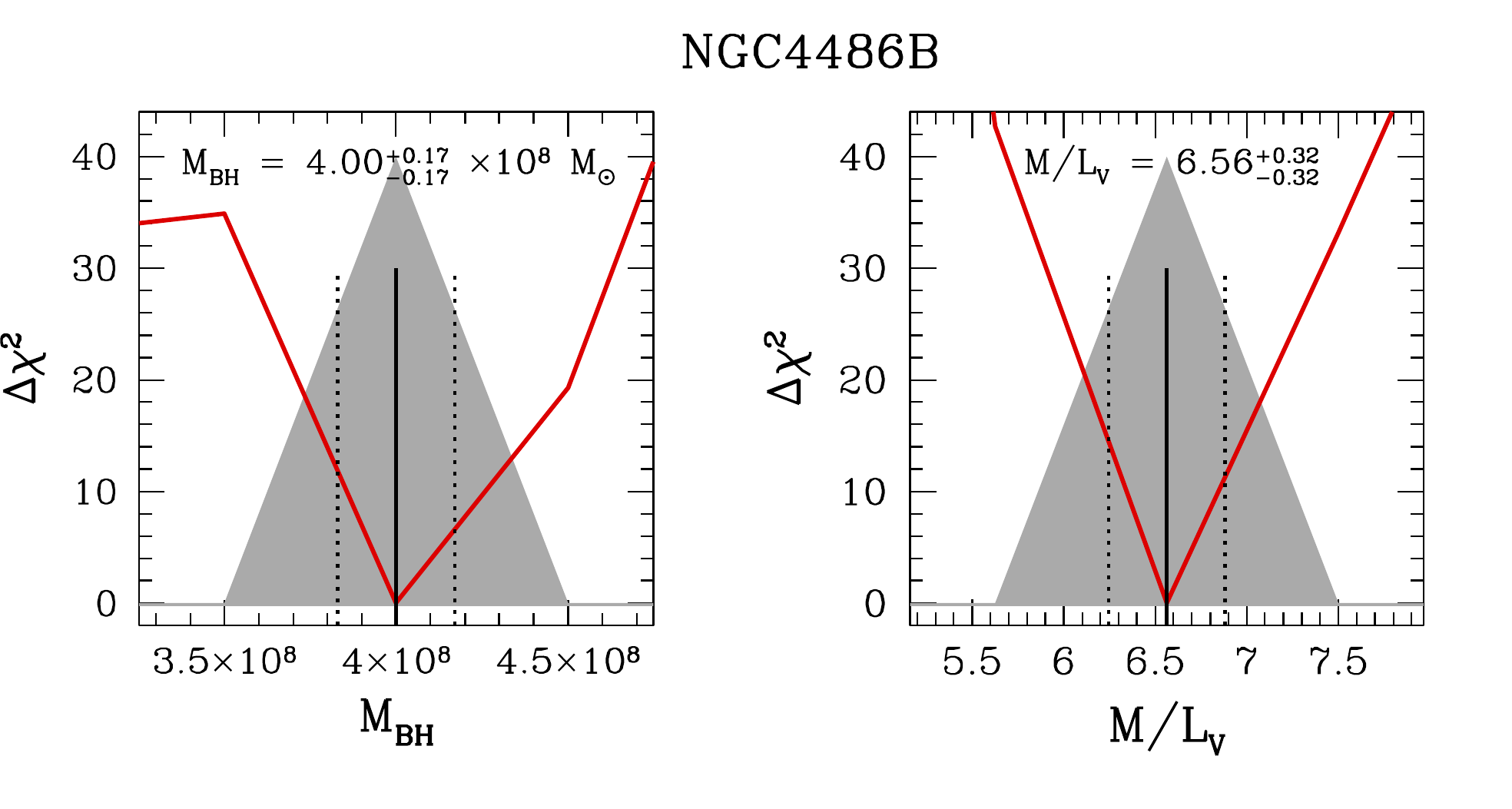}
\end{center}
\caption{Continued}
\end{figure*}
\addtocounter{figure}{-1}
\begin{figure*}
  \begin{center}
     \includegraphics[trim=1cm 15cm 0cm 4cm,clip,width=16cm]{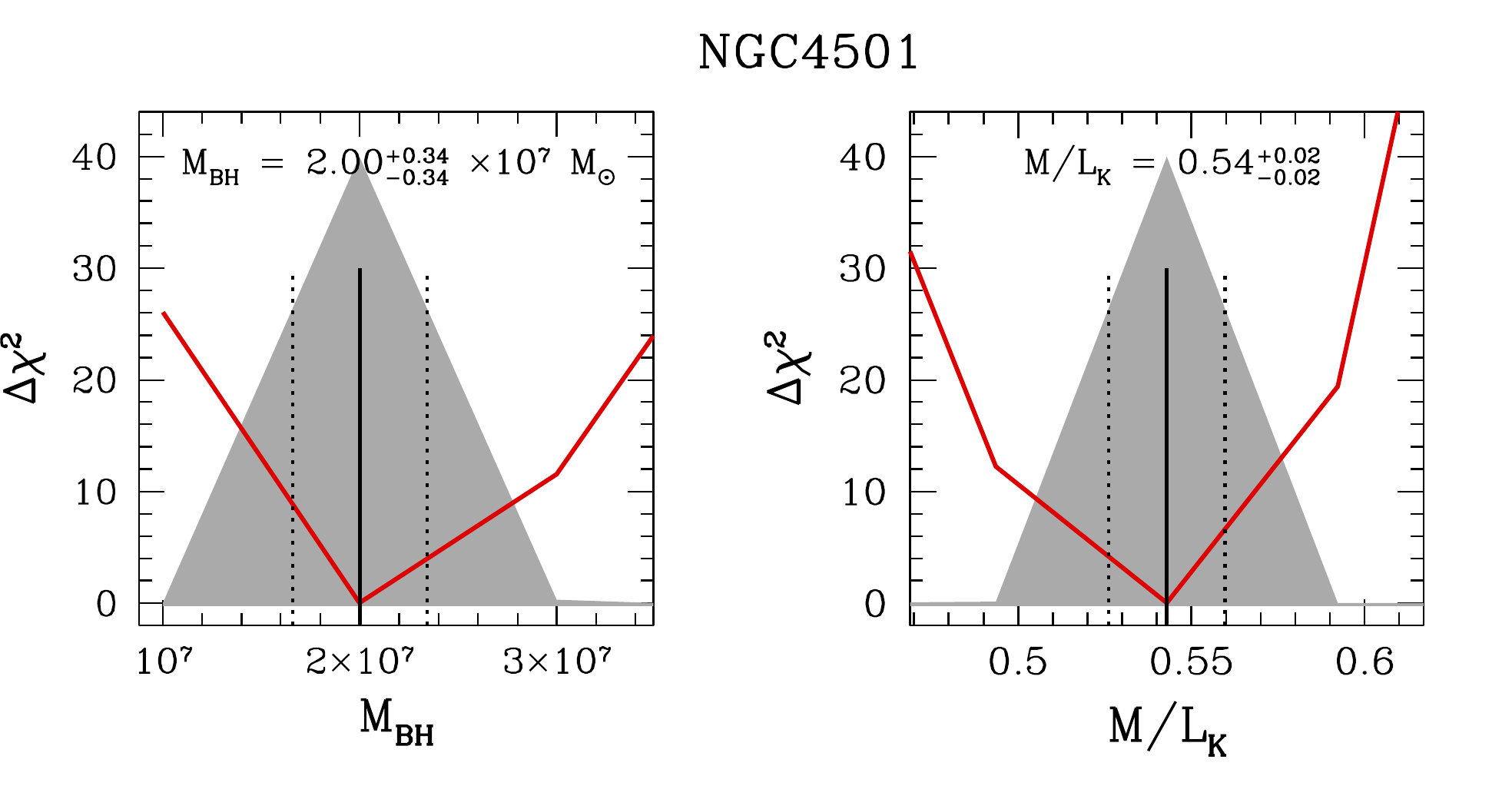}
    \includegraphics[trim=1cm 15cm 0cm 4cm,clip,width=16cm]{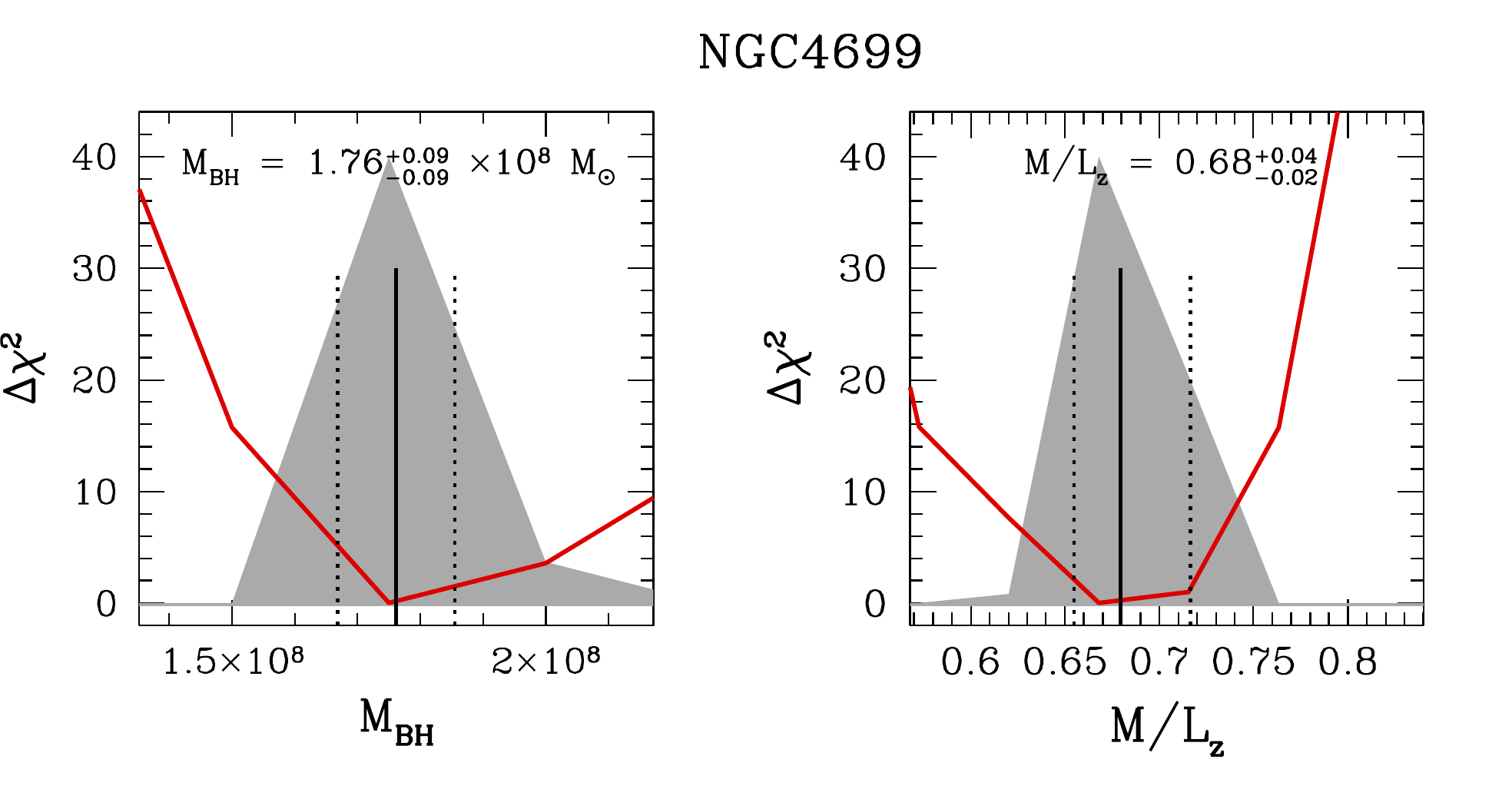}
  \end{center}
\caption{Continued}
\end{figure*}
\addtocounter{figure}{-1}
\begin{figure*}
  \begin{center}
    \includegraphics[trim=1cm 15cm 0cm 5cm,clip,width=16cm]{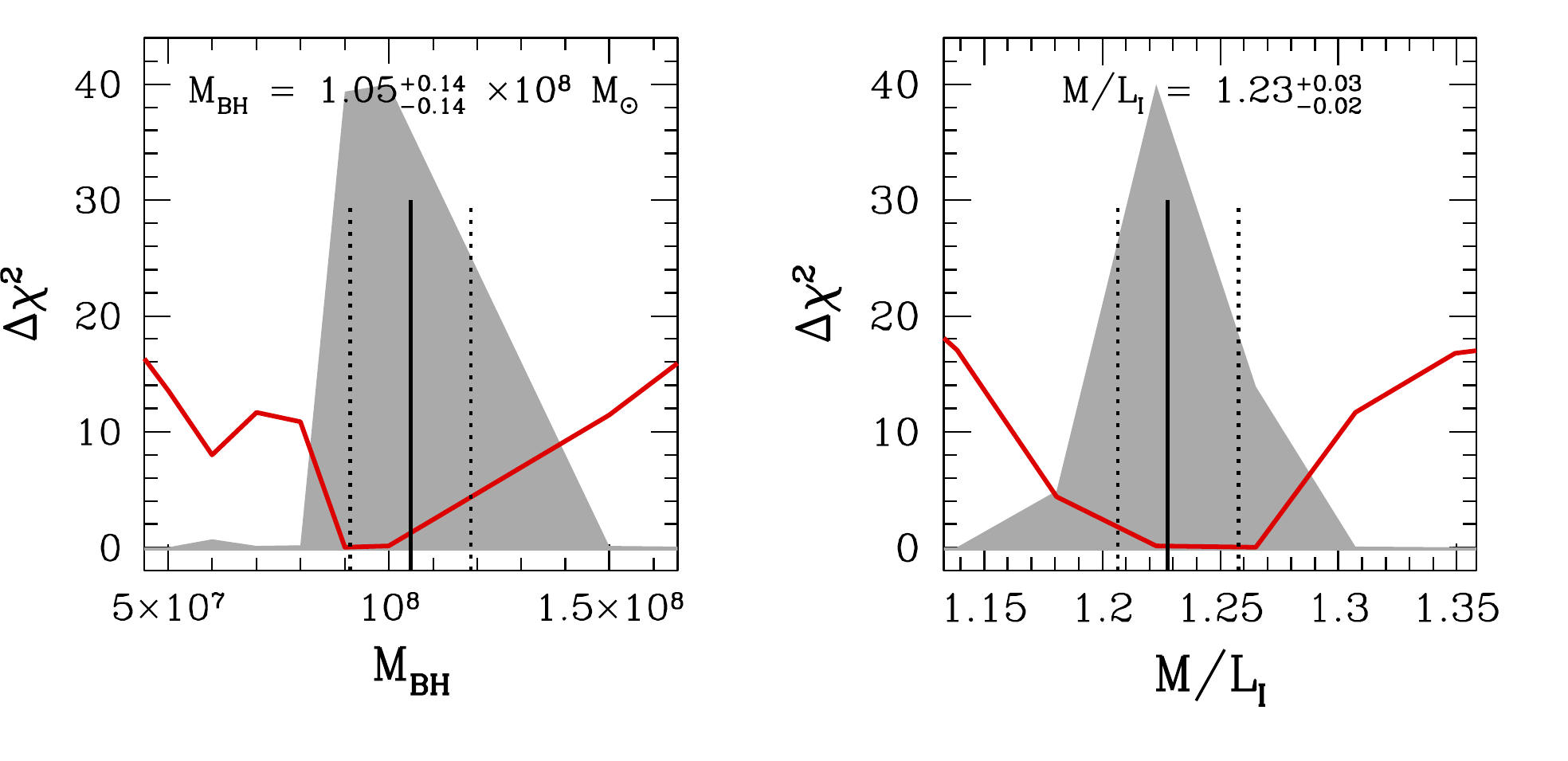}
  \end{center}
\caption{Continued}
\end{figure*}

\section*{Appendix C: $M/L$ from stellar kinematics}
\label{app_kinfit}

\begin{figure*}
  \begin{center}
    \includegraphics[trim=0cm 5cm 0cm 2cm,clip,width=16cm]{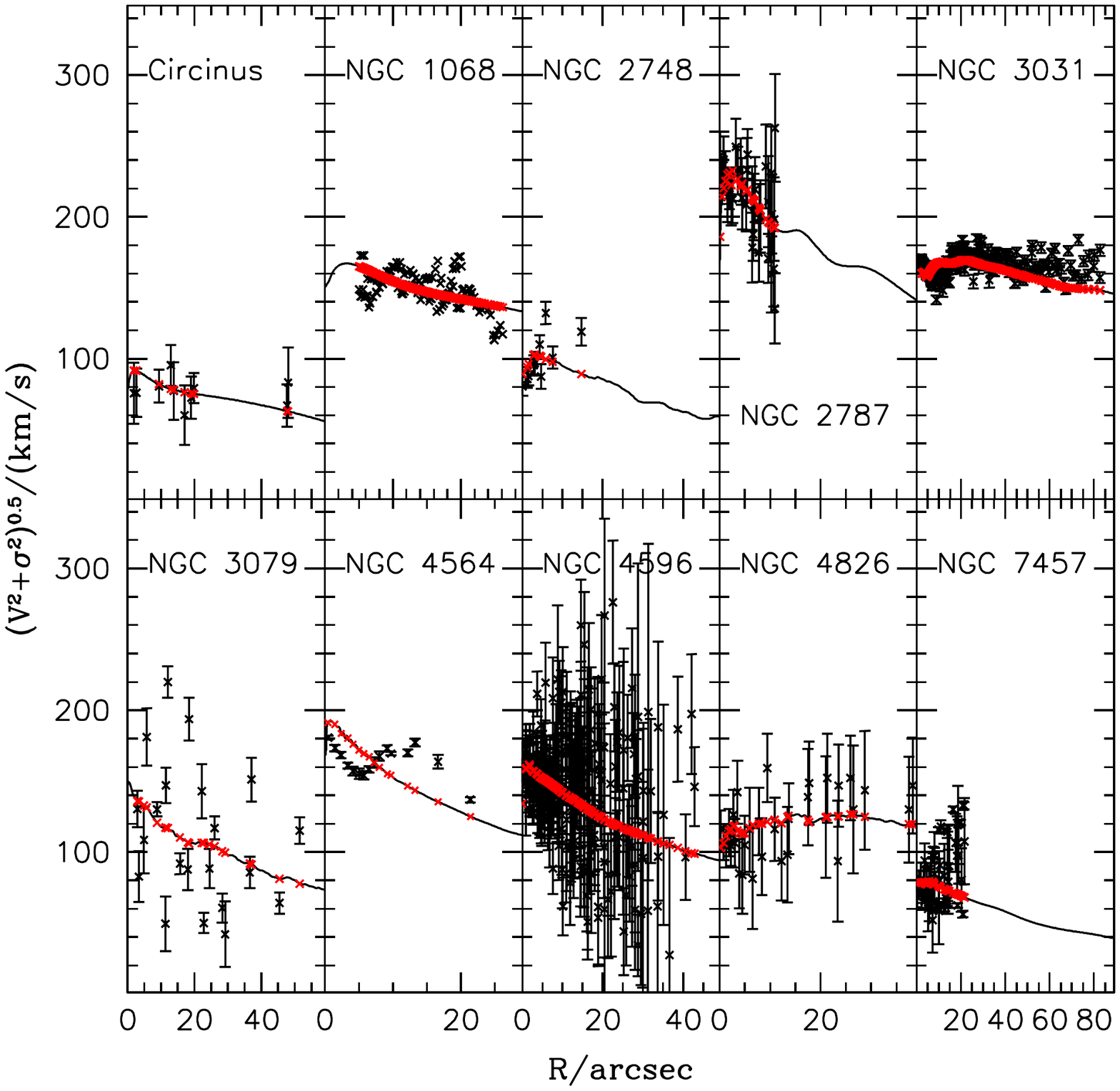}
  \end{center}
  \caption{The spherical Jeans fits to the ten galaxies with
    extended stellar kinematics, but without dynamical $M/L$.  The black
    crosses show the datapoints, the black line the model, and the red
    crosses the values of the model at the radii where data are
    available.\label{fig_kinprofile}}
\end{figure*}

For a number of galaxies in Table \ref{tab_ml} no dynamically
determined $M/L$ is available in the literature. 
Nevertheless, for all of these objects
stellar kinematics of some sort is available, 
either as stellar velocity and velocity dispersion
profiles or central velocity dispersions. In these cases we
determine the $M/L$ by solving the Jeans equations for a
self-consistent, isotropic non-rotating spherical system, using the
spherically deprojected surface brightness profiles determined in
Sect. \ref{sec_data}. If radially extended stellar kinematics is
available, we fit $\sigma_{kin}(R)=\sqrt{v(R)^2+\sigma(R)^2}$.
Fig. \ref{fig_kinprofile} shows the resulting fits for the ten
galaxies with extended kinematics. When the bulge and disk profiles
are of similar luminosity in the radial range where the stellar
kinematics is available, we solve the Jeans equations by summing the
two deprojected contributions and determining the $M/L$ value (equal for
both components) that minimizes $\chi^2$. The errors on $M/L$ are
computed by looking at $\chi^2=\chi^2_{min}+1$.  When only a
``central'' velocity dispersion is available, we select the $M/L$ that
predicts a line-of-sight velocity dispersion
$\sigma_{ap}=\sqrt{\langle\sigma^2L\rangle_{ap}/\langle
  L\rangle_{ap}}$, luminosity-averaged over the effective
aperture with radius $r_{ap}=\sqrt{slitwidth\times slitlength/\pi}$, 
matching the observed value. The percentage error on $M/L$ is
twice the percentage error on the observed velocity dispersion.

This approach is a drastic simplification of the appropriate dynamical
modeling of the galaxies, which would need to take into account the
geometry of the objects and the dynamical effects of the central black
hole and dark matter halos, not to mention the presence of bars. In
order to estimate how wrong our simple dynamical estimates of $M/L$ can
be, we apply both methods (i.e. $(M/L)_{kin}$ from the spherical Jeans
modeling of $\sqrt{V^2+\sigma^2}$, or $(M/L)_{ap}$ from modeling of the
single aperture velocity dispersion) to the sample of SINFONI galaxies
where we performed the full Schwarzschild modeling of the available
extended stellar kinematics, deriving $(M/L)_{best}$.  To compute
$M/L_{ap}$ we consider the $\sigma_{col}$ of Table \ref{tab_sigma},
obtained by fitting the SINFONI collapsed spectrum. Since the SINFONI
field of view is $3\times 3$ arcsec$^2$, the radius of the equivalent
aperture is $r_{ap}=3/\pi^{1/2}=1.7$ arcsec.

In Table \ref{tab_checkML} we report the ratios
$(M/L)_{kin}/(M/L)_{best}$ and $(M/L)_{ap}/(M/L)_{best}$. On average,
our simple methods tend to overestimate the true $M/L$ values by 40\%
with similar scatter ($\approx 0.15$ dex).  There are no trends with
$M_{Bu}$, $\sigma$, $r_h$ or $\rho_h$.

\begin{table}
\caption{The ratios $(M/L)_{kin}/(M/L)_{best}$ and $(M/L)_{ap}/(M/L)_{best}$ 
for the SINFONI sample of galaxies, without NGC 3489 (see App. C).}
\label{tab_checkML}
\begin{tabular}{lll}
\hline
\hline
Galaxy & $(M/L)_{kin}/(M/L)_{best}$ &  $(M/L)_{ap}/(M/L)_{best}$\\
\hline
     NGC0307           &       1.667           &       1.583          \\
     NGC1316           &       1.015           &      0.8354          \\
     NGC1332           &      0.9507           &       1.434          \\
     NGC1374           &       1.336           &       1.585          \\
     NGC1398           &        1.45           &        1.84          \\
     NGC1407           &       1.618           &       1.627          \\
     NGC1550           &       1.328           &       1.273          \\
     NGC3091           &       1.353           &       1.337          \\
     NGC3368           &       1.275           &       2.825          \\
     NGC3627           &        1.35           &       1.447          \\
     NGC3923           &        1.13           &       1.456          \\
     NGC4371           &       1.251           &       1.229          \\
     NGC4472           &       1.592           &       1.449          \\
    NGC4486a           &      0.9375           &        0.81          \\
    NGC4486b           &       1.101           &       1.609          \\
     NGC4501           &       1.358           &        1.23          \\
     NGC4699           &       1.467           &       1.451          \\
     NGC4751           &       1.493           &       1.345          \\
     NGC5018           &        1.04           &       0.665          \\
     NGC5328           &       1.455           &       1.706          \\
     NGC5419           &        1.33           &         1.2          \\
     NGC5516           &       1.386           &        1.09          \\
     NGC6861           &        1.19           &      0.8557          \\
     NGC7619           &       1.443           &        1.37          \\

\hline
\end{tabular}
\end{table}

\section*{Appendix D: Fits without NGC 2974, NGC 3079, NGC 3414, NGC 4151, NGC 4552, NGC 4621, NGC 5813, NGC 5846.}
\label{app_strictfits}

Tables \ref{tab_1dimstrict} and \ref{tab_2dimstrict} report the
results of the mono- and bivariate correlations involving black hole
masses, excluding the galaxies NGC 2974, NGC 3079, NGC 3414, NGC 4151,
NGC 4552, NGC 4621, NGC 5813, NGC 5846, for which only uncertain black
hole masses are available. No significant differences with the results reported 
for the full sample (see Tables \ref{tab_1dim} and \ref{tab_2dim}) are found.

\begin{deluxetable}{llllllllllll}
\tablewidth{0pt}
\tabletypesize{\tiny}
\tablecaption{One-dimensional correlations derived without 
the galaxies  NGC 2974, NGC 3079, NGC 3414, NGC 4151, NGC 4552, NGC 4621, NGC 5813, NGC 5846.\label{tab_1dimstrict}}
\tablehead{
\colhead{Fit} & \colhead{Sample} & \colhead{N} & \colhead{a} & \colhead{da} & \colhead{ZP} & \colhead{dZP} & \colhead{$\epsilon$} & \colhead{$d\epsilon$} & \colhead{rms} & \colhead{$r_S$}& \colhead{$P(r_S)$}
}
\startdata
      \msig\           &         All           &          88           &       5.175           &       0.274           &      -3.594           &        0.63           &       0.412           &        0.04           &       0.457           &      0.9111           &    7.33e-35          \\
                       &       CoreE           &          28           &       4.782           &       0.857           &      -2.499           &       2.102           &       0.351           &       0.068           &       0.379           &      0.6494           &    0.000185          \\
                       &  CorePowerE           &          42           &       4.696           &        0.48           &      -2.341           &       1.155           &       0.397           &       0.055           &       0.411           &      0.7838           &    8.33e-10          \\
                       & CorePowerEClass           &          64           &       4.521           &       0.342           &      -1.961           &        0.81           &       0.365           &       0.041           &       0.392           &      0.8667           &    2.17e-20          \\
                       & CorePowerEClassPC           &          70           &       4.864           &       0.333           &      -2.815           &       0.781           &       0.396           &       0.041           &       0.423           &       0.883           &    5.06e-24          \\
                       & CorePowerEClassPCSINFONI           &          22           &       4.979           &       0.717           &      -3.032           &       1.714           &       0.501           &       0.092           &       0.446           &      0.7134           &    0.000194          \\
\enddata
\tablenotetext{\ }{Column 1: Fit type; Column 2: Sample type, see Table
  \ref{tab_subsamples}; Column 3: number of data points; Column 4 and
  5: slope of the correlation and its error; Column 6 and 7:
  zero-point of the correlation and its errors; Column 8 and 9:
  intrinsic scatter and its errors; Column 10: measured scatter;
  Column 11 and 12: Spearman coefficient and its probability.}
\end{deluxetable}

\begin{deluxetable}{llllllllllllllll}
\tablewidth{0pt}
\rotate
\tabletypesize{\tiny}
\tablecaption{Two-dimensional correlations derived without 
the galaxies NGC 2974, NGC 3079, NGC 3414, NGC 4151, NGC 4552, NGC 4621, NGC 5813, NGC 5846.
\label{tab_2dimstrict}}
\tablehead{
\colhead{Fit} & \colhead{Sample} & \colhead{a} & \colhead{da} & \colhead{b} & \colhead{db} &\colhead{ZP} & \colhead{dZP} & \colhead{$\epsilon$} & \colhead{$d\epsilon$} &\colhead{rms} & \colhead{$P(b\neq 0)$} &\colhead{$\Delta cAIC$} & \colhead{$RP$} & \colhead{$r_S$}& \colhead{$P(r_S)$} 
}
\startdata
   \msigrho\           &         All           &       4.521           &       0.315           &       -0.17           &       0.047           &      -0.571           &       1.036           &       0.376           &       0.036           &       0.412           &      0.9999           &      -10.69           &       0.005           &     -0.4782           &    2.46e-06          \\
                       &       CoreE           &       4.409           &       1.138           &      -0.106           &       0.196           &      -0.757           &        3.94           &       0.356           &       0.069           &       0.367           &      0.7228           &       2.448           &         3.4           &      -0.295           &       0.127          \\
                       &  CorePowerE           &       3.906           &       0.576           &      -0.246           &       0.109           &       1.537           &        2.05           &       0.367           &       0.052           &       0.369           &      0.9865           &       -2.59           &       0.274           &     -0.4889           &     0.00102          \\
                       & CorePowerEClass           &       3.948           &       0.365           &      -0.179           &       0.054           &       0.917           &       1.165           &       0.328           &       0.037           &        0.35           &      0.9991           &      -8.611           &       0.013           &     -0.4624           &     0.00012          \\
                       & CorePowerEClassPC           &       4.019           &       0.355           &      -0.195           &       0.047           &       0.868           &       1.124           &       0.343           &       0.036           &        0.36           &      0.9999           &      -14.76           &       0.001           &     -0.5261           &    2.91e-06          \\
\enddata
\tablenotetext{\ }{Column 1: Fit type; Column 2: Sample type, see Table
  \ref{tab_subsamples};
Column 3 and 4: first variable slope of the correlation and its error; 
Column 5 and 6: second variable slope of the correlation and its error; 
Column 7 and 8: zero-point of the correlation and its errors; 
Column 9 and 10: intrinsic scatter and its errors; 
Column 11: measured scatter; 
Column 12: probability of the bivariate correlation;
Column 13 and 14: $\Delta cAIC$ value and $RP=exp(\Delta cAIC/2)$
(the relative probability of the mono- and bivariate solutions); they are computed
  matching the bivariate solutions of this table to the monovariate
  solutions of Table \ref{tab_1dimstrict} of the respective datasets. The
  pairings are: \msigrho\ with \msig, \mburho\ with \mbu, \msigr\ with
  \msig, \mbur with \mbu, \mbus\ with \msig, see
  Sect. \ref{sec_correlations}.
Column 15 and 16: Spearman coefficient and its probability.}
\end{deluxetable}

\end{document}